\documentclass[a4paper,12pt]{report}

\usepackage[all]{nowidow}

\usepackage{preamble}
\newcommand{\ie}{{i.e.~}}

\newcommand{\eg}{e.g.~}

\newcommand{\E}{E_{\mathrm{B}}}

\let\oldsqrt\sqrt
\def\sqrt{\mathpalette\DHLhksqrt}
\def\DHLhksqrt#1#2{%
\setbox0=\hbox{$#1\oldsqrt{#2\,}$}\dimen0=\ht0
\advance\dimen0-0.2\ht0
\setbox2=\hbox{\vrule height\ht0 depth -\dimen0}%
{\box0\lower0.4pt\box2}}

\newcommand{\mean}[1]{\left\langle #1 \right\rangle}
\newcommand{\order}[1]{\mathcal{O}\!\left(#1\right)}

\DeclareMathOperator{\erfc}{erfc}

\newcommand{\dd}{\mathrm{d}}
\newcommand{\ee}{e}

\newcommand{\sss}[1]{{\scriptscriptstyle{#1}}}

\newcommand{\SR}{{{}_\mathrm{SR}}}
\newcommand{\USR}{{{}_\mathrm{USR}}}

\newcommand{\uPl}{\mathrm{Pl}}
\newcommand{\uin}{\mathrm{in}}

\newcommand{\umax}{\mathrm{max}}
\newcommand{\uend}{\mathrm{end}}

\newcommand{\uc}{\mathrm{c}}

\newcommand{\lo}{\sss{\mathrm{LO}}}
\newcommand{\nlo}{\sss{\mathrm{NLO}}}
\newcommand{\nnlo}{\sss{\mathrm{NNLO}}}

\newcommand{\usssPl}{\sss{\uPl}}

\newcommand{\uNL}{\mathrm{NL}}

\newcommand{\calH}{\mathcal{H}}
\newcommand{\calP}{\mathcal{P}}

\newcommand{\eV}{\mathrm{eV}}
\newcommand{\keV}{\mathrm{keV}}
\newcommand{\MeV}{\mathrm{MeV}}
\newcommand{\GeV}{\mathrm{GeV}}

\newcommand{\cm}{\mathrm{cm}}

\newcommand{\km}{\mathrm{km}}
\renewcommand{\sec}{\mathrm{s}}

\newcommand{\Mpc}{\mathrm{Mpc}}
\newcommand{\K}{\mathrm{K}}

\newcommand{\phiend}{\phi_{\text{end}}}
\newcommand{\phiwell}{\phi_{\text{well}}}		
\newcommand{\phiuv}{\phi_{\text{uv}}}			

\newcommand{\phisto}{\Delta\phiwell}
\newcommand{\phicl}{\phi_{\mathcal{P}_{\zeta}|_{\mathrm{cl}}>1}}	
\newcommand{\N}{\mathcal{N}}

\newcommand{\Mp}{M_\usssPl}

\newcommand{\fnl}{f_\uNL}

\newcommand{\efolds}{$e$-folds}
\newcommand{\efold}{$e$-fold}

\newcommand{\beq}{\begin{equation}}
\newcommand{\eeq}{\end{equation}}
\newcommand{\bea}{\begin{equation}\begin{aligned}}
\newcommand{\eea}{\end{aligned}\end{equation}}

\newlength{\wsingfig}
\newlength{\wdblefig}
\newlength{\wquadfig}
\newlength{\wtriplefig}
\newcommand{\Eq}[1]{Eq.~(\ref{#1})}
\newcommand{\Eqs}[1]{Eqs.~(\ref{#1})}
\newcommand{\Fig}[1]{Fig.~{\ref{#1}}}
\newcommand{\Figs}[1]{Figs.~{\ref{#1}}}

\newcommand{\Sec}[1]{Sec.~\ref{#1}}
\newcommand{\Secs}[1]{Secs.~\ref{#1}}
\newcommand{\App}[1]{Appendix~\ref{#1}}

\begin{document}
\sloppy
\onehalfspacing

\begin{titlepage}

\newcommand{\HRule}{\rule{\linewidth}{0.5mm}} 

\center 


\textsc{\LARGE Institute of Cosmology and Gravitation, University of Portsmouth}\\[0.5cm] 


\HRule \\[0.4cm]
{ \huge \bfseries Inflation: a quantum laboratory on cosmological scales}\\[0.4cm] 
\HRule \\[1.5cm]
 

\begin{minipage}{0.5\textwidth}
\begin{flushleft} \large
\textbf{\LARGE Christopher Pattison}\\[0.5cm] 
\end{flushleft}
\end{minipage}
~
\begin{minipage}{0.4\textwidth}
\begin{flushright} \large
\emph{Supervisors:} \\
Prof. David Wands \\
Dr. Vincent Vennin \\
Dr. Hooshyar Assadullahi
\end{flushright}
\end{minipage}\\[2cm]


{\normalsize This thesis is submitted in partial fulfilment of\\
the requirements for the award of the degree of\\
Doctor of Philosophy of the University of Portsmouth.\\} 
\vspace{2cm}


\textsc{\large March 2020}\\[0.5cm] 

\vspace{1cm}

\includegraphics[width=4cm]{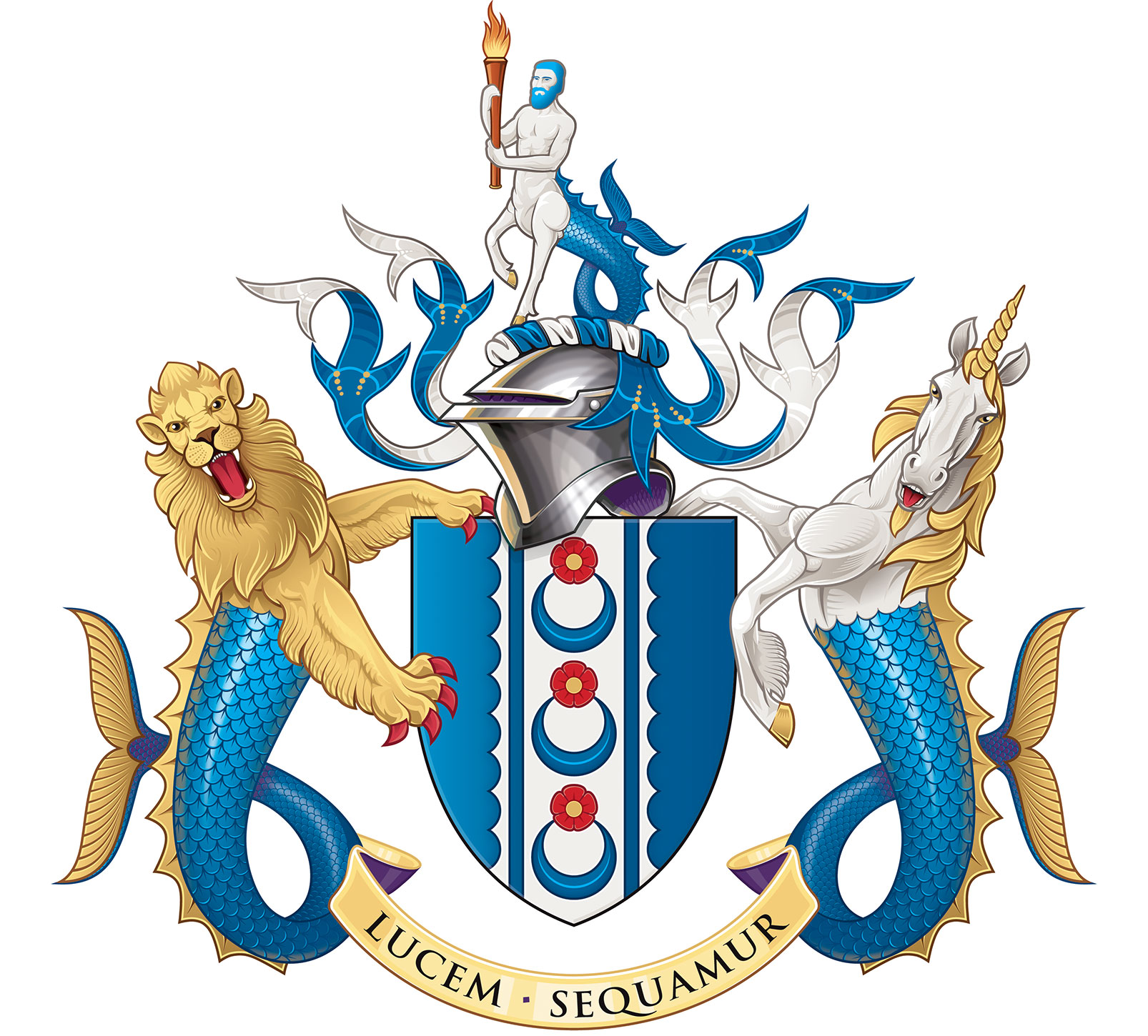}\hspace*{4.75cm}~%
   \includegraphics[scale=0.2]{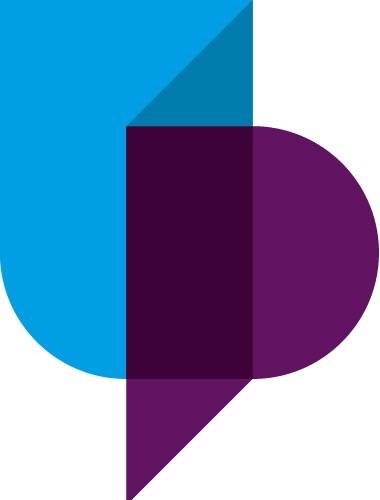} 

\vfill 

\end{titlepage}

\pagenumbering{roman}	

\phantomsection	
\addcontentsline{toc}{chapter}{Abstract}	
\chapter*{Abstract}	

\linespread{1.6}
This thesis is dedicated to studying cosmological inflation, which is a period of accelerated expansion in the very early Universe that is required to explain the observed anisotropies in the cosmic microwave background.
Inflation, when combined with quantum mechanics, also provides the over-densities that grow into the structure of the modern Universe.
Understanding perturbations during this period of inflation is important, and we study these perturbations in detail in this work. 

We will assume that inflation is driven by a single scalar field, called the inflaton.
When the shape of the potential energy is flat, the inflaton can enter a phase of ``ultra-slow-roll inflation''.
We study the stability of such a period of inflation, and find that it can be stable and long-lived, although it has a dependence on the initial velocity of the inflaton field. 
This is different to the slow-roll regime of inflation, which is always stable, but has no dependence on the initial velocity. 

In the second part of this thesis, we use the stochastic formalism for inflation in order to take account of the non-perturbative backreaction of quantum fluctuations during inflation. 
This formalism is an effective field theory for long wavelength parts of quantum fields during inflation, and hence is only valid on large scales. 
We use this formalism to study curvature fluctuations during inflation, and we derive full probability distributions of these fluctuations. 
This allows us to study the statistics of large fluctuations that can lead to the formation of rare objects, such as primordial black holes. 
In general, we find that when the quantum effects modelled by the stochastic formalism are correctly accounted for, many more primordial black holes can be formed than one would expect if these quantum effects were not taken into account. 

We finish by summarising our results and discussing future research directions that have opened up as a result of the work we have done. 
In particular, we mention future applications of the formalisms we develop using the stochastic techniques for inflation, and note that their applications can be broader than primordial black holes and they can be used to, for example, study other rare objects.

\newpage
\renewcommand{\contentsname}{Table of Contents}	
\pdfbookmark{Table of Contents}{contents}	
\tableofcontents


\newpage
\pdfbookmark{List of Figures}{figures}	
\listoffigures


\newpage
\phantomsection
\addcontentsline{toc}{chapter}{Declaration}
\chapter*{Declaration}
Whilst registered as a candidate for the above degree, I have not been registered for any other research award. The results and conclusions embodied in this thesis are the work of the named candidate and have not been submitted for any other academic award.

Chapter \ref{chapter:introduction} is an introductory chapter, written by myself and drawn from many references, as cited where appropriate.
Chapter \ref{chapter:USRstability} is primarily based on \href{https://iopscience.iop.org/article/10.1088/1475-7516/2018/08/048}{\textit{JCAP} $1808(2018) 048$}, while Chapter \ref{chapter:stochastic:intro} provides a combination of further introductory material and sections based on \href{https://iopscience.iop.org/article/10.1088/1475-7516/2019/07/031}{\textit{JCAP} $1907(2019) 031$}. Chapter \ref{chapter:stochastic:intro} also contains some new analytical results which are not yet published. 
Chapter \ref{chapter:quantumdiff:slowroll} is based on \href{https://iopscience.iop.org/article/10.1088/1475-7516/2017/10/046}{\textit{JCAP} $1710(2017)046$}, and Chapter \ref{chapter:USRstochastic} was based on work in progress which has since been expanded and submitted for publication.  

I am the first author of each publication that this thesis is based on, and in each of these I performed all of the analytic and numerical calculations, either originally or as checks for my co-authors.

\vspace{50pt}

\begin{center}
Word count: 50,001 words.

Ethical review code: 5D3F-6FD6-E158-5231-4FF5-FD1B-A724-6CF3
\end{center}

\newpage
\phantomsection
\addcontentsline{toc}{chapter}{Acknowledgements}
\chapter*{Acknowledgements}

First of all, I would like to thank my incredible PhD supervisors: Prof David Wands, Dr Vincent Vennin and Dr Hooshyar Assadullahi, for your remarkable wealth of knowledge, advice, and humour that you have shared with me throughout this time. 
I thank you for showing me the beauty of early universe physics, and for teaching me with kindness and patience, in a relaxed environment. 
You have inspired me to be a better researcher, and I could not have hoped for a better team of teachers and collaborators. 

I would like to thank my examiners Ed Copeland and Roy Maartens for the careful consideration they gave to this thesis, and for the enjoyable and fruitful discussions we had about this work.  

It is also a pleasure to thank all of my PhD colleagues for keeping me sane and helping me to achieve the work in this thesis. 
In particular, I would like to thank Natalie, Bill, and Mark, for your continued support and encouragement throughout my PhD. 
I would also like to thank everyone that I shared an office and department with throughout my time in Portsmouth, including my close friends Michael, Mike, Laura, Tays, Manu, The Sams, Jacob, and Gui - thank you for making every day fun, and inspiring me throughout my PhD.
I truly hope you all achieve your dreams.

Finally, and most importantly, I would like to thank Hannah, my parents Jan and Andy, and my sister Eve. 
Without your love and support I would not have been able to make it through the last three years, and I hope you know how much I love and appreciate you all. 
I also hope you are proud of the work you have helped me to produce here (even if this is the only page you read of this thesis!).
I would be lost without you. 

\newpage
\phantomsection
\addcontentsline{toc}{chapter}{Dedication}
\chapter*{Dedication}

Completing a PhD is always difficult, but I was lucky to pass through mine with a whole host of privileges that allowed me to focus entirely my work. I did not have to deal with racism. I did not have to deal with sexism. I did not have to deal with discrimination of any kind.

Many PhD students are forced to deal with these issues, both inside and outside of academia, throughout their PhD. This makes their PhD journey more complicated and difficult than I can imagine. 
It is our collective responsibility to ensure that the members of our society who feel these discriminations directly are not the only ones who are working to rectify them. If some of our community suffer, then we all feel the negative effects. 

Take care of your colleagues. Call out behaviour that should be consigned to the history books. We must make sure that the future of academia, and society, is fair and offers equal opportunities and support to everyone.
Black Lives Matter. 

\vspace{0.5cm}

\hdashrule{13cm}{1pt}{3mm}

\vspace{0.5cm}

\noindent This PhD thesis was completed and defended during the COVID-19 pandemic. It is dedicated to everyone who has suffered as a result of this disease.

\newpage
\phantomsection
\addcontentsline{toc}{chapter}{Dissemination}
\chapter*{Dissemination}
\LARGE\textbf{Publications}  \vspace{5mm}\\
\normalsize
 \textbf{C. Pattison}, V. Vennin, H. Assadullahi and D. Wands, \textit{Quantum diffusion during inflation and primordial black holes}, \href{https://iopscience.iop.org/article/10.1088/1475-7516/2017/10/046}{\textit{JCAP} $1710(2017)046$}, [\href{https://arxiv.org/abs/1707.00537}{1707.00537}] \\
\textbf{C. Pattison}, V. Vennin, H. Assadullahi and D. Wands, \textit{The attractive behaviour of ultra-slow-roll inflation}, \href{https://iopscience.iop.org/article/10.1088/1475-7516/2018/08/048}{\textit{JCAP} $1808(2018) 048$}, [\href{https://arxiv.org/abs/1707.00537}{1806.09553}]\\
\textbf{C. Pattison}, V. Vennin, H. Assadullahi and D. Wands, \textit{Stochastic inflation beyond slow roll}, \href{https://iopscience.iop.org/article/10.1088/1475-7516/2019/07/031}{\textit{JCAP} $1907(2019) 031$}, [\href{https://arxiv.org/abs/1905.06300}{1905.06300}]



\newpage

\pagenumbering{arabic}	

\chapter{Introduction}
\label{chapter:introduction}

In this chapter, we will review the standard model of cosmology and cosmological inflation, as well as perturbations during inflation and the effects they can have on observable quantities.
We will review the so-called Friedmann-Lema{\^i}tre-Robertson-Walker (FLRW) Universe, which describes an expanding spacetime that is homogeneous and isotropic, and discuss the Hot Big Bang model of the Universe, which describes the Universe since the initial singularity (some $13.7$ billion years ago), along with the problems of this model. 
Cosmological inflation will then be introduced, which was designed in the $1980s$ to solve the known problems of the Hot Big Bang model and also, when combined with quantum mechanics, inflation provides a mechanism to seed the large-scale structure we see in the Universe.
Finally, in this chapter, we will explain some modern problems that persist in cosmology, and then discuss the formation of primordial black holes. 
These objects form in the early universe, but after inflation has ended. 
A later part of this thesis is concerned with studying the effects of quantum diffusion during inflation; we find that this can have a large impact on the formation of primordial black holes. 
In this section, we only present an overview of the standard cosmological model, and more details can be found in a range of textbooks, see, for example, \cite{Kolb:1990vq, Peebles:1994xt, Liddle:1998ew, Liddle:2000cg, Dodelson:2003ft, Mukhanov:2005sc, Lyth:2009zz, Peter:2013avv}.

\section{Standard model of cosmology}

\subsection{The FLRW Universe}

Let us begin by discussing the Hot Big Bang Model of the Universe.
By implementing the cosmological principle of isotropy and homogeneity, the metric for spacetime can be simply written as the Friedmann-Lema{\^i}tre-Robertson-Walker (FLRW) metric, which is completely determined up to a single free function of cosmic time. 
This free function is the scale factor of the Universe, $a(t)$, which is often taken to be dimensionless (so that $a=1$ today, by convention), and given this function the FLRW metric is then 
\bea \label{eq:FLRWmetric}
\dd s^2 = g_{\mu \nu}\dd x^{\mu}\dd x^{\nu} = -\dd t^2 + a^2(t)\left[ \frac{\dd r^2}{1-Kr^2} + r^2\dd\theta^2 + r^2\sin^2\theta\dd\phi^2 \right]  \, ,
\eea
where $\mu,\nu$ run from $0$ to $3$, and the parameter $K$ describes the spatial curvature of the Universe and can take on the discrete values $K=1$ (closed universe), $K=0$ (flat Universe), or $K=-1$ (open Universe). 
Note that we are using natural units $c=\hbar=k_\mathrm{b} = 1$ here and throughout this thesis. 
In the metric \eqref{eq:FLRWmetric}, $t$ is cosmic time, $r$ is the comoving radial coordinate, and $\theta$ and $\phi$ are the comoving angular coordinates.
Note the simplicity of this metric, which is due to the symmetries of isotropy and homogeneity, and note that if the scale factor were to vary with space as well as time, then this metric would violate homogeneity. 

From the FLRW metric we can gain some physical understanding as to what the scale $a(t)$ represents. 
If we consider a constant $t$ hypersurface, and define the \textit{comoving} distance $L_\mathrm{com}$ between two points at fixed spatial coordinates $(r,\theta, \phi)$ to remain constant in the FLRW frame, then the \textit{physical} distance $L_\mathrm{phys}$ between these two points is $L_\mathrm{phys} = a(t)L_\mathrm{com}$. 
This means the scale factor $a(t)$ sets the physical expansion of spatial hypersurfaces in the FLRW metric \eqref{eq:FLRWmetric}

\begin{figure}
\centering
\includegraphics[width=0.75\columnwidth]{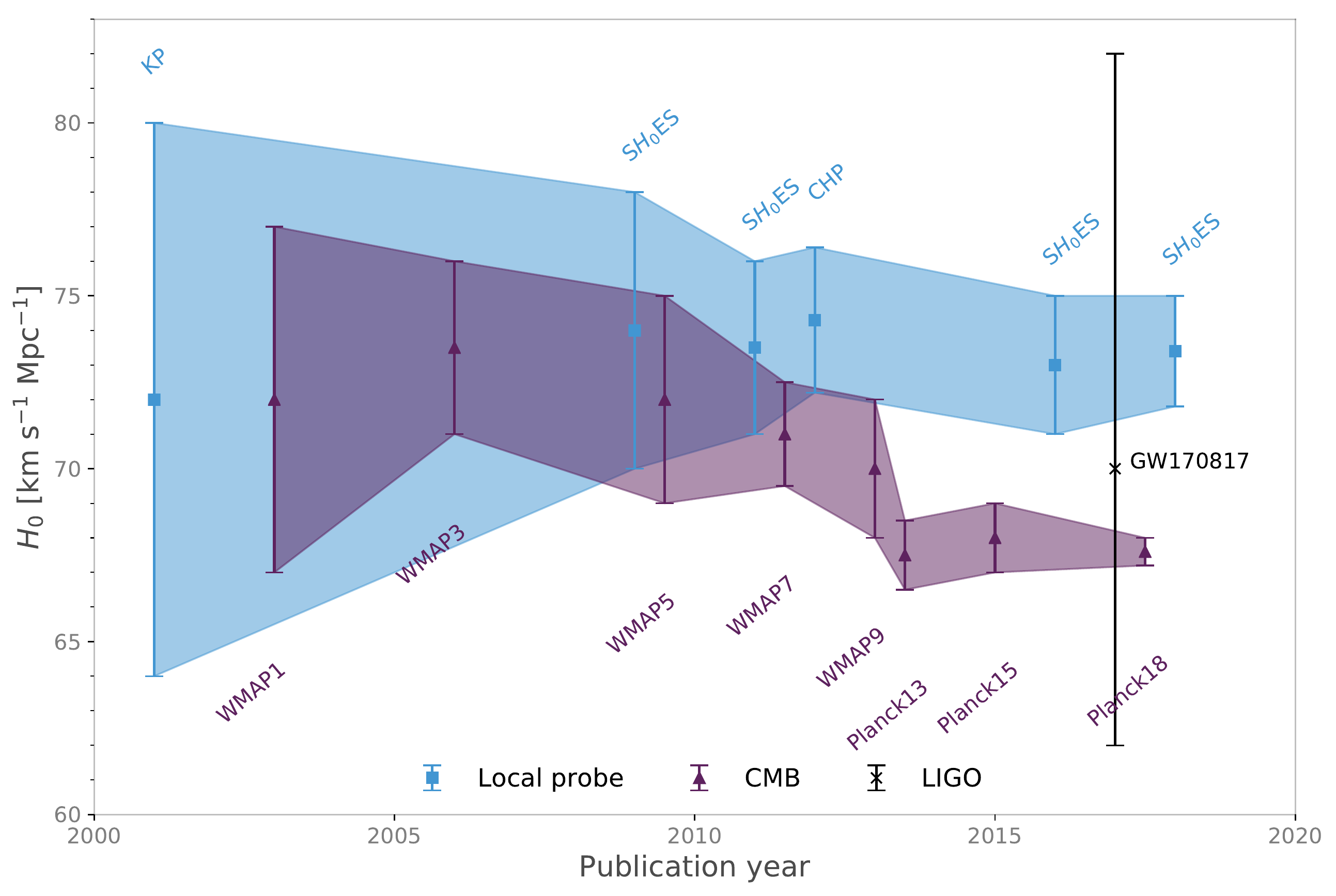}
\caption[$H_0$ tension over time]{$H_0$ measurements over time, demonstrating the emergence of a tension between local measurements and those inferred from CMB measurements using the $\Lambda$CDM model. ``KP" denotes the Hubble Space Telescope key project \cite{Freedman:2000cf}, ``CHP" is the Carnegie Hubble Program \cite{2012ApJ...758...24F}. WMAP \cite{Bennett_2013} and Planck \cite{Akrami:2018vks} are constraints from the cosmic microwave background (CMB), and S$H_0$ES are local supernovae constraints (eg, \cite{Riess:2016jrr}). The LIGO data point uses a single gravitational wave event to constrain $H_0$ \cite{Abbott:2017xzu}. \textsc{Image credit: Natalie Hogg \cite{H0GG}.}
\label{fig:h0tensionovertime}}
\end{figure}

In the FLRW metric we also have a simple, linear relationship betewen distance and velocity, known as the Hubble law. 
This can easily be seen by considering 
\bea 
v_\mathrm{phys} = \frac{\dd}{\dd t}L_\mathrm{phys} = \frac{\dot{a}}{a}L_\mathrm{phys} \equiv H L_\mathrm{phys} \, ,
\eea 
where a dot denotes a time derivative, and we have defined the Hubble parameter $H(t)=\dot{a}/{a}$ which sets the rate of expansion.
We denote the current value of $H(t)$ to be $H_0$, which we call the Hubble constant, and note that current measurements of its value are failing to converge on an agreed exact value for $H_0$ (see Ref. \cite{Bernal:2016gxb} for a review), see \Fig{fig:h0tensionovertime}.
While the original measurement of this constant by Edwin Hubble in 1929 was $500$ $\km \mathrm{ s}^{-1}\Mpc^{-1}$, modern measurements range between approximately $67$-$74$ $\km \mathrm{ s}^{-1}\Mpc^{-1}$. 
This tension is seen as important because the value of $H_0$ contains significant information about the content and history of the Universe, and hence resolving this tension is the subject of a great deal of research today.
The information encoded in the value of $H_0$ includes a characteristic time scale $H_0^{-1} \approx 4.551\times10^{17} \sec$ called the ``Hubble time", which ultimately sets the scale for the age of the Universe, and a characteristic length scale $H_0^{-1} \approx 1.364\times10^{26} \mathrm{m}$ called the ``Hubble radius", which ultimately sets the scale for the size of the observable Universe.

It is sometimes convenient to factor out the scale factor from the time coordinate of the FLRW metric, \ie remove the expansion from the time coordinate. 
This is done by defining the conformal time coordinate by $\dd t = a(\eta) \dd\eta$, and hence \eqref{eq:FLRWmetric} becomes 
\bea \label{eq:FLRWmetric:conformaltime}
\dd s^2 = a^2(\eta)\left[-\dd \eta ^2 +  \frac{\dd r^2}{1-Kr^2} + r^2\dd\theta^2 + r^2\sin^2\theta\dd\phi^2 \right]  \, .
\eea

\subsection{Einstein equations}

The dynamics of generic (\ie not necessarily FLRW) space-time metrics are described by the Einstein--Hilbert action
\bea \label{eq:action:einsteinhilbert}
S = \frac{\Mp^2}{2}\int \dd^4x\sqrt{-g}R \, ,
\eea 
where $\Mp = 1/\sqrt{8\pi G}\approx 2.435 \times 10^{18} \GeV$ in natural units of $c=1$ is the reduced Planck mass, $\dd^4x\sqrt{-g} = \dd^4x\sqrt{-\mathrm{det}\left(g_{\mu\nu}\right)}$ is the 4-dimensional volume element, and $R$ is the Ricci scalar which is defined as the contraction of the Ricci tensor $R=R^{\mu}{}_{\mu}$. 
One can vary this action with respect to the metric components $g_{\mu\nu}$ to obtain the vacuum solution 
\bea 
R_{\mu\nu} = 0 \, ,
\eea 
but it is perhaps more enlightening to first add gravitating matter and a cosmological constant $\Lambda$ to the action \eqref{eq:action:einsteinhilbert} in an attempt to describe a Universe that is more like our own. 
Doing so yields the action
\bea \label{eq:action:einsteinhilbert:matter}
S = \frac{\Mp^2}{2}\int \dd^4x\sqrt{-g}\left( R - 2\Lambda\right) + \int \dd^4x\sqrt{-g} \mathcal{L}_{\mathrm{matter}} \, ,
\eea 
where $\mathcal{L}_{\mathrm{matter}}$ is the Langrangian density of the gravitating matter of the universe. 

If one varies the matter sector of this equation with respect to $g_{\mu\nu}$ then it allows one to define the energy-momentum tensor $T^{\mu\nu}$  of the matter part as 
\bea \label{eq:def:energymomentumtensor}
-\frac{1}{2}\sqrt{-g}T_{\mu\nu} &\equiv \frac{\partial \sqrt{-g} \mathcal{L}_{\mathrm{matter}}}{\partial g^{\mu\nu} } = \frac{1}{2}\sqrt{-g}g_{\mu\nu}\mathcal{L}_{\mathrm{matter}} - \sqrt{-g}\frac{\partial \mathcal{L}_{\mathrm{matter}}}{\partial g^{\mu\nu} } \\
& = -\frac{1}{2}\sqrt{-g}\left(2\frac{\partial \mathcal{L}_{\mathrm{matter}}}{\partial g^{\mu\nu} } - g_{\mu\nu}\mathcal{L}_{\mathrm{matter}} \right) \, .
\eea 
Similarly, varying the first term of \eqref{eq:action:einsteinhilbert:matter} with respect to $g_{\mu\nu}$ lets us define the Einstein tensor as 
\bea \label{eq:def:einsteintensor}
G_{\mu\nu} + \Lambda g_{\mu\nu} \equiv R_{\mu\nu} + \frac{1}{2}Rg_{\mu\nu} + \Lambda g_{\mu\nu} =  \frac{2}{\sqrt{-g}\Mp^2}\frac{\partial \left[\sqrt{-g} \left( R-2\Lambda \right)\right]}{\partial g^{\mu\nu} } \, .
\eea 
Thus, varying the entire action \eqref{eq:action:einsteinhilbert:matter} and forcing this variation to vanish yields the well-known Einstein field equations
\bea \label{eq:einstein:fieldequation}
G_{\mu\nu} + \Lambda g_{\mu\nu} = \frac{1}{\Mp^2}T_{\mu\nu} \, ,
\eea 
where one can see that the Planck mass $\Mp$ acts as a coupling constant between the matter in the theory and the gravitaional sector $g_{\mu\nu}$, which follows because the matter sector does not feature $\Mp$ in its action (if it did the factors of $\Mp$ would cancel at the level of the equations of motion and it would not feature as a coupling constant). 

As an example, we can now explicitly calculate the tensors $T_{\mu \nu}$ and $G_{\mu\nu}$ when we consider an FLRW universe. 
One can plug the FLRW metric \eqref{eq:FLRWmetric} components into the definition of $G_{\mu\nu}$ \eqref{eq:def:einsteintensor} to find 
\bea 
& G_{00} = 3H^2 + \frac{3K}{a^2} \, , \\
& G_{ij} = -g_{ij}\left( H^2 + \frac{\ddot{a}}{a} + \frac{K}{a^2} \right) \, , \\
& G_{0i} = G_{i0} = 0 \, ,
\eea 
where the indicies $i,j$ correspond to the spatial indices and run from $1$ to $3$, and the index $0$ corresponds to the time component of the metric. 
These components are given explicitly in \App{appendix:flrwchristoffel}, along with the Christoffel symbols of the FLRW metric. 

In order to find the form of the energy momentum tensor, let us consider an observer that is moving with the matter fluid, \ie moving with respect to the rest frame and stationary with respect to the fluid frame, where we assume the matter is a perfect fluid with density $\rho$, pressure $P$, and (normalised) 4-velocity $u_\mu$.
For a perfect fluid, the energy-momentum tensor is given by
\bea 
T_{\mu\nu} = \left(\rho + P\right) u_\mu u_\nu + Pg_{\mu\nu} \, ,
\eea
where we note that for an observer at rest with respect to the fluid we have $u_0=-1$ and $u_i=0$, and hence $T_{\mu\nu}=\mathrm{diag}\left( \rho, P, P, P \right)$ for our perfect matter fluid. 
We note that, assuming that $\rho$ and $P$ do not vary spatially, this form of $T_{\mu\nu}$ is consistent with our global assumption of homogeneity and isotropy. 

If we now substitute the components of the metric \eqref{eq:FLRWmetric} and the $G_{\mu\nu}$, $T_{\mu\nu}$ tensors into the Einstein equations \eqref{eq:einstein:fieldequation}, we find two important equations in cosmology
\begin{align}
\label{eq:friedmannequation} H^2 &= \frac{\rho}{3\Mp^2} - \frac{K}{a^2} + \frac{\Lambda}{3} \, , \\
\label{eq:raychaudhuriequation} \frac{\ddot{a}}{a} &= -\frac{\rho+3P}{6\Mp^2} + \frac{\Lambda}{3} \, ,
\end{align}
which are called the Friedmann equation and Raychaudhuri equation, respectively 
\footnote{Note that one can also derive the Friedmann equation by varying the action \eqref{eq:action:einsteinhilbert:matter} with respect to the lapse function $A$ (see \Eq{eq:perturbedlineelement:FLRW} for the definition of the lapse function), and one can derive the Raychaudhuri equation by varying the action \eqref{eq:action:einsteinhilbert:matter} with respect to the scale factor $a$.}.

It is interesting to note that, in the absence of curvature and the cosmological constant ($K=\Lambda=0$), the Friedmann equation gives a direct relationship between the Hubble rate and its energy density, meaning that simply the presence of energy in the Universe will cause it to contract or expand. 
Similarly, if $\Lambda=0$, the Raychaudhuri equation tells us that the presence of energy (with or without pressure) stops the scale factor from being constant. 
In the context of what will follow in this thesis, it is important to note that the scale factor will accelerate ($\ddot{a}>0$) if $\rho + 3P <0$, which is possible for fluids with negative pressure (assuming $\rho$ is positive), and we will see that this is the case during inflation. 

From the energy-momentum tensor, if we implement the time component of the conservation equation $\nabla_\mu T^\nu{}^\mu=0$ (\ie take $\nu=0$), we also find that
\bea 
0 = \nabla_\mu T^{0\mu} 
&= -\dot{\rho} - 3\frac{\dot{a}}{a}\rho - 3\frac{\dot{a}}{a}P \, ,
\eea 
which can be simply rewritten as the continuity equation
\bea \label{eq:continuityequation}
\dot{\rho} + 3H(\rho + P) = 0 \, .
\eea 
The continuity equation \eqref{eq:continuityequation} has a simple solution when the equation of state parameter, defined as $w=\frac{P}{\rho}$, is constant, and these solutions are given by 
\bea \label{eq:rho:constantw}
\rho(a) = \rho_\mathrm{in}\left( \frac{a}{a_\mathrm{in}} \right)^{-3\left(1+w\right)} \, , 
\eea 
where $\rho_\mathrm{in}$ is the initial value of $\rho$ when $a=a_\mathrm{in}$.
Many cases of interest yield the simple solutions given by \eqref{eq:rho:constantw}.
For example, cold matter simply has $w_\mathrm{matter}=0$ and so we see that $\rho_\mathrm{matter} \propto a^{-3}$, which is to say that matter scales inversely with the volume of the spacetime that it is in. 
For radiation, we have $w_\mathrm{radiation}=1/3$, and so $\rho_\mathrm{radition} \propto a^{-4}$, so in an expanding spacetime, radiation dilutes with the volume increase of the spacetime ($\propto a^{-3}$) as well as with an additional redshift dependence ($\propto a^{-1}$) of each particles energy. 
In order to see how we can identify the effect of the curvature $K$ and the cosmological constant $\Lambda$ with a fluid, let us rewrite \eqref{eq:friedmannequation} as 
\bea \label{eq:rescaledFriedmann}
H^2 = \frac{1}{3\Mp^2}\left[ \rho_\mathrm{matter} + \rho_\mathrm{rad} + \rho_K + \rho_\Lambda \right] \equiv \frac{1}{3\Mp^2} \rho_\mathrm{tot}\, ,
\eea 
where we define $\rho_\mathrm{matter}$ to be the energy density of any matter and any other non-relativistic constituent present in the Universe, $\rho_\mathrm{rad}$ is the energy density of radiation, $\rho_K = -3\Mp^2K/a^{2}$ to be the energy density of curvature, $\rho_\Lambda = \Mp^2\Lambda$ to be the energy density of the cosmological constant, and $\rho_\mathrm{tot}$ to be the total energy density of the Universe. 
We see that $\rho_K \propto a^{-2}$, and so from \eqref{eq:rho:constantw} we can conclude that $w_K= -1/3$, and similarly $\rho_\Lambda \propto a^0$, and so we associate the cosmological constant to a fluid with equation of state $w_\Lambda=-1$. 
The constant values of $w$ and the forms of the energy density's dependence on the scale factor discussed above are summarised in Table \eqref{table:fluids}. 

\begin{table}[ht]
\centering
\begin{tabular}{ |p{4cm}|| m{4cm}|m{2cm}|m{2cm}|  }
 \hline
 \textbf{fluid} & \textbf{equation of state parameter $w$} & \textbf{$\rho(a)$} & \textbf{$a(t)$} \\
 \hhline{|=#=|=|=|}
 cold matter & $0$ & $\propto a^{-3}$ & $\propto t^{2/3}$ \\
 \hline
 radiation & $1/3$ & $\propto a^{-4}$ & $\propto t^{1/2}$ \\
 \hline
 spatial curvature & $-1/3$ & $\propto a^{-2}$ & $\propto t^{}$ \\
 \hline
 cosmological constant & $-1$ & $\propto a^{0}$ & $\propto \ee^{Ht}$ \\
 \hline
 scalar field & $-1+2\epsilon_1/3$  & $\propto a^{-2\epsilon_{1}}$ & $\propto t^{1/\epsilon_1}$ \\
 \hline
\end{tabular}
\caption[Properties of various cosmological fluids]{Equations of state for various fluids, along with the corresponding profiles for their density and scale factor.}
\label{table:fluids}
\end{table}

Rewriting the Friedmann equation as \eqref{eq:rescaledFriedmann} has another useful consequence, as it allows us to see that (assuming $a$ evolves monotonically) the right-hand side quickly becomes dominated by one of the fluids. 
This is the fluid with the smallest value of $w$ if space is expanding, and the fluid with the largest value of $w$ if space is contracting. 

Once $H$ is dominated by a single fluid, \eqref{eq:rescaledFriedmann} is integrable and we find
\bea \label{eq:scalefactor:time}
a(t) = \begin{cases} a_\mathrm{in}\left[ 1 \pm \frac{3}{2}(1+w)H_\mathrm{in}\left(t-t_\mathrm{in}\right) \right]^{\frac{2}{3(1+w)}} & \mathrm{if}\, w \neq -1 \\ 
a_\mathrm{in}\exp\left[ H_\mathrm{in}\left(t-t_\mathrm{in}\right) \right] & \mathrm{if}\, w = -1 
\end{cases}
\, ,
\eea
Here, the $\pm$ depends on whether space is expanding (take the $+$ sign) or contracting (take the $-$ sign).
For the rest of this thesis, only the case of expanding space will be considered.
For the values of $w$ discussed above, the corresponding profiles of $a(t)$ are displayed in Table \eqref{table:fluids}.

\begin{figure}
\centering
\includegraphics[width=0.9\columnwidth]{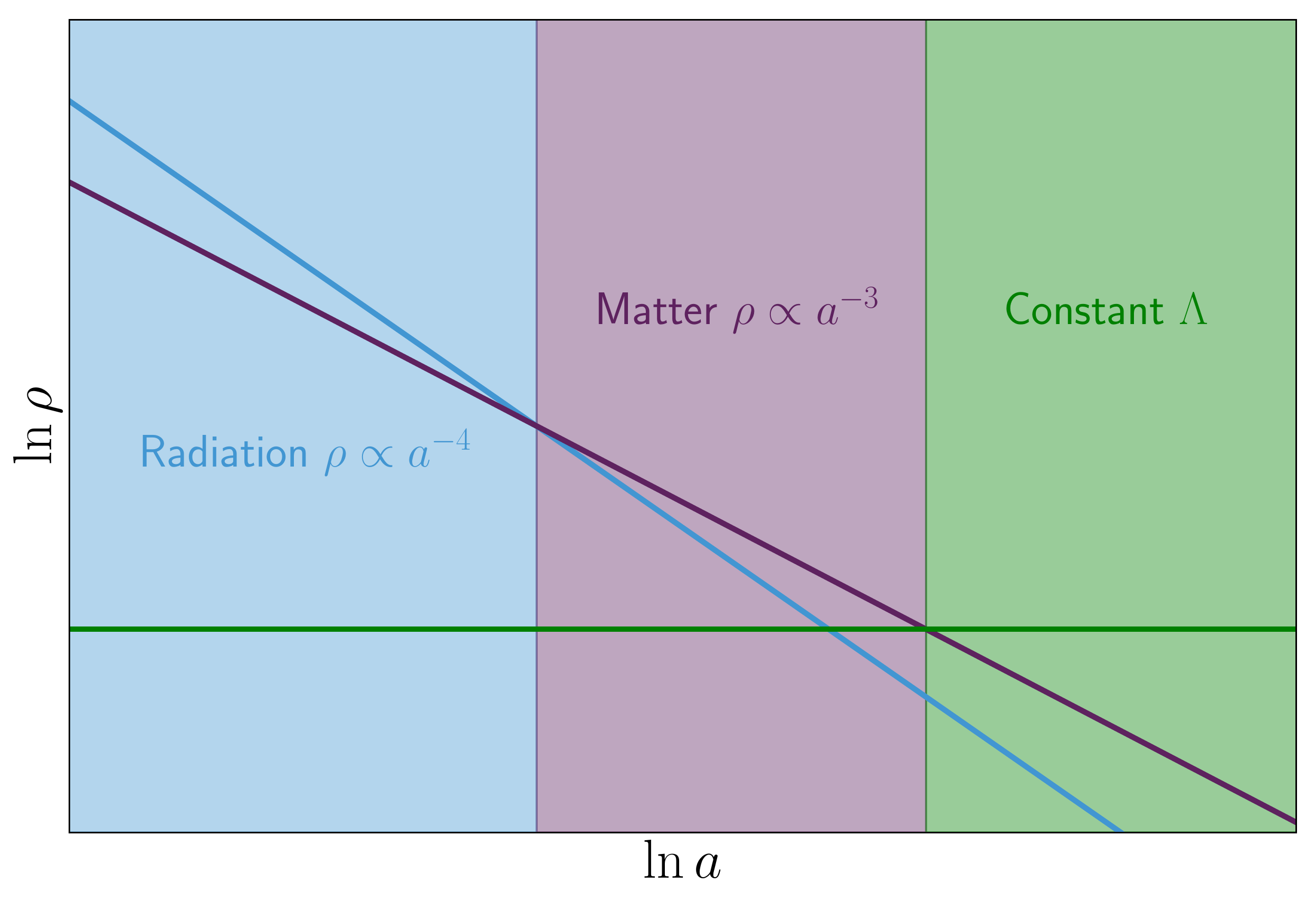}
\caption[Different epochs in the history of the Universe]{A simple illustration of the decay of energy density of the constituents of the Unvierse.
\label{fig:linesofdomination}}
\end{figure}

There a few interesting things that we can note when considering multiple fluids together. 
First of all, if one considers multiple fluids in the Freidmann equation \eqref{eq:friedmannequation}, that is we replace $\rho \to \sum_i \rho_i$, then we can consider all of the separate constituents of gravitating matter in the Universe which all dilute at different rates with the expansion of the Universe. 
In this way, we can estimate which constituent dominated the energy budget of the Universe throughout its history, as shown in \Fig{fig:linesofdomination}.
We see that the history of the Universe can be separated into three main epochs in which the total energy density $\rho_\mathrm{tot}$ of the Universe is dominated by a different constituent of the Universe. 
From the end of inflation until matter-radiation equality at $z_\mathrm{eq} \simeq 3402 $, the energy density of the Universe is dominated by radiation, and hence $\rho_\mathrm{tot} \propto a^{-4}$.
Then for redshifts $z_\mathrm{eq} < z < z_\mathrm{acc}$, where $z_\mathrm{acc} \simeq 0.6$,  matter (specifically dark matter) dominated the energy budget of the Universe and $\rho_\mathrm{tot} \propto a^{-3}$.
Finally, the current epoch of the Universe $z_\mathrm{acc} < z < 0$ is dominated by dark energy (assumed to be a cosmological constant), and hence $\rho_\mathrm{tot} \propto a^{0}$.
The evolution of each of these epochs is well understood, although during the transitions between epochs there are two equally important constituents of the Universe and hence the behaviour of $\rho_\mathrm{tot}$ is more complicated. 

Next, we can see that if we consider the Universe to be made up of multiple fields with independent equations of state $w_i=P_i/\rho_i$, then the continuity equation \eqref{eq:continuityequation} simply generalises to 
\bea \label{eq:continuityequation:multiplefluids}
\sum_i\left[\dot{\rho}_i + 3H(\rho_i + P_i)\right] = 0 \, .
\eea 
Finally, we note that if \eqref{eq:continuityequation:multiplefluids} is solved because each term vanishes individually, then this corresponds to the case of multiple fluids that are not interacting (so no energy transfers between the fields), and in this case (as we did before) we can solve for the total energy density $\rho_\mathrm{T}$ as 
\bea \label{eq:rho:constantw:multiplefluids}
\rho_\mathrm{T}(a) = \sum_{i} \rho_i^\mathrm{in}\left( \frac{a_i}{a_i^\mathrm{in}} \right)^{-3\left(1+w_i\right)} \, .
\eea

\subsection{Composition of the Universe}

\begin{figure}
\centering
\includegraphics[width=0.5\columnwidth]{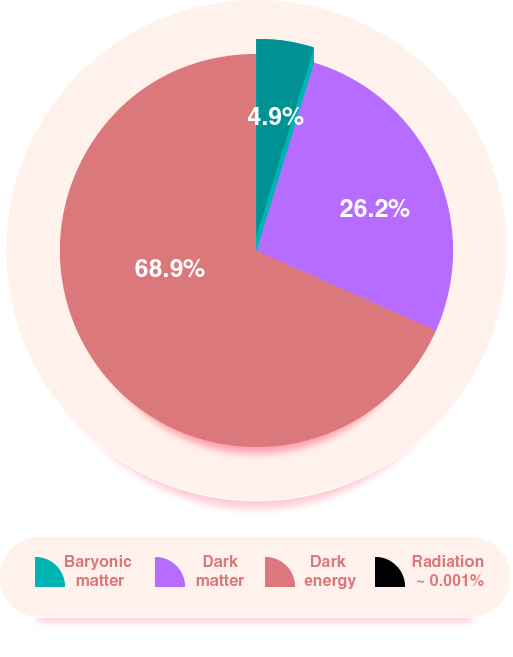}
\caption[Composition of the Universe]{Current composition of the Universe, according to the $\Lambda CDM$ model, where we take  $\Omega_k \simeq 0$. This is consistent with the latest measurements from the Planck satellite \cite{Aghanim:2018eyx}.
\label{fig:compositionUniverse}}
\end{figure}

It is possible to use \eqref{eq:rho:constantw:multiplefluids} to estimate the energy density of each constituent of the Universe at any time if we know the current values for their energy density (or if we know their values at any one time, but it is easiest to measure them at the current time).
This can be done as long as we assume each component has a constant equation of state and assuming there is no energy transferred between each component. 

Let us begin by defining the critical density of the universe to be 
\bea \label{eq:def:criticaldensity}
\rho_\mathrm{crit} = 3\Mp^2 H^2 \, ,
\eea 
and also defining the dimensionless quantities $\Omega_i = \rho_i/\rho_\mathrm{crit}$ for each fluid, where $\Omega_i$ describes the fraction of the Universe that is contained in each constituent.
We note at this point that $\Omega_K$ can be either positive or negative, while every other $\Omega_i$ is strictly positive (or zero). 
With these definitions in place, we can rewrite the Friedmann equation \eqref{eq:rescaledFriedmann} as 
\bea 
1 - \Omega_K = \Omega_\mathrm{matter} + \Omega_\mathrm{rad} + \Omega_\Lambda   \, ,
\eea 
where $\Omega_K$ has been singled out on the left-hand side due to it being the only one that can be positive or negative. 
Finally, we can implement the scaling of the scale factor \eqref{eq:rho:constantw:multiplefluids} for each constituent to show how each term evolves backwards (or forwards) in time from their current values, 
\bea \label{eq:sumofomegas}
1 - \frac{\Omega_K^0}{a^2} = \frac{\Omega_\mathrm{matter}^0}{a^3} + \frac{\Omega_\mathrm{rad}^0}{a^4} + \Omega_\Lambda^0   \, ,
\eea 
where a superscript `$0$' denotes it is its value today. 
Based on the latest observations \cite{Aghanim:2018eyx}, we now give the values of $\Omega_i^0$ for each known constituent of the Universe (see \cite{Aghanim:2018eyx} for a detailed description of how each measurement is made and for error bars). 

\textbf{Curvature:} There has not yet been a statistically significant detection of any non-zero curvature in the Universe, \ie all measurements are consistent with a flat Universe, and the latest value is $\Omega_K^0 = 0.0007 \pm 0.0019$,  at a $68\%$ confidence level.

We split the matter component into baryonic matter and cold dark matter, that is $\Omega_\mathrm{matter} = \Omega_\mathrm{BM} + \Omega_\mathrm{CDM}$, and explain each component separately:

\textbf{Baryonic matter:} This is the constituent of the universe that corresponds to ordinary matter (atoms, nuclei, etc) and this is dominated by cold baryons\footnote{a baryon is composite particle made of three quarks and held together by the strong nuclear force} such as protons and neutrons, which are significantly heavier than leptons\footnote{a lepton is an elementary particle with half integer spin that does not experience the strong force}, such as electons and neutrinos. 
Despite making up all of the matter we see around us, recent measurements show this constituent is a very small fraction of the total energy density of the Universe, with $\Omega_\mathrm{BM}^0 \simeq 0.049$.

\textbf{Dark matter:} This component of the matter in the Universe is necessary in order to explain many observations of our Universe (see \Sec{sec:darkmatter} for details), including galactic rotation curves and the CMB fluctuations seen by the Planck satellite. 
Dark matter has $w=0$, is pressureless, and does not interact electromagnetically (or if it does interact with the EM force, it must do so extremely weakly), hence the name ``dark matter". 
While the nature of dark matter is currently unknown, there are many candidates for dark matter, some of which are explored later in this thesis (see \Sec{sec:darkmatter}), including primordial black holes, which we will discuss in detail.
Despite the unknown nature of dark matter, it is much more abundant in the Universe than baryonic matter, with $\Omega_\mathrm{CDM}^0 \simeq 0.262$, so approximately $5$ times the energy density of ordinary matter is contained in dark matter. 

\textbf{Radiation:} Since radiation dilutes much faster than matter as the Universe expands, see \eqref{eq:rho:constantw} with $w=1/3$, the present energy density contained in radiation is much smaller than that contained in matter, with $\Omega_\mathrm{rad}^0 \simeq 9.23\times 10^{-5}$, and most of this is contained in photons from the CMB (neutrinos have a small mass and are hence non-relativistic in the current epoch). 

\textbf{Cosmological constant:} A cosmological constant with $w=-1$ is also necessary to explain several pieces of evidence that the current expansion of the Universe is accelerating, including supernovae observations, baryon acoustic oscillations and voids, and fitting the CMB observations. 
In fact, this dark energy fluid must be the dominant constituent of the Universe today, with $\Omega_\Lambda^0 \simeq0.6889 \pm 0.0056$, despite the fact that we do not know the exact nature of this dark energy fluid. 
However, it is likely to be a cosmological constant or similar, as the latest measurements \cite{Aghanim:2018eyx} for the equation of state are $w = -1.03 \pm 0.03$, at $68\%$ confidence level, which is consistent with a cosmological constant. 

The fractions of the energy density of the Universe contained in each component discussed here are summarised in \Fig{fig:compositionUniverse}.
Let us note that the majority of the energy density of the Universe is contained in fluids that we do not currently understand, and the nature of dark matter and dark energy are two of the most important open questions in modern cosmology.

\subsection{Timeline of the Universe}

\begin{figure}
\centering
\includegraphics[width=0.98\columnwidth]{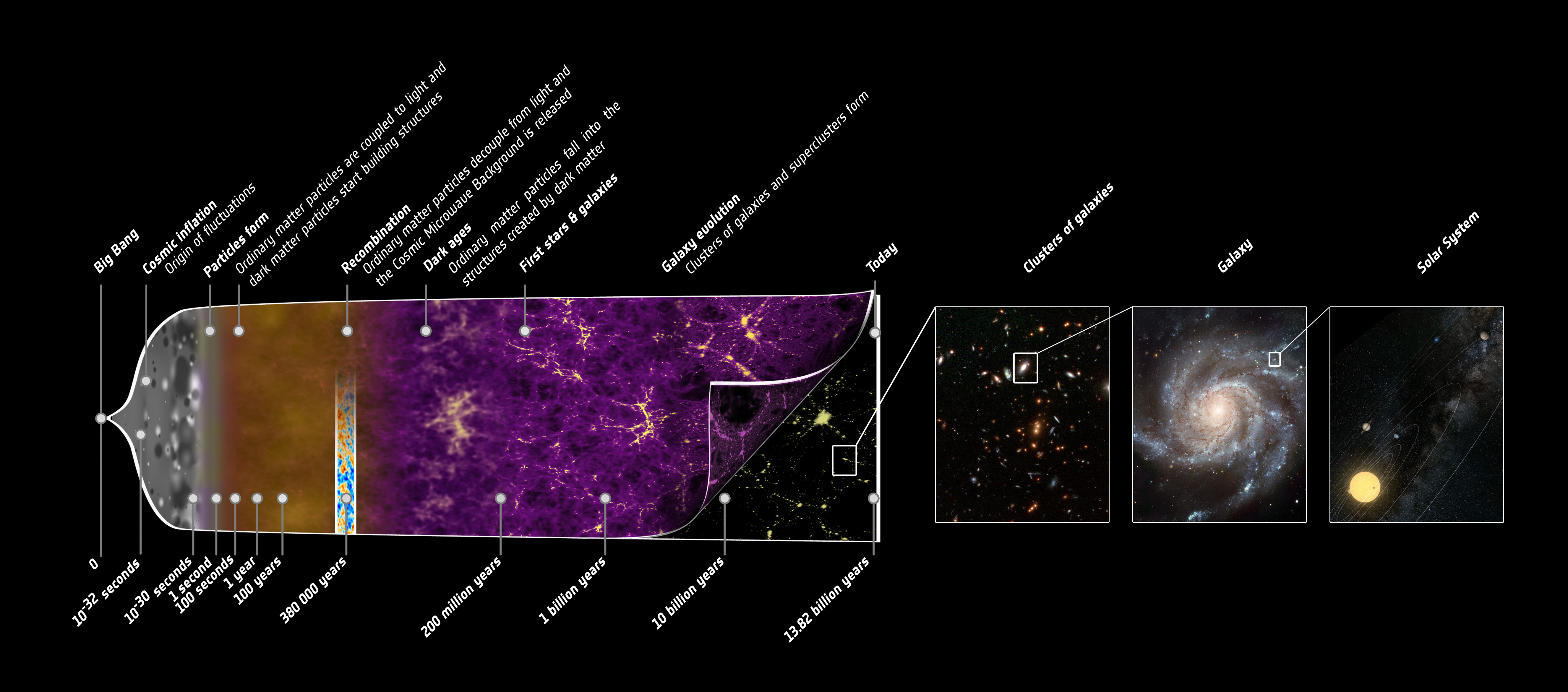}
\caption[History of the Universe]{An illustrated timeline of some of the key events in the evolution of the Universe, including the approximate times that they occured. \textsc{Image credit: NASA.}
\label{fig:historyofuniverse}}
\end{figure}

In this section we provide a brief description of some key events in the history of the Universe.
These events are summarised in \Fig{fig:historyofuniverse}, along with approximate time after the Big Bang that they took place. 
The initial singularity of the Big Bang picture is postulated because all photons travelling in an expanding Universe are such that their wavelength evolves as $a^{-1}$.
Since we also know that an initial black-body distribution remains a black-body at all time and its temperature decreases as $T \propto a^{-1}$, if we evolve this backwards in time we reach an initial singularity of infinite temperature at a finite time in the past (at $t=0$).
In the Hot Big Bang model of the Universe, this initial singularity is followed by a hot radiation dominated era which gradually cools. 
As the radiation cools and dilutes, it eventually becomes subdominant to the matter density of the Universe, and hence the radiation dominated era is followed by a matter dominated era. 
It is during this matter era that the familiar structures of the modern Universe form, including galaxies, stars and planets. 
Finally, as the matter dilutes with the expansion of the volume of spacetime, the cosmological constant (which has constant density and does not dilute) eventually dominates the energy density of the Universe and we enter the current era of accelerated expansion due to dark energy. 

Here, we give a brief description of some of the key events during this evolution of the Universe, with the times of their occurrence given relative to the Big Bang.
We begin our description of the events of the Universe after a period of inflation, which is a period of accelerated expansion at $t \lessapprox 10^{-35} \sec$, and is treated in great deal in \Sec{sec:intro:slowrollinflation} and is the main topic of this thesis. 
We assume that inflation leaves the Universe filled with a hot plasma that contains the fundamental particles of the Standard Model at $t \sim 10^{-35} \sec$ (possibly after a period of reheating). 

Once inflation is over, the Universe is radiation dominated and the first important event is the electroweak phase transition \cite{Kirzhnits:1972ut, Kirzhnits:1972iw, Dolan:1973qd, Weinberg:1974hy, Kirzhnits:1974as, Kirzhnits:1976ts,Dine:1992wr} at $t \sim 10^{-12} \sec$, when the energy drops below the vacuum expectation value of the Higgs field, around $246$ $\GeV$ \cite{2008PhLB..667....1A}. 
This phase transition broke the $SU(2)\times U(1)$ symmetry of the unified electroweak force into the $U(1)$ of the electromagnetic force we see today.
This marks the beginning of the ``quark epoch", in which the four fundamental forces were as they are now, but the temperature was too great to allow quarks to become bound together, \ie collisions between particles in the hot quark-gluon plasma were too energetic to allow quarks to combine. 

At $t\sim 10^{-6} \sec$, as the temperature drops below $\sim 938$ $\MeV$ (the rest energy of nucleons), quarks and gluons can bind together to form hadrons (either baryons or mesons\footnote{a meson is a composite particle made of a quark and an antiquark and held together by the strong nuclear force}) and anti-hadrons. 
This marks the end of the quark epoch and the beginning of the ``hadron epoch". 
The amount of matter compared to anti-matter created here must be large, since almost no anti-matter is observed in nature \cite{Sakharov:1967dj}, but it is not yet known how this asymmetry occured. 

As the energy of the Universe continues to drop, new hadron/anti-hadron pairs stop forming and most of the existing hadrons annihilate with anti-hadrons, creating high-energy photons, and by about $1 \sec$ after the Big Bang, almost all of the hadrons had been annihilated. 
If sufficiently large density fluctuations were seeded during inflation, then between the end of inflation and a few seconds after the Big Bang it is possible that primordial black holes formed. 
These black holes can have a very large range of masses, based on the time at which they formed, and at $t\sim 1 \sec$ these black holes would form with $10^5$ times the mass of the Sun. 

At $t\sim 1 \sec$, neutrinos decouple from matter and begin to freely stream through space, leaving a very low energy cosmic neutrino background (C$\nu$B) that still exists today (although it is almost impossible to directly detect\footnote{CMB observations are sensitive to the energy density of the C$\nu$B.}). 
This is analogous to the cosmic microwave background that is emitted much later, see below. 

Big Bang nucleosynthesis (BBN) takes place between $t\sim 10$ $\sec$ and $t\sim 20$ min \cite{Alpher:1948ve, Fuller:1987ue}, producing the nuclei for the lightest chemical elements and isotopes (BBN does not include the production of hydrogen-$1$ nuclei, which is just a single proton), including deuterium, helium, lithium and beryllium (although beryllium is unstable and later decays into helium and lithium), when the temperature of the Universe is $\sim 10$ $\MeV$ (the binding energy of nuclei). 
BBN ends at temperatures below $100$ $\keV$, when all of the deuterium has formed helium.
Note that although the nuclei for these elements form at this time, the energy of the Universe is still too high to allow electrons to bind to these nuclei. 
The observed amounts of each of these elements place constraints on the environment that BBN took place in and the time at which it happened \cite{Coc:2003ce, Pospelov:2010hj}
The short duration of BBN means only fast and simple processes can occur, and heavier elements only form later in the history of the Universe, though supernovae and kilonovae (see, for example, \cite{Kasen:2017sxr}).

At $t\sim 47,000$ years (redshift $z \sim 3400$), the energy density of matter begins to dominate over the energy density of radiation, and the Universe enters the matter dominated era. 
From this point on, perturbations are no longer erased by free-streaming radiation and hence structures can begin to form in the matter dominated era. 
The matter content of the Universe at this point is dominated by dark matter, although since the nature of dark matter is still unknown, the Hot Big Bang model does not give a unique explanation for its origin.
At $t\sim 100,000$ years, the temperature of the Universe is low enough for the first molecules (helium hydride) to form.

\begin{figure}
\centering
\includegraphics[width=0.98\columnwidth]{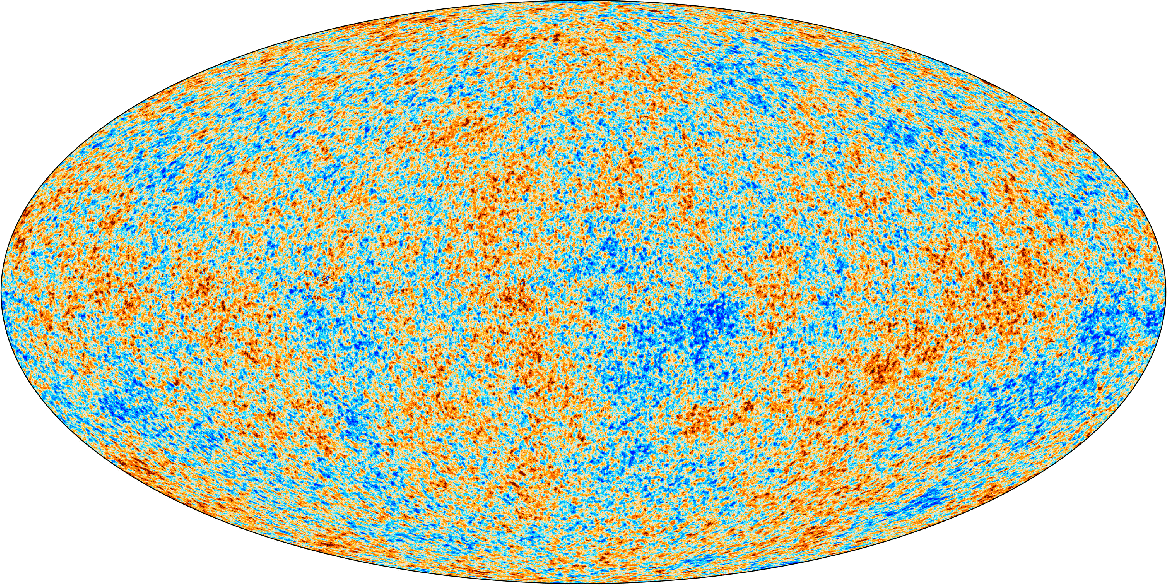}
\caption[Planck 2018 CMB temperature map]{Planck 2018 CMB temperature map. The temperature of the CMB is 2.725 K, with fluctions of just 0.01\%, \ie $1$ part in $10,000$. Image credit: ESA/Planck collaboration.
\label{fig:cmb2018}}
\end{figure}

At $t\sim 370,000$ years (redshift $z \sim 1080$), the Universe has cooled enough (to about $1$ $\eV$) that charged electrons can bind with protons to form the first neutral hydrogen atoms, in a phase called ``recombination'' (even though it is the first time these particles have combined). 
At this point the Universe becomes transparent for the first time \cite{Penzias:1965wn} and photons scatter off charged particles for the last time at the ``last scattering surface'', and can then freely stream through the Universe unimpeded and are still travelling today. 
Almost all protons and electrons in the Universe become bound in neutral atoms, and hence the mean free path for photons becomes very large, that is to say the photon-atom cross-section is much smaller than the photon-electron cross-section, and hence photons ``decouple'' from matter. 
This results in the cosmic microwave background (CMB), which is often called the ``oldest light in the Universe". 
The CMB reaches us with the same temperature distribution that it was emitted with (\ie a perfect black-body), but with its temperature redshifted by the expansion of spacetime between its emission and now, and the average temperature of the CMB today is $T_\mathrm{CMB} = 2.725 \K$.
The latest Planck image \cite{Akrami:2018vks} of the temperature fluctuations of the CMB is shown in \Fig{fig:cmb2018}, where the colours show that the CMB is homogeneous and isotropic up one part in $10^5$, and the deviations from homogeneity are key predictions of inflation. 
We note that when electrons and photons combine, it is more efficient for them to do so with the electron still in an excited state and then for the electron to transition to a lower energy state by releasing photons once the hydrogen forms, providing more photons for the CMB. 

After recombination, the Universe is no longer opaque to photons, but there are no light emitting structures, and hence this epoch is called the ``Dark Ages". 
As such, it is very difficult to make observations of anything from this epoch. 
There were only two sources of photons from this epoch, the CMB photons, and the rare $21 \cm$ spin line of neutral hydrogen, which is the spontaneous release of a $21 \cm$ photon from an electron dropping down an energy state. 
Although this $21 \cm$ emission is rare, the Universe was filled with neutral hydrogen, and so a detectable amount of this wavelength of light could have been emitted during the dark ages and there are currently many telescopes trying to detect this light, including the recent EDGES experiment \cite{Bowman:2018yin}. 

As perturbations grow during the matter era, the first stars (known as Population III stars) begin to form at $t \sim 100,000$ years, providing the first visible light after recombination and ending the Dark Ages. 
These early stars were made almost entirely of hydrogen (and some helium), and gradually began to fill the Universe with heavier elements as they evolved and went supernova. 

Galaxies then begin to form and in these early galaxies form objects that are energetic enough to ionise neutral hydrogen.
That is, objects that produce photons with enough energy to break apart the electron-proton bond of the hydrogen and leave separate charged particles, in a process that is called ``reionisation" \cite{1965ApJ...142.1633G, Becker:2001ee, Barkana:2000fd, 1987ApJ...322..597V}.
The objects thought to be able to produce these high energy photons include quasars, as well as the Population III stars and the formation of the early galaxies themselves. 
Reionisation took place between $t \sim 150$ millions years and $t \sim 1$ billion years, or at redshift $6 < z < 20$.
However, due to the large amount of expansion that the Universe had experienced since recombination, the Universe did not revert back to being opaque to photons after reionisation, \ie the mean free path of photons remains large, even today, due to scattering interactions remaining rare due to the dilution of matter. 

Entering the current epoch at $t \sim 9$ billion years, dark energy begins to dominate the Universe and the expansion of the Universe accelerates \cite{Riess:1998cb,Perlmutter:1998np}. 
During this epoch, the large scale structure of the Universe continues to develop to become the cosmic web we see today, full of clusters of galaxies, galaxies, stars and planetary systems.
The nature of dark energy is still unknown, and is a very active area of research in modern cosmology.

\section{Classical problems in the Hot Big Bang Model}

Above we have outlined the main events and features of the Hot Big Bang model of cosmology. 
Although this model is very compelling, it leaves several unanswered questions about the Universe when we compare this theory to observations, and this ultimately leads us to introduce an early phase of cosmological inflation into our view of the Universe.  
The three famous problems are known as the horizon, flatness, and monopole problems, although this is not an exhaustive list of known problems with this model and other questions raised by this model include the origin and nature of dark matter and dark energy, the nature of physics at the Big Bang singularity, etc.
Below, we will discuss the details of the horizon problem, as it is arguably the most fundamental problem, while simply noting the other two for historical reasons. 
Most modern cosmologists indeed treat these problems as the original motivation for inflation, but nowadays the main motivation for studying inflation is to seed the large scale structure of the Universe.

In this section, we will introduce the horizon problem, and in the next section we will show explicitly how this can be rectified by a period of accelerated expansion in the very early Universe.

\subsection{Horizon problem}

We begin by discussing the horizon problem \cite{1956MNRAS.116..662R, 1968ApJ...151..431M} of the Hot Big Bang model, which is a consequence of the finite travel time of light, and the existence of a causal horizon in the Hot Big Bang model. 
This causal horizon acts as a barrier between observable events and non-observable events, and no physical process can act on scales larger than the causal horizon. 
This means that we would, \textit{a priori}, expect the Universe to be inhomogeneous on scales larger than the causal horizon, as separate causal regions cannot ``talk" to each other in order to equilibriate. 
The causal horizon is defined as the furthest physical (proper) distance away that a photon received at a time $t_\mathrm{rec}$ could have originated. 

In this section, we will calculate the size of the causal horizon at the time of recombination and show that it is much smaller than the diameter of the last scattering surface, \ie the last scattering surface is made of many causally disconnected regions. 
Hence the extreme homogeneity of the CMB observations (see \Fig{fig:cmb2018}) requires either fine-tuning, an early period of inflation, or some other explanation. 

If a photon is emitted at time $t_\mathrm{em}$ and radius $r_\mathrm{em}$ and travels directly ($\dd\phi = \dd\theta = 0$) to an observer at $r=0$, then from \eqref{eq:FLRWmetric} we see that its trajectory, given by $\dd s^2=0$, is 
\bea 
\frac{\dd t}{a(t)} = \frac{\dd r}{\sqrt{1-Kr^2}} \, .
\eea 
If $K=0$ (as will be justified below), we can then solve for $r$ to get
\bea \label{eq:photonemissiondistance}
r(t) = r_\mathrm{em} - \int^{t}_{t_\mathrm{em}} \frac{\dd \hat{t}}{a(\hat{t})} \, ,
\eea 
and hence the physical distance between the photon and its point of emission is $d_\mathrm{phys}(t) = a(t)r(t)$.
In order to find the size of the horizon, we must maximise \eqref{eq:photonemissiondistance}, which will occur if the photon was emitted at the earliest possible time, so $t_\mathrm{em}=t_\mathrm{BB}$ is the time of the Big Bang, and received at the latest possible time, so $r(t_\mathrm{rec}) = 0$. 
Thus, \eqref{eq:photonemissiondistance} gives 
\bea 
r_\mathrm{em} = \int^{t_\mathrm{rec}}_{t_\mathrm{BB}} \frac{\dd t}{a(t)} \, ,
\eea
and hence the size of the horizon at a time $t_\mathrm{rec}$ is given by 
\bea \label{eq:horizonsize}
d_\mathrm{hor}\left( t_\mathrm{rec} \right) = a(t_\mathrm{rec})r_\mathrm{em} = a(t_\mathrm{rec})\int^{t_\mathrm{rec}}_{t_\mathrm{BB}} \frac{\dd t}{a(t)} \, .
\eea 
The size of the causal horizon at the last scattering surface is then found by taking $t_\mathrm{rec} = t_\mathrm{lss}$ in \Eq{eq:horizonsize}.

However, it is more useful to know the \textit{angular} size of the causal horizon at recombination as seen by an observer on Earth at time $t_0 > t_\mathrm{lss} $, denoted by $\Delta\Omega_{d_\mathrm{hor}}(t_0)$, as this is an observable quantity. 
If we let $\dd \Omega^2 = \dd \theta^2 + \sin\theta\dd\phi^2$, and assume that the last scattering surface is an instantaneous sphere of constant radius, \ie $\dd t = \dd r = 0$, then the FLRW metric \eqref{eq:FLRWmetric} reads $\dd s = a(t_\mathrm{lss})r_\mathrm{lss}\dd\Omega$.
Here, $r_\mathrm{lss} = \int^{t_0}_{t_\mathrm{lss}} \frac{\dd t}{a(t)}$ is the distance from Earth to the last scattering surface, and is found from \eqref{eq:photonemissiondistance} with $t=t_0$, $t_\mathrm{em}=t_\mathrm{lss}$ and $r(t_0)=0$.
Combining all of these results, we find that the angular size of the causal horizon of the last scattering surface as seen from Earth is 
\bea \label{eq:angularhorizon:lss}
\Delta\Omega_{d_\mathrm{hor}}(t_0) = \frac{d_\mathrm{hor}\left( t_\mathrm{lss} \right)}{a(t_\mathrm{lss})r_\mathrm{lss}} = \dfrac{\int^{t_\mathrm{lss}}_{t_\mathrm{BB}} \frac{\dd t}{a(t)}}{\int^{t_0}_{t_\mathrm{lss}} \frac{\dd t}{a(t)}} \, .
\eea 
This ratio of integrals can be solved by noting the relationship
\bea 
\int^{t_2}_{t_1} \frac{\dd t}{a(t)} = \frac{1}{a_0H_0}\int^{z_1}_{z_2}\frac{\dd z}{\sqrt{\sum_{i} \Omega_i^0 (1+z)^{3(1+w_i)}}} \, ,
\eea 
where $\omega_i$ refers to the energy density of each fluid present in the Universe, with equation of state $w_i$, as in \eqref{eq:sumofomegas}.
With this, the angular size of the causal horizon at last scattering can be found numerically to be 
\bea 
 \Delta\Omega_{d_\mathrm{hor}}(t_0) = \displaystyle\frac{\int^{\infty}_{z_\mathrm{lss}}\frac{\dd z}{\sqrt{\sum_{i} \Omega_i^0 (1+z)^{3(1+w_i)}}}}{\int^{z_\mathrm{lss}}_{0}\frac{\dd z}{\sqrt{\sum_{i} \Omega_i^0 (1+z)^{3(1+w_i)}}}} \simeq 0.0054 \, \mathrm{rad} \simeq 0.3\deg \, ,
\eea 
which corresponds to approximately $1/450,000$ of the sky. 
This means we should expect the last scattering surface to be comprised of around $450,000$ separate patches whose physical properties have, \textit{a priori}, no reason to be similar and can be completely different. 
However, observations tell us that the CMB is homogeneous and isotropic up to tiny fluctuations of order  $\delta T/T \simeq 10^{-5}$ across the whole sky \cite{Akrami:2018vks}, and hence we have a contradiction in the standard picture, and this problem is know as the ``horizon problem''.

Note that this is essentially an initial condition problem, because if one assumes that the conditions in each patch were identical before last scattering\footnote{Since the Universe at the time we need to set initial conditions is likely to be governed by quantum gravity, the nature of which we do not know, it may be argued that the horizon problem is a manifestation of our ignorance of quantum gravity and should not be considered a ``classical'' problem. However, in this case we cannot perform any calculations or make any predictions.}, then this problem vanishes. 
Another way to solve the horizon problem without fine-tuning the initial conditions is to introduce an early phase of cosmological inflation, as will be discussed in the next section.

\subsection{Flatness and monopole problems}

The flatness problem \cite{2010grae.book.....H} refers to the fact that observations of the Universe today are consistent with flatness at $95\%$ \cite{Akrami:2018vks}, while also noting that $\Omega_K=0$ is unstable in a radiation or matter dominated universe, and any initial curvature in the Universe should grow. 
This means that the flatness problem is another fine-tuning issue, since to explain the current observations we need to set the initial curvature of the Universe to be close to zero at an extremely precise level. 

This is, again, not a problem if one is happy to set such a fine-tuned initial configuration, but an early period of inflation also solves this problem. 
To see why, we simply need to note that curvature decays slower than the other fluids in the Universe (\ie slower than radiation and matter, see \Eq{eq:sumofomegas}), and hence if it is small today then it must have been even smaller in the past. 
If we introduce a phase in which the energy density decays slower than curvature, however, we can effectively dilute any curvature away in the early Universe. 
We know that $w_K=-1/3$, and so any phase dominated by a new fluid with $w<-1/3$ will see curvature decay faster than this new fluid. 
We will see shortly that this is precisely the case for an accelerating expansion phase. 
We also assume this new fluid decays directly into radiation and hence curvature is automatically subdominant to the radiation, provided the period of inflation lasted long enough (which turns out to be approximately $60$~\efolds~or more \cite{Wrase:2019sfp}).

For the rest of the thesis, we take $\Omega_K=0$, since from \eqref{eq:rescaledFriedmann} we see that the contribution from curvature grows slower than all other components as we go backwards in time ($a\to 0$), and hence we know that curvature has never dominated the energy content of the Universe.

The monopole problem is the apparent paradox between the predicted abundance of magnetic monopoles \cite{tHooft:1974kcl,Polyakov:1974ek} resulting from spontaneous symmetry breaking in Grand Unified Theories \cite{Georgi:1974sy,Pati:1974yy,Buras:1977yy} (GUT) as the universe cools below $M_\mathrm{GUT} \simeq 10^{16} \GeV$, and the observed number of magnetic monopoles in the Universe today. 
These GUT theories predict vast numbers of these monopoles being produced at high temperatures \cite{Guth:1979bh,Einhorn:1980ym} and they should still exist in the modern Universe in such numbers that they would be the dominant constituent of the Universe \cite{Zeldovich:1978wj,Preskill:1979zi}.
However, searches for these objects have failed to find any magnetic monopoles \cite{Rajantie:2012xh}, leading to the paradox of the monopole problem. 
Inflation offers a solution to this problem by effectively diluting the density of such monopoles to less than one per causal patch of the Universe. 
This does not exclude the possibility of these theories forming magnetic monopoles, but a long enough period of inflation explains the lack of observations of these objects (we require $N\gtrsim60$ \efolds,~assuming $w=-1$).


%
\section{Inflation}

Cosmological inflation \cite{Starobinsky:1980te, Sato:1980yn, Guth:1980zm, Linde:1981mu, Albrecht:1982wi, Linde:1983gd} is the leading paradigm for the very early Universe, in which space-time undergoes a period of accelerating expansion at very high energies (between $10^3$ and $10^{15}$ $\GeV$). 
While inflation provides a neat solution to the classical problems of the Hot Big Bang model, possibly a more important feature of cosmic inflation is its ability to seed the large-scale structure of the modern Universe when vacuum quantum fluctuations (of the gravitational and matter fields) are amplified and grow into the cosmic web~\cite{Mukhanov:1981xt, Mukhanov:1982nu, Starobinsky:1982ee, Guth:1982ec, Hawking:1982cz, Bardeen:1983qw}.
This provides an explanation for the observed homogeneity and isotropy of the Universe, and allows inflation to be a predictive theory, and measurements of these inhomogeneities provide knowledge about conditions during inflation. 
For example, inflation predicts that the spectrum of the cosmological fluctuations should be almost exactly scale invariant, that is to say their power is approximately equal on all spatial scales, and this is completely consistent with the latest observations \cite{Akrami:2018vks}. 

In fact, inflation is possibly the only case in physics where an effect based on General Relativity and Quantum Mechanics leads to predictions that, given our present day technological capabilities, can be tested experimentally. We note that many other possible explanations for the early universe have been suggested (see, for example, \cite{Alexander:2000xv,Martin:2001ue,Steinhardt:2001vw,Kallosh:2001ai,Magueijo:2003gj,Brandenberger:2009yt}), but inflation has outlasted them all and become an accepted part of modern cosmology. 
High precision data that can test the inflationary paradigm is now more readily available than ever, and will keep coming with missions planned for the next few decades that will continue to test inflation.
Recent observations from the Planck satellite \cite{Akrami:2018vks, Akrami:2018odb} (which build upon previous previous data from the WMAP satellite \cite{Bennett_2013, 2013ApJS..208...19H}) allow us to constrain inflation, while data about the smaller scales of the CMB is complemented by ground-based microwave telescopes such as the Atacama Cosmology Telescope \cite{Dunkley:2013vu,Sievers:2013ica} and the South Pole Telescope \cite{2014ApJ...782...74H, 2013ApJ...779...86S}, and dedicated ultra-sensitive polarization  experiments are planned for the future \cite{Baumann:2008aq,Crill:2008rd,2011JCAP...07..025K,Matsumura:2013aja,Andre:2013afa}.
Other observations that can test inflationary physics include polarisation measurements of the CMB, and $21$ cm telescopes (for example \cite{Bowman:2018yin}) that probe the dark ages of the Universe.
Direct detection of primordial gravitational waves, through future experiments like the space-based Laser Interferometer Space Antenna (LISA) \cite{Seoane:2013qna}, can also test the very early Universe \cite{Iacconi:2019vgc}, including inflation and the formation of primordial black holes. 
Probing the very early Universe is exciting because it provides an ultra-high energy quantum laboratory on cosmological scales, allowing us to test physics well beyond the scales accessible to Earth-based experiments such as the Large Hadron Collider (LHC). 

In this section, as well as demonstrating some of the motivation behind inflation, we will discuss the theory and tools that are typically used to study inflation, which will provide the foundations for the results presented later in this thesis.

\subsection{Solution to the Horizon Problem}

\begin{figure}
\centering
\includegraphics[width=0.98\columnwidth]{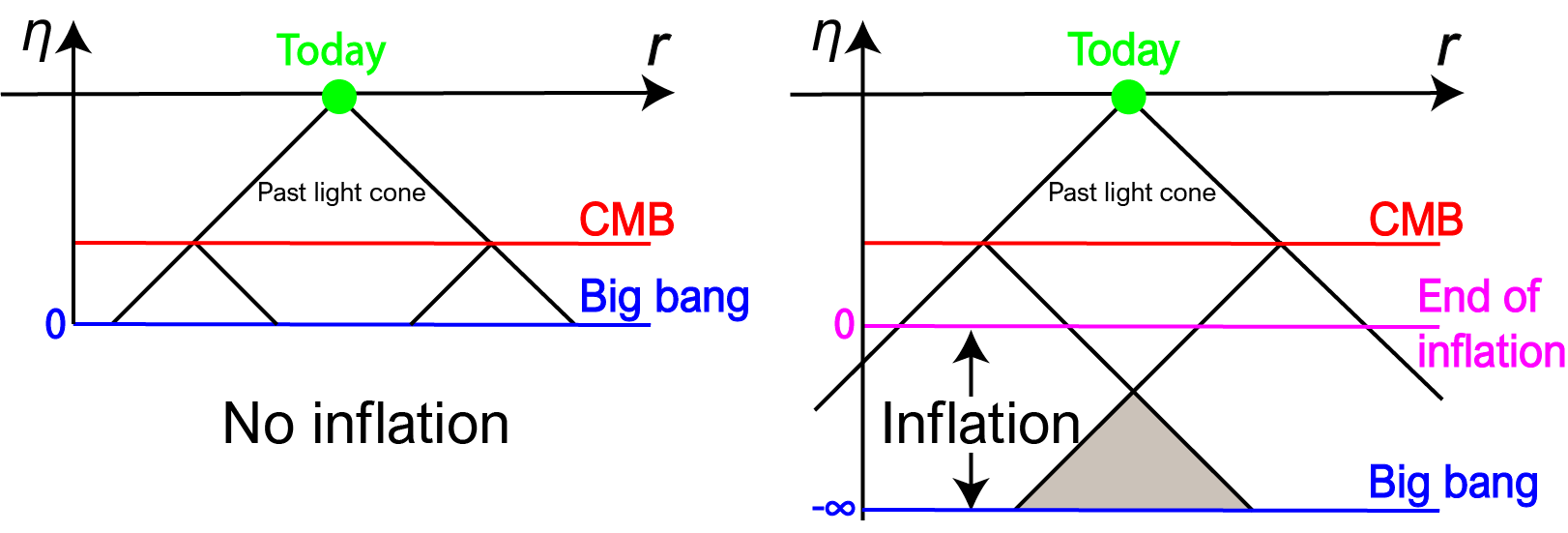}
\caption[Space-time conformal diagram with and without inflation]{Conformal diagrams in the $(\eta, r)$ plane, where light propagates in straight lines.
In the left panel we depict the standard Big Bang cosmology with no inflation, where the CMB consists of approximately $450,000$ causally disconnected patches. This means that the past light cones of these regions do not intersect before the initial singularity at $\eta=0$. In the right hand plot we show the conformal diagram when inflation is included and we extend conformal time to negative values, and any two patches on the CMB are causally connected as there past light cones intersect (denoted by the shaded region here).
\label{fig:conformaldiagram}}
\end{figure}

In order to understand how a period of inflation can solve the horizon problem, let us consider conformal space-time diagrams. 
By neglecting angular coordinates, the FLRW metric \eqref{eq:FLRWmetric:conformaltime}, written in conformal time, is simply given by 
\bea 
\dd s^2 = a^2(\eta) \left( -\dd\eta^2 + \dd r^2 \right) \, ,
\eea 
where we recall that conformal time is given by $\dd t = a \dd\eta $.
Thus, in this parameterisation, photons, which follow null geodesics defined by $\dd s^2 = 0$ and define the past light cones, follow the simple trajectories $\dd\eta = \dd r$.
The size of the causal horizon in terms of conformal time is then given by 
\bea \label{eq:horizon:conformaltime}
d_\mathrm{hor}\left( \eta \right) = a(\eta) \left( \eta - \eta_\mathrm{BB} \right) \, ,
\eea 
where $\eta_\mathrm{BB}$ is the value of conformal time at the Big Bang. 
Now, from \Eq{eq:scalefactor:time}, if the Universe is dominated by a single fluid with equation of state $w$, and we have $t_\mathrm{BB}=0$, then the scale factor is given by 
\bea 
a(t) = a_0 \left( \frac{t}{t_0} \right)^{\frac{2}{3(1+w)}} \, ,
\eea
where $a_0$ is a constant defined by \Eq{eq:scalefactor:time}, and which is positive for an expanding Universe and $w > -1$.
Hence conformal time, found by integrating $\dd t = a \dd\eta $, is given by 
\bea \label{eq:conformaltimeintermsofcosmictime}
\eta = \eta_0  \left( \frac{t}{t_0} \right)^{\frac{3w+1}{3(1+w)}} \, ,
\eea 
where we have defined 
\bea 
\eta_0 = \frac{3}{a_0 t_0}\frac{1+w}{3w+1} \, .
\eea 
Now, if $w>-\frac{1}{3}$, then when $t\to0$ we have $\eta \to 0$ from \Eq{eq:conformaltimeintermsofcosmictime}.
This tells us that $\eta_\mathrm{BB} = 0$, and hence the size of the horizon is $d_\mathrm{hor}\left( \eta \right) = a \eta$, which is finite and can lead to a horizon problem. 
This is the case represented by the left hand side of \Fig{fig:conformaldiagram}.

However, if $w<-\frac{1}{3}$, which we will see below is the case for an inflating universe, then when $t\to 0$, we have $\eta \to -\infty$ from \Eq{eq:conformaltimeintermsofcosmictime}. 
This means that $\eta_\mathrm{BB} = -\infty$, and hence the size of the causal horizon becomes infinite, see \Eq{eq:horizon:conformaltime}.
Since $w<-\frac{1}{3}$, the definition of $\eta_0$ means that $\eta$ is allowed to become negative, and the singularity at $a=0$ is only realised in the infinite past\footnote{In the infinite future, for $w<-1/3$, we have $a\to\infty$ in the infinite future $t\to 0$ (or $\eta\to 0$). This is only the case if we assume the inflation phase ($w<-1/3$) lasts indefinitely, but in practice inflation will end at some finite time $-1 \ll \eta_{\mathrm{end}} < 0$, and hence the surface $\eta\to0^{-}$ represents the end of inflation. This is shown in \Fig{fig:conformaldiagram}.}. 
This means that the horizon problem is eliminated in this case \cite{Guth:1980zm, Linde:1981mu}, which is represented in the right hand side of \Fig{fig:conformaldiagram}.

\subsection{Classical inflationary dynamics}

By definition, inflation is a period of accelerated expansion, hence the scale factor $a$ is growing at an increasing rate, so $\ddot{a} > 0$. 
The Raychauduri equation \eqref{eq:raychaudhuriequation} in the absence of a cosmological constant then tells us that 
\bea 
\rho + 3P < 0 \, , 
\eea 
which means that inflation will be realised by a fluid with negative pressure (since we require $\rho>0$ always).
More specifically, this tells us that, for inflation, the equation of state is $w < -1/3$.
Since inflation takes place at very high energies, the necessary formalism to describe the physics of this time is field theory, and hence a simple realisation of inflation is to consider the expansion to be driven by a real scalar field $\phi$, which we call the ``inflaton'' field, \ie we consider the energy density of the early Universe to be dominated by the inflaton. 
This is an assumption that is completely compatible with the flatness, isotropy and homogeneity of the early Universe that we observe. 
However, the physics of inflation cannot be tested terrestrially (\ie in a particle accelerator) because of the extremely high energies that it took place at, which are currently well beyond the limits of experimental probing on Earth. 
This means that the shape of the potential $V(\phi)$ is relatively unknown, other than the flatness requirement to ensure inflation is realised, allowing for many models of inflation to be suggested (and tested against observations). 

The action for a single scalar field $\phi$, the inflaton, minimally coupled to gravity is given by 
\bea \label{eq:4d:action}
S_\mathrm{infl} = \int \dd^4 x \sqrt{-g}\left[ \frac{\Mp^2}{2}R - \frac{1}{2}g^{\mu\nu}\partial_{\mu}\phi \partial_\nu\phi - V(\phi) \right] \, ,
\eea 
where the potential of the inflaton $V(\phi)$ is left unspecified for now, and the shape of the potential is still an area of extensive research and discussion today \cite{Obied:2018sgi,Palti:2019pca}. 
From this action, we can use the definition \eqref{eq:def:energymomentumtensor} to find the energy-momentum tensor of $\phi$ to be 
\bea \label{eq:energymomentumtensor:phi}
T_{\mu\nu} = \partial_{\mu}\phi \partial_\nu\phi - g_{\mu\nu}\left[ \frac{1}{2}g^{\rho\lambda}\partial_{\rho}\phi \partial_\lambda\phi + V(\phi) \right] \, ,
\eea 
and the equation of motion for $\phi$ is 
\bea 
\frac{\partial}{\partial \phi}\left(\sqrt{-g}\left[- \frac{1}{2}g^{\mu\nu}\partial_{\mu}\phi \partial_\nu\phi - V(\phi)\right]\right) = -\frac{g^{\mu\nu}}{\sqrt{-g}}\partial_\mu\left(\sqrt{-g}\partial_{\nu}\phi\right) - \frac{\partial V}{\partial\phi} = 0 \, .
\eea 
As explained previously, we assume a flat Universe ($K=0$, since curvature can \textit{never} have dominated the Universe) and in an FLRW Universe (see \eqref{eq:FLRWmetric}) this equation of motion becomes
\bea \label{eq:eom:scalarfield}
\ddot{\phi} + 3H\dot{\phi} + V_{,\phi} = 0 \, , 
\eea 
which is called the Klein--Gordon equation, and we have assumed that $\phi$ is homogeneous and we recall that a dot is a derivative with respect to cosmic time $t$ and a subscript ``$,\phi$'' denotes a derivative with respect to the field $\phi$. 
Since $\phi$ is a scalar field, and hence has only one degree of freedom, it has no anisotropic stress and can be identified as a perfect fluid, and therefore by comparing \eqref{eq:energymomentumtensor:phi} for flat space with $T_{\mu\nu}=\mathrm{diag}\left( \rho, P, P, P \right)$ for a perfect fluid, $\phi$ has energy density and pressure given by 
\begin{align}
\label{eq:phi:density} \rho &= \frac{\dot{\phi}^2}{2} + V(\phi) \\
\label{eq:phi:pressure} P &= \frac{\dot{\phi}^2}{2} - V(\phi) \, ,
\end{align}
and we can therefore write the equation of state of $\phi$ as 
\bea \label{eq:equationofstate:phi}
w = \frac{P}{\rho} = \frac{\frac{1}{2}\dot{\phi}^2 - V(\phi)}{\frac{1}{2}\dot{\phi}^2 + V(\phi)} \, .
\eea 
From \eqref{eq:phi:pressure}, we see that in order to have negative pressure, as is required to ensure inflation is realised, we need to have 
\bea \label{eq:inflationcondition}
V(\phi) > \frac{\dot{\phi}^2}{2} \, ,
\eea 
which mean that the potential energy must dominate over the kinetic energy if we have an inflating universe. 
Also note that with the forms of energy density and pressure given by \Eqs{eq:phi:density} and \eqref{eq:phi:pressure}, we can find the equation of motion \eqref{eq:eom:scalarfield} by plugging these into the continuity equation \eqref{eq:continuityequation}, and we find the Friedmann equation \eqref{eq:friedmannequation} for the inflaton $\phi$ is 
\bea \label{eq:friedmann:scalarfield}
H^2 = \frac{\frac{1}{2}\dot{\phi}^2 + V(\phi)}{3\Mp^2} \, .
\eea 
For given initial conditions on $\phi$ and $\dot{\phi}$ and any potential $V(\phi)$, the Klein--Gordon equation \eqref{eq:eom:scalarfield}, together with the Friedmann equation \eqref{eq:friedmann:scalarfield}, can be solved, although numerical methods must often be used, to give the dynamics of the inflaton. 
For a given potential, inflation will persist as long as $P < -\rho/3$, and when this condition fails, inflation will end and a phase of reheating is assumed to take place which fills the universe with the particles of the standard model.

\subsection{Slow-roll inflation}
\label{sec:intro:slowrollinflation}

While numerical solutions to \eqref{eq:eom:scalarfield} can be useful, it is helpful to consider cases when analytical solutions exist and this is the case when the inequality in \Eq{eq:eom:scalarfield} is extreme, \ie $V(\phi) \gg \dot{\phi}^2$.
From \Eqs{eq:phi:density} and \eqref{eq:phi:pressure}, this means that $P \simeq -\rho$ and hence the continuity equation \eqref{eq:continuityequation} gives $\dot{\rho} \simeq 0$, \ie the energy density of the inflaton field is approximately constant. 
In turn, the Friedmann equation \eqref{eq:friedmannequation} then tells us that $H=\dot{a}/a$ can be taken to be constant and hence the scale factor is given by 
\bea 
a(t) \simeq a_\mathrm{in}\exp\left[ H\left( t - t_\mathrm{in} \right) \right] \, .
\eea 
This means space-time is very close to de Sitter space (in which the $\simeq$ here is an exact equality), and we see that inflation does in fact give us exponential expansion. 
This motivates the study of solutions of \eqref{eq:eom:scalarfield} in this $V(\phi) \gg \dot{\phi}^2$ limit, which we call the ``slow-roll'' approximation because the kinetic energy is much smaller than the potential energy and hence the inflaton ``slowly rolls'' down its potential. 

In order to explicitly parameterise the deviations from de Sitter space, we introduce a set of hierarchical quantities called the ``slow-roll parameters''
\bea \label{eq:def:slowrollparameters}
\epsilon_{i+1} = \frac{\dd \ln \left|\epsilon_i\right|}{\dd N} = \frac{1}{\epsilon_i}\frac{\dd \left|\epsilon_i\right|}{\dd N} \, ,
\eea
where $\epsilon_0 = H_\mathrm{in}/H$, $H_\mathrm{in}$ is the value of the Hubble parameter at the initial time $t_\mathrm{in}$, $\dd N = \dd \ln a = H\dd t$, and $N$ is the number of ``\efolds'' of inflation, which we will often use as our time coordinate.
Typically, each $\epsilon_i$ is of the same order of magnitude and slow-roll inflation is defined by the condition $\epsilon_i \ll 1$ $\forall i>0$. 
For example, the first slow-roll parameter is 
\bea \label{eq:eps1exact}
\epsilon_1 = -\frac{\dot{H}}{H^2} = 3 \frac{\frac{1}{2}\dot{\phi}^2}{V(\phi) + \frac{1}{2}\dot{\phi}^2} \, , 
\eea 
where the second equality follows by using $\dot{H} = -\dot{\phi}^2/(2\Mp^2)$, which is found by inserting the Klein--Gordon equation in the time derivative of the Friedmann
equation.
In terms of the scale factor, $\epsilon_1$ can be written as 
\bea 
\epsilon_1 = 1 -\frac{a\ddot{a}}{\dot{a}^2} \, , 
\eea 
and hence the condition for inflation $\ddot{a} > 0$ is equivalent to $\epsilon_1 < 1$.
The slow-roll condition $V(\phi) \gg \dot{\phi}^2$ also ensures that $\epsilon_1 \ll 1$, and hence ensures inflation is easily realised. 
We also have, from \eqref{eq:equationofstate:phi}, that the equation of state for the inflaton is 
\bea 
w = -1 + \frac{2\epsilon_1}{3} \simeq -1 \, ,
\eea 
as previously stated in table \eqref{table:fluids}, and we see that the slow-roll parameter $\epsilon_1$ parameterises the deviation from the de Sitter value $w = -1$.

There are several nice consequences of the slow-roll assumption.
For example, implementing $V(\phi) \gg \dot{\phi}^2$ in \eqref{eq:friedmann:scalarfield} tells us that in slow roll we have the simplified Friedmann equation
\bea \label{eq:friedmann:slowroll}
H^2 \simeq \frac{V(\phi)}{3\Mp^2} \, ,
\eea 
which is valid at leading (zeroth) order in the slow-roll parameter $\epsilon_1$.
If we then consider the second slow-roll parameter $\epsilon_2$ (which tells us the relative change in $\epsilon_1$ in one $e$-fold) we find 
\bea \label{eq:eps2exact}
\epsilon_2 = \frac{\ddot{H}}{H\dot{H}} - 2\frac{\dot{H}}{H^2} = 6\left( \frac{\epsilon_1}{3} - \frac{V_{,\phi}}{3H\dot{\phi}} - 1 \right) \, ,
\eea 
which comes from \eqref{eq:def:slowrollparameters} and by noting that
\bea 
\ddot{H} = \frac{3H\dot{\phi}^2}{\Mp^2} + \frac{\dot{\phi}V_{,\phi}}{\Mp^2} \, , 
\eea 
which is found by using the Klein--Gordon equation and the time derivative of our previously found expression $\dot{H} = -\dot{\phi}^2/(2\Mp^2)$.
By considering the condition $\epsilon_2 \ll 1$, we see that, at leading order in slow roll, $\dot{\phi}_\mathrm{SR} \simeq -V'/(3H)$, which is equivalent to neglecting the second derivative term in the Klein--Gordon equation \eqref{eq:eom:scalarfield}.
This is a powerful result because it takes the equation of motion from second order to first order, and thus removes a dependence on the initial velocity of the field, and the kinetic energy is now entirely specified by the gradient of the potential.
The form of the potential entirely specifies the dynamics of the inflaton and there is just a single trajectory through phase space (we call this trajectory the ``slow-roll attractor''). 

We can calculate the length of time that slow-roll inflation lasts for by rewriting the slow-roll equation of motion with the number of \efolds~$N$ as the time variable, so 
\bea \label{eq:eom:slowroll:efolds}
\frac{\dd \phi_\mathrm{SR}}{\dd N} \simeq -\frac{V_{,\phi}}{3H^2} \, ,
\eea 
where we recall that $\dd N = H\dd t$ and that this is at leading order in slow roll (\ie we neglect terms that are $\propto \epsilon_1$). 
By first inserting the slow-roll Friedmann equation \eqref{eq:friedmann:slowroll}, this can be integrated to give 
\bea \label{eq:efolds:slowroll:LO}
\Delta N_\mathrm{SR} \equiv N_\mathrm{end} - N_\mathrm{in} = \int^{N_\mathrm{end}}_{N_\mathrm{in}} \dd N \simeq -\frac{1}{\Mp^2}\int ^{\phi_\mathrm{end}}_{\phi_\mathrm{in}} \frac{V}{V_{,\phi}} \dd \phi \, ,
\eea 
where $\phi_\mathrm{in}$ is the value of the inflaton at an inital time $N_\mathrm{in}$ and $\phi_\mathrm{end}$ is the field value at some end time $N_\mathrm{end}$. 
For a given potential $V(\phi)$, this expression can then be inverted to give $\phi(N)$. 
Recall that this is only valid at leading order in slow roll, and one can perturbatively include corrections to this slow-roll trajectory, and the limit of this expansion gives the slow-roll attractor in phase-space \cite{Remmen:2013eja}, which confirms that the slow-roll approximation is not only simple but is also physically motivated. 

In the slow-roll approximation, the hierarchy of slow-roll parameters \eqref{eq:def:slowrollparameters} can be rewritten in terms of the potential and its derivatives.
This can be done by noting that \eqref{eq:eom:slowroll:efolds}, together with the Friedmann equation \eqref{eq:friedmann:slowroll}, gives
\bea 
\frac{\dd}{\dd N} \simeq -\Mp^2 \frac{V_{,\phi}}{V}\frac{\dd}{\dd\phi} \, ,
\eea 
and hence the slow-roll parameters at leading order (LO) are given by 
\begin{align} 
\epsilon_0^{\mathrm{LO}} &\simeq H_\mathrm{in}\sqrt{\frac{3\Mp^2}{V}} \\
\epsilon_1^{\mathrm{LO}} &\simeq \frac{\Mp^2}{2}\left(\frac{V_{,\phi}}{V}\right)^2 \\
\epsilon_2^{\mathrm{LO}} &\simeq 2\Mp^2\left[ \left(\frac{V_{,\phi}}{V}\right)^2 - \frac{V_{,\phi\phi}}{V} \right] \\
\epsilon_3^{\mathrm{LO}} &\simeq \frac{2\Mp^4}{\epsilon_2^{\mathrm{SR}}}\left[ 2\left( \frac{V_{,\phi}}{V}\right)^4 - 3\frac{V_{,\phi\phi}V_{,\phi}^2}{V^3} +  \frac{V_{,\phi\phi\phi}V_{,\phi}}{V^2} \right] \, ,
\end{align}
and higher order parameters can continue to be calculated in the same way. 
We reiterate that these expressions are only valid in slow roll.
In this form, the first slow-roll parameter $\epsilon_1^{\mathrm{SR}}$ tells us that the potential of the inflaton needs to be sufficiently flat in order to support inflation, that is $\epsilon_1 \ll 1$ if 
\bea 
\frac{\dd V}{\dd\phi} \ll \frac{V}{\Mp} \, ,
\eea 
which hence provides a necessary (but not sufficient) condition for slow-roll inflation.
Beyond this required flatness, there is little known about the shape of the inflatons potential, and \textit{a priori} it can take a large range of different shapes. 
If one wants to derive the next-to-leading order expressions for the quantities given above, this can be done by noting the Friedmann equation can be written (exactly) as 
\bea 
H^2 = \frac{V(\phi)}{3\Mp^2\left(1-\frac{\epsilon_1}{3}\right)} \, .
\eea 
Combining this with \Eq{eq:friedmann:scalarfield}, gives us 
\bea 
\dot{\phi}^2 = 2V(\phi)\frac{\epsilon_1}{3 - \epsilon_1} \, ,
\eea 
and then we can write 
\bea 
\dd N = \pm \frac{1}{\Mp}\frac{\dd\phi}{\sqrt{2\epsilon_1}} \, .
\eea 
Using these relations, the next-to-leading-order in slow roll expressions can be obtained to be
\bea 
\epsilon_0^{\mathrm{NLO}} &= \epsilon_0^{\mathrm{LO}}\left( 1 - \frac{\epsilon_1^{\mathrm{LO}}}{6} \right) \\
\epsilon_1^{\mathrm{NLO}} &= \epsilon_1^{\mathrm{LO}}\left( 1 - \frac{\epsilon_2^{\mathrm{LO}}}{3}  \right) \\
\epsilon_2^{\mathrm{NLO}} &= \epsilon_2^{\mathrm{LO}}\left( 1 - \frac{\epsilon_2^{\mathrm{LO}}}{6} - \frac{\epsilon_3^{\mathrm{LO}}}{3} \right) \\
\epsilon_3^{\mathrm{NLO}} &= \epsilon_3^{\mathrm{LO}}\left(  1 - \frac{\epsilon_2^{\mathrm{LO}}}{3} - \frac{\epsilon_4^{\mathrm{LO}}}{3}  \right) \, ,
\eea 
and we note that we can continue to calculate more slow-roll parameters in the same way, and similarly we can calculate these parameters at higher and higher order.

\subsection{Inflationary perturbations} \label{sec:intro:infl:perturbations}

One of the huge successes of inflation is that, in addition to providing a solution for the classical Hot Big Bang problems, when combined with quantum mechanics it provides a natural explanation for the CMB anisotropies and the large-scale structure of the Universe. 
These deviations from homogeneity and isotropy arise from the vacuum quantum fluctuations of the coupled inflaton and gravitational fields, and are predicted to have an almost scale-invariant power spectrum, which matches observations \cite{Akrami:2018vks}. 
Since the slow-roll attractor of inflation is so strong, many models of inflation make the same prediction of an almost scale invariant power spectrum, and the deviations from scale invariance (\ie deviations from a massless field in de Sitter, where $H$ is approximately constant) probe the shape of the inflaton potential and characterise the deviations from flatness of the potential. 
As such, measurements of the CMB anisotropies allow us to constrain the inflationary potential $V(\phi)$.
In this section, we will discuss inflationary perturbations and demonstrate some key features of the predictions of (slow-roll) inflation. 
We will review the standard approach to inflationary perturbations here, while in Chapter \ref{chapter:stochastic:intro} we introduce the stochastic formalism for inflationary perturbations, which seeks to also include the non-perurbative backreaction effects of field fluctuations of the background equations. 

Beyond homogeneity and isotropy, we can expand the metric about the flat FLRW line element \eqref{eq:FLRWmetric:conformaltime}
\bea \label{eq:perturbedlineelement:FLRW}
\dd s^2 &= a^2(\eta)\left\{ -(1+2A)\dd \eta^2 + 2\partial_{i}B\dd x^{i}\dd \eta + \left[ (1-2\psi)\delta_{ij} + 2\partial_i \partial_j E \right]\dd x^i \dd x^j \right\} \, ,
\eea 
where $a$ is the scale factor, and $A$, $B$, $\psi$ and $E$ are scalar fluctuations. 
Here, $A$ is called the lapse function perturbation and represents a fluctuation in the proper time interval with respect to the coordinate time interval. 

By perturbing the Klein--Gordon equation and the Einstein equations according to \Eq{eq:perturbedlineelement:FLRW}, and rewriting the resultant equation in Fourier space ($\nabla^2 \to -k^2$), one finds the equation of motion for scalar perturbations in an FLRW metric.
At linear order, and for a given comoving wavenumber $k$, this is given by~\cite{Gordon:2000hv, Malik:2008im}
\bea \label{eq:pertubations:general}
\ddot{\delta\phi_{\bm{k}}} + 3H\dot{\delta\phi_{\bm{k}}} + \left(\frac{k^2}{a^2}+V_{,\phi\phi}\right)\delta\phi_{\bm{k}} = -2V_{,\phi}A_{\bm{k}} + \dot{\phi}\left[ \dot{A_{\bm{k}}} + 3\dot{\psi_{\bm{k}}} + \frac{k^2}{a^2}\left(a^2\dot{E_{\bm{k}}}-aB_{\bm{k}}\right)\right] .
\eea 

The metric perturbations that feature in the right-hand side of \Eq{eq:pertubations:general} satisfy the Einstein field equations \eqref{eq:einstein:fieldequation}, and in particular the energy and momentum constraints
\begin{align}
\label{eq:energyconstraint:arbgauge} 3H\left(\dot{\psi_{\bm{k}}}+HA_{\bm{k}}\right) + \frac{k^2}{a^2}\left[\psi_{\bm{k}} + H\left(a^2\dot{E_{\bm{k}}}-aB_{\bm{k}}\right)\right] &= -\frac{1}{2\Mp^2}\left[ \dot{\phi}\left(\dot{\delta\phi_{\bm{k}}}-\dot{\phi}A_{\bm{k}}\right)+V_{,\phi}\delta\phi_{\bm{k}}\right], \\
\label{eq:momentumconstraint:arbgauge} \dot\psi_{\bm{k}} + H A_{\bm{k}} &= \frac{\dot\phi}{2\Mp^2}  \delta\phi_{\bm{k}} \, ,
\end{align}
which come from the $00$ (energy) and $0i$ (momentum) components respectively.
Introducing the Sasaki--Mukhanov variable~\cite{Sasaki:1986hm, Mukhanov:1988jd}
\bea \label{eq:def:Q}
Q_{\bm{k}} = \delta\phi_{\bm{k}} + \frac{\dot{\phi}}{H}\psi_{\bm{k}} \, ,
\eea
and using \Eqs{eq:energyconstraint:arbgauge} and \eqref{eq:momentumconstraint:arbgauge} to eliminate the metric perturbations, \Eq{eq:pertubations:general} can be rewritten as 
\bea \label{eq:pertubations}
\ddot{Q}_{\bm{k}} + 3H\dot{Q}_{\bm{k}} + \left[ \frac{k^2}{a^2} + V_{,\phi\phi} - \frac{1}{a^3\Mp^2}\frac{\dd}{\dd t}\left( \frac{a^3}{H}\dot{\phi}^2 \right) \right]Q_{\bm{k}} = 0 \, .
\eea 

Note that \eqref{eq:pertubations} takes a simple and familiar form in the spatially flat gauge, which corresponds to the choice $\psi=0$, and in which case we define $v_{\bm{k}} = aQ_{\bm{k}}$, which has equation of motion
\bea \label{eq:vk:sasakimukhanov}
v_{\bm{k}}'' + \left( k^2 - \frac{z''}{z} \right) v_{\bm{k}} = 0 \, ,
\eea 
where a prime denotes a derivative with respect to conformal time $\eta$, and we have defined 
\bea \label{eq:def:z}
z = a \sqrt{2 \epsilon_{1}}\Mp \, .
\eea 
\Eq{eq:vk:sasakimukhanov} is known as the Sasaki--Mukhanov equation, and it is particularly useful because $v_{\bm{k}}$ is related to the gauge-invariant curvature perturbation $\zeta_{\bm{k}}$ (see \Eq{eq:def:zetacg}) through 
\bea \label{eq:def:v:zeta}
v_{\bm{k}} = z\zeta_{\bm{k}} \, ,
\eea 
with $z$ defined as above. 
This can easily be seen to take the form of a harmonic oscillator by defining 
\bea 
\omega^2(\eta, k) = k^2 - \frac{z''}{z} \, ,
\eea 
and hence \Eq{eq:vk:sasakimukhanov} is $v_{\bm{k}}'' + \omega^2(\eta, k) v_{\bm{k}} = 0$, \ie a harmonic oscillator with frequency $\omega$.
Let us note that in the case of constant $\epsilon_1$, we simply have $\omega^2(\eta, k) = k^2 - a''/a$. 
Since, during inflation, $a\propto \exp(Ht)$, this further simplifies to $\omega^2(\eta, k) = k^2 - 2/\eta^2$.
This allows us to describe the different behaviours of this variable at different times. 
At early times, $|\eta| \gg |k|$, and hence $\omega^2 \simeq k^2$, which means that the Sasaki--Mukhanov variable oscillates with constant frequency and $v_{\bm{k}} \propto \cos(k\eta)$. 
On the other hand, at late times ($|\eta| \ll |k|$) we have $\omega^2 \simeq - 2/\eta^2$, and hence
$v_{\bm{k}} \propto -1/\eta + \eta^2 \simeq -1/\eta$, as discussed in more detail below. 

One can show that, in full generality $z''/z=\calH^2(2-\epsilon_1+3\epsilon_2/2-\epsilon_1\epsilon_2/2+\epsilon_2^2/4+\epsilon_2\epsilon_3/2)$, where $\mathcal{H} = a'/a$ is the conformal Hubble parameter. 
For future use however, instead of working with the second and third slow-roll parameters, it will be more convenient to work with the field acceleration parameter
\bea
\label{eq:intro:def:f} f&=-\frac{\ddot{\phi}}{3H\dot{\phi}} = 1+\frac{1}{3H\dot{\phi}}V_{,\phi}
\eea
and the dimensionless mass parameter
\bea
\label{mu:def}
\mu &= \frac{V_{,\phi\phi}}{3H^2} \, ,
\eea
in terms of which
\bea \label{eq:z''overz:general}
\frac{z''}{z} &= \calH^2 \left( 2 + 5\epsilon_{1} - 3\mu - 12f\epsilon_{1} + 2\epsilon_{1}^2 \right) ,
\eea 
see  \App{appendix:MSequation}. 

\subsection*{Solution in the slow-roll limit}
\label{appendix:solving:MSequation}

At leading order in slow roll, the slow-roll parameters can simply be evaluated at the Hubble-crossing time $\eta_{*}\simeq -1/k_*$, since their time dependence is slow-roll suppressed, \ie $\epsilon_{1} = \epsilon_{1*} + \mathcal{O}(\epsilon^2)$, etc. At that order, we begin by writing 
\bea \label{eq:z:nu}
\frac{z''}{z} \equiv \frac{\nu^2 - \frac{1}{4}}{\eta^2} \, ,
\eea 
where $\nu^2 =9/4 + 3\epsilon_{1*} + 3\epsilon_{2*}/2$ can be taken as constant at leading order in slow roll, and we have used the fact that $\mathcal{H}\simeq -\frac{1}{\eta}\left( 1 + \epsilon_{1*} \right)$.
At that order, the Sasaki--Mukhanov equation has the generic solution
\bea \label{eq:v:generalsolution:slowroll}
v_{\bm{k}}\left(\eta\right) &= \sqrt{-\eta} \left[ A J_{\nu}\left(-k\eta \right) + BY_{\nu} \left( -k\eta \right) \right] \\
&= \sqrt{-\eta} \left[ \alpha H^{(1)}_{\nu}\left(-k\eta \right) + \beta H^{(2)}_{\nu} \left( -k\eta \right) \right] \, ,
\eea 
where we recall that conformal time $\eta$ runs from $-\infty$ to $0$ during inflation. 
In \Eq{eq:v:generalsolution:slowroll}, $J_{\nu}$ is the Bessel function of the first kind,  $Y_{\nu}$ is the Bessel function of the second kind, $A$, $B$, $\alpha$ and $\beta$ are constants, and the second line follows from 
\begin{align}
H_{\nu}^{(1)} &= J_{\nu} + iY_{\nu}\, , \\
H_{\nu}^{(2)} &= J_{\nu} - iY_{\nu} \, ,
\end{align}
where $H_{\nu}^{(1)}$ is the Hankel function of the first kind, and $H_{\nu}^{(2)}$ is the Hankel function of the second kind.

In order to fix the constants $\alpha$ and $\beta$, we need to set our initial conditions.
However, this is not a simple task, since the mode functions in \eqref{eq:vk:sasakimukhanov} have a time-dependent frequency and hence defining a vacuum state is difficult.
However, we avoid this problem by noticing that in the sub-Hubble (early time) limit we have $|k\eta| \gg 1 $ and we can neglect this time dependence and asymptotically define a ground state, known as the Bunch--Davies vacuum, which serves as our inital condition and is given by \cite{Mukhanov:2007zz}
\bea 
\lim_{\eta \to - \infty} v_{\bm{k}}(\eta) = \frac{\ee^{-ik\eta}}{\sqrt{2k}}  \, .
\eea 
We implement this initial condition by making use of the following asymptotic behaviour for the Hankel functions 
\bea 
\lim_{k\eta \to -\infty} H_{\nu}^{(1)}\left(-k\eta \right) &= \sqrt{\frac{2}{\pi}}\frac{1}{\sqrt{-k\eta}} \ee^{ik\eta} \ee^{i \frac{\pi}{2} \left(\nu + \frac{1}{2}\right)} \\
\lim_{k\eta \to -\infty} H_{\nu}^{(2)}\left(-k\eta \right) &= \sqrt{\frac{2}{\pi}}\frac{1}{\sqrt{-k\eta}} \ee^{- ik\eta} \ee^{-i \frac{\pi}{2} \left(\nu + \frac{1}{2}\right)} \, .
\eea 
Thus 
\bea 
\lim_{k\eta \to -\infty} v_{\bm{k}}\left(\eta\right) &= \sqrt{\frac{2}{\pi k}} \left[ \alpha \ee^{i \frac{\pi}{2} \left(\nu + \frac{1}{2}\right)} \ee^{ik\eta} + \beta \ee^{-i \frac{\pi}{2} \left(\nu + \frac{1}{2}\right)}\ee^{-ik\eta} \right] = \frac{\ee^{-ik\eta}}{\sqrt{2k}}  \, .
\eea 
By comparing the two expressions in this equation, we conclude that $\alpha = 0$ and $\beta = \frac{\sqrt{\pi}}{2}$ (where the irrelevant phase factor $\ee^{-i \frac{\pi}{2} \left(\nu + \frac{1}{2}\right)}$ is dropped). 
Thus the Bunch--Davies modes at first order in slow roll are
\bea 
v_{\bm{k}} \left( \eta \right) = \frac{\sqrt{-\pi\eta}}{2} H_{\nu}^{(2)}\left(-k\eta \right) \, .
\eea 

We now have the slow-roll result for $v_{\bm{k}}$, and can use it to find some interesting physical quantities.
For example, we can use the fact that $v_{\bm{k}}=z\zeta_{\bm{k}}$ to calculate the curvature perturbation $\zeta_{\bm{k}}$. 
For this, we need to find the behaviour of $z$ in the slow-roll regime. 
This is done by solving \eqref{eq:z:nu}, which has general solution
\bea 
z = C_{1} (-\eta)^{\frac{1}{2} - \nu} + C_{2}(-\eta)^{\frac{1}{2} + \nu} \, .
\eea
Now note that $\frac{1}{2} - \nu \approx -1$ and $\frac{1}{2} + \nu \approx 2$, so since $\eta$ is increasing from $-\infty$ to $0$, we have that $\eta^{-1}$ is the growing mode, and $\eta^2$ is the decaying mode, and hence in the late-time expansion and at leading order in slow roll we take 
\bea \label{eq:z:slowroll}
z \propto (-\eta)^{\frac{1}{2} - \nu} \, .
\eea
More specifically, we have 
\bea \label{eq:z:morespecific}
z = a_*\sqrt{2\epsilon_{1*}}\Mp\left( \frac{\eta}{\eta_*} \right)^{\frac{1}{2}-\nu} \, ,
\eea 
where $\eta_*$ is some reference time such that $z = a_*\sqrt{2\epsilon_{1*}}\Mp$ when $\eta=\eta_*$.
At late times, the second Hankel function behaves as
\bea \label{eq:hankel:superhorizon}
\lim_{-k\eta \to 0} H^{(2)}_{\nu} \left( - k\eta \right) = \frac{i}{\pi} \Gamma (\nu) \left( -\frac{k\eta}{2} \right)^{-\nu} \, ,
\eea 
where $\Gamma(\nu)$ in the Gamma function, and hence in this limit 
\bea \label{eq:vk:latetime}
v_{\bm{k}} \left( \eta \right) = \frac{i\Gamma(\nu)}{2}\sqrt{\frac{-\eta}{\pi}}\left( \frac{-k\eta}{2} \right)^{-\nu} \, .
\eea 
Then, in the super-horizon limit, we have
\bea 
\zeta_{\bm{k}} &= \frac{v_{\bm{k}}}{z} = \frac{i}{2\Mp}\left( \frac{k_*}{k} \right)^{\frac{3}{2}}\frac{(-\eta_*)^{\frac{1}{2}}}{a_*\sqrt{\epsilon_{1*}}} = \mathrm{constant} \, ,
\eea 
where we have also use $\eta_* = -1/k_*$.
This demonstrates that at late time in slow roll, the curvature perturbation $\zeta$ is constant on super-horizon scales. 

We can also calculate the power spectrum of curvature fluctuations, $P_{\zeta}$, using
\bea 
P_{\zeta} = z^{-2} P_{v} \, ,
\eea 
where we note that $P_{v} \equiv |v_{\bm{k}}|^2$. 
Using this, and \Eq{eq:z:morespecific}, we find
\bea 
P_{\zeta} = \frac{\pi}{4}\frac{(-\eta_*)^{1-2\nu}}{a_*^2\epsilon_{1*}} \frac{\left( -\eta \right)^{2\nu}}{2\Mp^2} \left| H^{(2)}_{\nu} \left( - k\eta \right) \right|^2 
\eea 
Then, in the super-horizon limit, $k\eta \to 0$, we use \eqref{eq:hankel:superhorizon} to find
\bea 
P_{\zeta} = \frac{\Gamma^2(\nu)}{2\Mp^2a_*^2 \epsilon_{1*}}\left( \frac{2}{k} \right)^{2\nu} \frac{(-\eta_*)^{1-2\nu}}{4\pi} \, ,
\eea 
and hence 
\bea 
\mathcal{P}_{\zeta} &\equiv \frac{k^3}{2\pi^2}P_{\zeta} &= \frac{\Gamma^2(\nu)}{16\pi^3\Mp^2} k^{3-2\nu}\frac{(-\eta_*)^{1-2\nu}2^{2\nu}}{a_*^2 \epsilon_{1*}} \, ,
\eea
which is noticeably independent of conformal time $\eta$, and hence on the large scales that this expression is valid on, curvature perturbations are constant, confirming what is claimed above. 
At leading order in slow roll (\ie $\nu=3/2$), this simplifies nicely to
\bea \label{eq:intro:powerspectrum}
\mathcal{P}_{\zeta} =  \left( \frac{H_*}{2\pi}\right)^2 \left( \frac{\dd N}{\dd \phi}\right)^2 \, ,
\eea
where each function is evaluated at horizon crossing ($k_*=a_*H_*$), and we have used the identity 
\bea 
2\Mp^2 \epsilon_1 = \frac{\dd \phi}{\dd N} \, .
\eea 
We usually evaluate the power spectrum around some pivot scale, where the pivot scale $k_*$ is usually taken to be
the scale best constrained by observations (see \cite{Aghanim:2018eyx}).
We can also calculate the spectral index $n_s$, which is defined as
\bea
n_{s} - 1 &= \frac{\dd \ln \mathcal{P}_{\zeta}}{\dd \ln k} &= 3 - 2\nu  \, ,
\eea 
from which we can see that if we take $\nu=\frac{3}{2}$ ($\epsilon_{1}=\epsilon_{2}=0$), we have exact scale invariance, and hence the slow-roll parameters also parameterise the amount of scale dependence one has in the power spectrum. 
Note also that, by definition, $\epsilon_1 \geq 0$, and inflation occurs for $\epsilon_1<1$, so if we take inflation to end at $\epsilon_1=1$ then it is natural to assume that $\epsilon_1 $ increases during inflation and thus $\epsilon_2>0$ in slow roll. 
Together, these conclusions predict, that, in slow roll $n_s <1$ is slightly ``red'', rather than exactly scale invariant or ``blue'' ($n_s>1$), which is consistent with current observations which exclude exact scale invariance at more than a $5\sigma$ confidence level, see \Sec{sec:inflationaryconstraints}.


Note that once the slow-roll approximation breaks down, the Sasaki--Mukhanov equation is hard to solve because one cannot take the slow-roll parameters to be approximately constant.
This is one of the reasons that the slow-roll limit is so well-studied. 
Further methods need to be developed in order to solve beyond slow-roll, see \App{app:usr:epscorrections} for a discussion on this.

As well as scalar perturbations, we can also consider vector and tensor perturbations to the spacetime metric during inflation.
While vector perturbations decay during inflation (due to the conservation of angular momentum), tensor perturbations can be studied in the same way as scalar perturbations. 
There is one subtlety that arises here, which is the fact that there are two separate helicities for tensor perturbations, denoted $h_+$ and $h_-$. 
The two helicities come from the fact that we impose two requirements, namely that tensor perturbations are transverse $\partial_i h_{ij} = 0$ and trace-free $h^i{}_i =0$.

Following the same reasoning as for scalar perturbations, we arrive at an analogous equation to \Eq{eq:vk:sasakimukhanov} for tensor perturbations, namely
\bea 
h_{\pm}'' + 2\mathcal{H} h_{\pm}' + k^2h_{\pm} = 0 \, .
\eea 
where $\mathcal{H} = a'/a$. 
By proceeding in the same way as we did for scalars, and allowing for the two helicities, we can compute the power spectrum of tensor perturbations to be 
\bea \label{eq:def:power:tensors}
\mathcal{P}_{h}(k) \simeq \frac{8}{\Mp^2}\left( \frac{H}{2\pi}\right)^2 \simeq \frac{2 V}{3 \pi^2\Mp^4} \, ,
\eea 
where functions here should be evaluated at the scale $k$, usually taken to be at horizon crossing $k_* = a_*H_*$.
Finally, we can define a parameter called the tensor-to-scalar ratio as 
\bea \label{eq:def:tensorscalarratio}
r = \frac{\mathcal{P}_{h}(k)}{\mathcal{P}_{\zeta}(k)} \simeq 16\epsilon_1 \, ,
\eea 
which can be constrained by cosmological data such as the CMB, and where the second equality is valid in slow roll.


%
\section{Modern problems in cosmology}

While we have outlined a well understood timeline for the evolution of the Universe above, there are many aspects of these events that are not well understood. 
As such, it is fair to conclude that our understanding of fundamental physics is incomplete. 
In this section, we will discuss several of these open problems of modern cosmology and explore their origin and possible explanations. 
The list of problems that we discuss is not an exhaustive list of all the questions that still exist, and examples of open problems that we do not discuss in detail here include the cosmological constant problem (why is the measured vacuum energy so different to the predicted value from theory?) \cite{Weinberg:1988cp}, and the origin of matter-antimatter asymmetry \cite{Sakharov:1967dj}.

%
\subsection{Inflationary constraints and model selection} \label{sec:inflationaryconstraints}

We have seen that inflation offers natural solutions to the classical problems of the Hot Big Bang model and also provides the seeds for the large-scale structure of the Universe, but in theory many possible models of inflation exist and using observations to select the physically relevant ones is an important open question in cosmology. 

\begin{figure}
\centering
\includegraphics[width=0.98\columnwidth]{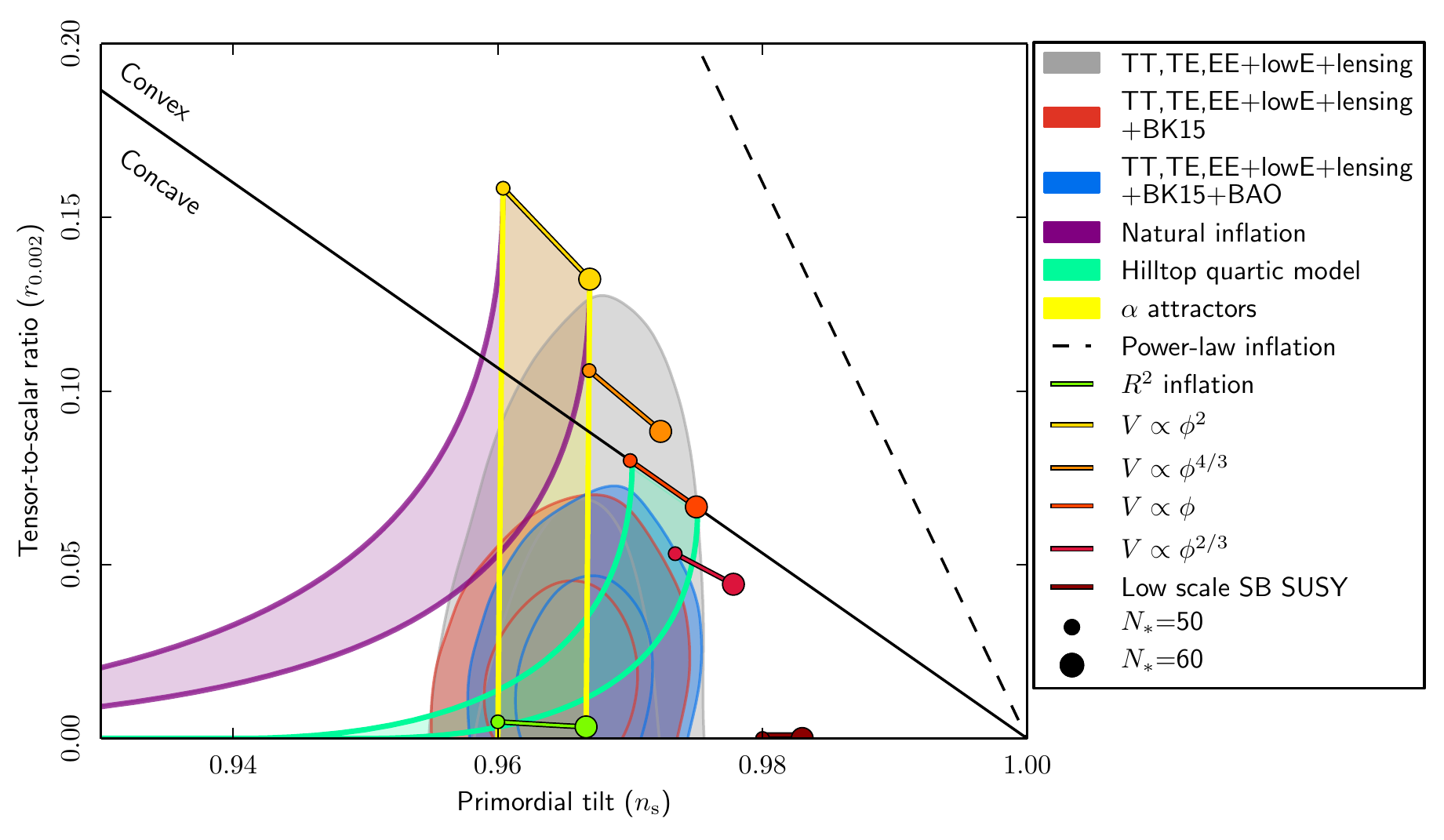}
\caption[Planck 2018 inflationary constraints]{Planck 2018 marginalised joint $68\%$ and $95\%$ CL regions (assuming $\dd n_\mathrm{s}/\dd \ln k = 0$) for the spectral index $n_\mathrm{s}$ and the tensor-to-scalar ratio $r$ at $k=0.002 \Mpc^{-1}$ from Planck alone and in combination with BK15 or BK15+BAO data, compared to the theoretical predictions of selected (slow-roll)  inflationary models. \textsc{Image credit: Planck collaboration \cite{Akrami:2018odb}}.
\label{fig:planckinflation}}
\end{figure}

As before, we assume inflation is driven by a single scalar field and discuss the current constraints on such inflationary models and which models are currently favoured by data. 
This discussion is mainly based on \Fig{fig:planckinflation}, which is taken from \cite{Akrami:2018odb} and uses the final data release of the Planck satellite in 2018. 

Planck observed the CMB in nine wavelength bands, ranging from $1 \cm$ to $\frac{1}{3}$mm \cite{Akrami:2018vks}, which corresponds to wavelengths from microwaves to the very-far-infrared. 
The wavelength range, along with with high sensitivity and small angular resolution allowed Planck to map the CMB in more detail than ever before. 
At the range of scales that the CMB is sensitive to, the Planck mission found no evidence of slow-roll violation for inflation \cite{Akrami:2018odb}, and hence the models discussed here are assumed to be in slow roll. 

The Planck observations for the spectral index $n_\mathrm{s}$ and the tensor-to-scalar ratio $r$ at $k=0.002 \Mpc^{-1}$ can be compared to theoretical prediction from potential models of inflation to see which models fit the data well. 
Assuming a $\Lambda CDM$ cosmology, combining Planck data with BAO data, the constraint on the spectral index is $n_\mathrm{s} = 0.966 \pm 0.0038$ at $68\%$ confidence level.
By also using $B$-mode polarisation data from BICEP2/Keck field (BK 15) \cite{Ade:2018gkx}, the tensor-to-scalar ratio is constrained to be $r_{0.002} < 0.056$ at $95\%$ confidence level.

To compare with this data, a selection of models are chosen, including several monomial potentials, $R^2$ inflation \cite{Starobinsky:1980te}, $\alpha$ attractors \cite{Kallosh:2013yoa, Ferrara:2013rsa}, and natural inflation \cite{Freese:1990rb, Adams:1992bn}, and $n_\mathrm{s}$ and $r$ are calculated in these models.
These calculations are each done at first order in slow roll, are performed at a scale $k = 0.002 \Mpc^{-1}$, and include an uncertainty in the number of \efolds~of $50 < N_* < 60$.
The comparison between theory and data is shown in \Fig{fig:planckinflation}, and we make some comments about the conclusions here. 

First of all, we see that monomial potentials of the form $V(\phi) \propto \phi^p$ are strongly ruled out for $p>2$, and once $B$-mode data is included even potentials with $1<p<2$ are disfavoured compared to, say, the $R^2$ model of inflation. 
Out of the models considered here, $R^2$ inflation is the best fit to the data, while $\alpha$ attractors can also fit the data well, but can do so because of an additional degree of freedom in the model, compared to $R^2$ inflation. 
While natural inflation is now disfavoured by the Planck 2018 data, the Hilltop quartic model \cite{Boubekeur:2005zm} provides a very good fit for a large swathe of its parameter space. 

However, while this data reduces the number of models that are physically viable, there are still many models that are consistent with the data. 
Indeed, models such as $\alpha$ attractors can tune their parameters so that they can satisfy almost all current and future observations. 
This means that improved data will not rule out models like this, but can be used to constrain the values of their parameters. 
While more complicated Bayesian analysis for model selection in light of the Planck data is possible (see for example \cite{Martin:2013nzq}), it is unlikely that any future data will be able to select a unique, viable model for inflation, although improved data will indeed continue to rule out models \cite{Martin:2014rqa} and shrink the parameter space of viable models.

\subsection{Dark energy}

\begin{figure}
\centering
\includegraphics[width=0.49\columnwidth]{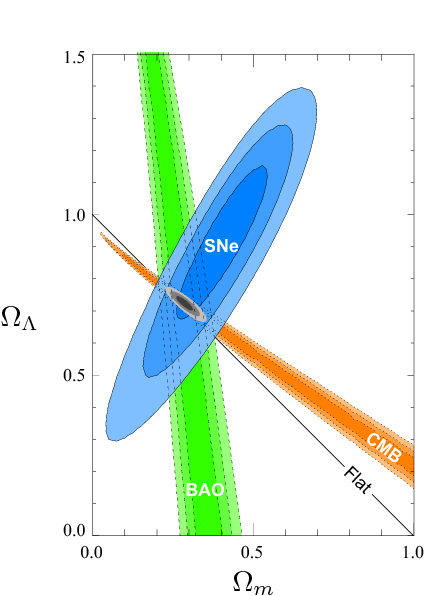}
\caption[Evidence for dark energy]{Evidence for dark energy. Shown are a combination of observations of the cosmic microwave background (CMB), supernovae (SNe) and baryon acoustic oscillations (BAO). \textsc{ Image credit: Supernova Cosmology Project \cite{Kowalski:2008ez}}. 
\label{fig:darkenergyevidence}}
\end{figure}

With a plethora of evidence for a dark energy component (see, for example, \cite{Caldwell:2009ix}) of the Universe that dominates at late (current) times, it is now widely accepted that this component exists, although the nature of dark energy remains elusive. 
The leading paradigm for dark energy is that of a cosmological constant $\Lambda$ (see \cite{Martin:2012bt} for a review on the cosmological constant), \ie a perfect fluid with an equation of state $w=-1$.
The evidence for dark energy is summarised in \Fig{fig:darkenergyevidence}, which shows a combination of evidence from CMB, supernovae (SNe) and baryon acoustic oscillations (BAO), and their agreement in the $\Omega_\Lambda-\Omega_m$ plane.
As stated previously, we see evidence for $\Omega_\Lambda \simeq 0.69$ \cite{Aghanim:2018eyx}, and conclude that dark energy is the dominant component of the Universe today. 

\begin{figure}
\centering
\includegraphics[width=0.49\columnwidth]{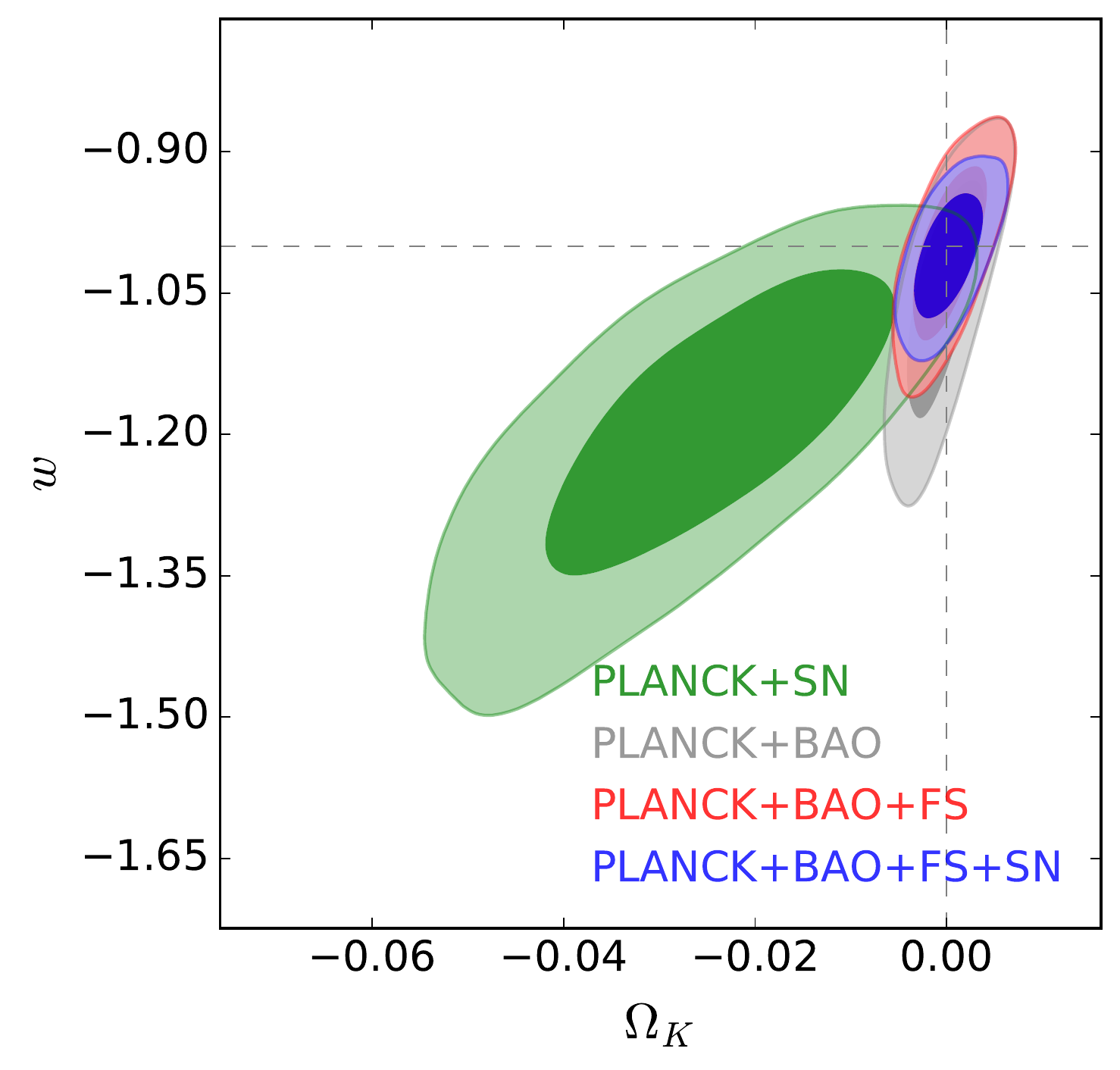}
\includegraphics[width=0.49\columnwidth]{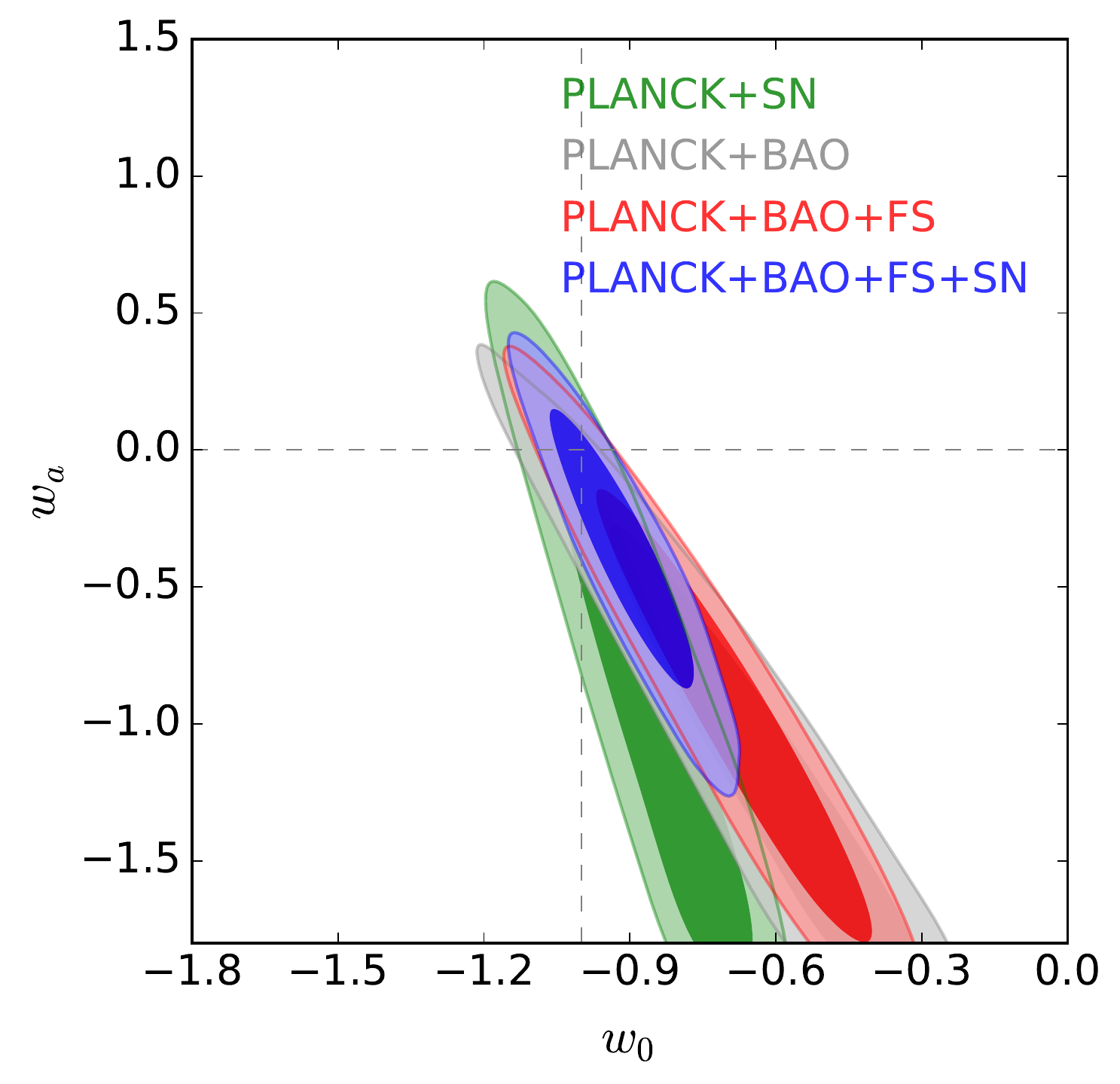}
\caption[Evidence for zero curvature and a cosmological constant]{Evidence for zero curvature and a cosmological constant. Parameter constraints for the owCDM (left) and w0waCDM (right) cosmological models, comparing the results from BAO and BAO+FS to those with JLA SNe. 
\textsc{Image credit: BOSS \cite{Alam:2016hwk}}. \label{fig:curvature}}
\end{figure}

Beyond the cosmological constant, a popular alternative is to assume a dark energy fluid with an equation of state that varies with time.
For example, one possible parameterisation is to allow $w$ to vary as a function of the scale factor $a$ as 
\bea \label{eq:darkenergy:parameterisation}
w(a) = w_0 + w_a(1-a) \, ,
\eea 
where $w_0$ and $w_a$ are constants.
Note that a cosmological constant would correspond to $w_0= -1$ and $w_a=0$.
In \Fig{fig:curvature}, we show results from the Baryon Oscillation Spectroscopic Survey (BOSS), part of the Sloan Digital Sky Survey III, which in the left panel constrains the curvature and a non-varying dark energy equation of state to be consistent with $0$ and $-1$ respectively, and in the right panel constrains the extended model \eqref{eq:darkenergy:parameterisation} for a varying equation of state and finds consistency with a cosmological constant. 

To demonstrate the similarities between dark energy and inflation, let us note that late-time accelerated expansion can be achieved with a scalar field, which we call the ``quintessence field" $Q$, which behaves just like the inflaton except that it dominates at late times rather than early times. 
In this case, the equation of state for the quintessence field is given by 
\bea \label{eq:equationofstate:quintessence}
w_Q = \frac{\frac{1}{2}\dot{Q}^2 - V(Q)}{\frac{1}{2}\dot{Q}^2 + V(Q)} \, ,
\eea 
and so has an equation of state that varies in the same way as the inflaton, see \Eq{eq:equationofstate:phi}. 
This explanation for dark energy differs from the cosmological constant because the scalar field is dynamical over time, while a constant $\Lambda$, by definition, does not. 


These different models of dark energy may be constrained through improved measurements of the equation of state $w$ and its evolution (or lack thereof) and how this affects observations such as the CMB and the matter power spectrum, or through observations the dark energy speed of sound, although this is currently much less constrained \cite{Linton:2017ged}. 
Let us also note that the late time acceleration of the Universe may be due to a large-scale modification to gravity that we do not yet understand \cite{Clifton:2011jh,Joyce:2016vqv,Wright:2019qhf, Kenna-Allison:2019tbu}.
Currently, there is not sufficient data to rule out or isolate a best candidate for dark energy, meaning that we currently lack a fundamental understanding of the dominant constituent of our Universe.
As such, there is much research activity on dark energy at this time and its nature remains an open question \cite{Copeland:2006wr}.

\subsection{Dark matter} \label{sec:darkmatter}

There is a significant amount of evidence \cite{Freese:2008cz} that approximately $80\%$ of the matter in the Universe is contained in non-baryonic matter that does not interact via electromagnetism, but does interact with gravity, making it very hard to detect. 
We use the term ``dark matter'' to refer to this hypothetical matter, and additionally, classify dark matter as ``cold'' or ``hot'' depending on its typical velocity, with cold dark matter (which moves with non-relativistic velocity) currently favoured by observations, as the dark matter must cluster to form the large scale structure of the universe. 

The need for dark matter was first noticed in the $1930$'s with the work of Lundmark and Zwicky \cite{2009GReGr..41..207E}, who noted that galaxies in the Coma cluster were moving too quickly to be explained by the visible matter in the cluster, leading to the hypothesis that something massive and dark must be providing additional gravitational pull. 
Later, in the $1970$'s, Rubin et al \cite{Rubin:1970zza} found more evidence for dark matter when studying galactic rotation curves, leading to the dark matter paradigm becoming widely accepted. 
For a more detailed review of the history of dark matter, see Ref. \cite{Bertone:2016nfn}.

In this section, we outline some of the evidence for dark matter, and then discuss some of the leading candidates for this elusive constituent of the Universe. 

\subsubsection{Evidence for dark matter:}

We will first discuss some of the evidence for dark matter. 
We include, arguably, some of the most convincing evidence for dark matter, but note that more evidence exists for dark matter, for example Lyman-$\alpha$ forest observations \cite{2009MNRAS.399L..39V}. 

\begin{figure}
\centering
\includegraphics[width=0.49\columnwidth]{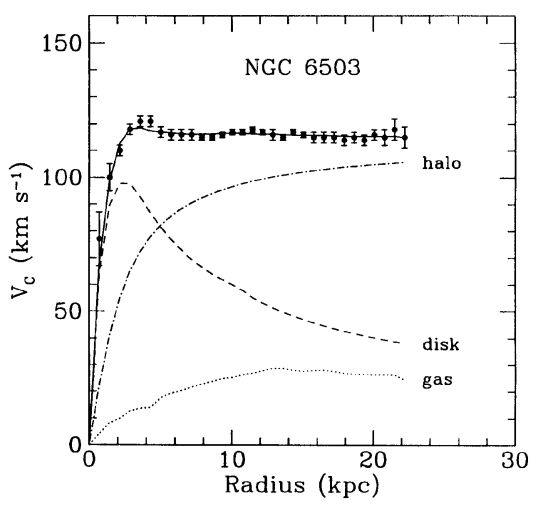}
\caption[Galaxy rotation curve for NGC $6503$]{Galactic rotation curve for NGC $6503$, showing the disk and gas contribution, as well as an additional dark matter halo contribution that is needed to match the data. \textsc{Image credit: Katherine Freese \cite{Freese:2017idy}}
\label{fig:flatrotationcurve}}
\end{figure}

\textbf{Galaxy rotation curves:}
This piece of evidence refers to measuring the rotation velocities of visible stars and gas within disk galaxies (elliptical galaxies have random motion within them) as a function of distance from the centre of the galaxy. 
For a spiral galaxy, the density of visible matter decreases with radius from the centre of the galaxy, and hence the usual dynamics of Kepler's Second Law (\ie Newtonian dynamics) predict the velocity of stars further from the centre to orbit the galaxy slower than those in the centre where gravity should be stronger due to the higher density of mass. 
This is what we observe in systems such as our solar system.
However, rather than decreasing in this way, rotation curves of galaxies are observed to be flat, out to large radii from the galaxy centre \cite{Corbelli:1999af}. 
These flat rotation curves have been seen in all galaxies that have been studied, including the Milky Way \cite{Bhattacharjee:2013exa}, and the visible stars and gas observed in these galaxies cannot provide the force to speed up these orbits sufficiently.
If one postulates that these galaxies contain large amounts of additional, unseen (dark) matter, then these rotation curves can be explained. 
This fit is demonstrated in \Fig{fig:flatrotationcurve}, where the galactic rotation curve for NGC $6503$ is shown. 
The matter of the stars and gas contained in the disk alone cannot fit the observed velocity, but the existence as a large ``dark matter halo'', making up over $95\%$ of the mass of the galaxy provides the gravity to explain these rotation curves. 

Note that rotation curves can only be observed as far out as there is visible matter or neutral hydrogen, which does not allow us to trace the full extent of the dark matter halo.
This limitation is not shared with gravitational lensing probes of dark matter discussed next. 

\textbf{Gravitational lensing:}
As a consequence of GR, massive objects such as galaxy clusters can warp spacetime and act as a lens for distant objects, \ie a lens between an observer and, say, a distant quasar will warp the image of the quasar seen by the observer. 
The more massive the lens, the more extreme the warping is, so observing more extreme lensing events means the lens is more massive.
This phenomenon can be used to probe the existence and distribution of mass in the Universe, even if that mass is ``dark'', and many lensing observations confirm the existence of dark matter, both in galaxies and clusters of galaxies.

For example, strong lensing observations of the Abell 1689 cluster \cite{Tyson:1998vp} provide mass measurements for the cluster, which are again much higher than the mass in visible matter in the cluster. 
Weak lensing can also be used map underlying dark matter halos in the Universe, and for instance SDSS used weak lensing to identify the fact that galaxies (including the Milky Way) are much more massive than the visible light suggests, leading to the requirement of a large dark matter halo for each galaxy \cite{AdelmanMcCarthy:2005se}.

\begin{figure}
\centering
\includegraphics[width=0.98\columnwidth]{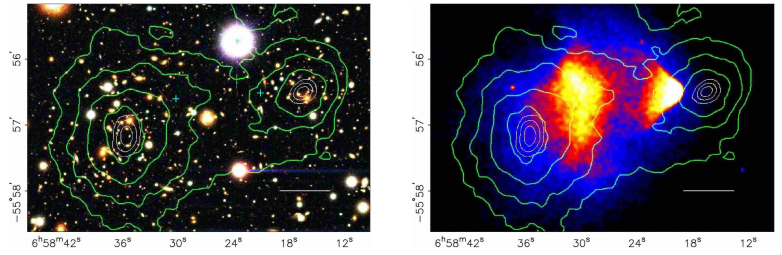}
\caption[Mass distribution for the Bullet Cluster]{Left panel: optical images from the Magellan telescope with contours of spatial distribution of mass, from gravitational lensing. Right panel: the same contours plotted over Chandra x-ray data that traces hot plasma in the galaxy. We see that most of the matter in not found in the visible plasma, which felt the friction of the collision and slowed down, but is in a dark matter halo further out which feels gravity but not electromagnetic friction and hence traveled further in the merger. \textsc{Image credit:\cite{Clowe:2006eq}} 
\label{fig:bulletcluster}}
\end{figure}

\textbf{Bullet Cluster:}
The Bullet Cluster \cite{Clowe:2006eq} is a cluster of galaxies formed from the collision of two smaller galaxy clusters. 
Gravitational lensing of background galaxies of the Bullet Cluster allow us to map the mass of the Bullet Cluster, which can then be compared to electromagnetic observations of the visible matter in the cluster. 
In \Fig{fig:bulletcluster}, we can see that the majority of the mass in the cluster is not located where the bulk of the baryons are observed, providing evidence of two distinct types of matter - baryons and dark matter - that behave differently in the collision. 
The baryons interact both electromagnetically and gravitationally, and feel a ``friction'' in the collision that slows them down and causes them to cluster more easily in the centre. 
However, the dark matter feels no friction and moves through the collision more easily, only being bound by gravity, and forms a much larger halo. 

\textbf{CMB:}
There is also evidence for dark matter in the anisotropies of the CMB.
Since dark matter and baryons behave differently (even though they are both matter), the CMB is sensitive to both of these types of matter in different ways. 
In the early Universe, baryons are ionised and interact strongly with radiation (through Thompson scattering), while dark matter is neutral, and so only interacts gravitationally (and via its effect on the density and velocity of baryons) with the CMB. 
Hence, the perturbations of dark matter and perturbations of baryons evolve differently and leave different imprints on the CMB. 

\begin{figure}
\centering
\includegraphics[width=0.98\columnwidth]{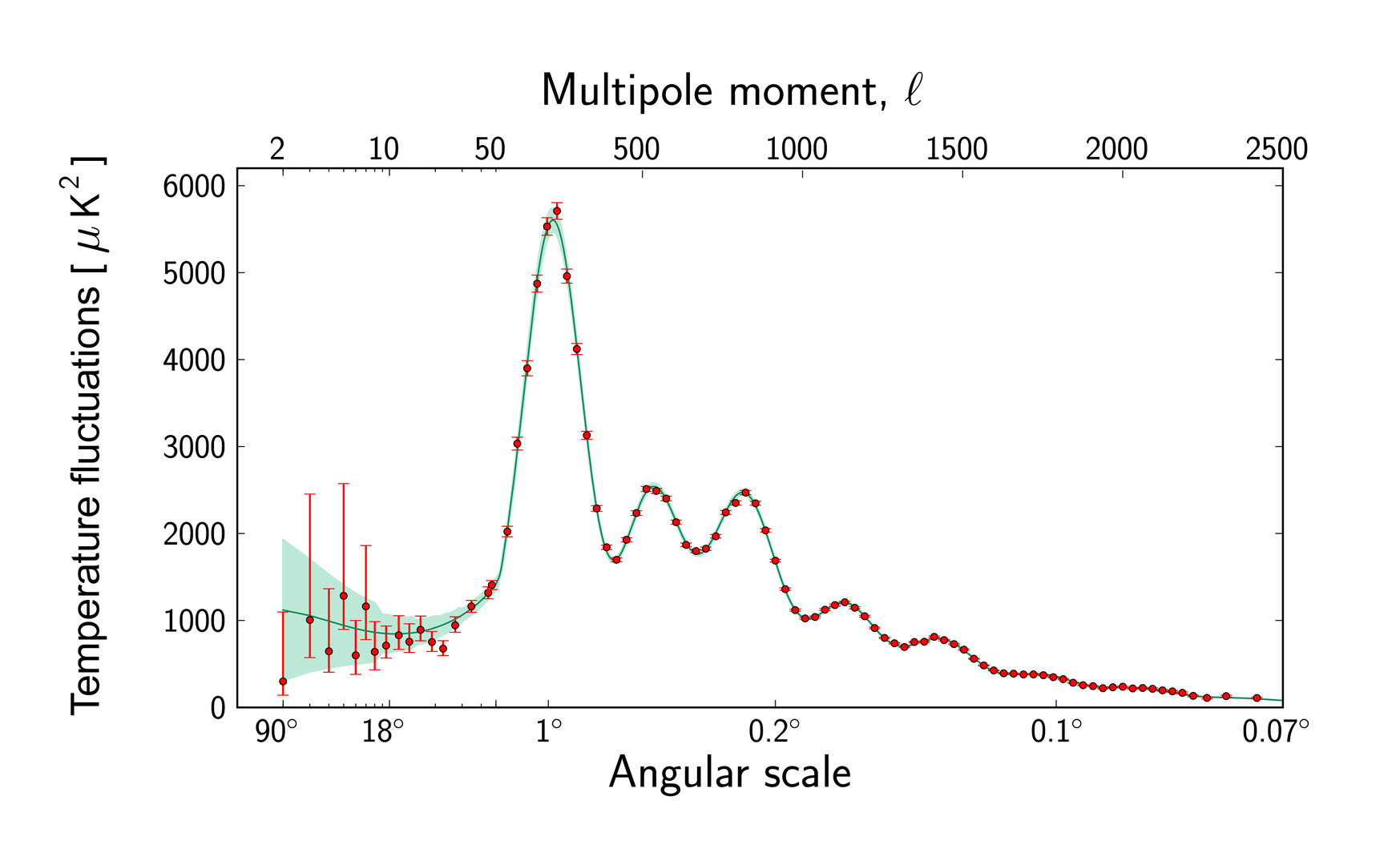}
\caption[Planck power spectrum]{Planck's power spectrum of temperature fluctuations in the cosmic microwave background. The fluctuations are shown at different angular scales on the sky. Red dots with error bars are the Planck data. The green curve represents the standard model of cosmology, $\Lambda$CDM. The peak at 1 degree is consistent with a flat geometry of the universe, the height of the second peak with $5\%$, and the second and third peaks with $26\%$ dark matter. 
\textsc{Image credit: ESA and the Planck Collaboration\cite{Aghanim:2019ame}}
\label{fig:planckpowerspectrum}}
\end{figure}

The photons of the CMB underwent oscillations that ``froze in'' at the time of decoupling, and the angular scale and heights of the peaks and troughs of these oscillations provide a powerful probe of cosmological parameters, including the total energy density, the baryonic fraction, and the dark matter component, as shown in \Fig{fig:planckpowerspectrum}. 
We can use the sound horizon to probe the geometry of the Universe. 
If the Universe is spatially flat, then the angular scale of the first Doppler peak is predicted to be found at $1\deg$ \cite{Freese:2017idy}, which is precisely what we see in the power spectrum in \Fig{fig:planckpowerspectrum}, suggesting that the Universe is indeed flat.
The height of the second peak of the power spectrum corresponds to ordinary baryonic matter making up $\sim 5\%$ of the energy density of the Universe, while the third peak tells us that $\sim 26\%$ of the energy density is in dark matter. 
It is difficult to reproduce this evidence for dark matter with competing explanations, such as modified Newtonian dynamics (MOND) \cite{Skordis:2005xk}. 
Additionally, a few small galaxies have been observed that contain \textit{no} dark matter \cite{vanDokkum:2018vup}, which would not be possible if dark matter came from modifications to gravity

Finally, let us note that without dark matter, the large scale structure of the Universe could not have formed by the present time (and we would fail to exist).
Before recombination, the baryons were coupled to photons (the Universe is ionised) and so both photons and baryons stream away and stop structure forming. 
However, the dark matter is not coupled to radiation and can hence collapse and begin to form structure. 
This provides potential wells that baryons can later fall into after recombination and form the structure we see today. 
This also provides evidence that the dark matter is cold, rather than hot. 
Relativistic dark matter would stream in the same way that photons did, and hence non-relativistic cold dark matter has become the accepted paradigm. 

\subsubsection{Candidates for dark matter:}

While there is much evidence for a cold dark matter component to the Universe, we know about the nature of dark matter. 
While a breakdown of our theories of gravity on galactic scales is a possible explanation for the observations of dark matter, here we will mention a few of the leading candidates for ``particle''  dark matter, although we note that this is not an exhaustive list. 

\textbf{WIMPS:} Weakly interacting massive particles (WIMPs) are a popular candidate for cold dark matter. 
A WIMP is a hypothetical elementary particle that only interacts via gravity and the weak nuclear force (and possibly any other unknown force beyond the standard model that is as weak, or weaker, than the weak force).
WIMPs in the mass range of $\sim 100$ $\GeV$ can offer the correct dark matter abundance that is observed today \cite{Jungman:1995df,Steigman:2012nb}, and the ``WIMP miracle'' refers to the fact that the supersymmetry (SUSY) extension of the standard model of particle physics predicts a particle in this mass range that only interacts via the weak force (and gravity).
Thus, a stable SUSY particle is a long-proposed WIMP candidate for dark matter. 

While the WIMP is a popular candidate for dark matter, both indirect and direct searches for WIMPs have failed to have any positive detections, and the Large Hadron Collider (LHC) has also failed to find evidence for SUSY. 
Indirect searches for WIMPs, such as the Fermi-LAT gamma ray telescope \cite{Ackermann:2013yva}, look for excess gamma rays that can be produced when WIMPs interact with themselves. 
Direct detectors attempt to observe the effects of a WIMP collision with a heavy nucleus in a sensitive system set up on Earth. 
For example, noble gas scintillators produce a pulse of light when the nucleus of an atom interacts with a WIMP. 
The XENON1T detector \cite{Aprile:2017iyp} uses $3.5$ tons of liquid xenon (even larger experiments are planned for the future), although no detections of dark matter WIMPs have yet occured. 

\textbf{Axions:}
The axion is a hypothetical elementary field (and hence, particle) which was proposed by Peccei and Quinn in $1977$ \cite{1977PhRvL..38.1440P,1977PhRvD..16.1791P} as a solution to the ``strong CP problem'' of quantum chromodynamics (QCD).
An axion with a sufficiently low mass offers a suitable candidate for cold dark matter, as they may have been produced in large numbers in the early Universe. 
A comprehensive review of axions can be found in Ref. \cite{DiLuzio:2020wdo}.
The strong CP problem \cite{Cheng:1987gp,Peccei:2006as,Kim:2008hd} refers to the apparent discrepancy between theory and observation of QCD phenomena. 
The theory of QCD allows for charge-parity (CP) violation, \ie the equations governing the strong nuclear force are not necessarily invariant under change and parity inversion\footnote{Note that CP violation is predicted and observed in the weak nuclear force.}.
However, observations have not yet found any evidence for this CP violation existing within QCD. 
For example, if CP violation occurs in QCD, then the neutron should exhibit an electric dipole field, but no such electric field has yet been observed (see, for example, \cite{Pendlebury:1984zz}), suggesting that the neutron does not have this field (or that is is at least $10^9$ times weaker than a CP violating QCD theory predicts) and hence QCD does not appear to violate CP symmetry. 

The axion arises from a proposed solution of the strong CP problem. 
The Lagrangian for QCD can be written as 
\bea 
\mathcal{L}_\mathrm{QCD} = \mathcal{L}_\mathrm{inv} - \theta \mathcal{L}_\mathrm{vio} \, ,
\eea 
where $\mathcal{L}_\mathrm{inv}$ is a Langrangian contribution that is invariant under CP symmetry, and $\mathcal{L}_\mathrm{vio}$ is a contribution that violates CP symmetry\footnote{This term appears in the Lagrangian because QCD does not have a unique vacuum, but instead has infinitely many lowest energy states.}, and $\theta$ is a new physical parameter. 
Thus, one solution to the strong CP problem would be if $\theta \simeq 0$, which would be the case if any of the standard model quarks were massless (which is contradicted by many measurements of the quark masses \cite{2014ChPhC..38i0001O}), or if $\theta$ is fine-tuned to $0$.
A more ``natural'' solution, suggested by Peccei and Quinn, is to promote $\theta $ to a new dynamical field, which over time reduces in value in order to minimise the energy of the vacuum. 
The new $\theta$ field can then be quantised to give a particle, named the axion, which turns out to have no spin, no electric charge and have a mass of $\sim 10^{-11} M_e$, where $M_e$ is the mass of the electron. 

\begin{figure}
\centering
\includegraphics[width=0.98\columnwidth]{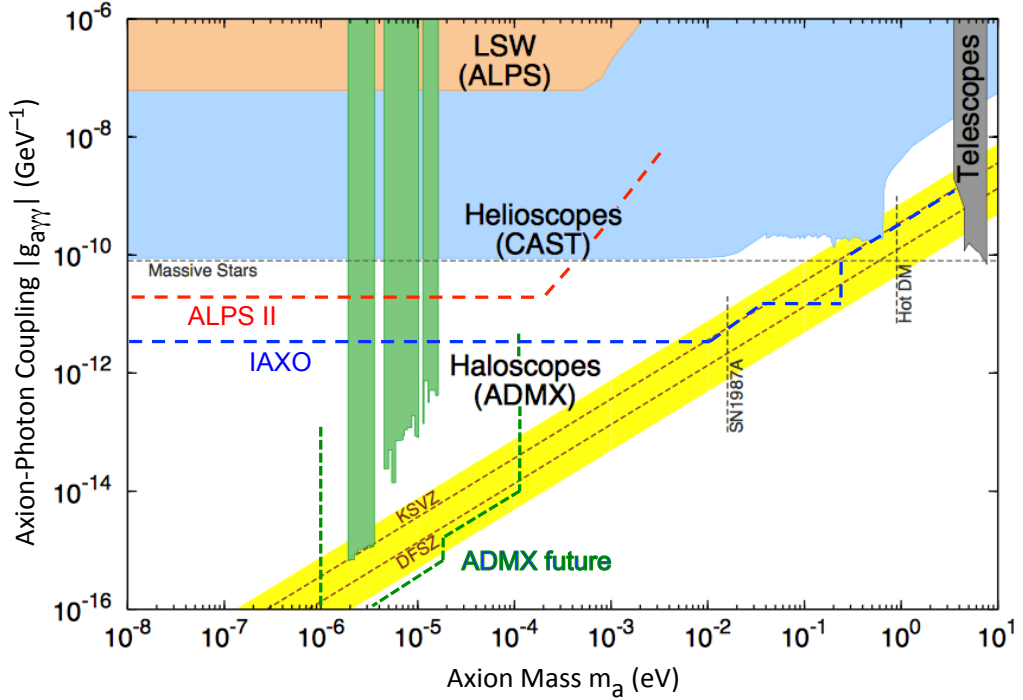}
\caption[Constraints on properties of the axion]{Existing limits on the photon coupling of axions and axion-like particles and the projected coverage of ongoing upgrades for these experiments. 

\textsc{Image credit: \cite{Graham:2015ouw}}
\label{fig:axionconstraints}}
\end{figure}

Axions also only interact very weakly through the strong and weak forces, and through gravity, making detection difficult. 
Many experiments, such as the CERN Axion Solar Telescope (CAST) \cite{Aalseth:2002qf}, are currently searching for axions by searching for X-rays that may be produced when axions pass through strong magnetic fields. 
While no detections have yet been made, this may mean that axions are simply lighter or more weakly interacting than expected (see \Fig{fig:axionconstraints} for recent constraints on axion properties), and axions remain a popular dark matter candidate, as well as being the leading solution to the strong CP problem. 

\textbf{MaCHOs, including primordial black holes:}
Finally, we will discuss massive compact halo objects (MaCHOs) as a candidate for dark matter. 
MaCHOS include objects such as brown dwarfs, neutron stars and primordial black holes (PBHs), the last of which we will focus on here. 
As a dark matter candidate, MaCHOs are appealing because they do not require any extension to the standard model of particle physics, as they are typically made of the known constituents of the Universe (baryons, or collapsed particles in the case of PBHs) just in a form that is very difficult to detect. 
Section \ref{sec:pbhs} focuses on PBHs, and therefore a full discussion of these objects will be performed there.

\subsection{Origin of supermassive black holes}

\begin{figure}
    \centering 
    \includegraphics[width=0.9\linewidth]{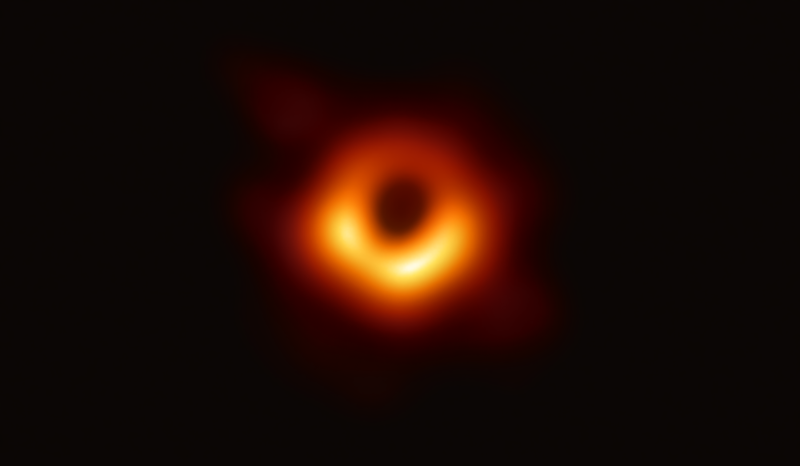}
    \caption[Event Horizon Telescope image of M87 black hole]{The first direct image of a black hole, taken by the Event Horizon Telescope in 2019. The image shows light bending around the supermassive black hole at the centre of the galaxy M87, which is $6.5$ billion times the mass of the Sun. 
    \textsc{Image credit: Event Horizon Telescope Collaboration} \cite{Akiyama:2019cqa}}
    \label{fig:eventhorizon}
\end{figure}

A supermassive black hole (SMBH) is an especially large black hole, with masses typically ranging from $10^{6} M_\odot$ to $10^{10} M_\odot$.
The centre of every massive galaxy is expected to contain a supermassive black hole, and the existence of such black holes has been confirmed through the direct imaging of the supermassive black holes in the M87 galaxy, see \Fig{fig:eventhorizon}, and through observations of the stellar orbits around the Milky Way's own supermassive black hole, Sagittarius A* \cite{doi:10.1146/annurev.astro.39.1.309}.
The most massive known SMBH is named TON $618$ and has an approximate mass of $6.6\times10^{10}M_\odot$ \cite{Shemmer_2004}. 
Accretion of gas onto supermassive black holes powers quasars, objects so massive and bright that they can outshine entire galaxies. 

While the existence of SMBHs is known, the origin of such massive objects is still an open question \cite{Li:2006ti,Pelupessy:2007mt}. 
The earliest black holes to form through stellar collapse can form from the death of populations III stars, which may be up to a few hundred solar masses, and can form by approximately redshift $15$ \cite{NaokiYOSHIDA2019PJA9501B-02}. 
It is thought that these first stellar black holes can then accrete matter and merge with one another throughout the rest of the age of the Universe, eventually becoming the SMBHs we see today. 
However, there are issues with this formation route for SMBHs. 
In particular, quasars have been observed at high redshifts\footnote{Currently, the most distant known quasar is named ULAS J1342+0928 \cite{Banados:2017unc}, and has a mass of $\sim 800\times10^6 M_\odot$. It is at redshift $z=7.54$, when the Universe was approximately $690$ million years old.} $z>7$, and stellar black holes would not have had time to grow to these masses at such high redshifts in this way.
This is because there is a limit to how fast a body can accrete matter, known as the Eddington limit.
This is because the in-falling matter heats up and produces an outward flow of radiation, and if this radiation becomes too powerful it can carry the material surrounding the black hole away, shutting down accretion.

Note that there are other possible formation mechanisms for SMBHs, such as the collapse of ``quasi-stars'' \cite{Begelman:2006db} in pregalactic halos, or the collapse of ``dark stars'' \cite{Spolyar:2007qv,Freese:2008ct} at reshifts $z>10$. 
Both of these methods can produce black holes at an earlier time than the collapse of population III stars, which can then accrete matter to become SMBHs.

An alternative route to forming SMBH is through PBHs formed in the very early Universe. 
This idea has two advantages over forming SMBH through stellar collapse and growth through subsequent accretion and mergers. 
Firstly, PBHs can form at larger masses than black holes formed from the death of stars, and for example PBHs formed around $1\sec$ after the Big Bang may have a mass of up to $\sim 10^5 M_\odot$ \cite{Garcia-Bellido:2017fdg}. 
Secondly, PBHs have a longer time to grow to supermassive sizes, even at high redshifts, making the Eddington limit less impactful on their status as potential seed for SMBHs. 
This is a second open problem in cosmology that PBHs may offer a solution to, and we study these objects in more detail in the following section.

\section{Primordial black holes}
\label{sec:pbhs}
%


\begin{figure}
    \centering 
    \includegraphics[width=0.95\linewidth]{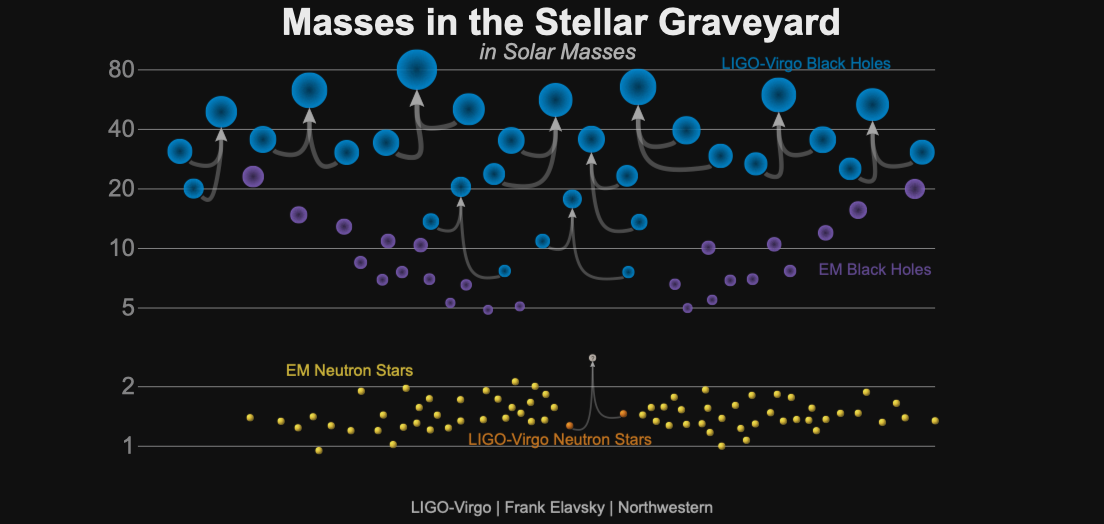}
    \caption[Known black holes in the Universe]{Known black holes in the Universe from the first and second observing runs of LIGO, plus black holes discovered through X-ray observations, and known neutron stars in the Universe. 
    \textsc{Image credit
    : LIGO/Frank Elavsky/Northwestern}}
    \label{fig:blackholes:LIGO}
\end{figure}

In this section, we will discuss primordial black holes (PBHs), and motivate why these objects are currently the subject of much research.
There is compelling evidence that black holes exist, following the LIGO/Virgo detections of gravitational waves from merging black holes (see \Fig{fig:blackholes:LIGO}), and the direct imaging of the event horizon of the supermassive black hole in the centre of the M87 galaxy (see \Fig{fig:eventhorizon}) .
While black holes in the modern Universe form through the supernovae and subsequent collapse of massive stars, the early Universe was much denser than the Universe now and so called PBHs could form from causal patches of radiation exceeding a critical curvature, or density, relative to the rest of the Universe.
This forms a black hole with roughly the mass of the causal horizon that collapsed \cite{Garcia-Bellido:2017fdg}, and the resultant PBH is expected to form in isolation with no accretion disk or associated system. 
In the standard picture, which we adopt here, this collapse happens in the first few seconds after inflation and during the radiation dominated period of the Universe's history\footnote{It is possible that an early matter dominated period in the Universe's history may affect the formation of PBHs (see, eg, \cite{Harada:2016mhb}), but here we only consider PBHs forming in the radiation era. }. 
The curvature fluctuations that seed these PBHs collapse come from vacuum quantum fluctuations during inflation.

PBHs offer possible solutions to some of the modern problems in cosmology discussed above, namely they offer a plausible candidate to make up some or all of the dark matter\footnote{Note, however, that recent work suggests that the existence of PBHs may be inconsistent with the existence of WIMPs \cite{Adamek:2019gns}.} (see, for example, \cite{Carr:2016drx,Green:2016xgy,Clesse:2017bsw}), and also offer candidates to seed the SMBHs observed at high redshifts. 
In addition to this, PBHs may provide some the black holes that LIGO has observed \cite{Bird:2016dcv,Sasaki:2016jop} and the formation and merger of PBHs could also provide some of the stochastic gravitational waves background that LISA will observe. 

During the inflationary epoch, vacuum quantum fluctuations were amplified to become large-scale cosmological perturbations that seeded the cosmic microwave background (CMB) anisotropies and the large-scale structure of our Universe \cite{Mukhanov:1981xt, Mukhanov:1982nu, Starobinsky:1982ee, Guth:1982ec, Hawking:1982cz, Bardeen:1983qw}.
In the range of scales accessible to CMB experiments~\cite{Ade:2015xua, Ade:2015lrj}, these perturbations are constrained to be small, at the level $\zeta\simeq 10^{-5}$ until they re-enter the Hubble radius during the radiation era, where $\zeta$ is the scalar curvature perturbation. 
At smaller scales however, they may be sufficiently large so that when they re-enter the Hubble radius, they overcome the pressure forces and collapse to form PBHs~\cite{Hawking:1971ei, Carr:1974nx, Carr:1975qj}. 
In practice, PBHs form when the mean curvature perturbation in a given Hubble patch exceeds a threshold denoted $\zeta_\uc\simeq 1$~\cite{Zaballa:2006kh, Harada:2013epa} (see Ref. \cite{Young:2014ana} for an alternative criterion based on the density contrast rather than the curvature perturbation). 

The abundance of PBHs is usually stated in terms of the mass fraction of the Universe contained within PBHs at the time of formation, $\beta$. 
If the coarse-grained curvature perturbation $\zeta_{\mathrm{cg}}$ follows the probability distribution function (PDF) $P(\zeta_{\mathrm{cg}})$, $\beta$ is given by~\cite{1975ApJ...201....1C}
\bea
\label{eq:def:beta}
\beta\left(M\right) = 2\int_{\zeta_\uc}^\infty P\left(\zeta_{\mathrm{cg}}\right) \dd \zeta_{\mathrm{cg}}\, .
\eea
Here, $\zeta_{\mathrm{cg}}$ is obtained from keeping the wavelengths smaller than the Hubble radius at the time of formation, 
\bea
\label{eq:def:zetacg}
\zeta_{\mathrm{cg}}(\bm{x}) = \left(2\pi\right)^{-3/2}\int_{k>aH_{\mathrm{form}}}\dd {\bm{k}}\zeta_{\bm{k}} e^{i\bm{k}\cdot\bm{x}}\, ,
\eea
where $a$ is the scale factor, $H\equiv \dot{a}/a$ is the Hubble scale, and a dot denotes differentiation with respect to cosmic time. In \Eq{eq:def:beta}, $M$ is the mass contained in a Hubble patch at the time of formation~\cite{Choptuik:1992jv, Niemeyer:1997mt, Kuhnel:2015vtw}, $M=3\Mp^2/H_{\mathrm{form}}$, where $\Mp$ is the reduced Planck mass.

Observational constraints on $\beta$ depend on the masses PBHs have when they form.
For masses between $10^9\mathrm{g}$ and $10^{16} \mathrm{g}$, the constraints mostly come from the effects of PBH evaporation on big bang nucleosynthesis and the extragalactic photon background\footnote{Interestingly, there is another suggested method of PBH detection. If a PBH with mass around $5\times 10^{-19} M_\odot$ ($\sim 2\times 10^{14}$g, and a radius similar to that of a proton), were to pass through the Earth, it would have almost no impact on the Earth, but would deposit $\sim 10^9$ joules of Hawking radiation into the Earth. 
This would leave detectable traces in crystalline material in Earth's crust, and hence offer an alternative detection method for PBHs \cite{Khriplovich:2007ci}.}, and typically range from $\beta<10^{-24}$ to $\beta<10^{-17}$. 
Heavier PBHs, with mass between $10^{16} \mathrm{g}$ and $10^{50} \mathrm{g}$, have not evaporated yet and can only be constrained by their gravitational and astrophysical effects (such as the mircolensing of quasars), at the level $\beta<10^{-11}$ to $\beta<10^{-5}$ (see Refs. \cite{Carr:2009jm, Carr:2017jsz, Carr:2020gox} for summaries of constraints, see in particular Fig. 10 of \cite{Carr:2020gox} for recent constraints on PBHs as dark matter).

Compared to the CMB anisotropies that allow one to measure $\zeta$ accurately in the largest $\sim 7$ \efolds~of scales in the observable Universe, PBHs only provide upper bounds on $\beta(M)$, and hence on $\zeta$. However, these constraints span a much larger range of scales and therefore yield valuable additional information. 
This is why PBHs can be used to constrain the shape of the inflationary potential beyond the $\sim 7$ \efolds~that are accessible through the CMB. 

\begin{figure}
    \centering
    \includegraphics[width=0.8\linewidth]{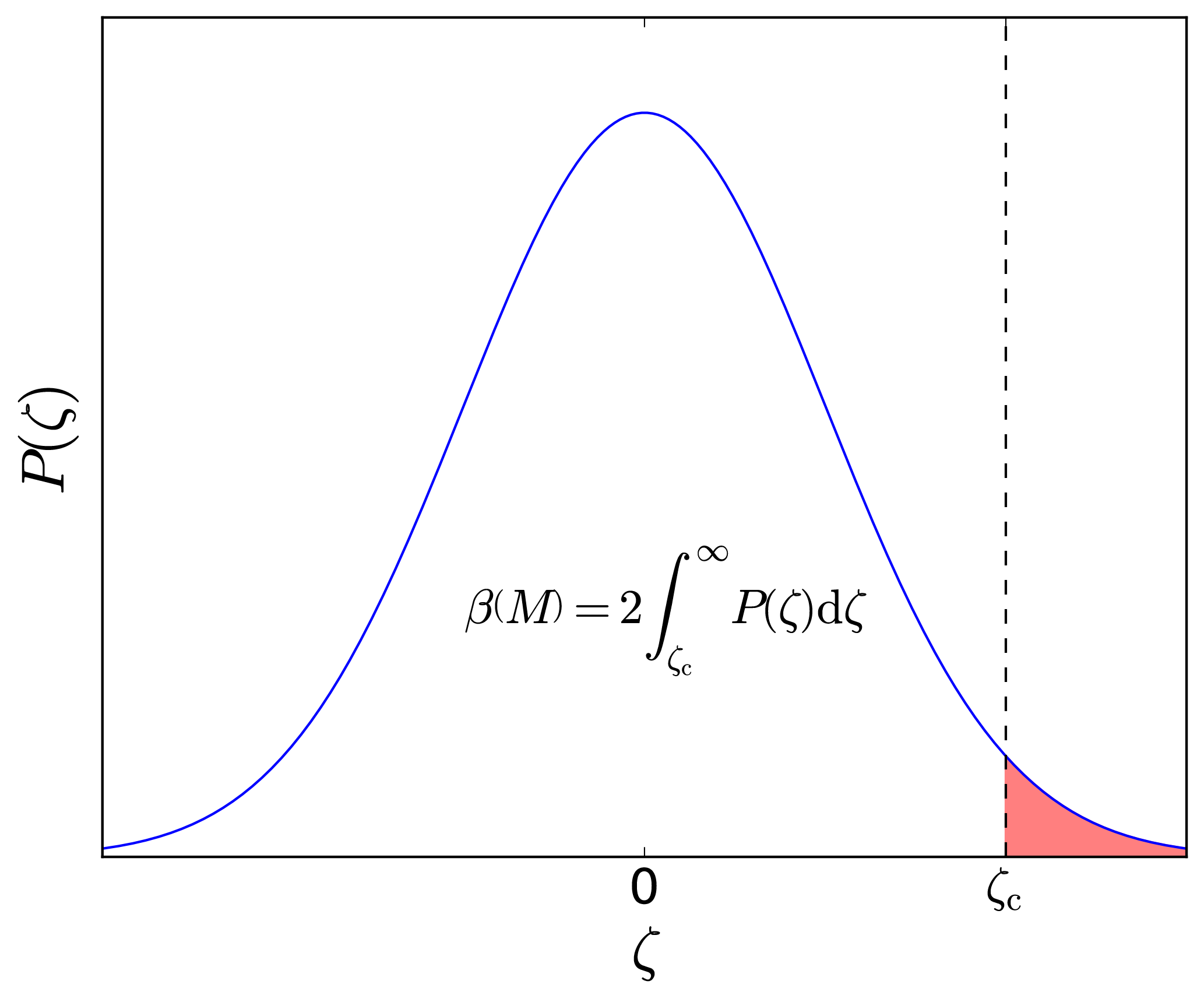}
    \caption[Mass fraction of the Universe contained in primordial black holes]{Sketch of how the mass fraction of the Universe contained in PBHs, denoted $\beta$, is obtained. One integrates the tail of the PDF of curvature fluctuation above some critical value $\zeta_\mathrm{c}$, above which PBHs are expected to form.}
    \label{fig:integral:sketch}
\end{figure}

In practice, one usually assumes $P(\zeta_{\mathrm{cg}})$ to be a Gaussian PDF, see \Fig{fig:integral:sketch}, with standard deviation given by the integrated power spectrum $\left\langle \zeta_{\mathrm{cg}}^2 \right\rangle = \int_{k}^{k_\uend} \calP_\zeta(\tilde{k})\dd \ln \tilde{k}$, where $k$ is related to the time of formation through $k=aH_{\mathrm{form}}$, and where $k_\uend$ corresponds to the wavenumber that exits the Hubble radius at the end of inflation. Combined with \Eq{eq:def:beta}, this gives rise to
\bea \label{eq:beta:erfc}
\beta\left(M\right) = \erfc\left[\frac{\zeta_\uc}{\sqrt{2  \int_{k}^{k_\uend} \calP_\zeta(\tilde{k})\dd \ln \tilde{k}}}\right]\, ,
\eea
where $\erfc$ is the complementary error function, $M$ is the mass contained in the Hubble volume, and $2\pi/k$ is the comoving Hubble length when the black holes form. In the limit $\beta \ll 1$, this leads to $\int_{k}^{k_\uend} \calP_\zeta(\tilde{k})\dd \ln \tilde{k} \simeq \zeta_\uc^2/(-2\ln\beta)$. 
Assuming the power spectrum to be scale invariant, one has $\int_{k}^{k_\uend} \calP_\zeta(\tilde{k})\dd \ln \tilde{k} \simeq \calP_\zeta \ln(k_\uend/k)  \simeq \calP_\zeta \Delta N$, where $\Delta N = \Delta \ln a$ is the number of \efolds~elapsed between the Hubble radius exit times of $k$ and $k_\uend$ during inflation. This leads to
\bea 
\label{eq:powerconstraint:standard}
\calP_\zeta \Delta N \simeq - \frac{\zeta_\uc^2}{2\ln\beta}\, .
\eea
For instance, with $\zeta_\uc=1$, the bound $\beta<10^{-22}$ leads to the requirement that $\calP_\zeta \Delta N< 10^{-2}$. 
This can be translated into constraints on the inflationary potential $V=24\pi^2\Mp^4 v$ and its derivative $V_{,\phi}$ with respect to the inflaton field $\phi$ using the slow-roll formulae~\cite{Mukhanov:1985rz, Mukhanov:1988jd} 
\bea
\label{eq:classicalPower}
\calP_\zeta = \frac{2v^3}{\Mp^2 {v_{,\phi}}^2}\, ,\quad 
\Delta N=\int_{\phi_\uend}^\phi\frac{v}{\Mp^2v_{,\phi}}   \dd \tilde{\phi} \, .
\eea
However, as we shall later see, producing curvature fluctuations of order $\zeta\sim \zeta_\uc\sim 1$ or higher precisely corresponds to the regime where quantum diffusion dominates over the classical field dynamics over a typical time scale of one \efold. 

Let us note in particular that the assumptions here rely on two things.
Firstly, the above considerations rely on the use of a Gaussian PDF for $P(\zeta_{\mathrm{cg}})$, which is only valid in the regime where quantum diffusion provides a subdominant correction to the classical field dynamics during inflation.
This is discussed in great detail in Chapter \ref{chapter:quantumdiff:slowroll}, and there we explore what happens when this condition is violated, but we remain in a slow-roll regime. 
Secondly, we are applying the classical slow-roll formulae for the curvature power spectrum $\calP_\zeta$ and number of \efolds~$\Delta N$, and hence we rely on the slow-roll assumptions holding throughout PBH formation.
Producing large curvature perturbations is more likely precisely when the power spectrum $\calP_\zeta$ becomes large. 
From \eqref{eq:classicalPower} it is easy to see that this can be achieved when $v_{,\phi} \to 0$, but this may cause problems for the slow-roll conditions, and much current research focuses on producing PBHs outside of slow roll \cite{Biagetti:2018pjj,Ezquiaga:2018gbw,Firouzjahi:2018vet,Byrnes:2018txb,Braglia:2020fms}. 

There are straightforward ways that we can see that slow-roll violation can easily produce the conditions needed for PBH production.
If $V'\to 0$, then the slow-roll equation of motion $3H\dot{\phi}_\mathrm{SR} + V_{,\phi} = 0$ becomes simply $\dot{\phi}_\mathrm{SR}=0$, and there are no dynamics. 
Also, in the case that $V_{,\phi}=0$, $\epsilon_2$ is no longer necessarily small, and hence there is no reason to neglect the second derivative term in the Klein--Gordon equation \eqref{eq:eom:scalarfield} in the first place. 
We treat this as a break down of the slow-roll equations. 

In order to explicitly see that this can lead to large curvature perturbations during inflation, let us rewrite the Sasaki--Mukhanov equation \eqref{eq:vk:sasakimukhanov} in terms of the curvature perturbation $\zeta_{\bm{k}} = v_{\bm{k}}/z$, which gives the equation of motion 
\bea \label{eq:eom:zeta}
\frac{\dd^2\zeta_{\bm{k}}}{\dd N} + \left( 3 - \epsilon_1 + \epsilon_2\right) \frac{\dd\zeta_{\bm{k}}}{\dd N} + \left(\frac{k}{aH}\right)^2 \zeta_{\bm{k}} = 0 \, ,
\eea 
where we have used the fact that $\dd N = aH\dd \eta$ and $\dd z/\dd N = (1+\epsilon_2/2)z$. 
This equation can be solved on super-horizon scales ($k\ll aH$), and the asymptotic equation can be integrated once to give 
\bea 
\frac{\dd \zeta_{\bm{k}}}{\dd N}\bigg|_{k\ll aH} = C_3 \ee^{-\int \left( 3 - \epsilon_1 + \epsilon_2 \right)\dd N } =  C_2 \ee^{ -3N + \ln H - \ln \epsilon_1  } \, ,
\eea 
and again to give the solution
\bea 
\zeta_{\bm{k}}\big|_{k\ll aH} = C_1 + C_2 \int \ee^{ -3N + \ln H - \ln \epsilon_1  }\dd N \, , 
\eea 
and hence we see the existence of two ``modes'' that govern the behaviour of $\zeta_{\bm{k}}$ in the super-horizon limit.
The first of these modes is constant and corresponds to the ``adiabatic'' mode, while the second mode evolves and either grows or decays, depending on the sign of $3 - \epsilon_1 + \epsilon_2 $.
This second mode corresponds to ``non-adiabatic'' (or ``entropy'') perturbations which can cause the curvature perturbation $\zeta$ to evolve, even after horizon crossing. 
If the curvature perturbation were to grow, then its power spectrum
\bea 
\mathcal{P}_\zeta = \frac{k^3}{2\pi^2} | \zeta_{\bm{k}} |^2\bigg|_{k\ll aH} \, ,
\eea 
can become large and we expect many more PBHs to form. 

In the standard slow-roll picture, the non-constant, non-adiabatic mode is exponentially suppressed because $\epsilon_1$, $\epsilon_2 \ll 1$, meaning that the curvature perturbation very quickly becomes constant after horizon crossing, meaning we can simply evaluate the power spectrum at horizon crossing ($k=aH)$. 
However, beyond slow roll, the evolving mode of the curvature perturbation can grow on super-horizon scales if the second slow-roll parameter changes sign and we have $\epsilon_2 < -3 + \epsilon_1$. 
This growth will also cause the power spectrum to grow, and hence we have the possibility of producing many black holes once we leave the slow-roll regime. 

Combining these two considerations, we see that we expect many PBHs to be formed when we have a very flat potential $V_{,\phi} \to 0$ and we violate slow roll and have $\epsilon_2 < -3 + \epsilon_1$.
This is precisely the case for ``ultra-slow-roll inflation'', in which we have a potential that is either exactly flat or very close to flat, and typically the second slow-roll parameter is large, with $\epsilon_2 \simeq -6$.

This motivates us to study ultra-slow-roll (USR) inflation. 
In the next chapter (\ie in Chapter \ref{chapter:USRstability}), we perform a classical analysis of the stability of USR inflation, in which we perform a full phase space analysis of USR. 
The stability of USR is an important feature to understand if one wants to study PBH production, as a long-lived (\ie stable) period of USR inflation may lead to many more PBH being produced than a short-lived regime may produce. 
Later in this thesis, after introducing the formalism of ``stochastic inflation'' in Chapter \ref{chapter:stochastic:intro}, we will study the quantum effects modelled by this formalism in a USR regime in Chapter \ref{chapter:USRstochastic}.



\newpage

\chapter{Attractive behaviour of ultra-slow-roll inflation} \label{chapter:USRstability}

As we have seen in the previous chapter, the production of primordial black holes may require the violation of slow-roll inflation. 
As such, in this chapter, we discuss the regime of ultra-slow-roll inflation, where the dynamics of the inflaton field are friction dominated, and we study in detail the dynamics of this regime. 
It is often claimed that the ultra-slow-roll regime of inflation is a non-attractor and/or transient, and here we study this claim in a classical context, \ie without considering quantum diffusion.
We carry out a phase-space analysis of ultra-slow roll in an arbitrary potential, $V(\phi)$, and show that while standard slow roll is always a dynamical attractor (whenever it is a self-consistent approximation), ultra-slow roll is stable for an inflaton field rolling down a convex potential with $\Mp V_{,\phi\phi}>|V_{,\phi}|$ (or for a field rolling up a concave potential with $\Mp V_{,\phi\phi}<-|V_{,\phi}|$). 
In particular, when approaching a flat inflection point, ultra-slow roll is always stable and a large number of \efolds~may be realised in this regime. 
However, in ultra-slow roll, $\dot{\phi}$ is not a unique function of $\phi$ as it is in slow roll and dependence on initial conditions is retained. 
We illustrate our analytical results with numerical examples.

This chapter is based on the publication \cite{Pattison:2018bct}, and is arranged as follows. 
In \Sec{sec:d&d} we briefly review the necessary machinery of inflation and the slow roll and ultra-slow roll (USR) regimes of the inflaton dynamics. In \Sec{sec:stability} we perform a phase-space analysis of ultra-slow-roll inflation and derive a necessary and sufficient condition for stability in USR. We then demonstrate how this condition can be applied to some simple examples in \Sec{sec:examples}, and provide some brief conclusions in \Sec{sec:conclusions}.

\section{Introduction}
\label{sec:intro}

We have seen that the simplest, and therefore perhaps the most natural, realisation of inflation is that of a single scalar field called the inflaton, which we denote $\phi$. 
The classical equation of motion of $\phi$ in an FLRW cosmology is given by the Klein-Gordon equation
\bea
\ddot{\phi} + 3 H\dot{\phi} + V_{,\phi}(\phi) = 0\, ,
\label{eq:eom:kleingordon}
\eea
where a dot denotes a derivative with respect to cosmic time $t$, $V_{,\phi}$ is the derivative of the potential with respect to the inflaton field value, and $H=\dot{a}/a$ is the Hubble expansion rate which satisfies \eqref{eq:friedmann:scalarfield}
\bea
\label{eq:friedmann}
H^2 = \frac{1}{3\Mp^2} \left( V+\frac{\dot{\phi}^2}{2} \right) \, .
\eea
Generally, it is not possible to solve this system of equations analytically, and so approximations are often made to simplify the dynamics. 

The most common approximation is that of slow-roll (SR) inflation, which we have reviewed in detail in \Sec{sec:intro:slowrollinflation}.
In this regime, one takes $\dot{\phi}^2 \ll V(\phi)$, so the energy budget of the inflaton is potential dominated and \Eq{eq:friedmann} reads $H^2\simeq V/(3\Mp^2)$. Under this approximation, one also neglect the acceleration term $\ddot{\phi}$ in the Klein-Gordon equation~(\ref{eq:eom:kleingordon}), and hence the equation of motion becomes first order,
\bea 
\label{eq:eom:KG:sr}
\dot{\phi}_\SR\simeq -\frac{V_{,\phi}}{3H} \, .
\eea

However, there are regimes in which this approximation is not a valid one.
For instance, one can imagine cases where the potential of the inflaton becomes very flat, so $V_{,\phi}(\phi) \to 0$, and SR begins to break down. 
A simple example of this is a potential with a flat inflection point at which $V_{,\phi} =V_{,\phi\phi}= 0$.
Under the SR approximation, \Eq{eq:eom:KG:sr} would give us $\dot{\phi} = 0$ at the inflection point, and so the inflaton rolling down this potential would come to a complete stop at the flat point of the potential.
In practice, we may expect the residual (although small) kinetic energy to carry the field through the inflection point, contrary to the SR prediction.
When SR is violated in this way, a phase of so-called ``ultra-slow-roll" (USR)~\cite{Inoue:2001zt, Kinney:2005vj}, or ``friction dominated", inflation takes place. 

Let us take a moment at this point to clear up the nomenclature surrounding ``ultra-slow-roll" inflation and similar scenarios, as this can sometimes be unclear in the literature. 
``Ultra-slow roll" is the name we use to describe the situation when the potential of the inflaton is very flat, so that $V_{,\phi} \simeq 0$, and hence from the Klein-Gordon equation~(\ref{eq:eom:kleingordon}) 
\bea
\label{eq:eom:KG:usr}
\ddot{\phi}_\USR\simeq-3H\dot{\phi}_\USR\, .
\eea
``Constant-roll" inflation~\cite{Martin:2012pe, Motohashi:2014ppa, Karam:2017rpw, Yi:2017mxs, Morse:2018kda} is intended to be a generalisation of ultra-slow-roll inflation, and is simply defined as a regime where $-\ddot{\phi}/(3H\dot{\phi}) = \mathrm{constant}$, and this constant is not necessarily equal to $1$.
However, often when constant-roll inflation is considered, ultra-slow roll is actually a singular point of the equations studied, and so the analysis does not include ultra-slow roll. One must thus be careful when calling constant-roll inflation a generalisation of ultra-slow roll. Another situation where SR is violated is ``fast-roll" inflation \cite{Linde:2001ae}, where the effective mass of the inflaton is of the same order as $H$ and all three terms in the Klein-Gordon equation~(\ref{eq:eom:kleingordon}) are of comparable magnitude (in the ``ultra-fast roll'' limit, the friction term is subdominant).

In this chapter, we are interested in the stability properties of USR inflation. 
While SR inflation is known to be a dynamical attractor~\cite{Salopek:1990jq, Liddle:1994dx, Vennin:2014xta, Grain:2017dqa}, it is often claimed in the literature that USR is always non-attractive~\cite{Cai:2017bxr, Dimopoulos:2017ged, Anguelova:2017djf, Biagetti:2018pjj, Morse:2018kda}. 
However, this conclusion is obtained from investigating constant-roll inflation, which is supported only by a very specific class of potentials and which, as already pointed out, only reduces to USR in a (singular) limit. 
This is why we carry out a generic analysis of USR that does not assume a specific potential.
We will derive a simple criterion on the potential for when it is stable.

\section{Inflaton dynamics and definitions}
\label{sec:d&d}
We begin by discussing the formalism and language we will use to construct and define USR, and explain how it differs from SR.
As introduced before, we shall make use of the Hubble flow parameters \eqref{eq:def:slowrollparameters}, where we recall that inflation requires $\epsilon_1 < 1$, along with the field acceleration parameter, 
\bea  \label{eq:def:f}
f&=-\frac{\ddot{\phi}}{3H\dot{\phi}} = 1+\frac{1}{3H\dot{\phi}}V_{,\phi} \, ,
\eea 
which quantifies the relative importance of the acceleration term compared with the friction term in the Klein-Gordon equation~(\ref{eq:eom:kleingordon}).
The field acceleration parameter can be expressed in terms of the first two Hubble-flow parameters, as can be seen from combining \Eqs{eq:eps1exact} and \eqref{eq:eps2exact} with \Eq{eq:def:f},
\bea
\label{eq:f:epsilon}
f = \frac{2\epsilon_1-\epsilon_2}{6}\, .
\eea

Since $f$, like the slow-roll parameters, is a function of $\phi$ and $\dot{\phi}$, phase space (which is usually parametrised by $\phi $ and $\dot{\phi}$) can also be parametrised by $\phi$ and $f$. This will prove useful in the following. To this end, let us express $\dot{\phi}$ in terms of $\phi$ and $f$, which can be done by combining \Eqs{eq:friedmann} and~(\ref{eq:f:epsilon}),
\bea
\label{eq:phidot:f:phi}
\dot{\phi}^2 = V\left[\sqrt{1+\frac{2\Mp^2}{3\left(f-1\right)^2}\left(\frac{V_{,\phi}}{V}\right)^2}-1\right]\, .
\eea
Thus by combining \Eqs{eq:eps1exact} and~(\ref{eq:phidot:f:phi}), one can write the first Hubble flow parameter as
\bea
\label{eq:eps1:f:phi}
\epsilon_1 =  3\frac{\sqrt{1+\frac{2\Mp^2}{3\left(1-f\right)^2}\left(\frac{V_{,\phi}}{V}\right)^2}-1}{\sqrt{1+\frac{2\Mp^2}{3\left(1-f\right)^2}\left(\frac{V_{,\phi}}{V}\right)^2}+1}\, .
\eea
The condition for inflation to take place, $\epsilon_1<1$, then reads
\bea
\label{eq:inflation:condition}
\frac{\Mp}{\left\vert 1-f \right\vert }\left\vert \frac{V_{,\phi}}{V}\right\vert < \frac{3}{\sqrt{2}}\, .
\eea
We will often work in the quasi-de Sitter (quasi-constant-Hubble) approximation which corresponds to $\epsilon_1\ll1$, and hence $[\Mp/\vert 1-f\vert]\vert {V_{,\phi}}/{V} \vert\ll1$.

\subsection{Slow-roll inflation}

Slow-roll inflation corresponds to the regime where all Hubble flow parameters are much smaller than one, \ie $\vert \epsilon_n\vert \ll 1$ for $n\geq 1$ and we have seen that this means that the dynamical system boils down to
\bea
\label{eq:slowroll}
H_\SR^2 \simeq \frac{V}{3\Mp^2}\,  \quad {\rm and} \quad 3H\dot\phi_\SR \simeq - V_{,\phi} \, ,
\eea
and hence $|f|\ll 1$ in slow roll.
In this limit, $\dot{\phi}$ is determined completely by the gradient of the potential and a single trajectory is selected out in phase space since $\dot{\phi}_\SR$ has no dependence on initial conditions. One notices that while SR is usually defined as $\vert\epsilon_n\vert \ll 1$ for \textit{all} $n\geq 1$, the above system only relies on $\epsilon_1\ll 1$ and $\vert \epsilon_2 \vert \ll 1$.

Since $\dot{\phi}$ is an explicit function of $\phi$ through \Eq{eq:slowroll}, any phase space function can be written as a function of $\phi$ only. For the first Hubble-flow parameter and the field acceleration parameter, substituting \Eq{eq:slowroll} into \Eqs{eq:eps1exact} and \eqref{eq:def:f}, one obtains
\bea
\epsilon_{1\SR} &\simeq \frac{\Mp^2}{2}\left(\frac{V_{,\phi}}{V}\right)^2 \,,
\eea
\bea
f_\SR &\simeq \frac{\Mp^2}{3} \left[\frac{V_{,\phi\phi}}{V} - \frac{1}{2}\left(\frac{V_{,\phi}}{V}\right)^2\right]\, .
\label{eq:f:SR}
\eea
For a given inflationary potential $V(\phi)$, the existence of a regime of SR inflation can thus be checked by verifying that  the potential slow-roll parameters $\epsilon_V$ and $\eta_V$, defined as
\bea
\label{eq:epsV}
\epsilon_{V} \equiv \frac{\Mp^2}{2}\left(\frac{V_{,\phi}}{V}\right)^2 
\quad {\rm and} \quad
\eta_V \equiv \Mp^2 \frac{V_{,\phi\phi}}{V} \, ,
\eea
remain small,
\bea
\label{eq:sr:consistency}
 \epsilon_{V} \ll 1 \quad {\rm and} \quad |\eta_V |\ll1 \,.
\eea

\subsection{Ultra-slow roll inflation}
\label{sec:Def:USR}

In the ultra-slow roll regime it is the driving term, corresponding to the gradient of the potential, that is neglected in the Klein-Gordon equation \eqref{eq:eom:kleingordon}, rather than the field acceleration. 
This corresponds to the relative field acceleration $f\approx 1$ in \Eq{eq:def:f}, leading to \Eq{eq:eom:KG:usr}. 
Note that if $\phi$ follows the gradient of its potential, then $\dot{\phi} V_{,\phi} = \dot{V}<0$, and conversely $\dot{\phi} V_{,\phi}>0$ if the field evolves in the opposite direction. 
From \Eq{eq:def:f}, one can then see that $f<1$ corresponds to situations where the inflaton rolls down its potential and $f>1$ to cases where the field climbs up its potential.

Integrating \Eq{eq:eom:KG:usr} leads to the USR solution
\bea
\label{eq:USR:phidot:N}
\dot{\phi}_{\USR} \propto \ee^{-3N} \, .
\eea
This is the USR limit in which one takes $V_{,\phi} = 0$, but we shall see later that other solutions exist approaching USR when $V_{,\phi}$ does not exactly vanish.
Instead of being driven by $V_{,\phi}$ as in the SR case \eqref{eq:eom:KG:sr}, here the time derivative $\dot{\phi}_{\textrm{USR}}$ is exponentially decreasing with the number of \efolds. If we also assume quasi-de Sitter ($\epsilon_1\ll1$), the above can be integrated as
\bea
\label{eq:USR:traj}
\phi_\USR - \phi_{\USR,*} \simeq \frac13 \frac{\dot{\phi}_{\USR,*}}{H_*}  \left[ 1-\ee^{-3\left(N-N_*\right)} \right]  \, ,
\eea
where the star denotes some reference time. 
Thus the USR solution may be thought of as the free or transient response of the scalar field in an expanding FRLW cosmology. 
It is independent of the shape of the potential, but depends instead on the initial value of the field and its time derivative, $\phi_*$ and $\dot\phi_*$.

Despite the different background evolution, linear fluctuations of a massless field about ultra-slow-roll inflation have the same scale-invariant form as during slow-roll inflation \cite{Seto:1999jc,Leach:2001zf,Kinney:2005vj}. 
This is a striking example of the invariance of field perturbations under ``duality'' transformations \cite{Wands:1998yp,Biagetti:2018pjj}. 

The condition under which USR takes place reads $\vert f-1 \vert \ll 1$, which implies that $3H \vert\dot{\phi}\vert \gg \vert V_{,\phi} \vert$. 
Clearly this is possible for any finite potential gradient so long as we have a sufficiently large field kinetic energy.
However, in order to have inflation we also need to have $\epsilon_{1} < 1$, which from \Eq{eq:inflation:condition} corresponds to
\bea
\label{eq:consistency:usr}
\epsilon_{V} < \frac94 \left(1-f \right)^2
 \, ,
\eea
where $\epsilon_{V}$ is the first potential slow-roll parameter given in \Eq{eq:epsV}. 
The quasi-de Sitter approximation $\epsilon_{1} \ll 1$ simply corresponds to $\epsilon_{V} \ll (1-f)^2$. Comparing this relation with \Eq{eq:sr:consistency}, one can see that USR inflation requires a potential that is even flatter than what SR imposes at the level of $\epsilon_V$ (hence the name ``ultra''-slow roll, which is otherwise not so apt since SR and USR are disjoint regimes), but that no constraint is required on $\eta_V$, \ie on the second derivative of the potential and hence there is no constraints on $\epsilon_2$\footnote{For USR we typically have $\epsilon_2 \simeq -6$.}.

In the following, we thus distinguish two regimes: USR, which corresponds to $\vert f-1\vert \ll 1$, and USR \emph{inflation}, which corresponds to $\sqrt{\epsilon_V}\ll\vert f-1\vert \ll 1$.

\section{Stability analysis}
\label{sec:stability}

USR is often referred to as a transient or non-attractor solution during inflation~\cite{Cai:2017bxr, Dimopoulos:2017ged, Anguelova:2017djf, Biagetti:2018pjj, Morse:2018kda}. 
This is because of results in constant-roll models~\cite{Martin:2012pe, Motohashi:2014ppa, Karam:2017rpw, Yi:2017mxs, Morse:2018kda}, where the field acceleration parameter $f$ defined in \Eq{eq:def:f} is taken to be a constant. In the Hamilton-Jacobi formalism, this corresponds to taking $H(\phi)\propto \exp (\pm\sqrt{3f/2}\phi/\Mp)$, and the potentials that support such a phase of constant roll can be obtained from $V=3\Mp^2H^2-2\Mp^4 H'^2$. 
In these potentials, the constant-roll solution is only one possible trajectory in phase space and one can study its stability. 
One finds that the constant-roll solution is an attractor if $f<1/2$~\cite{Motohashi:2014ppa}. This excludes the USR limit $f\simeq 1$, which could lead to the incorrect conclusion that USR is always unstable. 
However this result only applies to the family of potentials mentioned above. 
Moreover, it is singular in the limit $f \rightarrow 1$ since combining the equations above, one finds $V\equiv \mathrm{constant}$ in that case, for which $f=1$ is the only solution so nothing can be concluded about its attractive or non-attractive behaviour.

This motivates us to go beyond these considerations and to study the phase-space stability of USR in a generic potential.

\subsection{Dynamical equation for the relative field acceleration}

Since the field acceleration parameter, $f$, quantifies the importance of the acceleration term in the Klein-Gordon equation \eqref{eq:eom:kleingordon}, it essentially parameterises whether we are in SR ($\vert f \vert \ll 1$) or USR ($\vert f-1 \vert \ll 1$).
As such, knowing the evolution of $f$ tells us which regime we are in and when we transition from one to the other, and will allow us to study the stability of the two regimes.
As such, we seek a dynamical equation for $f$.

We begin by recasting \Eq{eq:eom:kleingordon} with $\phi$ as the ``time" variable, which reduces the equation to a first-order differential equation, namely
\bea 
\label{eq:kgphi}
\frac{\dd\dot{\phi}}{\dd\phi} +3H + \frac{V_{,\phi}}{\dot{\phi}} = 0 \, ,
\eea 
where we assume that $\phi$ evolves monotonically with time, and hence acts as a ``clock'' for inflation. 
\Eq{eq:kgphi} can be rewritten as
\bea
\label{eq:phiddot:f:phi}
\frac{\dd}{\dd\phi}  \left(\dot{\phi}^2\right) = -2 V_{,\phi}\frac{f}{f-1}\, ,
\eea
and, combined with \Eq{eq:phidot:f:phi}, this leads to an equation for the evolution of $f$,
 \bea
 \frac{\dd f}{\dd \phi} = \frac{3}{2\Mp^2}\frac{V}{V_{,\phi}}\left(f-1\right)^2\left(f+1\right)\left[\sqrt{1+\frac{2\Mp^2}{3\left(f-1\right)^2}\left(\frac{V_{,\phi}}{V}\right)^2}-\frac{1-f}{1+f}\right]-\left(1-f\right)\frac{V_{,\phi\phi}}{V_{,\phi}}\, .
 \label{eq:f:dynamical}
 \eea
This can be written in terms of the potential slow-roll parameters \eqref{eq:epsV} as
\bea
\frac{\dd f}{\dd \phi} = \frac{3}{2\Mp}\frac{\left(f-1\right)^2\left(f+1\right)}{\sqrt{2\epsilon_{V}}}\left[\sqrt{1+\frac{4\epsilon_{V}}{3\left(f-1\right)^2}}-\frac{1-f}{1+f}\right]-\frac{\left(1-f\right)\eta_{V}}{2\Mp\sqrt{2\epsilon_{V}}}\, .
\label{eq:f:dynamical:srparams}
\eea
Note that this equation is exact and does not make any assumption about the smallness or otherwise of the slow-roll parameters.

\subsection{Slow-roll limit}

We shall begin our stability analysis by considering the slow-roll case.
If we expand the right-hand side of \Eq{eq:f:dynamical:srparams} to first order in the potential slow-roll parameters \eqref{eq:epsV} and take $f$ to be of first order in the slow-roll parameters as suggested by \Eq{eq:f:SR}, one obtains
\bea
\frac{V_{,\phi}}{V}\frac{\dd f}{\dd \phi} \simeq \frac{1}{2}\left(\frac{V_{,\phi}}{V}\right)^2 - \frac{V_{,\phi\phi}}{V}+\frac{3}{\Mp^2}f\, .
\label{eq:f:dynamical:SR}
\eea
We see that the right-hand side of \Eq{eq:f:dynamical:SR} vanishes for the slow-roll solution \eqref{eq:f:SR}, which is consistent with the fact that $f$ is first order in the slow-roll parameters and $  V_{,\phi}/V \dd f/\dd \phi  \simeq \dd f/\dd N$ is second order in slow roll. 

The stability of the slow-roll solution can then be studied by considering a deviation from \Eq{eq:f:SR} parametrised by
\bea
f \simeq f_{\mathrm{SR}}+\Delta\, .
\label{eq:f:fSR:epsilon}
\eea
In this expression, $f_{\mathrm{SR}}$ is given by \Eq{eq:f:SR} plus corrections that are second order in slow roll and $\Delta$ describes deviations from slow roll that are nonetheless first order in slow-roll parameters or higher. For instance, we imagine that initially, one displaces $f$ from the standard slow-roll expression given in \Eq{eq:f:SR} (\eg by adding another linear combination of some slow-roll parameters) and study how this displacement evolves in time. 
By substituting \Eq{eq:f:fSR:epsilon} into \Eq{eq:f:dynamical:SR}, one obtains
\beq
\frac{V_{,\phi}}{V}\frac{\dd \Delta}{\dd \phi} \simeq \frac{3}{\Mp^2}\Delta\, ,
\eeq
which at leading order in slow roll, using \Eq{eq:slowroll}, can easily be solved to give
\bea
\Delta \simeq \Delta_\uin \exp\left[-3\left(N-N_\uin\right)\right]\, ,
\eea
which is always decreasing as inflation continues. This shows that SR is a stable attractor solution whenever the consistency conditions \eqref{eq:sr:consistency} are satisfied. This is of course a well-known result~\cite{Salopek:1990jq, Liddle:1994dx} but it is interesting to see how it can be formally proven in the formalism employed in this work.

\subsection{Ultra-slow-roll limit}
\label{sec:stability:USR}

In the ultra-slow-roll limit we have $f=1$, which we can readily see is a fixed point of \Eq{eq:f:dynamical} for any potential. We can therefore carry out a generic stability analysis of this fixed point that is valid for any potential. The results will be illustrated with two specific models in \Sec{sec:examples}.

The strategy is to linearise \Eq{eq:f:dynamical} around $f=1$ by parameterising 
\bea
f=1-\delta\, ,
\eea
where we assume $\vert \delta \vert \ll 1$ in order to study small deviations from USR, and from \Eq{eq:def:f} we see that $\delta = -V_{,\phi}/(3H\dot{\phi})$.
The only ambiguity is in the argument of the square root in \Eq{eq:f:dynamical}, that reads $1+\epsilon_V/(6\delta^2)$, since both $\epsilon_V$ and $\delta$ are small numbers. However, from \Eq{eq:consistency:usr} and the discussion below it, one recalls that inflation requires $\epsilon_V<9\delta^2/4$, and $\epsilon_V\ll \delta^2$ ensures quasi de-Sitter inflation $\epsilon_1\ll 1$. 
This is why \Eq{eq:f:dynamical} should be expanded in the USR \emph{inflation} limit $\sqrt{\epsilon_V}\ll\vert\delta\vert\ll 1$,\footnote{\label{footnote:USR:Non:Inflation}An expansion in the USR \emph{non-inflating} limit, $\vert \delta \vert \ll \sqrt{\epsilon_V}$ and $\vert \delta \vert \ll 1$, can also be performed along similar lines.  At linear order in $\delta$, \Eq{eq:f:dynamical} gives rise to
\bea
\frac{\dd \delta}{\dd \phi} \simeq \left[-\frac{\sqrt{6}}{\Mp}\mathrm{sign}\left(V_{,\phi}\delta \right)+\frac{V_{,\phi\phi}}{V_{,\phi}}\right]\delta\, .
\label{eq:f:dynamical:USR:noninflating}
\eea
If the field follows the gradient of its potential, one obtains the stability condition
\bea
\frac{V_{,\phi\phi}}{\left\vert V_{,\phi} \right\vert }>\frac{\sqrt{6}}{\Mp}\, ,
\label{eq:USR:noninflating:stability:condition}
\eea
and conversely, if the field climbs up the potential, one gets $V_{,\phi\phi}/|V_{,\phi}|<-\sqrt{6}/\Mp$. The solution to \Eq{eq:f:dynamical:USR:noninflating} reads
\bea
\delta \simeq \delta_\uin \frac{V_{,\phi}(\phi)}{V_{,\phi}\left(\phi_\uin\right)}\exp\left(\sqrt{6}\frac{|\phi-\phi_\uin|}{\Mp}\right) \, .
\label{eq:delta:sol:USRnoninflating}
\eea
To determine how $\epsilon_1$ varies, one can plug \Eq{eq:delta:sol:USRnoninflating} into \Eq{eq:eps1:f:phi}. One finds that if the field follows the gradient of its potential, then $\epsilon_1$ decreases if $\epsilon_V<3$ and increases otherwise, and it always decreases if the field climbs up its potential. When $\epsilon_1$ decreases, it may become smaller than one at some point, and a phase of USR \emph{inflation} starts, whose stability properties are discussed in the main text.} which gives rise to
\bea
\label{eq:f:dynamical:USR:new}
\frac{\dd\delta}{\dd\phi}\simeq -\frac{3}{\Mp^2}\frac{V}{V_{,\phi}}\delta^2+\frac{V_{,\phi\phi}}{V_{,\phi}}\delta\, .
\eea
The right-hand side of this expression is proportional to $\delta-\eta_V/3$, so which term dominates depends on the magnitude $\vert\delta\vert$ with respect to $\vert\eta_V\vert$. Since $\delta$ must be larger than $\sqrt{\epsilon_V}$, two possibilities have to be distinguished.

\subsubsection{Case $\eta_V^2<\epsilon_V$}
\label{sec:stability:USR:inflation:case1}

In this case the condition for USR inflation, $\sqrt{\epsilon_V}\ll\vert\delta\vert$, guarantees that $\vert\delta\vert \gg \eta_V$ and the first term on the right-hand side of \Eq{eq:f:dynamical:USR:new} dominates,
\bea
\label{eq:f:dynamical:USR:case1}
\frac{\dd\delta}{\dd\phi}\simeq -\frac{3}{\Mp^2}\frac{V}{V_{,\phi}}\delta^2\, .
\eea
As explained at the beginning of \Sec{sec:Def:USR}, if the field follows the gradient of its potential then we have $f<1$ and $\delta>0$. 
If $V_{,\phi}>0$ and $\phi$ decreases with time, then from \Eq{eq:f:dynamical:USR:case1} $\delta$ increases with time and USR is unstable. 
If $V_{,\phi}<0$ and $\phi$ increases with time, then again $\delta$ increases with time and USR is still unstable. 
Conversely, if we have $f>1$ and $\delta<0$, so that the field climbs up the potential, if $V_{,\phi}>0$ then $\phi$ increases with time and so does $\delta$, so USR is unstable, and if $V_{,\phi}<0$ then $\phi$ decreases with time and USR is still unstable.
Note that if one substitutes $V_{,\phi} = -3H\dot{\phi}\delta$ into \Eq{eq:f:dynamical:USR:case1}, then these same conclusions can also be reached by again considering the signs of $\delta$ and $\dot{\phi}$ for any case of interest. 

We conclude that USR inflation is always unstable in that case.  The example discussed in \Sec{sec:Staro} corresponds to this situation.

\subsubsection{Case $\epsilon_V<\eta_V^2$}

In this case, which term dominates in \Eq{eq:f:dynamical:USR:new} depends on the magnitude of $\delta$. However, strictly speaking, a stability analysis of the fixed point $\delta\sim 0$ should only deal with its immediate neighbourhood, \ie with the smallest possible values of $\vert \delta \vert$ which in this case are smaller than $\vert \eta_V \vert$ (notice that if $\vert \eta_V \vert \gtrsim 1$ then $\epsilon_V<\eta_V^2$ becomes true for all  $\sqrt{\epsilon_V}\ll \vert \delta \vert \ll 1$). The second term in \Eq{eq:f:dynamical:USR:new} then dominates and one has
\bea
\label{eq:f:dynamical:USR:case2}
\frac{\dd\delta}{\dd\phi}\simeq \frac{V_{,\phi\phi}}{V_{,\phi}} \delta\, .
\eea
A similar discussion as in the previous case can be carried out, by first considering the situation where the field follows the gradient of its potential, so $f<1$ and $\delta>0$. 
If $V_{,\phi}>0$ and $\phi$ decreases with time then $\vert \delta\vert$ decreases if $V_{,\phi\phi}>0$. 
If $V_{,\phi}<0$ and $\phi$ increases with time then $\vert \delta\vert$ decreases under the same condition 
\bea
V_{,\phi\phi}>0\, .
\eea
Thus we conclude that USR is stable for a scalar field rolling down a convex potential.
Conversely, we find that USR is stable for a scalar field rolling up a concave potential,
$V_{,\phi\phi}<0$. 

The fact that $\vert \delta\vert$ decreases with time is a necessary condition for USR inflation stability but not a sufficient one, since one also has to check that $\vert\delta\vert$ remains much larger than $\sqrt{\epsilon_V}$, \ie that the system remains inflating. To this end, let us notice that \Eq{eq:f:dynamical:USR:case2} can be integrated and gives
\bea
\delta \simeq \delta_\uin \frac{V_{,\phi}(\phi)}{V_{,\phi}\left(\phi_\uin\right)} \, .
\label{eq:delta:sol}
\eea
This confirms that $\vert \delta \vert$ decreases with time when $\vert V_{,\phi} \vert$ decreases. 
Note that, given $\delta=-V_{,\phi}/3H\dot\phi$, this solution corresponds to $H\dot{\phi} =$constant, which differs from the ultra-slow-roll limit \Eq{eq:USR:phidot:N}.
Substituting \Eq{eq:delta:sol} into \Eq{eq:eps1:f:phi} (expanded in the $\sqrt{\epsilon_V}\ll\vert\delta\vert\ll 1$ limit), one obtains
\bea
\label{eq:USR:stable:eps1:appr}
\epsilon_1\simeq\epsilon_{1,\uin}\left( \frac{V_\uin}{V}\right)^2\, ,
\eea
where a subscript ``in'' denotes a quantities value at the initial field value $\phi_{\mathrm{in}}$. 
Therefore, $\epsilon_1$ increases if the field follows the gradient of its potential and decreases otherwise. Whether or not this increase can stop inflation in the former case depends on the potential. 
If the relative variations of the potential are bounded this may never happen if $\epsilon_1$ has a sufficiently small value initially. 

One can also use \Eq{eq:delta:sol} to compute the number of \efolds~spent in the USR regime. 
Since $\dd N/\dd\phi = H/\dot{\phi} = -3H^2\delta/V_{,\phi}$ where we have used the definition~(\ref{eq:def:f}), in the quasi de-Sitter limit where $H^2\simeq V/(3\Mp^2)$, one obtains $\dd N/\dd \phi = -V\delta/(V'\Mp^2)$. Making use of \Eq{eq:delta:sol}, this gives rise to $\dd N/\dd \phi = - V \delta_\uin/[V_{,\phi}(\phi_\uin) \Mp^2]$, and hence
\bea
\label{eq:DeltaN:USR}
\Delta N_\USR = -\frac{\delta_\uin}{\Mp^2V_{,\phi}(\phi_\uin)}\int_{\phi_\uin}^\phi V(\tilde{\phi})\dd\tilde{\phi}\, .
\eea
This should be compared with the slow-roll formula $\Delta N_\SR = - 1/\Mp^2\int_{\phi_\uin}^\phi V(\tilde{\phi})/V_{,\phi}(\tilde{\phi})\dd\tilde{\phi} $, which shows that in general fewer \efolds~are realised between two given field values in the USR regime than in standard slow roll. 
From this slow-roll formula it is even clear that $\Delta N_\SR$ can become infinite if there is a flat point in the potential such that $V/V_{,\phi}$ is not integrable as $V_{,\phi} \to 0$. 
However, the USR formula \eqref{eq:DeltaN:USR} is always integrable and finite, even when one crosses a flat inflection point of the potential.

Note that when we find USR to be a local attractor ($|\delta|$ decreases), it is so for sufficiently small values of $|\delta|<\vert \eta_V \vert$ only. If $|\delta| > \vert \eta_V \vert$ initially, then the first term on the right-hand side of \Eq{eq:f:dynamical:USR:new} dominates even if $\epsilon_V<\eta_V^2$ and the analysis of section~\Sec{sec:stability:USR:inflation:case1} shows that USR becomes unstable. 
In this case trajectories diverge from USR ($f\sim 1$) to approach the standard slow roll ($f\ll1$) for $\epsilon_V\ll1$ and $|\eta_V|\ll1$. This shows that, if $\epsilon_V<\eta_V^2 \ll 1$, the boundary between the SR and the USR basins of attraction is located around the line $\vert\delta\vert\sim\vert\eta_V\vert$. This will be checked explicitly in the example presented in \Sec{sec:example:inflectionPoint}.
This is similar to bifurcation behaviour that has previously been observed for inflection point quintessence \cite{Chang:2013cba}.

In summary, we find that if the inflaton rolls down its potential, USR inflation is stable if $V_{,\phi\phi}>0$ and $\eta_V^2>\epsilon_V$, which can be combined into the condition
\bea
\label{eq:USR:stable:criterion}
\eta_V>\sqrt{\epsilon_V}\, ,
\eea
and continues to inflate provided $V/V_\uin$ remains larger than $\sqrt{\epsilon_{1,\uin}}$.

\section{Examples}
\label{sec:examples}

Let us now illustrate the stability analysis performed in the previous section with two examples. In the first one, the potential has a discontinuity in its slope which produces a transient regime of USR inflation. In the second one, the potential has a flat inflection point around which the inflaton field evolves in the USR regime.

\subsection{Starobinsky inflation}
\label{sec:Staro}
\begin{figure}[t]
\begin{center}
\includegraphics[width=0.65\textwidth]{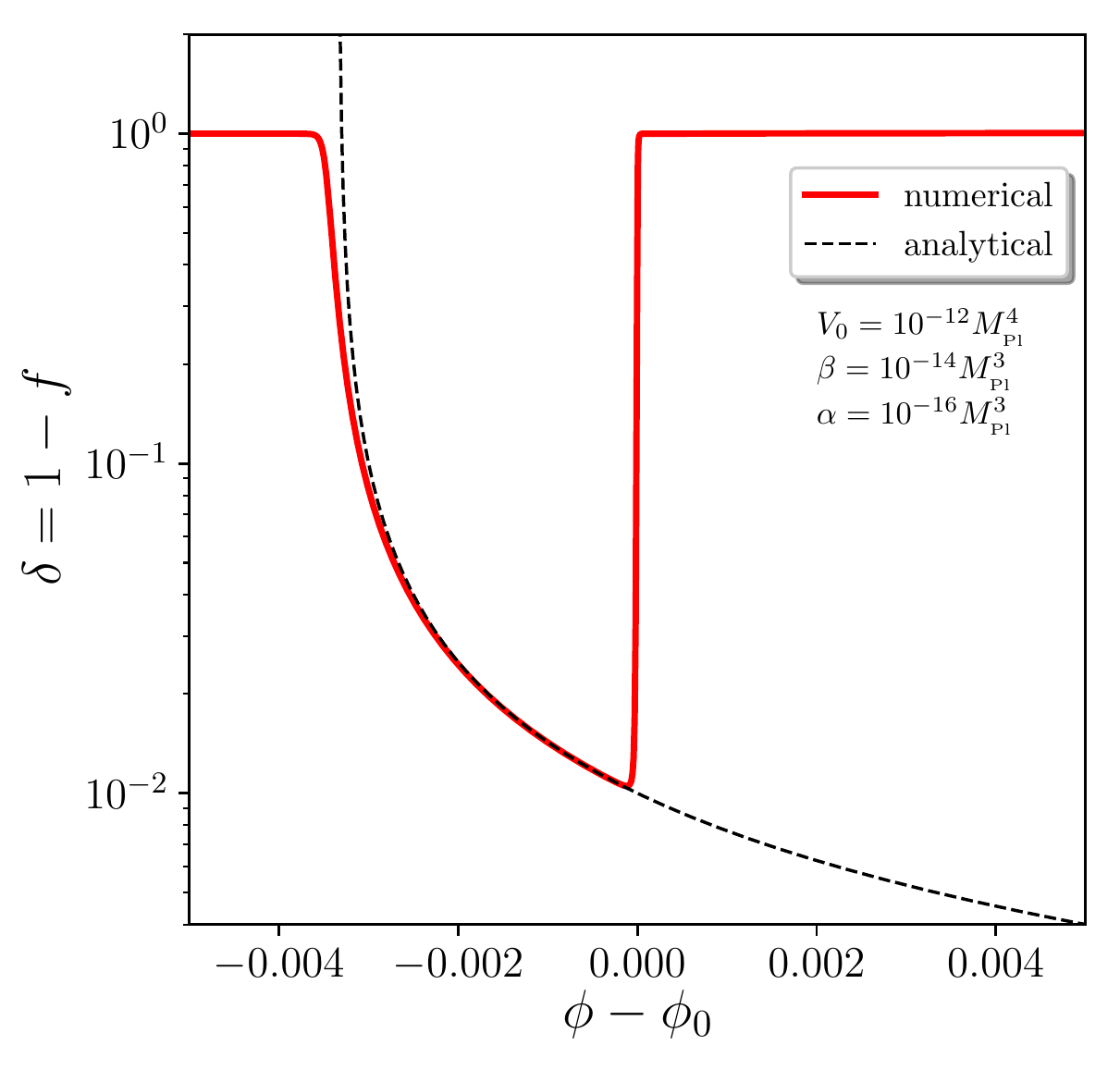}
\caption[Field acceleration parameter for the Starobinsky model of inflation]{Field acceleration parameter $\delta=1-f$ in the Starobinsky model~(\ref{eq:def:pot:strobinsky}) as a function of the field value. Before crossing the discontinuity point, a regime of SR inflation takes place where $\delta\simeq 1$. Right after crossing $\phi=\phi_0$, $\delta$ drops to small values which signals the onset of a USR phase of inflation, that quickly transitions towards a new SR phase. The solid red curve is obtained from numerically solving \Eqs{eq:eom:kleingordon} and~(\ref{eq:friedmann}) and making use of \Eq{eq:def:f}, while the black dashed curve corresponds to the analytical approximation~(\ref{eq:Staro:delta:appr}). One can check that it provides a good fit to the numerical result when $\vert\delta\vert\ll 1$.}
\label{fig:f:Staro}
\end{center}
\end{figure}
Let us first analyse the Starobinsky piece-wise linear model~\cite{Starobinsky:1992ts}, where the potential is made up of two linear segments with different gradients,
\bea
\label{eq:def:pot:strobinsky}
V(\phi) = \begin{cases}
V_0 + \alpha \left( \phi-\phi_0 \right) &\text{for $\phi < \phi_0$}\\
V_0 + \beta \left( \phi-\phi_0 \right) &\text{for $\phi > \phi_0$} 
\end{cases}
\, ,
\eea
where $\beta>\alpha>0$.

Starting with $\phi>\phi_0$, the inflaton quickly relaxes to the slow-roll attractor for $V_0\gg\beta \Mp$ (corresponding to $\epsilon_V\ll1$), where, according to \Eq{eq:slowroll}, $3H\dot{\phi}\simeq-\beta$. Right after crossing $\phi=\phi_0$ where the gradient of the potential is discontinuous, $\dot{\phi}$ is still given by the same value (since the equation of motion~(\ref{eq:eom:kleingordon}) for $\phi$ is second order, $\dot{\phi}$ is continuous through the discontinuity point) but the value of $V_{,\phi}$ is now different, such that $f$ given by \Eq{eq:def:f} reads
\bea
\label{eq:Staro:fminus}
f^{-}= 1+\frac{V_{,\phi}^-}{3(H\dot{\phi})^-} = 1+\frac{V_{,\phi}^-}{3(H\dot{\phi})^+}\simeq 1-\frac{V_{,\phi}^-}{V_{,\phi}^+}
= 1-\frac{\alpha}{\beta}
\, .
\eea
In this expression, a superscript ``$-$'' (or ``$+$'') means that the quantity is evaluated at $\phi\rightarrow \phi_0$ with $\phi<\phi_0$ (or $\phi>\phi_0$, respectively). If $\alpha\ll \beta$, $f_-\simeq 1$ and a phase of USR is triggered. 

The analysis of \Sec{sec:stability:USR} revealed that the stability of USR inflation depends on whether $\epsilon_V$ is smaller or larger than $\eta_V^2$. In the present model, since $\eta_V$ exactly vanishes, one necessarily falls in the later case, \ie the case discussed in \Sec{sec:stability:USR:inflation:case1} where it was shown that USR inflation is always unstable. Let us also notice that in the Starobinsky model, \Eq{eq:f:dynamical:USR:case1} can be integrated analytically, and making use of \Eq{eq:Staro:fminus} for the initial condition, one finds
\bea
\label{eq:Staro:delta:appr}
\delta\simeq \frac{\alpha}{\beta+\frac{3V_0}{\Mp^2}\left(\phi-\phi_0\right)}
\, .
\eea
Since $\phi$ decreases as a function of time, $\delta$ increases, and this confirms that USR is unstable in the Starobinsky model.

These considerations are numerically checked in \Fig{fig:f:Staro}. One can see that when the inflaton field crosses the discontinuity point at $\phi=\phi_0$, a phase of USR inflation with small values of $\delta$ starts, which \Eq{eq:Staro:delta:appr} accurately describes. This regime is however unstable and when the inflaton field crosses the value
\bea
\phi_{\USR\rightarrow\SR}=\phi_0-\frac{\Mp^2\left(\beta-\alpha\right)}{3V_0}
\, ,
\eea 
$\delta\simeq 1$ and the system relaxes back to SR. Making use of \Eq{eq:USR:traj}, one can also estimate the number of \efolds~spent in the USR regime between the field values $\phi_0$ and $\phi_{\USR\rightarrow\SR}$, and one finds
\bea
N_\USR\simeq\frac{1}{3}\ln\left(\frac{\beta}{\alpha}\right)\, .
\eea
The number of USR \efolds~is therefore of order a few or less in this model.

\subsection{Cubic inflection point potential}
\label{sec:example:inflectionPoint}

Let us now consider the case where the potential contains a flat inflection point at $\phi=0$, around which it can be expanded as
\bea 
\label{eq:pot:inflectionPoint:cubic}
V(\phi) = V_0 \left[1+\left(\frac{\phi}{\phi_0}\right)^3\right] \, .
\eea
One could parametrise the potential with a higher odd power of the field, say $V\propto 1+(\phi/\phi_0)^5$, but this would not change the qualitative conclusions that we draw below. 
The potential~(\ref{eq:pot:inflectionPoint:cubic}) has a flat inflection point at $\phi=0$, where $V_{,\phi}=V_{,\phi\phi}=0$. 
In the slow-roll regime, it takes the inflaton an infinitely long time to reach the inflection point, which it never crosses.
However in the USR regime the inflaton can traverse the inflection point in a finite time, which we estimate below. 

As explained in \Eq{eq:sr:consistency}, SR inflation requires $\epsilon_V\ll 1$ and $\vert\eta_V\vert\ll 1$, where the potential slow-roll parameters \eqref{eq:epsV} are here given by
\bea
\label{eq:InflectionPoint:SRpotParam}
\epsilon_V &= \frac92 \frac{\Mp^2}{\phi_0^2} \frac{(\phi/\phi_0)^4}{\left[ 1+(\phi/\phi_0)^3 \right]^2} \,, \\
\eta_V &= 6 \frac{\Mp^2}{\phi_0^2} \frac{\phi/\phi_0}{1+(\phi/\phi_0)^3} \,. 
\eea
Let us first focus on the part of the potential located before the inflection point, \ie at $\phi>0$. The parameter $\epsilon_V$ vanishes at $\phi=0$ and at $\phi\rightarrow\infty$, and in between it reaches a maximum at $\phi=2^{1/3}\phi_0$ where its value is $\epsilon_{V,\umax}=2^{1/3}\Mp^2/\phi_0^2$. The parameter $\eta_V$ has a similar behaviour, with a maximum at $\phi=2^{-1/3}\phi_0$ where its value is $\eta_{V,\umax}=2^{5/3}\Mp^2/\phi_0^2$. 

Two regimes need therefore to be distinguished: (i) if $\phi_0\gg \Mp$, SR inflation can be realised for all $\phi>0$, while (ii) if $\phi_0\ll \Mp$, SR inflation only takes place at sufficiently large ($\phi\gg\Mp$) or sufficiently small ($\phi\ll \phi_0^3/\Mp^2$) field values. After crossing the inflection point at  $\phi=0$, the potential decreases towards zero and the potential slow-roll parameter, $\epsilon_V$, diverges, signalling the end of inflation, so we restrict our analysis to the field values $\phi>-\phi_0$.

USR inflation can be studied making use of the results of \Sec{sec:stability:USR}, where it was shown that USR inflation is stable if $\eta_V>\sqrt{\epsilon_V}$, see \Eq{eq:USR:stable:criterion}. Together with \Eq{eq:InflectionPoint:SRpotParam}, this gives rise to the USR stability condition
\bea
\label{eq:InflectionPoint:USR:stability:Condition}
0<\phi<2\sqrt{2}\Mp\, .
\eea
We shall now study the two regimes $\phi_0\gg\Mp$ and $\phi_0\ll\Mp$ separately.

\subsubsection{Case $\phi_0\gg \Mp$}
\begin{figure}
\begin{center}
\includegraphics[width=0.45\textwidth, height=6cm]{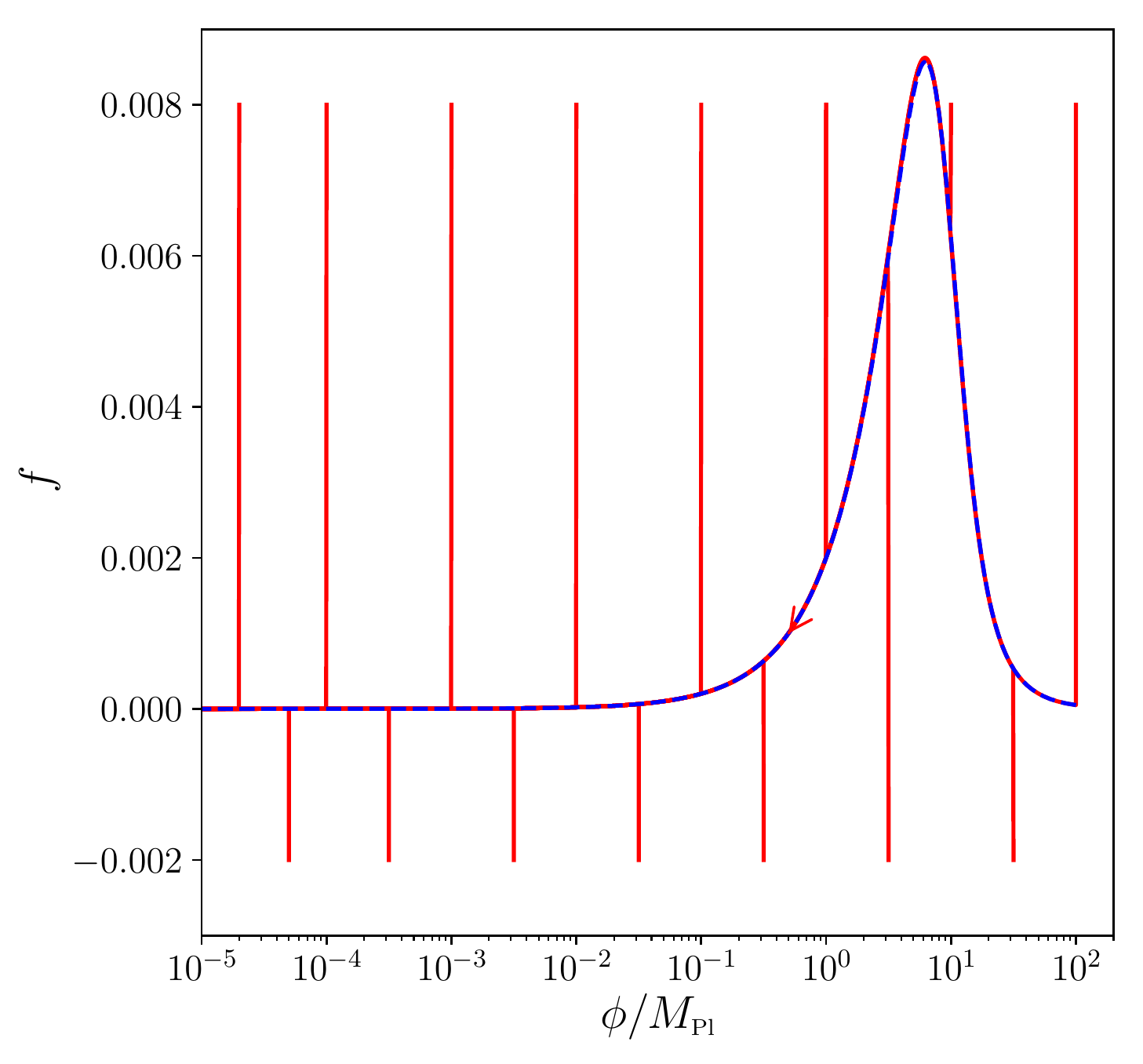} 
\includegraphics[width=0.45\textwidth, height=6.1cm]{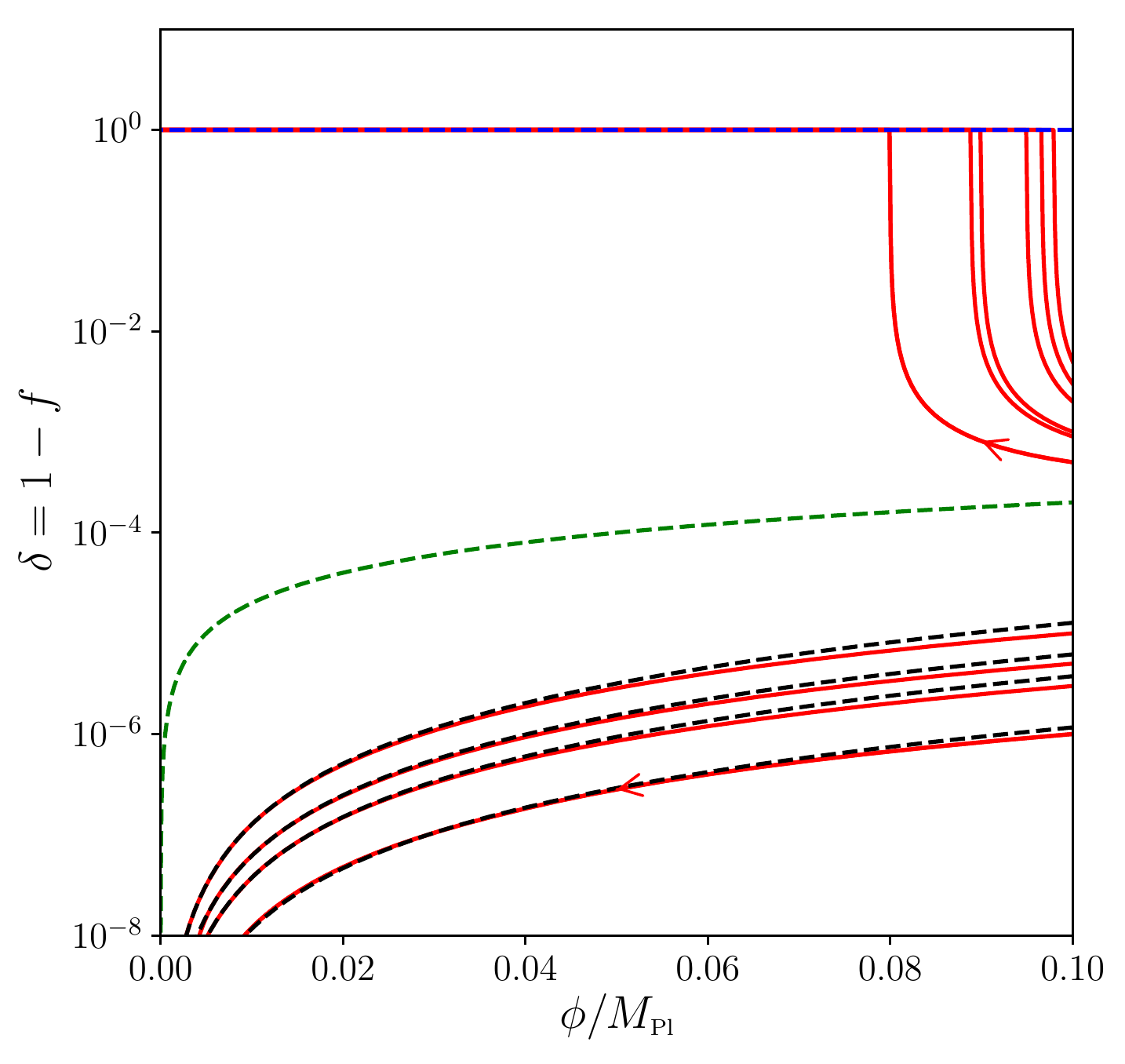} 
\caption[Field acceleration parameter for the cubic inflection point potential with $\phi_0=10\Mp$]{Field acceleration parameter in the cubic inflection point model~(\ref{eq:pot:inflectionPoint:cubic}) as a function of the field value, for $\phi_0=10\Mp$ and $V_0=4.2\times 10^{-11}$. The red lines correspond to numerical solutions of \Eq{eq:f:dynamical} and the dashed blue line stands for the slow-roll limit~(\ref{eq:f:SR}). The left panel zooms in on the region $f\simeq 0$ where one can see that SR is an attractor. The right panel uses a logarithmic scale on $1-f=\delta$, such that it zooms in on the USR regime $f\simeq 1$. If $\delta$ is initially smaller than $\vert \eta_V \vert/3$, represented with the dashed green line, the trajectories evolve towards $\delta=0$, otherwise they evolve to reach the SR attractor. The black dotted lines correspond to the analytical USR approximation~(\ref{eq:delta:sol}).
}
\label{fig:InflectionPoint:fsol:phi0GTMp}
\end{center}
\end{figure}
In this case, as already mentioned, SR inflation is an attractor over the entire range $\phi>0$ (until inflation stops when $\phi$ approaches $-\phi_0$). This implies that if one starts from an initial field value that is larger than the USR stability upper bound given in \Eq{eq:InflectionPoint:USR:stability:Condition}, $\phi=2\sqrt{2}\Mp$, the system relaxes towards the SR attractor (SR is the only stable attractor at $\phi>2\sqrt{2}\Mp$) and stays in SR until the end of inflation. In this scenario, even though USR inflation is also a local attractor at $\phi<2\sqrt{2}\Mp$, the inflaton field never drives a phase of USR inflation. 

The only way to get a period of USR inflation is therefore to start with $\phi<2\sqrt{2}\Mp$. There, as explained in \Sec{sec:stability:USR}, USR inflation is stable and its basin of attraction is bounded by the condition $\sqrt{\epsilon_V}<|\delta|<\eta_V$.

These considerations are numerically verified in \Fig{fig:InflectionPoint:fsol:phi0GTMp}. In the left panel, the SR region $\vert f\vert\ll 1 $ is displayed, where one can check that the numerical solutions of \Eq{eq:f:dynamical} (red curves) all converge towards the SR attractor~(\ref{eq:f:SR}) (dashed blue curve). In the right panel, a logarithmic scale is used on $1-f$, which allows one to zoom in on the USR region $f\simeq 1$. Since the initial values for $\phi$ satisfy \Eq{eq:InflectionPoint:USR:stability:Condition}, one can check that the trajectories with $\delta<\eta_V$ converge towards USR, while the ones for which $\delta>\eta_V$ approach the SR attractor. This confirms that the boundary between the two basins of attraction is located around the line $|\delta|=|\eta_V|$. The analytical approximation~(\ref{eq:delta:sol}) is displayed with the black dotted lines and one can check that it provides a good fit to the numerical result in the USR regime.

Let us finally estimate the number of \efolds~that is typically realised in the USR inflating regime. In the stability range~(\ref{eq:InflectionPoint:USR:stability:Condition}) of USR inflation, the potential is dominated by its constant piece since $\phi_0\gg\Mp$. The first Hubble-flow parameter is therefore roughly constant during the USR epoch, see \Eq{eq:USR:stable:eps1:appr}. Starting USR inflation at $\phi_\uin\sim \Mp$ with $\delta=\delta_\uin$, \Eq{eq:delta:sol} implies that $\delta$ goes back to its initial value $\delta_\uin$ at around $\phi\sim -\Mp$.  Plugging these values into \Eq{eq:DeltaN:USR}, one obtains
\bea
\label{eq:FlatInflectionPoint:NUSR:phi0GTMp}
\Delta N_\USR\simeq \frac{2\delta_\uin}{3}\left(\frac{\phi_0}{\Mp}\right)^3.
\eea
This shows that, in the regime $\phi_0\gg \Mp$, a large number of USR~\efolds~can be realised.
However, we should note that this number remains finite, contrary to what happens in the slow-roll regime where it takes an infinite time to cross the inflection point, as already mentioned.

\subsubsection{Stochastic diffusion}
This large number of USR inflationary \efolds~is however derived under the assumption that the field behaves classically all the way down to the inflection point, while stochastic diffusion is expected to play a role when the potential becomes very flat. Let us estimate how this changes the above result.
Stochastic diffusion and its impacts on the dynamics will be studied in detail in both slow roll and ultra-slow roll, in chapters \ref{chapter:quantumdiff:slowroll} and \ref{chapter:USRstochastic} respectively, and here we simply provide a brief discussion relevant to the example under consideration. 

Starting from $\phi_\uin=\Mp$ and $\delta=\delta_\uin$ as explained above, \Eq{eq:delta:sol} leads to $\delta(\phi)\simeq \delta_\uin(\phi/\Mp)^2$ (where we assume $\phi<\Mp$). Then, making use of \Eq{eq:DeltaN:USR}, if the field behaved in a purely classical manner, the number of \efolds~realised between $\phi$ and $-\phi$ would be given by $\Delta N_\USR(\phi)\simeq 2\delta_\uin/3 (\phi_0/\Mp)^3(\phi/\Mp)$.

On the other hand, if the field was only driven by stochastic noise, its equation of motion would be given by~\cite{Starobinsky:1986fx} $\dd\phi/\dd N=H/(2\pi)\xi$, where $\xi$ is a white Gaussian noise with vanishing mean and unit variance, such that $\langle \xi(N) \xi(N') \rangle = \delta(N-N')$.  Assuming that $H$ is roughly constant, this leads to $\langle \phi^2 \rangle = H^2/(2\pi)^2 N$, hence the typical number of \efolds~required for the inflaton field value to go from $\phi$ to $-\phi$ is given by $\Delta N_{\mathrm{sto}}= 48\pi^2\phi^2\Mp^2/V_0$. Notice that this can also be obtained using the ``first-passage-time techniques'' developed in Ref. \cite{Vennin:2015hra}, and which we will discuss in \Sec{sec:firstpassage}. Setting a reflective boundary condition at $\phi$ and an absorbing one at $-\phi$, one finds that the mean number of \efolds~required to reach $-\phi$ starting from $\phi$ exactly coincides with the expression given here for $\Delta N_{\mathrm{sto}}$.

Since $\Delta N_\USR$ scales as $\phi$ and $\Delta N_{\mathrm{sto}}$ as $\phi^2$, two regimes need to be distinguished. When $\phi>\phi_{\mathrm{sto}}$, where 
\bea
\frac{\phi_{\mathrm{sto}}}{\phi_0} = \frac{\delta_\uin}{72\pi^2}\left(\frac{\phi_0}{\Mp}\right)^2\frac{V_0}{\Mp^4}
\eea
is the solution of $\Delta N_\USR(\phi_{\mathrm{sto}})=\Delta N_{\mathrm{sto}}(\phi_{\mathrm{sto}})$, one has $\Delta N_\USR<\Delta N_{\mathrm{sto}}$, which means that classical USR is more efficient at driving the field than stochastic diffusion, hence the dynamics of the field are essentially classical. When $\phi<\phi_{\mathrm{sto}}$ on the other hand, stochastic diffusion takes over, which means that the part of the potential where $-\phi_{\mathrm{sto}}<\phi<\phi_{\mathrm{sto}}$ is dominated by quantum diffusion.  

This is why, for classical USR to take place, one needs to impose $\phi_{\mathrm{sto}}<\Mp$, which means that the potential energy cannot be too large,
\bea
\frac{V_0}{\Mp^4}\ll 72\pi^2\left(\frac{\Mp}{\phi_0}\right)^3
\eea
(recall that $\phi_0\gg \Mp$ so this is not necessarily guaranteed). When this is the case, the number of \emph{classical} USR infationary \efolds~is given by $\Delta N_{\USR {}_{,\mathrm{class}}} = \Delta N_{\USR}(\Mp)-\Delta N_{\USR}(\phi_{\mathrm{sto}} )$, where $\Delta N_{\USR}(\Mp)$ was given in \Eq{eq:FlatInflectionPoint:NUSR:phi0GTMp}, and one obtains
\bea
\Delta N_{\USR{}_{,\mathrm{class}}} = \frac{2\delta_\uin}{3}\left(\frac{\phi_0}{\Mp}\right)^3 \left[1 - \frac{\delta_\uin}{72\pi^2}\left(\frac{\phi_0}{\Mp}\right)^3 \frac{V_0}{\Mp^4}\right]\, .
\eea
If the parameter $\phi_0$ is chosen such that $1\ll \phi_0/\Mp \ll (72\pi^2V_0/\Mp^4)^{-1/3}$, this number can still be very large and a sustained phase of classical USR inflation takes place.
\subsubsection{Case $\phi_0\ll \Mp$}
\begin{figure}
\begin{center}
\includegraphics[width=0.45\textwidth, height=6cm]{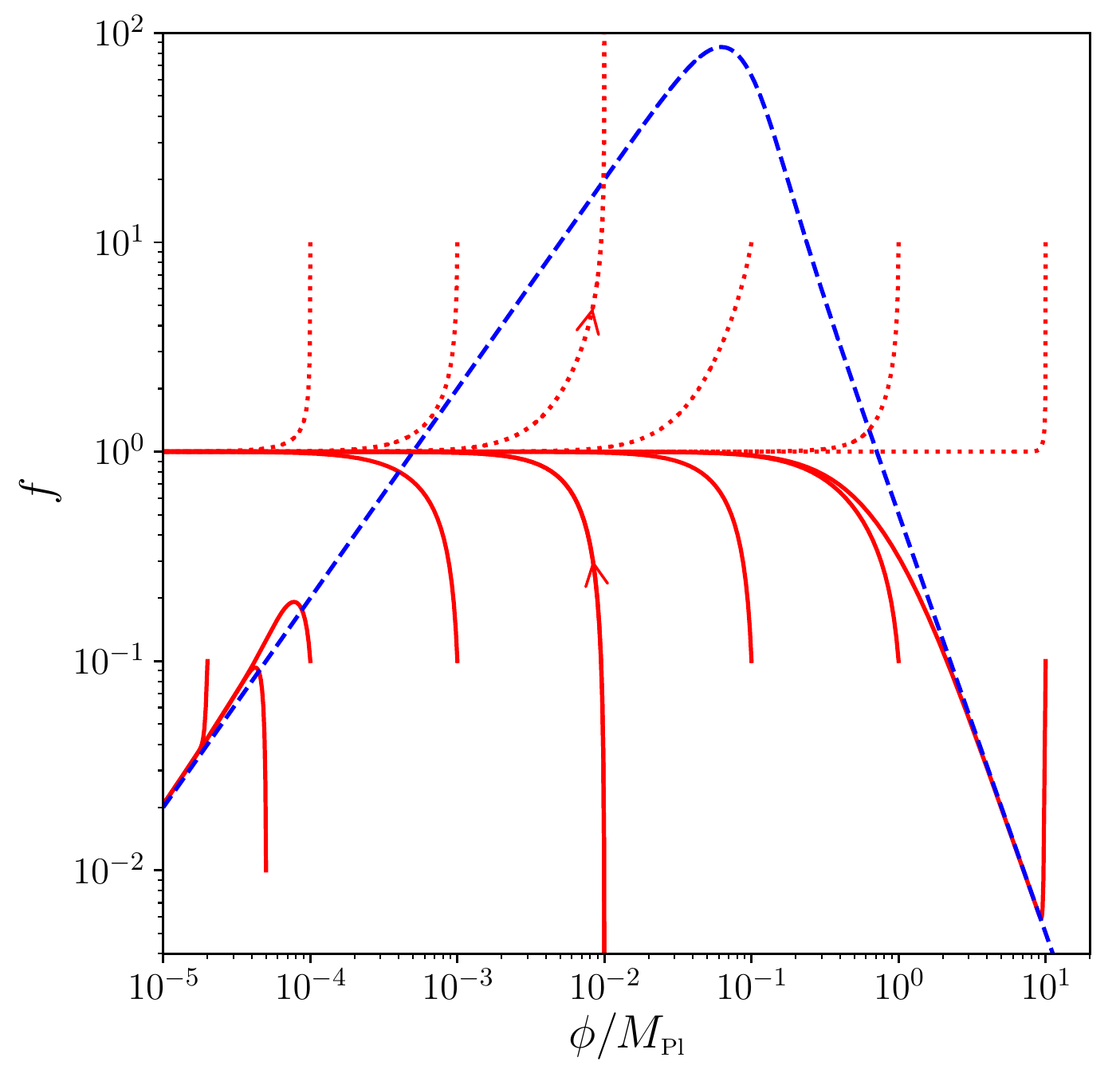} 
\includegraphics[width=0.45\textwidth, height=6.1cm]{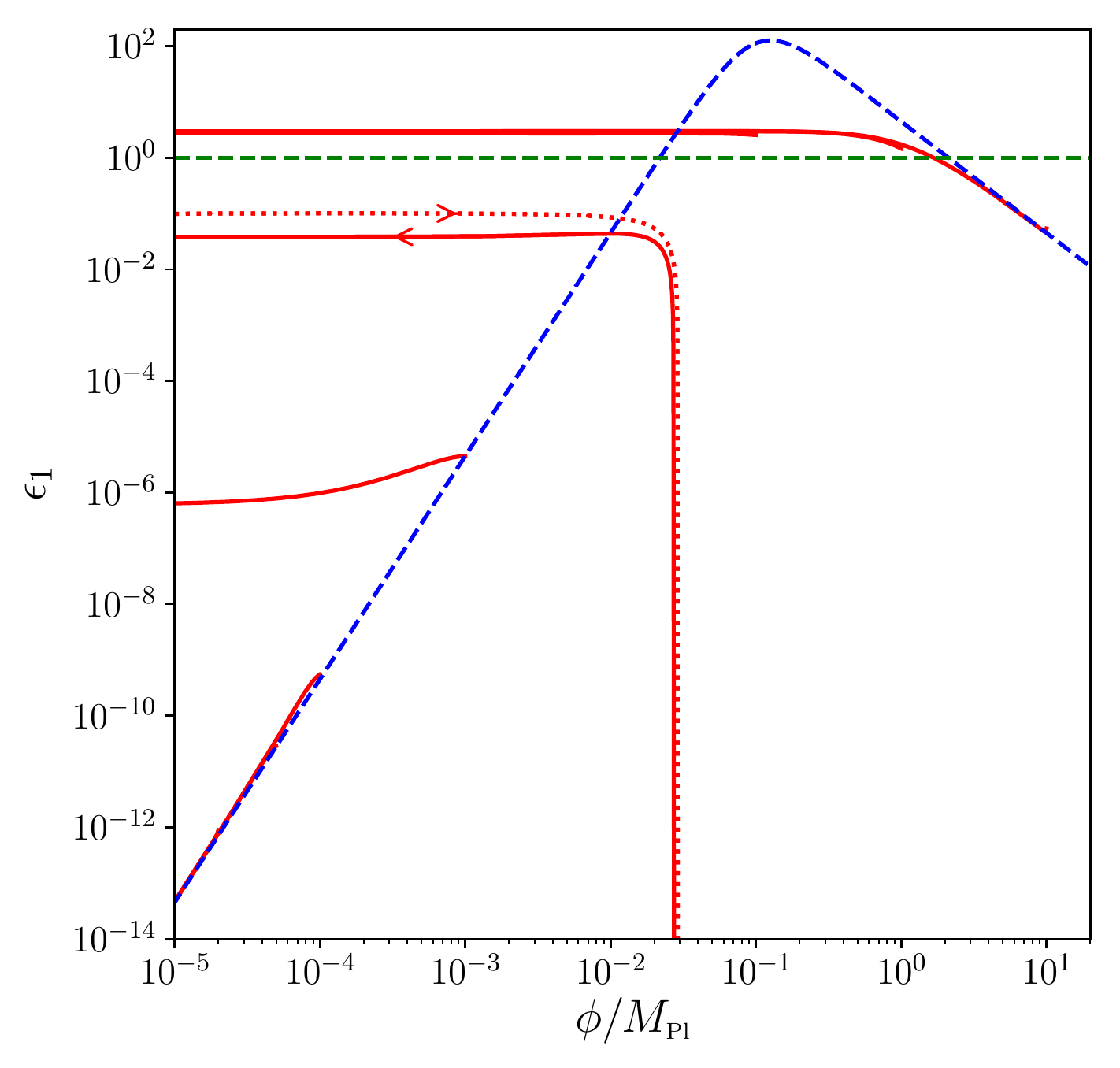} 
\caption[Field acceleration and first slow-roll parameter for the cubic inflection point model with $\phi_0=0.1\Mp$]{Left panel: Field acceleration parameter in the cubic inflection point model~(\ref{eq:pot:inflectionPoint:cubic}) as a function of the field value, for $\phi_0=0.1\Mp$ and $V_0=4.2\times 10^{-11}$, with the same conventions as in \Fig{fig:InflectionPoint:fsol:phi0GTMp}. The solid part of the red curves correspond to when $f<1$ and $\phi$ decreases with time, while the dotted parts are for $f>1$ and $\phi$ increases (as indicated by the arrows). Right panel: first Hubble-flow parameter $\epsilon_1$ as a function of the field value for the same solid trajectories and the dotted trajectory with an arrow in the left panel. The dashed green line stands for $\epsilon_1=1$ below which inflation proceeds. The trajectories that have both $\epsilon_1 \ll 1$ and $f \to 1$ drive a phase of USR inflation. 
}
\label{fig:InflectionPoint:fsol:phi0LTMp}
\end{center}
\end{figure}
In this case, SR inflation can only occur at $\phi\gg\Mp$ or $\phi\ll\phi_0^3/\Mp^2$. One therefore has three regions: if $\phi\gg\Mp$, SR is the only attractor, if $\phi_0^3/\Mp^2\ll\phi\ll\Mp$, USR is the only attractor, and if $\phi\ll \phi_0^3/\Mp^2$, both SR and USR are local attractors. These three regimes can be clearly seen in the left panel of \Fig{fig:InflectionPoint:fsol:phi0LTMp}, where the same colour code as in \Fig{fig:InflectionPoint:fsol:phi0GTMp} is employed. In the right panel of \Fig{fig:InflectionPoint:fsol:phi0LTMp}, the first Hubble-flow parameter is displayed for the same trajectories. The solid curves have $f<1$ for which $\phi$ decreases with time and the dotted curves have $f>1$ for which $\phi$ increases with time. We shall now discuss each of these three regimes in more detail.

Firstly, if one starts with an initial value of $\phi$ that is super Planckian, one quickly reaches the SR attractor. Then when $\phi$ becomes of order $\Mp$, $f_\SR$ becomes of order one which signals the breakdown of SR and one leaves the SR line to settle down to $f\simeq 1$, \ie in the USR regime. However, as can be seen on the right panel of \Fig{fig:InflectionPoint:fsol:phi0LTMp}, the first Hubble-flow parameter converges towards $\epsilon_1\simeq 3$, so inflation stops around $\phi\simeq\Mp$ and does not resume afterwards. In this case, for $\phi<\Mp$, we have USR but not USR inflation, and this non-inflating USR regime is stable due to the considerations of footnote~\ref{footnote:USR:Non:Inflation}.

Secondly, if one starts with an initial field value between $\phi_0^3/\Mp^2$ and $\Mp$ and with $\dot\phi<0$ (rolling down the potential), the field converges towards USR, since it is the only stable solution. This is the case for the trajectory with $f<1$ on which an arrow has been added in \Fig{fig:InflectionPoint:fsol:phi0LTMp}. Let us recall that the dotted part of the trajectory corresponds to $f>1$ and the inflaton climbs up its potential ($\dot\phi>0$), until its velocity changes vanishes at which point $f$ diverges and $\dot\phi$ changes sign. The inflaton then rolls down its potential starting from very negative values for $f$ (solid part of the curve) and quickly reaches USR. On the right panel of \Fig{fig:InflectionPoint:fsol:phi0LTMp} one can see that $\epsilon_1$ is roughly constant in the rolling down phase, which is consistent with \Eq{eq:USR:stable:eps1:appr} since the potential is dominated by its constant piece $V\simeq V_0$ when $\phi\ll\phi_0$.
\begin{figure}
\begin{center}
\includegraphics[width=0.65\textwidth]{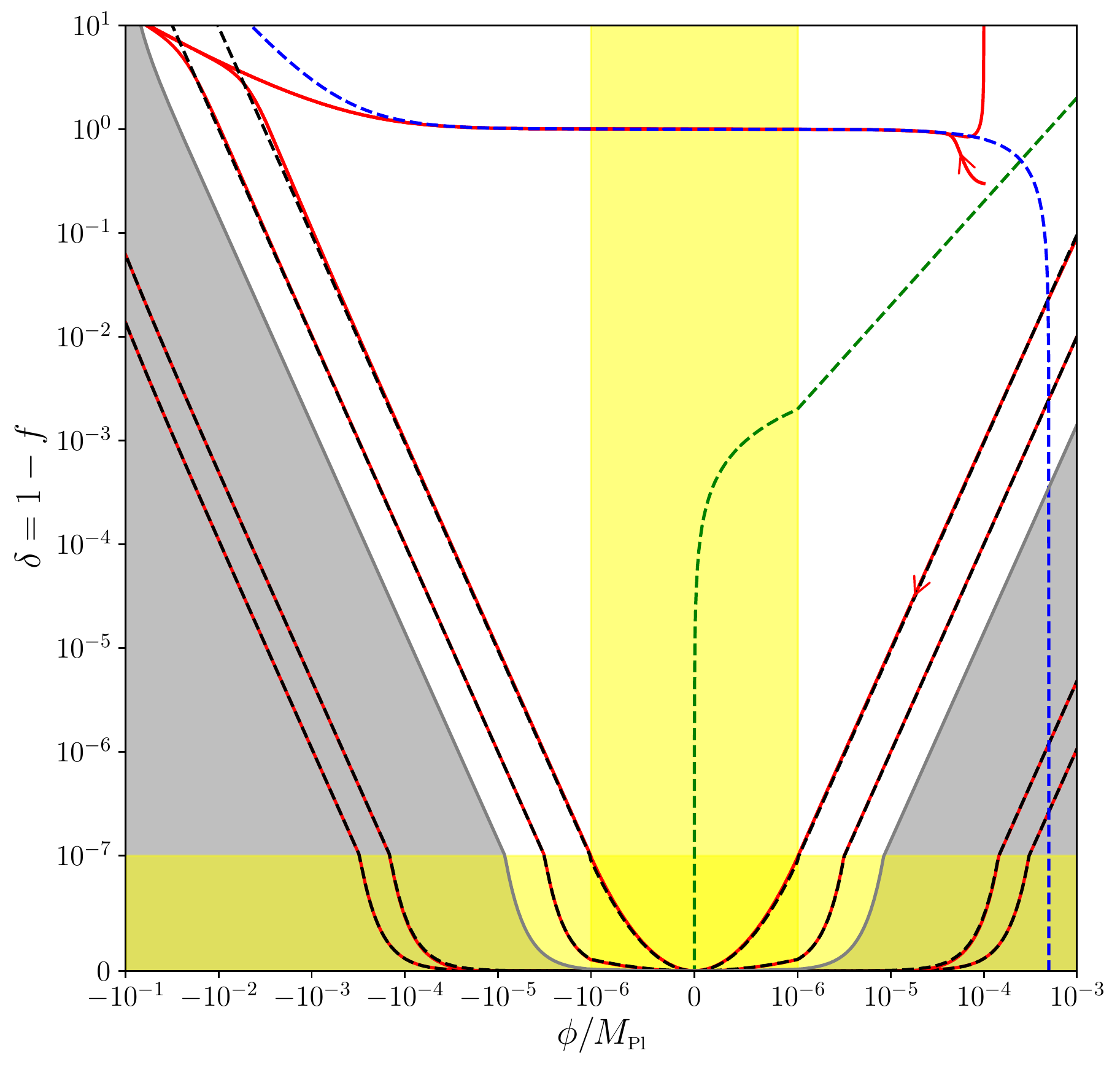} 
\caption[Field acceleration for the cubic inflection point model for the parameter range $-\phi_0<\phi<\phi_0^3/\Mp^2$]{Same as in the left panel of \Fig{fig:InflectionPoint:fsol:phi0LTMp} for the region $-\phi_0<\phi<\phi_0^3/\Mp^2$. Yellow shading denotes regions when the axis scale is linear rather than logarithmic, and the grey shaded region is where inflation in not happening $\epsilon_1>1$, as given by \eqref{eq:inflation:condition}. The black dashed lines stand for the analytical USR inflation approximation~(\ref{eq:delta:sol}) in the inflating part, and to the USR \emph{non}-inflation approximation~(\ref{eq:delta:sol:USRnoninflating}) in the non-inflating part (the field excursion being sub Planckian, the two behaviours are very much similar). The dashed green line stands for $\delta=\vert\eta_V\vert/3$, which in the $\phi>0$ region corresponds to the boundary between the SR and the USR basins of attraction. For $\phi<0$, only SR is an attractor which explains why those trajectories that reach the USR attractor in the $\phi>0$ region have $\delta$ increasing with time in the $\phi<0$ region where USR is unstable. 
}
\label{fig:InflectionPoint:fsol:phi0LTMp:phiLTphi0power3}
\end{center}
\end{figure}

Lastly, if one starts with $\phi\ll \phi_0^3/\Mp^2$, one either reaches the SR attractor if $|\delta|>|\eta_V|$ or the USR attractor if $|\delta|<|\eta_V|$. This can be more clearly seen in \Fig{fig:InflectionPoint:fsol:phi0LTMp:phiLTphi0power3}, where the whole region $-\phi_0<\phi<\phi_0^3/\Mp^2$ is displayed. One can check that the inflating USR approximation~(\ref{eq:delta:sol}) provides a good approximation to the numerical solutions of \Eq{eq:f:dynamical} in the inflating part of phase space for those trajectories that reach the USR attractor, while the \emph{non}-inflating USR approximation~(\ref{eq:delta:sol:USRnoninflating}) correctly describes the non-inflating trajectories (in the grey shaded region of the plot). Here, because the field excursion is sub-Planckian (since $\phi_0\ll\Mp$), these two behaviours are almost identical. As in the right panel of \Fig{fig:InflectionPoint:fsol:phi0GTMp}, one can also check that the line $\vert\delta\vert\sim\vert\eta_V\vert$ correctly delimitates the boundary between the two basins of attraction when $\phi>0$. 
If $\phi<0$, USR becomes unstable and only SR remains as an attractor, which explains why $\delta$ increases with time for those trajectories that reached the USR attractor before crossing the flat inflection point. However, one should note that those trajectories do not have time to reach the SR attractor before the potential becomes too steep and SR is violated.
Similarly, for $\phi > \phi_0^3/\Mp^2$ the potential is too steep and SR is not a valid approximation.

Let us also estimate the number of \efolds~that is typically realised in the USR inflating regime. USR is stable in the range~(\ref{eq:InflectionPoint:USR:stability:Condition}). However, for $\phi>\phi_0$, the potential is not dominated by its constant piece so $\epsilon_1$ can substantially increase because of \Eq{eq:USR:stable:eps1:appr}. Whether or not USR \emph{inflation} is maintained depends on the initial value of $\epsilon_1$ (see the right panel of \Fig{fig:InflectionPoint:fsol:phi0LTMp}) and to avoid this initial condition dependence, let us consider the case where we start USR inflation around $\phi\sim\phi_0$. Starting with $\delta=\delta_\uin$, \Eq{eq:delta:sol} implies that $\delta$ goes back to its initial value $\delta_\uin$ at around $\phi\sim -\phi_0$.  Making use of \Eq{eq:DeltaN:USR}, this gives rise to
\bea
\Delta N_\USR \simeq \frac{2\delta_\uin}{3}\left(\frac{\phi_0}{\Mp}\right)^2\, .
\eea
This shows that, in the regime $\phi_0\ll\Mp$, the number of \efolds~realised in the USR regime is necessarily small, contrary to the case $\phi_0\gg\Mp$, see \Eq{eq:FlatInflectionPoint:NUSR:phi0GTMp}.
\begin{figure}
\begin{center}
\includegraphics[width=0.49\textwidth]{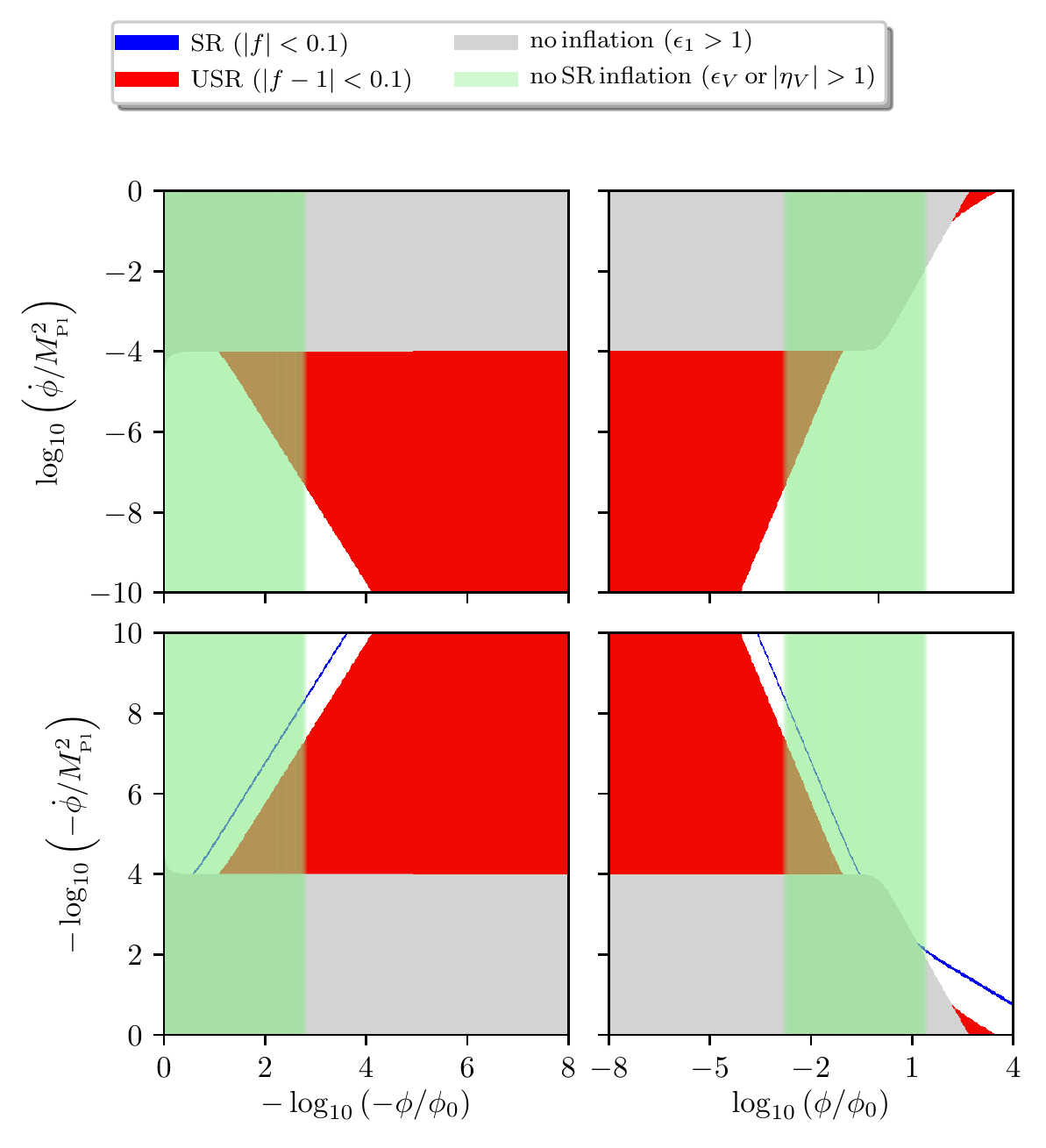} 
\includegraphics[width=0.49\textwidth]{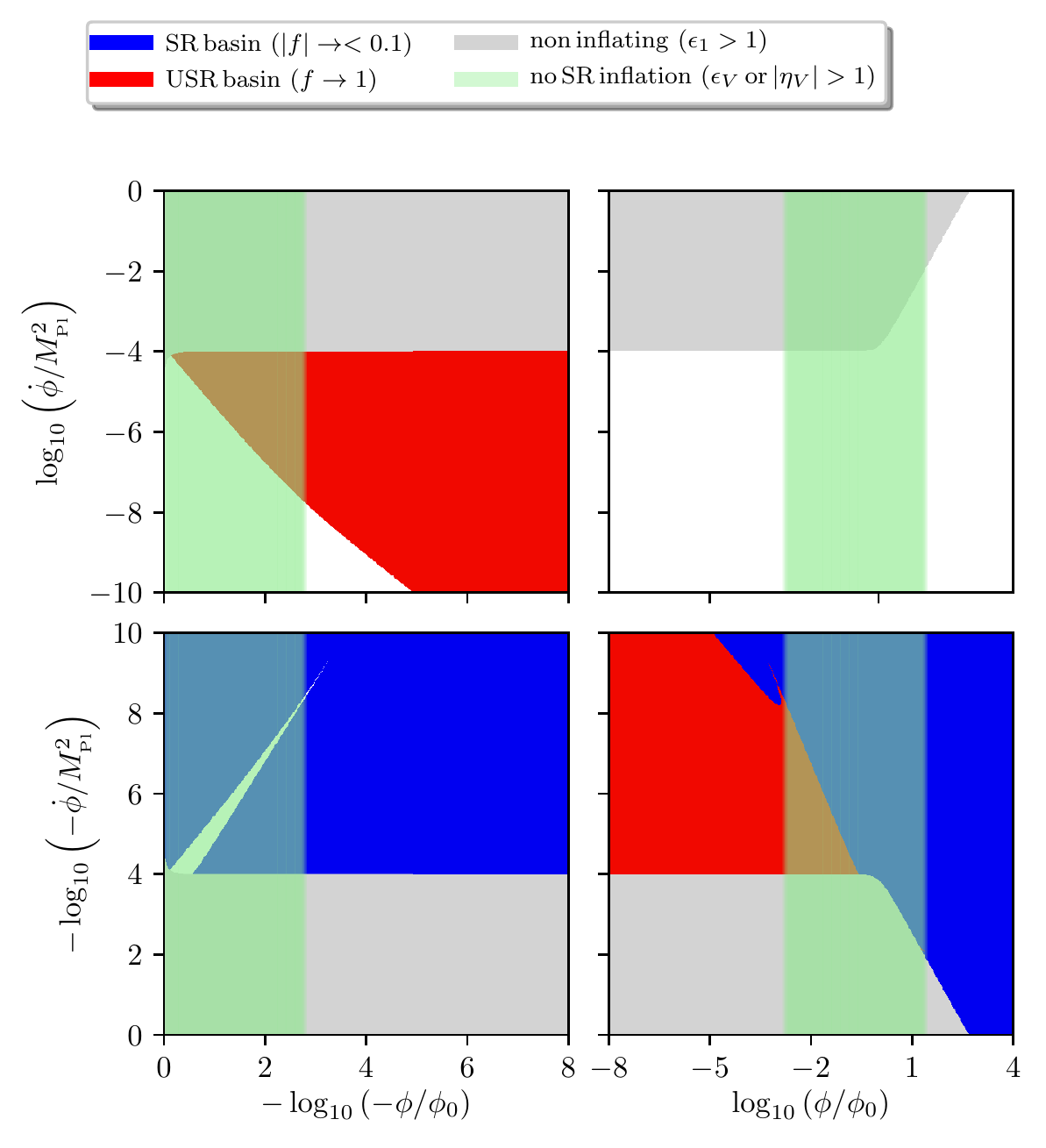} 
\caption[Existence of and basins of attraction for slow roll and ultra-slow roll-regimes in the cubic inflection point model for $\phi_0=0.1\Mp$]{Regions in phase space for the cubic inflection point model~(\ref{eq:pot:inflectionPoint:cubic}) with $V_0=4.2\times 10^{-11}$ and $\phi_0=0.1\Mp$ where SR and USR solutions exist (left panel), and (right panel) basins of attraction for SR ($f<1$ and $|f|$ decreasing) and USR ($|1-f|$ decreasing).}
\label{fig:phasespace_phi0eq0p1}
\end{center}
\end{figure}

\begin{figure}
\begin{center}
\includegraphics[width=0.49\textwidth]{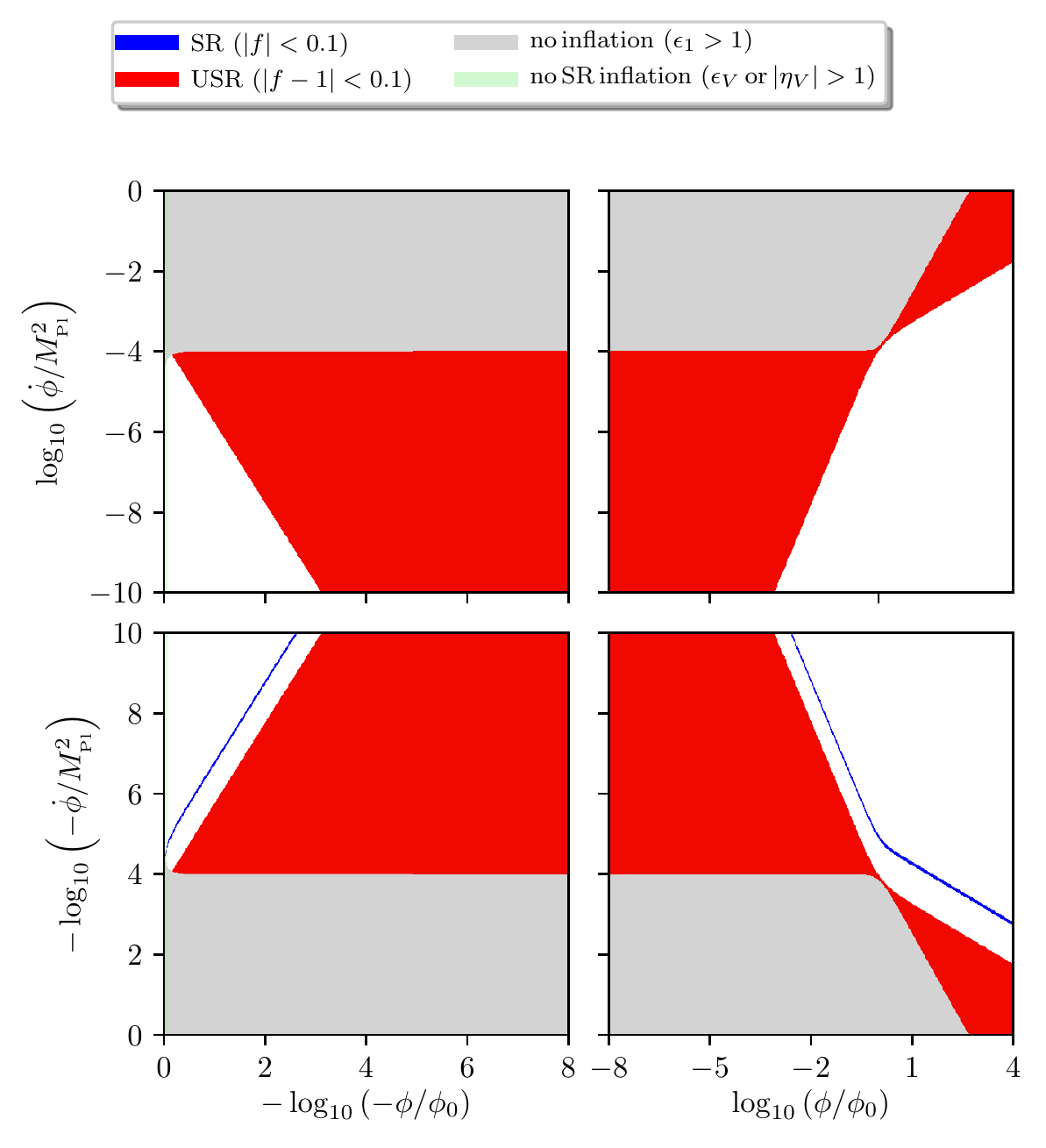} 
\includegraphics[width=0.49\textwidth]{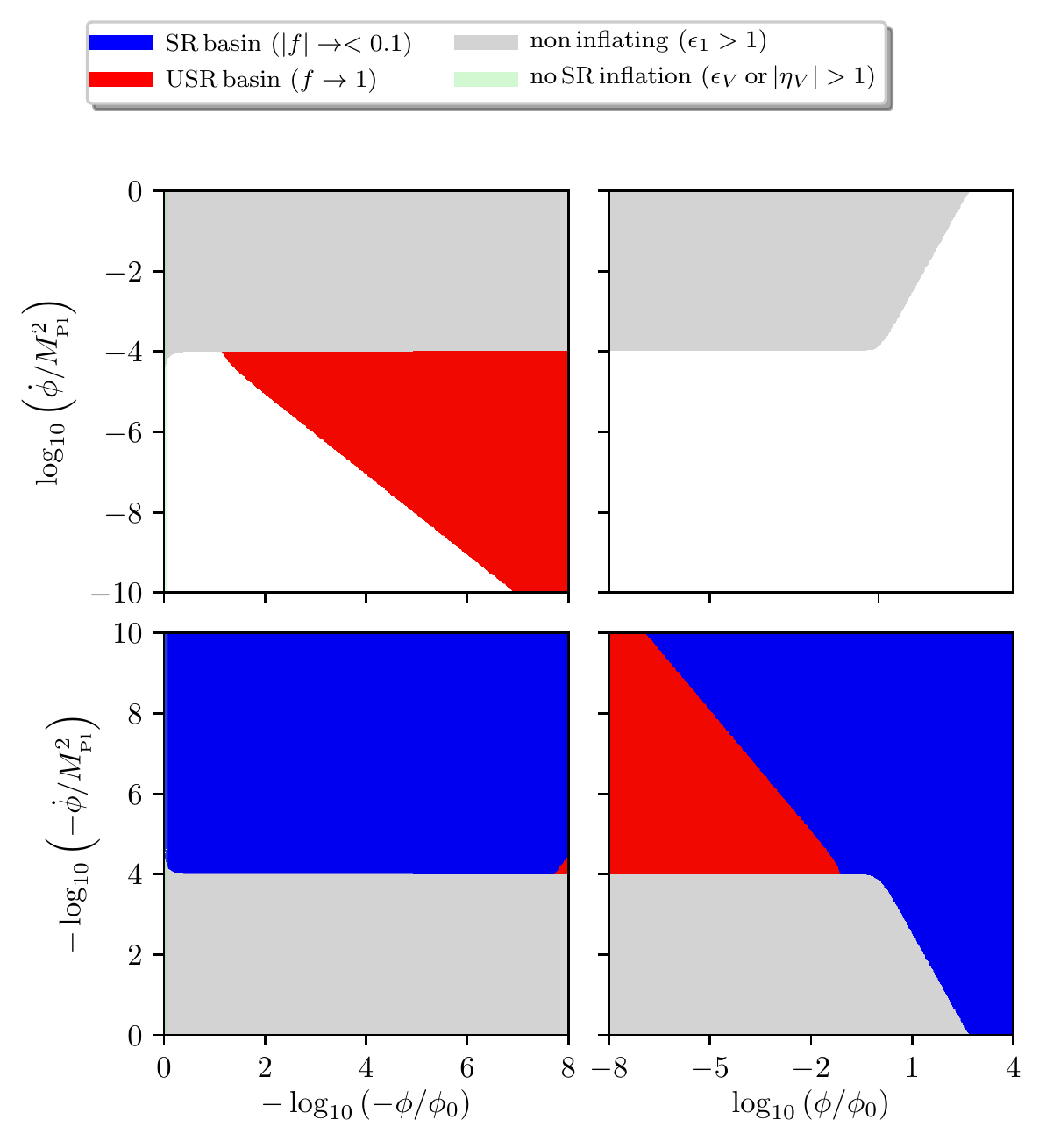} 
\caption[Existence of and basins of attraction for slow roll and ultra-slow-roll regimes in the cubic inflection point model for $\phi_0=10\Mp$]{Regions in phase space for the cubic inflection point model~(\ref{eq:pot:inflectionPoint:cubic}) with $V_0=4.2\times 10^{-11}$ and $\phi_0=10\Mp$ where SR and USR solutions exist (left panel), and regions of stability for SR and USR (right panel).}
\label{fig:phasespace_phi0e10}
\end{center}
\end{figure}
Finally we plot in \Figs{fig:phasespace_phi0eq0p1} and~\ref{fig:phasespace_phi0e10} the phase space $(\phi,\dot\phi)$ for the cubic inflection point model~(\ref{eq:pot:inflectionPoint:cubic}), for $\phi_0=0.1\Mp$ and $\phi_0=10\Mp$ respectively. In the left panels, the blue region corresponds to SR solutions (defined as $|f|<0.1$) and the red region to USR (defined as $|1-f|<0.1$). In the right panels we show the basins of attraction of SR and USR, defined by the behaviour of $f$.

One can see that SR corresponds to a thin line in phase space while USR spans a larger region. This is due to the fact that in USR inflation, there is no unique USR trajectory in phase space and solutions retain a dependence on initial conditions as can be seen \eg in \Eq{eq:delta:sol}. This is not the case for SR that singles out a unique phase-space trajectory, see \Eq{eq:f:SR}. Note also that SR solutions only exist in the quadrants where the field velocity is aligned with the potential gradient while USR exists in every quadrant.

The right-hand plots show the basins of attraction of SR (\ie where $f<1$ and $|f|$ decreases) and USR (where $|1-f|$ decreases). When $\phi>0$, if the field goes up the potential ($\dot{\phi} >0$) then  USR is unstable, and there is no SR regime and hence no SR basin of attraction either. If the field rolls down the potential ($\dot{\phi}<0$), when $\phi\gg \Mp$ or $\phi<0$ we see that only SR is an attractor as discussed above, and when $0<\phi\ll\Mp$ both SR and USR can be attractors. When $\phi<0$ and the field goes up the potential, USR is an attractor in some region of the phase space.
This corresponds to initial conditions where the field arrives at the inflection point with an almost vanishing velocity and inflates in the USR regime.

\section{Summary}
\label{sec:conclusions}

In this chapter we have shown that, contrary to what is sometimes suggested in the literature, ultra-slow-roll inflation can be a classical stable attractor in phase space.
We have seen that ultra-slow-roll inflation ($|\delta|\ll1$) is stable for a scalar field rolling down a convex potential ($\dot{V}<0$ and $V_{,\phi\phi}>0$) if the condition~(\ref{eq:USR:stable:criterion}) is fulfilled, which in terms of the potential function $V(\phi)$ reads
\bea
\label{eq:USR:stable:criterion:conclusion}
\Mp V_{,\phi\phi}>\left\vert V_{,\phi} \right\vert\, .
\eea
Conversely, standard slow roll ($|f|\ll1$) is always an attractor whenever the slow-roll consistency conditions \Eq{eq:sr:consistency} are satisfied.

In practice, however, in models with a long-lived USR epoch, the classical description may eventually break down when quantum fluctuations become more efficient at driving the field than its (decreasing) residual velocity. 
We have estimated when this happens, and shown that classical USR can still be long lived. 
When the quantum and classical evolutions become comparable then the resulting primordial density perturbations after inflation become large, which, as discussed previously, can lead to the formation of primordial black holes \cite{Hawking:1971ei,Carr:1975qj, GarciaBellido:1996qt} (see ~\cite{Ezquiaga:2018gbw, Ozsoy:2018flq} for explicit models similar to the inflection point potential discussed in this chapter). 
In such cases we need a non-perturbative formalism to describe the cosmological evolution on large scales, such as stochastic inflation. 
In the next chapter, we will review and develop the formalism of stochastic inflation, before going on to apply this formalism to both slow roll and ultra-slow roll, in chapters \ref{chapter:quantumdiff:slowroll} and \ref{chapter:USRstochastic}, respectively.


\newpage

\chapter{Stochastic inflation} \label{chapter:stochastic:intro}

We have seen in \Sec{sec:intro:infl:perturbations} how deviations from homogeneity and isotropy can be studied in the standard, perturbative description of inflation. 
In this chapter we describe, and develop new results for, the stochastic formalism for inflation~\cite{Starobinsky:1986fx, Nambu:1987ef, Nambu:1988je, PhysRevD.39.2245, Nakao:1988yi, Nambu:1989uf, Mollerach:1990zf, Linde:1993xx, Starobinsky:1994bd}, which goes beyond the standard treatment by also including the non-linear backreaction of perturbations on the background dynamics. 
The stochastic formalism is an effective field theory for the large-scale modes of the inflaton field, which is coarse-grained at scales larger than the Hubble radius, while sub-Hubble degrees of freedom are integrated out and treated as a classical, stochastic noise. 

One can intuitively understand the derivation of this theory as simply adding a stochastic noise term to the classical equations of motion. 
For example, in slow roll, the classical equation $\dd\phi/\dd N = -V_{,\phi}/(3H^2)$ becomes a Langevin equation of the form
\bea
\frac{\dd \phi}{\dd N} = -\frac{V_{,\phi}}{3H^2} + \frac{H}{2\pi}\xi\left( N \right) \, .
\eea
The right-hand side of this equation has two terms, the first of which involves $V_{,\phi}$ and is a classical drift term, and the second term involves $\xi$ which is a Gaussian white noise such that $\left\langle \xi \left( N \right) \right\rangle = 0$ and $\left\langle \xi \left( N \right) \xi \left( N' \right)\right\rangle = \delta\left( N - N' \right)$, and which makes the dynamics stochastic.
Indeed, when light fields are coarse grained at a fixed, non-expanding, physical scale that is larger than the Hubble radius during the whole period of inflation, one can show that their dynamics indeed become classical and stochastic. 


Parts of this chapter are based on Ref. \cite{Pattison:2019hef}, and this chapter is organised as follows: 
In \Sec{sec:stochastic}, we derive the general Langevin equations of a field and its momentum in stochastic inflation, before describing the requirements of this stochastic formalism to be valid in \Sec{sec:stochastic:requirements}. 
Next, in \Sec{sec:stochbeyondSR:paper}, we explicitly show that these requirements are valid in physical cases, and demonstrate this in three examples: slow roll, ultra-slow roll, and Starobinsky's piece-wise linear model for inflation. 
In \Sec{sec:stochasticdeltaN}, we explain the $\delta N$ formalism, which can be used to calculate perturbations during inflation while taking into account the non-perturbative backreaction effects of quantum fluctuations. 
Finally, in \Sec{sec:firstpassage}, we use the techniques of ``first passage time analysis'' to calculate the first few moments of the probability distribution of curvature fluctuations when stochastic effects are included. 
In this final section, the calculations are done in the framework of DBI inflation, which is motivated from string theory and includes non-canonical kinetic terms, and this is a new calculation in the literature (although the canonical calculation is easily recovered).

\section{Deriving the Langevin equations} \label{sec:stochastic}
%


Recall that, in this thesis, we consider a single inflaton field $\phi$ with potential $V(\phi)$ for simplicity, but our results can easily be extended to multiple-field setups. 
We also restrict our analysis to scalar fluctuations only, which are found from expanding about the flat FLRW line element, as shown in \Eq{eq:perturbedlineelement:FLRW}.

For now we consider a general system under no assumptions (in particular we \textit{do not} assume slow roll at this point), and hence the homogeneous background field $\phi$ and its conjugate momentum $\pi$ are two independent dynamical variables and stochastic inflation needs to be formulated in the full phase space (see also Ref. \cite{Grain:2017dqa}). 
This can be done by deriving the Hamiltonian equations\footnote{This can also be obtained from the Klein--Gordon equation written with $N$ as the time variable} from the action \eqref{eq:4d:action},
\begin{align} 
\label{eq:conjmomentum} \frac{\partial \hat{\phi}}{\partial N} &= \hat{\pi} \, ,\\
\label{eq:KG:efolds} \frac{\partial\hat{\pi}}{\partial N} &+ \left(3-\epsilon_{1}\right)\hat{\pi} + \frac{V_{,\phi}(\hat{\phi})}{H^2} = 0 \, ,
\end{align}
where we recall that $N=\ln a$ is the number of \efolds, $\epsilon_{1}$ is the first slow-roll parameter, $H\equiv \dot{a}/a$ is the Hubble parameter, a dot denotes derivatives with respect to cosmic time $t$ and a subscript $_{,\phi}$ means derivative with respect to the field $\phi$. 
At this stage, $\hat{\phi}$ and $\hat{\pi}$ are quantum operators, as stressed by the hats. 
The Hubble parameter is related to the field phase-space variables through the Friedmann equation \eqref{eq:friedmann:scalarfield}.

When linear fluctuations are added to the homogenous field and its conjugate momentum, they can be split according to 
\begin{align}
\hat{\phi} &= \hat{\bar{\phi}} + \hat{\phi}_{\mathrm{s}} \label{eq:decomp:phi}\, , \\
\hat{\pi} &= \hat{\bar{\pi}} + \hat{\pi}_{\mathrm{s}} \label{eq:decomp:pi}\, ,
\end{align}
where $\hat{\phi}_{\mathrm{s}}$ and $\hat{\pi}_{\mathrm{s}}$ are the short-wavelength parts of the fields that can be written as a Fourier decomposition 
\begin{align}
    \hat{\phi}_{\mathrm{s}} &= \int_{\mathbb{R}^3}\frac{\dd \bm{k}}{\left(2\pi\right)^{\frac{3}{2}}}W\left( \frac{k}{\sigma aH}\right) \left[ \ee^{-i\bm{k}\cdot\bm{x}}\phi_{\bm{k}}(N)\hat{a}_{\bm{k}} + \ee^{i\bm{k}\cdot\bm{x}}\phi_{\bm{k}}^{*}(N)\hat{a}^{\dagger}_{\bm{k}}  \right] \, ,\\
        \hat{\pi}_{\mathrm{s}} &= \int_{\mathbb{R}^3}\frac{\dd \bm{k}}{\left(2\pi\right)^{\frac{3}{2}}}W\left( \frac{k}{\sigma aH}\right) \left[ \ee^{-i\bm{k}\cdot\bm{x}}\pi_{\bm{k}}(N)\hat{a}_{\bm{k}} + \ee^{i\bm{k}\cdot\bm{x}}\pi_{\bm{k}}^{*}(N)\hat{a}^{\dagger}_{\bm{k}}  \right] \, .
\end{align}
In these expressions, $\hat{a}^{\dagger}_{\bm{k}}$ and $\hat{a}_{\bm{k}}$ are creation and annihilation operators, $\phi_{\bm{k}}$ and $\pi_{\bm{k}}$ are field fluctuations with wavenumber $\bm{k}$, and $W$ is a window function that selects out modes such that $k/(\sigma aH) > 1$, where $\sigma\ll 1$ is the coarse-graining parameter. 
Specifically, $W$ is such that $W\simeq 0$ for $k/(\sigma aH) \ll 1$ and $W\simeq1$ for $k/(\sigma aH) \gg 1$.
For example, the simplest choice of window function, and the one we shall take later on, is a Heaviside function $W(x) = \theta(x-1)$, where $x=k/(\sigma aH)$.

The coarse-grained fields $\bar{\phi}$ and $\bar{\pi}$ thus contain all wavelengths that are much larger than the Hubble radius, $k< \sigma aH$. 
They stand for the local background values of the fields, which are continuously perturbed by the small wavelength modes as they emerge from $\hat{\phi}_{\mathrm{s}}$ and $ \hat{\pi}_{\mathrm{s}}$ and cross the coarse-graining radius. 
The perturbation of the coarse-grained field by the small-scale modes leaving the Hubble radius comes from the time dependence of the argument of the window function $W$, and we do not consider any couplings between super-Hubble and sub-Hubble modes (couplings such as this would, however, need to be taken into account beyond linear order in perturbation theory).
Our aim is to find Langevin equations (which are effective equations of motions with the small-wavelength parts of the fields $\hat{\phi}_{\mathrm{s}}$ and $\hat{\pi}_{\mathrm{s}}$ integrated out) for these long-wavelength parts of the quantum fields.

Inserting the decompositions \eqref{eq:decomp:phi} and \eqref{eq:decomp:pi} into the classical equations of motion \eqref{eq:conjmomentum} and \eqref{eq:KG:efolds}, to linear order in the short-wavelength parts of the fields, the equations for the long-wavelength parts become
\begin{align}
\frac{\partial \hat{\bar{\phi}}}{\partial N} &= \hat{\bar{\pi}} + \hat{\xi}_{\phi}(N) \label{eq:conjmomentum:langevin:quantum} \, ,\\
\frac{\partial \hat{\bar{\pi}}}{\partial N} &= -\left(3-\epsilon_{1}\right)\hat{\bar{\pi}} - \frac{V_{,\phi}(\hat{\bar{\phi}})}{H^2} +\hat{\xi}_{\pi}(N) \label{eq:KG:efolds:langevin:quantum} \, ,
\end{align}
where the source functions $\hat{\xi}_{\phi}$ and $\hat{\xi}_{\pi}$ are given by
\begin{align}
    \hat{\xi}_{\phi} &= -\int_{\mathbb{R}^3}\frac{\dd \bm{k}}{\left(2\pi\right)^{\frac{3}{2}}}\frac{\dd W}{\dd N}\left( \frac{k}{\sigma aH}\right) \left[ \ee^{-i\bm{k}\cdot\bm{x}}\phi_{\bm{k}}(N)\hat{a}_{\bm{k}} + \ee^{i\bm{k}\cdot\bm{x}}\phi_{\bm{k}}^{*}(N)\hat{a}^{\dagger}_{\bm{k}}  \right]\, , \\
    \hat{\xi}_{\pi} &= -\int_{\mathbb{R}^3}\frac{\dd \bm{k}}{\left(2\pi\right)^{\frac{3}{2}}}\frac{\dd W}{\dd N}\left( \frac{k}{\sigma aH}\right) \left[ \ee^{-i\bm{k}\cdot\bm{x}}\pi_{\bm{k}}(N)\hat{a}_{\bm{k}} + \ee^{i\bm{k}\cdot\bm{x}}\pi_{\bm{k}}^{*}(N)\hat{a}^{\dagger}_{\bm{k}}  \right] \, .
\end{align}
Note that in the stochastic formalism, background quantities such as $H$ and $\epsilon_1$ are only evaluated as function of the coarse-grained fields (eg through the Friedmann equation) and not the short wavelength perturbations, which means that these short wavelength modes are taken to be ``test perturbations''.

Now, if the window function is taken to be a Heaviside function, then the two-point correlation functions of the sources are given by \cite{Grain:2017dqa} 
\bea 
\langle 0| \hat{\xi}_{\phi}(N_1)\hat{\xi}_{\phi}(N_2)|0\rangle & 
\kern-0.1em = \kern-0.1em
 \frac{1}{6\pi^2}\frac{\dd k_\sigma^3(N)}{\dd N}\bigg|_{N_1}\left\vert\phi_{\bm{k}_\sigma}(N_1) \right\vert^2 \delta\left(N_1 - N_2\right) \kern-0.1em ,\\
\langle 0| \hat{\xi}_{\pi}(N_1) \hat{\xi}_{\pi}(N_2)|0\rangle &
\kern-0.1em = \kern-0.1em
\frac{1}{6\pi^2}\frac{\dd k_\sigma^3(N)}{\dd N}\bigg|_{N_1}\left\vert \pi_{\bm{k}_\sigma}(N_1) \right\vert^2 \delta\left(N_1 - N_2\right) \kern-0.1em, \\
\langle 0| \hat{\xi}_{\phi}(N_1)\hat{\xi}_{\pi}(N_2)|0\rangle &
\kern-0.1em = \kern-0.1em \langle 0| \hat{\xi}_{\pi}(N_1) \hat{\xi}_{\phi}(N_2)|0\rangle^* 
\kern-0.1em = \kern-0.1em
 \frac{1}{6\pi^2}\frac{\dd k_\sigma^3(N)}{\dd N}\bigg|_{N_1}\phi_{\bm{k}_\sigma}(N_1) \pi^*_{\bm{k}_\sigma}(N_1)  \delta\left(N_1 \kern-0.1em - \kern-0.1em N_2\right)  \kern-0.1em,
 \label{eq:noise:correlators}
\eea 
where $k_\sigma \equiv \sigma a H$ is the comoving coarse-graining scale. 
In order to evaluate these correlation functions more explicitly, one needs to calculate the amplitudes of the field fluctuations.
This leads us to note that the specific gauge in which one solves for the field fluctuations is now important, as the amplitude of the fluctuations is a gauge-dependent quantity. 
This point is discussed in detail in \Sec{sec:uniformexpansion}. 

The important step is now to view the source functions as random Gaussian noises rather than quantum operators\footnote{The mapping between $\xi_\phi$ and $\xi_\pi$ (which are, strictly speaking, quantum operators) and their classical noise counterparts is found by identifying their quantum expectation values with stochastic moments, that is by imposing the relationship 
\bea 
\left< \xi_i^p( N_1) \xi_j^q( N_2) \right> = \left< 0 \vert T\left[\xi_i^p(N_1) \xi_j^q(N_2)\right] \vert 0 \right> \, ,
\eea
where $i$ and $j$ stand for $\phi$ or $\pi$, $p$ and $q$ are natural numbers, and where the $T$-product is the time ordering product. 
Note that in the left hand side of the above, the angle brackets stand for stochastic average while in the right hand side, they denote the bra and the ket of the vacuum state $\vert 0 \rangle$.}, correlated according to \Eqs{eq:noise:correlators}, and to interpret \Eqs{eq:conjmomentum:langevin:quantum} and~(\ref{eq:KG:efolds:langevin:quantum}) as stochastic Langevin equations for the random field variables $\bar{\phi}$ and $\bar{\pi}$, 
\begin{align}
\frac{\partial {\bar{\phi}}}{\partial N} &= {\bar{\pi}} + {\xi}_{\phi}(N) \label{eq:conjmomentum:langevin} \, ,\\
\frac{\partial {\bar{\pi}}}{\partial N} &= -\left(3-\epsilon_{1}\right){\bar{\pi}} - \frac{V_{,\phi}({\bar{\phi}})}{H^2(\bar{\phi},\bar{\pi})} +{\xi}_{\pi}(N) \label{eq:KG:efolds:langevin} \, ,
\end{align}
where we have removed the hats to stress that we now work with stochastic quantities rather than quantum operators. 
These are the general Langevin equations for stochastic inflation. 

Before we discuss the general criteria that one must satisfy in order to use the stochastic formalism, let us make a few notes about the formalism described here. 

Firstly, due to the presence of the term $\delta(N_1-N_2)$ in each correlator in \eqref{eq:noise:correlators}, we see that the noise terms $\hat{\xi}_{\phi}$ and $\hat{\xi}_{\pi}$ are ``white noises'', meaning that their value at time $N_1$ is uncorrelated with their value at any other time $N_2$. 
This means that they describe a Markovian process, and the noises have no ``memory'' of their history. 
However, this property is a direct consequence of our choice of the Heaviside window function. 
If we were to choose a smooth window function, we would find ``coloured noises'', which are non-Markovian and their values at different times are correlated \cite{Casini:1998wr,Matarrese:2003ye,Liguori:2004fa}. 
While this may have physical implications, because a coloured noise can provide an additional source of non-Gaussianity \cite{Hu:1992ig} and affect the shape of the power spectrum \cite{Liguori:2004fa}, we continue with a Heaviside window function for simplicity. 

Secondly, in order to obtain the noise correlators \eqref{eq:noise:correlators}, we have evaluated them at the same point $\bm{x}$ in space, so there is no $\bm{x}$-dependence in these expression. 
This is done both for simplicity of the expressions given, and because this is the case of physical interest for perturbations in stochastic inflation. 

Finally, let us demonstrate how to calculate the noise correlators in some specific cases. 
Here, we will focus on $\left< \xi_\phi \xi_\phi \right>$ in \eqref{eq:noise:correlators}, and simply note that the calculation of the other moments follows in the same way. 
Let us first consider the case of a de Sitter background. 
To calculate the auto-correlation of $\xi_\phi$, we need to calculate the field fluctuation $\phi_{\bm{k}}$, which in the notation of \Eq{eq:def:Q} is $\delta\phi_{\bm{k}}$, by solving the Sasaki--Mukhanov equation \eqref{eq:vk:sasakimukhanov}. 
By implementing the Bunch--Davies initial condition, as in \Sec{sec:intro:infl:perturbations}, we find that in the super-Hubble limit $k\gg aH$ that we are interested in, the field fluctuation is given by 
\bea \label{eq:deSitter:fieldfluctuation}
(\sigma aH)^3 \vert\phi_{\bm{k}}\vert^2_{k=\sigma aH} \simeq \frac{H}{4\pi^2}\Gamma(\nu)2^{2\nu}\sigma^{3-2\nu} \, ,
\eea 
for $\sigma\ll 1$, where $\nu= \frac{3}{2}\sqrt{1-\frac{4}{9H^2}\frac{\dd^2 V}{\dd \phi^2}}$, and $\Gamma$ is the Euler Gamma function. 
Hence, we find 
\bea 
\langle \hat{\xi}_{\phi}(N_1)\hat{\xi}_{\phi}(N_2)\rangle = \left(\frac{H}{2\pi}\right)^2 \left(\frac{\sigma}{2}\right)^{3-2\nu}\frac{4\Gamma^2(\nu)}{\pi}\delta(N_1-N_2) \, .
\eea 
If we additionally impose that we are considering a light field, so $\frac{\dd^2 V}{\dd \phi^2} \ll H^2 $, then $\nu \simeq 3/2$ and 
\bea 
\langle \hat{\xi}_{\phi}(N_1)\hat{\xi}_{\phi}(N_2)\rangle = \left(\frac{H}{2\pi}\right)^2 \delta(N_1-N_2) \, ,
\eea 
along with $\xi_{\pi} = 0$.
We can thus combine the Langevin equations \eqref{eq:conjmomentum:langevin:quantum} and \eqref{eq:KG:efolds:langevin:quantum} and write them as 
\begin{align}
\frac{\partial^2 {\bar{\phi}}}{\partial N^2} &= -3\frac{\partial {\bar{\phi}}}{\partial N} - \frac{V_{,\phi}}{H^2} + 3\frac{H}{2\pi}\xi(N) \, ,
\end{align}
where we have redefined the noise term as $\xi_\phi = H/(2\pi)\xi$, where $\xi$ is a Gaussian white noise. 
Furthermore, in slow roll, we can neglect the second derivative term to obtain the Langevin equation 
\begin{align}
\frac{\partial {\bar{\phi}}}{\partial N} = -\frac{V_{,\phi}}{3 H^2} + \frac{H}{2\pi}\xi(N) \, .
\end{align}

In the following section, we shall consider the conditions required for this stochastic formalism to be a consistent approach.

\section{Requirements for the stochastic approach} \label{sec:stochastic:requirements}

In order for the stochastic approach outlined above to be valid, we require three main conditions to be met:

\begin{itemize}
\item {\em quantum-to-classical transition}

This requirement is what allows us to describe the coarse-grained field as a classical stochastic quantity, since we replace quantum operators by stochastic fields, but we note that this is a non-trivial procedure. 
For instance, stochastic variables always commute while quantum operators do not. 
From the last of \Eqs{eq:noise:correlators}, one can see that this notably implies the imaginary part of $\phi_{\bm{k}}\pi^*_{\bm{k}}$ to be negligible compared to its real part. 
During inflation, cosmological perturbations are placed in a two-mode highly squeezed state on large scales and indeed experience such a quantum-to-classical transition~\cite{Polarski:1995jg, Lesgourgues:1996jc, Kiefer:2008ku}. 
The only requirement for this to happen is the dominance of a growing mode over a decaying mode, which is guaranteed as long as perturbations get amplified outside the Hubble radius. 
Note that this requirement is always satisfied and does not rely on slow roll or any other approximation to be valid.

Furthermore, the importance of this first requirement should be taken with a grain of salt. 
Hermitian two-point functions involving the field and its conjugate momentum, or any higher-order correlator involving only one phase-space variable, can be well reproduced by a stochastic description regardless of the amount of quantum squeezing~\cite{Martin:2015qta, Grain:2017dqa}. 
There are also a class of observables called ``improper''~\cite{2005PhRvA..71b2103R}, the expectation values of which can never be reproduced by a stochastic theory, even in the large-squeezing limit (giving rise \eg to Bell inequality violations~\cite{Martin:2016tbd, Martin:2017zxs}). 

The quantum-to-classical transition is therefore a delicate concept, which however does not rely on the slow-roll approximation, and hence when we later study the stochastic formalism beyond slow roll, this condition will not hinder the use of a stochastic formalism outside the slow-roll regime. 

\item {\em separate universe approach}

Since the spatial gradients in the Langevin equations~(\ref{eq:conjmomentum:langevin}) and~(\ref{eq:KG:efolds:langevin}) are neglected, one assumes that, on super-Hubble scales, each Hubble patch evolves forward in time independently of the other patches, and under a locally FLRW metric. 
This is the so-called separate universe picture~\cite{Salopek:1990jq, Sasaki:1995aw, Wands:2000dp, Lyth:2003im, Rigopoulos:2003ak, Lyth:2005fi}, or quasi-isotropic \cite{Lifshitz:1960, Starobinsky:1982mr, Comer:1994np, Khalatnikov:2002kn} picture. 
The validity of this approximation beyond slow roll has recently been questioned in Ref. \cite{Cruces:2018cvq}, and in \Sec{sec:separateuniverse}, we will show why it is in fact still valid.

\item {\em use of the uniform-$N$ gauge}

In order to derive the Langevin equations~(\ref{eq:conjmomentum:langevin}) and~(\ref{eq:KG:efolds:langevin}), only the field variables have been perturbed according to \Eqs{eq:decomp:phi} and~(\ref{eq:decomp:pi}), and not the entries of the metric. 
In particular, the lapse function, \ie $A$ in the notation of \Eq{eq:perturbedlineelement:FLRW}, has been neglected. This implies that the Langevin equations are written in a specific gauge, namely the one where the time coordinate is fixed. 
Since we work with the number of \efolds~as the time variable, this corresponds to the uniform-$N$ gauge. 
In \Eqs{eq:noise:correlators}, the field perturbations $\phi_{\bm{k}}$ and $\pi_{\bm{k}}$ must therefore be calculated in that same gauge. 
However, it is common to compute the field perturbations in the spatially-flat gauge, since in that gauge, they are directly related to the gauge-invariant curvature perturbation, which is quantised in the Bunch-Davies vacuum. 
One must therefore compute the correction to the noise amplitude that comes from translating the field fluctuations in the spatially-flat gauge to the uniform-$N$ gauge, and this is what is done in detail in \Sec{sec:uniformexpansion}. 

Let us note that one could work with a different time coordinate, hence in a different gauge (for instance, working with cosmic time $t$ would imply working in the synchronous gauge). 
This is not a problem as long as one computes gauge-invariant quantities in the end, such as the curvature perturbation $\zeta$. However, since $\zeta$ is related to the fluctuation in the number of \efolds~in the so-called ``stochastic-$\delta N$ formalism''~\cite{Fujita:2013cna, Vennin:2015hra}, we find it convenient to work with the number of \efolds~as a time variable. 
Another reason is that, as will be shown in \Sec{sec:uniformexpansion}, in the slow-roll regime, the spatially-flat gauge coincides with the uniform-$N$ gauge (but not, say, with the synchronous gauge), which makes the gauge correction identically vanish, and which explains why it is usually recommended~\cite{Vennin:2015hra} (but not compulsory) to work with $N$ as a time variable. 
\end{itemize}

\section{Stochastic inflation beyond slow roll} \label{sec:stochbeyondSR:paper}

Having identified three main requirements for the stochastic formalism for inflation to be valid, we note it is well known that these conditions are all satisfied in the slow-roll limit of inflation, but they have not been studied in detail beyond this approximation. 
We therefore perform a systematic analysis of these conditions, and confirm that stochastic inflation is valid well beyond the slow-roll regime.

As noted above, the quantum-to-classical transition of super-Hubble fluctuations has no relation to slow roll of any other regime, and is therefore not studied any further here.
However, we demonstrate the validity of the separate universe approach to evolving long-wavelength scalar field perturbations beyond slow roll, and also calculate the gauge correction to the amplitude of the stochastic noise that is discussed above.
We show that this gauge correction vanishes in the slow-roll limit, but we also explain how to calculate them in general, and explicitly compute the correction in different cases, including ultra-slow roll and the Starobinsky model that interpolates between slow roll and ultra-slow roll, and find the corrections to be negligible in practice. 
This confirms the validity of the stochastic formalism for studying quantum backreaction effects in the very early universe beyond slow roll.

This study is motivated by recent works where situations in which non-slow-roll stochastic effects are at play have been highlighted~\cite{Garcia-Bellido:2017mdw, Germani:2017bcs, Firouzjahi:2018vet, Biagetti:2018pjj, Ezquiaga:2018gbw}. 
For instance, if the inflationary potential features a very flat section close to the end of inflation, large curvature perturbations could be produced that later collapse into primordial black holes. 
If such a flat portion exists, it may be associated with both large stochastic diffusion~\cite{Pattison:2017mbe} and deviations from slow-roll, \eg the ultra-slow-roll regime~\cite{Inoue:2001zt, Kinney:2005vj}, which we have already seen can be stable in some cases~\cite{Pattison:2018bct}. 
This explains the need for implementing the stochastic inflation programme beyond slow roll, which is the aim of this section.

\subsection{Separate universes}
\label{sec:separateuniverse}

We first consider the separate universe approach, which is valid when each causally-disconnected patch of the universe evolves independently, obeying the same field equations locally as in a homogeneous and isotropic (FLRW) cosmology. 
Combining \Eqs{eq:conjmomentum} and~(\ref{eq:KG:efolds}), it is straightforward to recover the  Klein--Gordon equation for a homogeneous field in an FLRW cosmology, $\phi(t)$, which is given by \eqref{eq:eom:scalarfield}.
In this section, we derive the equation of motion for linear fluctuations about a homogeneous scalar field from perturbations of the background FLRW equations of motion, i.e., the separate universe approach.
We show that the resultant equation of motion matches the known result from cosmological perturbation theory, which is given by \Eq{eq:pertubations}, at leading order in a spatial gradient expansion, with or without slow roll.

\subsubsection{Perturbed background equations}

In order to easily relate the field fluctuation $\delta\phi$ to the Sasaki--Mukhanov variable, one usually chooses to work in the spatially-flat gauge where $\psi=0$, and hence $Q=\delta\phi$ according to \Eq{eq:def:Q}.
Here, we will show how to perturb the background equations in that gauge (see \Sec{sec:PertBackEOM:SFG}), but also in the uniform-$N$ gauge that is used in stochastic inflation (see \Sec{sec:PertBackEOM:UNG}). 

Let us perturb the quantities appearing in \Eq{eq:eom:scalarfield}, according to 
\bea \label{eq:perturbkleingordon}
&\phi \to \phi + \delta\phi \, , &\dd t \to (1+A) \dd t \, ,
\eea 
where $A$ is the lapse function introduced in \Eq{eq:perturbedlineelement:FLRW}. Let us stress that the lapse function needs to be perturbed, otherwise one is implicitly working in a synchronous gauge (where $A=0$), which in general differs from the spatially-flat and uniform-$N$ gauges, and this leads to inconsistencies~\cite{Cruces:2018cvq}. Inserting \Eq{eq:perturbkleingordon} into \Eq{eq:eom:scalarfield} gives rise to
\bea \label{eq:perturbedKG}
\ddot{\delta\phi} + \left(3H + \frac{\dot{\phi}^2}{2\Mp^2H}\right)\dot{\delta\phi} \: + &\left(\frac{\dot{\phi}}{2\Mp^2H}V_{,\phi} + V_{,\phi\phi}\right)\delta\phi
\\
- &\dot{\phi}\dot{A} - \left(2\ddot{\phi}+3H\dot{\phi}+ \frac{\dot{\phi}^3}{2\Mp^2H}\right)A = 0 \, ,
\eea 
where we have also used 
\bea \label{eq:deltaH} 
\delta H = \frac{V_{,\phi}\delta\phi + \dot{\phi}\dot{\delta\phi} - \dot{\phi}^2A}{6\Mp^2 H}  
\eea 
that comes from perturbing the Friedmann equation \eqref{eq:friedmann} under \Eq{eq:perturbkleingordon}.

\subsubsection{Spatially-flat gauge}
\label{sec:PertBackEOM:SFG}

In the spatially-flat gauge, the lapse function can readily be rewritten in terms of the field perturbation by imposing the momentum constraint~\eqref{eq:momentumconstraint:arbgauge}, which simplifies to 
\bea
\label{eq:flatA}
A = \frac{\dot{\phi}}{2\Mp^2 H} \delta \phi \,.
\eea
Substituting this relation into \Eq{eq:perturbedKG} gives rise to 
\bea \label{eq:perturbedKG:simplified}
\ddot{\delta\phi} + 3H\dot{\delta\phi} + \left[ V_{,\phi\phi} - \frac{1}{\Mp^2a^3}\frac{\dd}{\dd t}\left(\frac{a^3}{H}\dot{\phi}^2\right) \right] \delta\phi = 0 \, .
\eea 
Comparing \Eq{eq:perturbedKG:simplified}, obtained from the perturbed background equations, with \Eq{eq:pertubations}, obtained in linear perturbation theory in the spatially-flat gauge where $Q=\delta\phi$, we see that the two are consistent in the super-Hubble limit where $k\ll a H$. 

It is important to note that the local proper time in each patch is perturbed with respect to the cosmic time, $t$, in the background in the presence of a non-zero lapse perturbation, $A$. As can be seen from \Eq{eq:flatA} the perturbation $A$ vanishes in the spatially-flat gauge in the slow-roll limit, $\dot\phi\to0$, and the local proper time in this limit coincides with the background cosmic time. Beyond slow roll one must consistently account for local variations in the proper time interval in different patches if one wants to relate the separate universe equations to the perturbation equations written in terms of a global (background) cosmic time. This will be the aim of \Sec{sec:uniformexpansion}.

\subsubsection{Uniform-$N$ gauge}
\label{sec:PertBackEOM:UNG}
Let us introduce the expansion rate of $t=$ constant hypersurfaces
\bea
\theta={n^\mu}_{;\mu}\, ,
\eea
where $n^\mu$ is the unit time-like vector, orthogonal to the constant-time hypersurfaces.
It is related to the metric perturbations in \Eq{eq:perturbedlineelement:FLRW} according to~\cite{Malik:2008im} 
\bea \label{eq:def:expansion}
\theta = \frac{3}{a}\left( \mathcal{H} - \mathcal{H}A - \psi' + \frac{1}{3}\nabla^2\sigma \right) \, ,
\eea 
where $\mathcal{H} = a'/a$ is the conformal Hubble parameter, a prime is a derivative with respect to conformal time $\eta$ defined through $\dd t = a \dd \eta$, and $\sigma = E'-B$ is the shear potential.
From the perturbed expansion rate $\theta$, one can define a perturbed integrated expansion up to first order in the metric perturbations
\bea 
\tilde{N} &= \frac{1}{3}\int \theta (1+A) \dd t 
= N - \psi + \frac{1}{3}\nabla^2\int\sigma\dd\eta \, .
\eea 
The last term in the right-hand side can be re-written in terms of $\E\equiv \int \sigma\dd\eta$, which corresponds to $E$ in the hypersurface-orthogonal threading where $B=0$. From now on, we work in such a spatial threading. This gives rise to
\bea  \label{eq:def:deltaN}
\delta N = -\psi + \frac{1}{3}\nabla^2\E \, ,
\eea 
i.e., the perturbation of the trace of the spatial metric on constant-time hypersurfaces.
Note, in particular, that in the spatially-flat gauge where $\psi=B=0$, we have $\delta N|_{\psi=0} = \frac{1}{3}\nabla^2\E |_{\psi=0}$. 

The uniform-$N$ gauge used in the Langevin equations \eqref{eq:conjmomentum:langevin} and \eqref{eq:KG:efolds:langevin} is defined by keeping the integrated expansion unperturbed across all patches of the universe, \ie $\delta N = 0$. From \Eq{eq:def:deltaN}, this imposes a direct relationship between $\psi$ and $E$, namely $\psi = \frac{1}{3}\nabla^2E_B$.
In the uniform-$N$ gauge, we note that the perturbation equation \eqref{eq:pertubations:general} can be written as
\bea 
\label{eq:pert:uniformN}
\ddot{\delta\phi_{\bm{k}}} + 3H\dot{\delta\phi_{\bm{k}}} + \left( \frac{k^2}{a^2} + V_{,\phi\phi} \right)\delta\phi_{\bm{k}} &= \dot{\phi}\dot{A_{\bm{k}}} - 2V_{,\phi}A_{\bm{k}}  \\
&= \dot{\phi}\dot{A_{\bm{k}}} + \left(2\ddot{\phi} + 6H\dot{\phi}\right)A_{\bm{k}} \, .
\eea
This can be recast in a form similar to the perturbed background equation~\eqref{eq:perturbedKG}, namely
\bea 
\label{eq:KGuniformN}
\ddot{\delta\phi_{\bm{k}}} + \left(3H + \frac{\dot{\phi}^2}{2\Mp^2H}\right)\dot{\delta\phi_{\bm{k}}} &+ \left( \frac{\dot{\phi}}{2\Mp^2H}V_{,\phi} +V_{,\phi\phi}\right)\delta\phi_{\bm{k}}
 \\
& - \dot{\phi}\dot{A_{\bm{k}}} - \left( 2\ddot{\phi}+3H\dot{\phi} + \frac{\dot{\phi}^3}{2\Mp^2H} \right)A_{\bm{k}} = \Delta_{\bm{k}}
\eea
where the difference between \Eqs{eq:perturbedKG} and \eqref{eq:pert:uniformN} is quantified as
\bea \label{eq:difference}
\Delta_{\bm{k}} = \frac{\dot{\phi}}{H}\left\lbrace \frac{1}{2\Mp^2}\left[\dot{\phi}\left(\dot{\delta\phi}_{\bm{k}} - \dot{\phi}\dot{A}_{\bm{k}}\right) + V_{,\phi}\delta\phi_{\bm{k}}\right] + 3H^2A_{\bm{k}}\right\rbrace - \frac{k^2}{a^2}\delta\phi_{\bm{k}} \, .
\eea 
If we now impose the energy constraint \eqref{eq:energyconstraint:arbgauge} in the uniform-$N$ gauge, and recalling that since we choose $B=0$, $\psi = \frac{1}{3}\nabla^2E$, one can show that
\bea \label{eq:UN:diff}
\Delta_{\bm{k}} = - \frac{k^2}{a^2}\left( \delta\phi_{\bm{k}} + \frac{\dot\phi}{H}\psi_{\bm{k}} \right) = - \frac{k^2}{a^2}Q_{\bm{k}} \, ,
\eea 
see \Eq{eq:def:Q}.
Hence, since we neglect $k^2/a^2$ terms in the large-scale limit, the perturbation equations and the perturbed background equations become identical on large scales. We conclude that the separate universe approach, describing the evolution of long-wavelength perturbations about an FLRW background in terms of locally FLRW patches, is valid in both the spatially-flat and uniform-$N$ gauges. This result does not rely on slow roll; we only require that we can neglect gradient terms on super-Hubble scales. 
\subsubsection{Arbitrary gauge}
Let us finally see how the above arguments can be formulated without fixing a gauge. 
It is instructive to collect together metric perturbation terms in the Klein-Gordon equation from the full linear perturbation theory, \Eq{eq:pertubations:general}, which describe the perturbation of the local expansion rate \eqref{eq:def:expansion}
\bea
\delta\theta_{\bm{k}} = - 3\dot\psi_{\bm{k}} - \frac{k^2}{a^2}\left(a^2\dot{E_{\bm{k}}}-aB_{\bm{k}}\right) - 3HA_{\bm{k}} \,.
\eea
Re-writing the perturbed Klein-Gordon equation \eqref{eq:pertubations:general} in terms of $\delta\theta_{\bm{k}}$ we obtain
\bea
\label{eq:pertubations:general2}
\ddot{\delta\phi_{\bm{k}}} + 3H\dot{\delta\phi_{\bm{k}}} + \left(\frac{k^2}{a^2}+V_{,\phi\phi}\right)\delta\phi_{\bm{k}} 
= \left(2\ddot\phi+3H\dot\phi\right)A_{\bm{k}} + \dot{\phi} \dot{A_{\bm{k}}} 
- \dot{\phi}
\delta\theta_{\bm{k}}
\,.
\eea 
Finally, combining \Eq{eq:pertubations:general2} with the background equation \eqref{eq:eom:scalarfield} and rewriting the time derivatives in terms of the local proper time rather than the coordinate time, $\partial/\partial\tau\equiv(1-A)\partial/\partial t$, one obtains
\bea
\frac{\partial^2}{\partial\tau^2}(\phi+\delta\phi) + \theta\frac{\partial}{\partial\tau}(\phi+\delta\phi) + V_{,\phi}(\phi+\delta\phi) = \frac{\nabla^2}{a^2} (\delta\phi) \,.
\eea
Thus we see that the perturbed Klein-Gordon equation~\eqref{eq:pertubations:general} from cosmological perturbation theory in an arbitrary gauge has exactly the same form, up to first order in the inhomogeneous field and metric perturbations and up to spatial gradient terms of order $\nabla^2\delta\phi$, as the Klein-Gordon equation for a homogeneous scalar field in an FLRW cosmology, \Eq{eq:eom:scalarfield}, where we identify the local proper time, $\tau$, with the coordinate time, $t$, in an FLRW cosmology and the local expansion rate, $\theta/3$, with the Hubble rate, $H$, in an FLRW cosmology.
However to relate these local quantities to a global background coordinate system we need to fix a gauge. This cannot be determined by the local FLRW equations but requires to use additional constraint equations from the cosmological perturbation theory, as demonstrated in the preceding sub-sections for the spatially-flat and uniform-$N$ gauges.

\subsection{Gauge corrections to the noise} 
\label{sec:uniformexpansion}

In the previous section, it was explained that the field fluctuations, which determine the noise correlators through \Eq{eq:noise:correlators}, are usually calculated in the spatially-flat gauge, where the field perturbations coincide with the Sasaki--Mukhanov variable. However, we have seen that the local time in the spatially-flat gauge is in general perturbed with respect to the global time. As stressed in \Sec{sec:stochastic}, the Langevin equations (\ref{eq:conjmomentum:langevin}) and~(\ref{eq:KG:efolds:langevin}) are written in terms of the number of \efolds, i.e., the integrated expansion, $N$, is used as a time variable. If we are to use the number of \efolds\ as a local time coordinate and also as a global coordinate, relating the stochastic distribution of field values in many different patches at a given time, then we need to work in the uniform-$N$ gauge.
Thus one needs to gauge transform the field fluctuations calculated in the spatially-flat gauge to the uniform-$N$ gauge before evaluating \Eq{eq:noise:correlators}, and in this section, we explain how this can be done.
\subsubsection{Gauge transformations}
Let us denote quantities in the uniform-$N$ gauge with a tilde, \ie $\widetilde{\delta N}=0$. The transformations from the spatially-flat to the uniform-$N$ gauge can be written by means of a gauge transformation parameter $\alpha$ (that will be determined below), according to~\cite{Malik:2008im}
\begin{align} 
\label{eq:transform:phi} \delta\phi &\to \widetilde{\delta\phi} = \delta\phi + \phi'\alpha\, , \\
\label{eq:transform:psi}\psi &\to \widetilde{\psi} = \psi - \mathcal{H}\alpha\, , \\
\E &\to \widetilde{\E} = \E + \int\alpha\dd\eta \, .
\end{align}
Combining these transformation rules with \Eq{eq:def:deltaN}, the perturbed integrated expansion transforms as 
\bea \label{eq:transform:deltaN}
\delta N \to \widetilde{\delta N} = \delta N + \mathcal{H}\alpha + \frac{1}{3}\nabla^2\int\alpha \dd \eta \, .
\eea 
By definition, $\widetilde{\delta N}=0$, so one is lead to
\bea \label{eq:integraleq:alpha}
\delta N\Big|_{\psi=0} + \mathcal{H}\alpha  + \frac{1}{3}\nabla^2\int\alpha\dd\eta = 0 \, .
\eea
Taking the derivative of this expression with respect to conformal time, one obtains a differential equation for the gauge transformation parameter $\alpha$, namely
\bea \label{eq:alpha:diff}
3\mathcal{H}\alpha' + \left(3\mathcal{H}'+ \nabla^2\right)\alpha = S
 \, ,
\eea
where the source term reads
\bea
S = - 3 \delta N'\Big|_{\psi=0} = - \nabla^2 \sigma \Big|_{\psi=0} \,.
\eea

Two remarks are then in order. First, the source standing on the right-hand side of \Eq{eq:alpha:diff} remains to be calculated. In \Sec{sec:Nad:press}, we will show that for a scalar field it is related to the non-adiabatic pressure perturbation, and we will explain how it can be calculated. In \Sec{sec:source}, we will provide the general solution to \Eq{eq:alpha:diff}. Second, once $\alpha$ is determined, the field fluctuations in the uniform-$N$ gauge can be obtained from those in the spatially-flat gauge via \Eq{eq:transform:phi}. The noise correlators~(\ref{eq:noise:correlators}) also involve the fluctuation in the conjugate momentum, so this needs to be transformed into the uniform-$N$ gauge as well. However, precisely since $N$ is unperturbed in the uniform-$N$ gauge, one simply has
\bea 
\label{eq:GaugeTransf:pi}
\widetilde{\delta \pi} & = \frac{\dd \widetilde{\delta\phi}}{\dd N} \, ,
\eea 
and $\widetilde{\delta \pi}$ can be inferred from $\widetilde{\delta\phi}$ by a straightforward time derivative.
\subsubsection{Non-adiabatic pressure perturbation}
\label{sec:Nad:press}
Let us now show that the source function, $S(\eta)$ of \Eq{eq:alpha:diff}, 
coincides with the non-adiabatic pressure perturbation for a scalar field. This will prove that if inflation proceeds along a phase-space attractor (such as slow roll), where non-adiabatic pressure perturbations vanish, the source function vanishes as well; in this case \Eq{eq:alpha:diff} is solved by $\alpha=0$, and there are no gauge corrections.

Let us start by recalling the expressions for the energy constraint in an arbitrary gauge~\cite{Malik:2008im}
\begin{align} \label{eq:energyconstraint}
3\mathcal{H}\left(\psi' + \mathcal{H}A\right) - \nabla^2\left(\psi + \mathcal{H}\sigma\right)
&= 
-\frac{a^2}{2\Mp^2}\delta\rho \, . 
\end{align}
Combining this with the momentum constraint \eqref{eq:momentumconstraint:arbgauge} gives 
\bea \label{eq:deltarhodeltaphi}
\nabla^2(\psi+\mathcal{H}\sigma) &= \frac{a^2}{2\Mp^2}
\delta\rho_{\mathrm{com}}
\, ,
\eea
where the comoving density perturbation for a scalar field is given by 
\bea \label{eq:def:comdensity}
\delta\rho_{\mathrm{com}} = \delta\rho - 
\frac{\rho'}{\phi'}\delta\phi \, .
\eea
This in turn can be related to the non-adiabatic pressure perturbation~\cite{Malik:2008im}
\bea \label{eq:def:deltaPnad}
\delta P_{\mathrm{nad}} = -\frac{2a^2}{3\mathcal{H}\phi'}V_{,\phi}\delta\rho_{\mathrm{com}} \, .
\eea
In particular, in the spatially-flat gauge where $\psi=0$, \Eq{eq:deltarhodeltaphi} becomes
\bea \label{eq:energyconstraint:Pnad}
\mathcal{H}\nabla^2\sigma|_{\psi=0}
= -\frac{3\mathcal{H}\phi'}{4\Mp^2V_{,\phi}} \delta P_{\mathrm{nad}} \, .
\eea
Thus the source term $S$ on the right-hand side of \Eq{eq:alpha:diff} reads
\bea
S = \frac{3\phi'}{4\Mp^2V_{,\phi}} \delta P_{\mathrm{nad}}\,,
\eea  
and it vanishes if the non-adiabatic pressure perturbation is zero, which is the case whenever inflation proceeds along a phase-space attractor, $\phi'=\phi'(\phi)$, such as during slow roll.

In order to find a general expression for $S(\eta)$, one can use the (arbitrary gauge) expression for $\delta\rho$ for a scalar field~\cite{Malik:2008im},
\bea \label{eq:constraint:rho}
\delta\rho = \frac{\phi'\delta\phi' - \phi'^2A}{a^2} + V_{,\phi}\delta\phi \, ,
\eea
and combine it with \Eq{eq:deltarhodeltaphi} to obtain
\bea 
\nabla^2(\psi+\mathcal{H}\sigma) &= \frac{a^2}{2\Mp^2}\left[\left(3H\dot{\phi} + V_{,\phi}\right)\delta\phi + \dot{\phi}\delta\dot{\phi} - \dot{\phi}^2A\right] \, .
\eea 
Hence, in terms of the field fluctuations in the spatially-flat gauge and using \Eq{eq:flatA} for the perturbed lapse function, one finds 
\bea
S = -\frac{1}{2\Mp^2\mathcal{H}}\left[\left(3\mathcal{H}\phi' + a^{2}V_{,\phi} - \frac{\phi'^3}{2\Mp^2\mathcal{H}} \right)Q + \phi'Q'\right]\, .
\eea
Introducing the second slow-roll parameter $\epsilon_2\equiv \dd\ln\epsilon_1/\dd N$, the source function can be rewritten in the simpler form
\bea \label{eq:sourcefunction:general}
S = \frac{Q\sqrt{2\epsilon_1}}{2\Mp} \mathrm{sign}(\dot{\phi}) \left(  \calH\frac{\epsilon_{2}}{2} - \frac{Q'}{Q}\right) \, .
\eea
\subsubsection{General solution}
\label{sec:source}
When written in Fourier space, the differential equation~\eqref{eq:alpha:diff} for $\alpha_k$ has the general solution
\bea \label{eq:alphaintegral1:general}
\alpha_k &= \frac{1}{3\mathcal{H}}\int^{\eta}_{\eta_0} S_k(\eta')\exp\left[ \frac{k^2}{3}\int^{\eta}_{\eta'}\frac{\dd\eta''}{\mathcal{H}(\eta'')} \right]\dd\eta' \, .
\eea 
In this expression, $\eta_0$ is an integration constant that defines the slicing relative to which the expansion is measured. In what follows, we will consider situations in which an attractor is reached at late times. Since, in such a regime, the gauge correction vanishes (given that the non-adiabatic pressure perturbation does), we will take $\eta_0$ in the asymptotic future, \ie $\eta_0 = 0^-$.

Finally, in \Eq{eq:sourcefunction:general} the Sasaki--Mukhanov variable, $Q$, needs to be determined, which can be done by solving the Sasaki--Mukhanov equation
\bea \label{eq:MSequn}
v_k'' + \left(k^2 - \frac{z''}{z}\right)v_k = 0 \, ,
\eea
where $v_k = aQ_k$ and $z \equiv a\sqrt{2\epsilon_{1}}\Mp $. One can show that, in full generality $z''/z=\calH^2(2-\epsilon_1+3\epsilon_2/2-\epsilon_1\epsilon_2/2+\epsilon_2^2/4+\epsilon_2\epsilon_3/2)$, where we have introduced the third slow-roll parameter $\epsilon_3\equiv \dd\ln\epsilon_2/\dd\ln N$. 
Recall that, in terms of the field acceleration parameter $f=-\ddot{\phi}/(3H\dot{\phi})$ and the dimensionless mass parameter $\mu=V_{,\phi\phi}/(3H^2)$, we can write $z''/z$ as 
\bea 
\frac{z''}{z} &= \calH^2 \left( 2 + 5\epsilon_{1} - 3\mu - 12f\epsilon_{1} + 2\epsilon_{1}^2 \right) \, ,
\eea 
as given in \Eq{eq:z''overz:general}.

The following three sections will be devoted to three case studies, for which \Eq{eq:MSequn} will be solved and \Eq{eq:alphaintegral1:general} will be evaluated in order to derive the gauge corrections to the stochastic noise correlators in the uniform-$N$ gauge with respect to those in the spatially-flat gauge. In all cases, we will find that at the order of the coarse-graining parameter at which the stochastic formalism is derived, these gauge corrections can be neglected.

\subsection{Case study 1: slow roll}
\label{sec:slowroll}

Let us first apply the programme sketched above to the case of slow-roll inflation. As argued before, the presence of a dynamical attractor in that case makes the non-adiabatic pressure perturbation vanish, hence we should not find any gauge correction to the field fluctuations in the uniform-$N$ gauge and thus to the correlators for the noise. This is therefore a consistency check of our formalism. 

At leading order in slow roll, the slow-roll parameters can simply be evaluated at the Hubble-crossing time $\eta_{*}\simeq -1/k$, since their time dependence is slow-roll suppressed, \ie $\epsilon_{1} = \epsilon_{1*} + \mathcal{O}(\epsilon^2)$, etc. At that order, \Eq{eq:MSequn} is solved according to 
\bea \label{eq:MS:SRsolution}
v_k = \frac{\sqrt{-\pi\eta}}{2}\mathrm{H}_{\nu}^{(2)}\left(-k\eta\right) = aQ_k \, ,
\eea
where $\mathrm{H}_{\nu}^{(2)}$ is the Hankel function of the second kind and $\nu \equiv 3/2+\epsilon_{1*}+\epsilon_{2*}/2$, see \Sec{appendix:solving:MSequation}. Since the coarse-graining parameter is such that $\sigma\ll 1$, the mode functions in \Eq{eq:noise:correlators} need to be evaluated in the super-Hubble regime, \ie when $-k\eta \ll 1$. One can therefore make use of the asymptotic behaviour
\bea \label{eq:Hankel2:largescale}
H_{\nu}^{(2)}(-k\eta) \simeq \frac{i\Gamma(\nu)}{\pi}\left(\frac{2}{-k\eta}\right)^{\nu}\left[1 + \frac{1}{4(\nu-1)}\left(-k\eta\right)^{2} +\order{k^4\eta^4}\right] \, .
\eea
On the other hand, at first order in slow roll, the scale factor can also be expanded, and one finds \cite{Vennin:2014xta}
\bea \label{eq:scalefactor:slowroll}
a = -\frac{1}{H_{*}\eta}\left[1 + \epsilon_{1*} - \epsilon_{1*}\ln{\left(\frac{\eta}{\eta*}\right)}+\mathcal{O}(\epsilon^2)\right] \, ,
\eea
where we have used the expression $\mathcal{H}\simeq -\frac{1}{\eta}\left( 1 + \epsilon_{1*} \right)$ derived in  \App{sec:z''/z:SR}. Combining the two previous equations then leads to
\bea \label{eq:Q'overQ:SR}
\frac{Q'_k}{Q_k} \simeq \frac{\frac{3}{2}+\epsilon_{1*}-\nu}{\eta} + \frac{\frac{7}{2}-\nu}{4(\nu-1)}k^2\eta \, ,
\eea
which is valid at next-to-leading order both in the slow-roll parameters and in $k\eta$. With the expression given above for $\nu$, one can see that $Q'_k/Q_k\simeq -\epsilon_2/(2\eta)$ at leading order in $k\eta$. Since, at leading order, $\calH\simeq -1/\eta$, the two terms in the right-hand side of \Eq{eq:sourcefunction:general} exactly cancel, and the source function $S_k$ vanishes. This confirms that the gauge corrections are indeed suppressed in that case.

In fact, the first contribution to the gauge correction comes from the decaying mode, and for completeness we now derive its value. Plugging the previous expressions into \Eq{eq:sourcefunction:general} leads to
\bea \label{eq:sourcefunction:SR}
S_k &= \frac{i}{2}\frac{H_*}{\Mp}\sqrt{k\epsilon_{1*} }\eta\left(-k\eta\right)^{-\epsilon_{1*}-\frac{\epsilon_{2*}}{2}}  \mathrm{sign}\left(\dot{\phi}\right) \, .
\eea
One can then insert \Eq{eq:sourcefunction:SR}, along with $\mathcal{H} = -(1+\epsilon_{1*})/\eta$, into \Eq{eq:alphaintegral1:general}, and derive the gauge transformation parameter from the spatially-flat gauge to the uniform-$N$ gauge in the large-scale and slow-roll limit, 
\bea \label{eq:alphaLO:SR}
\alpha_k = \frac{iH_{*}\sqrt{\epsilon_{1*}}}{12\Mp}k^{-\frac{5}{2}}\left(-k\eta\right)^{3-\epsilon_{1*}-\frac{\epsilon_{2*}}{2}}  \mathrm{sign}\left(\dot{\phi}\right) \, .
\eea
In the uniform-$N$ gauge, according to \Eq{eq:transform:phi}, the field fluctuation thus reads
\bea 
\label{eq:GaugeCorr:SR}
\widetilde{\delta\phi}_k &= Q_k\left[ 1 - \frac{\epsilon_{1*}}{6}\left(-k\eta\right)^2 \right] ,
\eea
and its deviation from $Q$ is therefore both slow-roll suppressed and controlled by the amplitude of the decaying mode. Since it needs to be evaluated at the coarse-graining scale $k_\sigma = \sigma a H$ in \Eq{eq:noise:correlators}, the relative gauge correction to the correlations of the noises scales as $\epsilon_1\sigma^2$, which can be neglected since the stochastic formalism assumes $\sigma\rightarrow 0$.
\subsection{Case study 2: ultra-slow roll}
\label{sec:USR}
Let us now consider the case of ultra-slow-roll (USR) inflation~\cite{Inoue:2001zt, Kinney:2005vj, Pattison:2018bct}, where the dynamics of $\phi$ is friction dominated and the gradient of the potential can be neglected in the Klein--Gordon equation~\eqref{eq:eom:scalarfield}, which becomes
\bea 
\ddot{\phi} + 3H\dot{\phi} \simeq 0 \, .
\eea
This gives rise to $\dot{\phi}_{\mathrm{USR}} \propto \ee^{-3N}$, hence $\dot{\phi}=\dot{\phi}_\uin+3H(\phi_\uin-\phi)$. The phase-space trajectory thus carries a dependence on initial conditions that is not present in slow roll, which explains why ultra-slow roll is not a dynamical attractor while slow roll is. We therefore expect the non-adiabatic pressure perturbation not to vanish in ultra-slow roll, which may lead to some non-trivial gauge corrections. In ultra-slow roll,  the field acceleration parameter $f$ introduced in \Eq{eq:def:f} is close to one (while it is close to zero in slow roll), so $\delta\equiv 1-f$ quantifies how deep in the ultra-slow-roll regime one is. In the limit where $\delta=0$,  $\dot{\phi}_{\mathrm{USR}} \propto \ee^{-3N}$ gives rise to $\epsilon_1^{\mathrm{USR}}\propto \ee^{-6N}/H^2$, hence
\bea \label{eq:eps:USR}
\epsilon_n^{\mathrm{USR}} = \begin{cases}
-6 + 2\epsilon_{1} &\text{if $n$ is even}\\
2\epsilon_{1} &\text{if $n>1$ is odd} \, .
\end{cases}
\eea 
The even slow-roll parameters are therefore large in ultra-slow roll. When $\delta$ does not strictly vanish, these expressions can be corrected, and for the second and the third slow-roll parameters, one finds
\begin{align}
\label{eq:eps2:USR}
\epsilon_{2}&= -6  + 2\epsilon_{1}+6\delta \,, \\
\epsilon_{3}&= 2\epsilon_{1} - \frac{\dd \delta}{\dd N}\frac{6}{6 - 2\epsilon_{1} - 6\delta} \, ,
\end{align}
which are exact formulas. One can then calculate
\bea \label{eq:ddeltadt:USR}
\frac{\dd \delta}{\dd N} &=  -\mu + 3\delta - 3\delta^2 + \delta\epsilon_{1} \, ,
\eea 
where $\mu$ is the dimensionless mass parameter defined in \Eq{mu:def}. For small $\delta$ and $\epsilon_{1}$, one then has
\bea 
\epsilon_{3}^{\mathrm{USR}} \simeq 2\epsilon_{1} + \mu - 3\delta + \mu\left( \frac{2\epsilon_{1}+6\delta}{6} \right) \, .
\eea 
There is no reason, \textit{a priori}, that $\mu$ needs to be small, and hence these corrections can be large for models with $V_{,\phi\phi}\neq 0$.
Note also that \Eq{eq:ddeltadt:USR} provides a criterion for the stability of ultra-slow roll, which is stable when the right-hand side of this equation is negative, in agreement with the results of Ref. \cite{Pattison:2018bct}. 

Let us now derive the gauge corrections in ultra-slow roll. We perform a calculation at leading order in $\epsilon_1$, $\delta$ and $\mu$, but in \App{app:usr:epscorrections} the calculation is extended to next-to-leading order in $\epsilon_1$, and it is shown that the result derived below is indeed valid. At leading order, one simply has $z''/z\simeq 2 \calH^2$, hence \Eq{eq:MSequn} is solved according to
\bea \label{eq:v:nu=3/2}
v_k &= \frac{1}{\sqrt{2k}}\ee^{-ik\eta}\left( 1 - \frac{i}{k\eta}\right) \, .
\eea 
Since $a = -1/(H_{*}\eta)$ at leading order, this gives rise to
\bea \label{eq:Q'overQ:USR:nu=3over2}
\frac{Q'_k}{Q_k} = \frac{-ik^2\eta}{k\eta-i} \, ,
\eea 
and the source function~\eqref{eq:sourcefunction:general} reads
\bea \label{eq:source:USR:nu=3/2}
S_k &= \frac{H_{*}}{2\Mp}\sqrt{\frac{\epsilon_{1}}{k}}\ee^{-ik\eta}\left( 3 - \frac{3i}{k\eta} + ik\eta \right) \mathrm{sign}\left(\dot{\phi}\right) \, .
\eea
Since $\epsilon_{1}\simeq \epsilon_{1*}(a/a_*)^{-6}$, the gauge transformation parameter $\alpha$ can be obtained from \Eq{eq:alphaintegral1:general} and is given by
\bea \label{eq:alpha:USR:nu=3/2}
\alpha_k = \frac{iH_{*}\sqrt{\epsilon_{1*}}}{6\Mp}{k^{-\frac{5}{2}}}(k\eta)^4\mathrm{sign}\left(\dot{\phi}\right) \left[  1+ \mathcal{O}(k\eta)^2\right] \, .
\eea
Comparing this expression with \Eq{eq:alphaLO:SR}, one can see that the gauge correction decays even faster than in the slow-roll regime, hence is even more suppressed. This is because, although slow roll is a dynamical attractor while ultra-slow roll is not, the field velocity (hence the conjugate momentum) decays very quickly in ultra-slow roll, and this also damps away one of the two dynamical degrees of freedom. Finally, the gauge transformation \eqref{eq:transform:phi} gives rise to
\bea \label{eq:deltaphi:USRtransform}
\widetilde{\delta\phi_k} &= Q_k\left[ 1 + \frac{\epsilon_{1*}}{3}\left(-k\eta\right)^6\right]\, .
\eea
The relative corrections to the noises correlators scale as $\epsilon_{1*}\sigma^6$ and can therefore be neglected, even more accurately than in slow roll. 
\subsection{Case study 3: Starobinsky model}
\label{sec:starobinsky}
In the two previous sections, we have shown that the gauge corrections to the noise correlators are negligible both in slow-roll and in ultra-slow-roll inflation. In this section, we consider a model that interpolates between these two limits, namely the Starobinsky piece-wise linear model~\cite{Starobinsky:1992ts}. This allows us to study a regime that is neither slow roll nor ultra-slow roll, but for which the early-time (ultra-slow roll) and the late-time (slow roll) limits are under control. 

The Starobinsky model is based on a potential made up of two linear parts with different gradients defined by the dimensionless parameters $a_+\gg a_- > 0$:
\bea 
V(\phi) = \begin{cases} V_{0}\left(1+a_+\frac{\phi}{\Mp}\right) & \mathrm{for}\, \phi> 0 \\ 
V_{0}\left(1+a_-\frac{\phi}{\Mp}\right) & \mathrm{for}\, \phi< 0 
\end{cases}
\, .
\eea
In order to ensure both parts of the potential are able to support slow-roll inflation,
we require $a_{\pm}\ll 1$. 

The dynamics of the inflaton, as it evolves across the discontinuity in the potential gradient at $\phi=0$, can be split into three phases. The first phase, which we label $\mathrm{SR}_{+}$, is a slow-roll phase for $\phi>0$ and $\dot\phi<0$. When the inflaton crosses $\phi=0$, it then starts down the $\phi<0$ part of the potential with an initial velocity inherited from the first slow-roll phase $\mathrm{SR}_{+}$ that is much larger than the slow-roll velocity for $\phi<0$. The second phase thus starts in an ultra-slow-roll regime and is denoted USR. It corresponds to the field range $\phi_{\mathrm{USR}\to\mathrm{SR}}<\phi<0$. Finally, the inflaton relaxes back to slow roll for $\phi<\phi_{\mathrm{USR}\to\mathrm{SR}}$, and we call this third phase $\mathrm{SR}_{-}$. 

\begin{figure}
    \centering
    \includegraphics[width=0.485\textwidth]{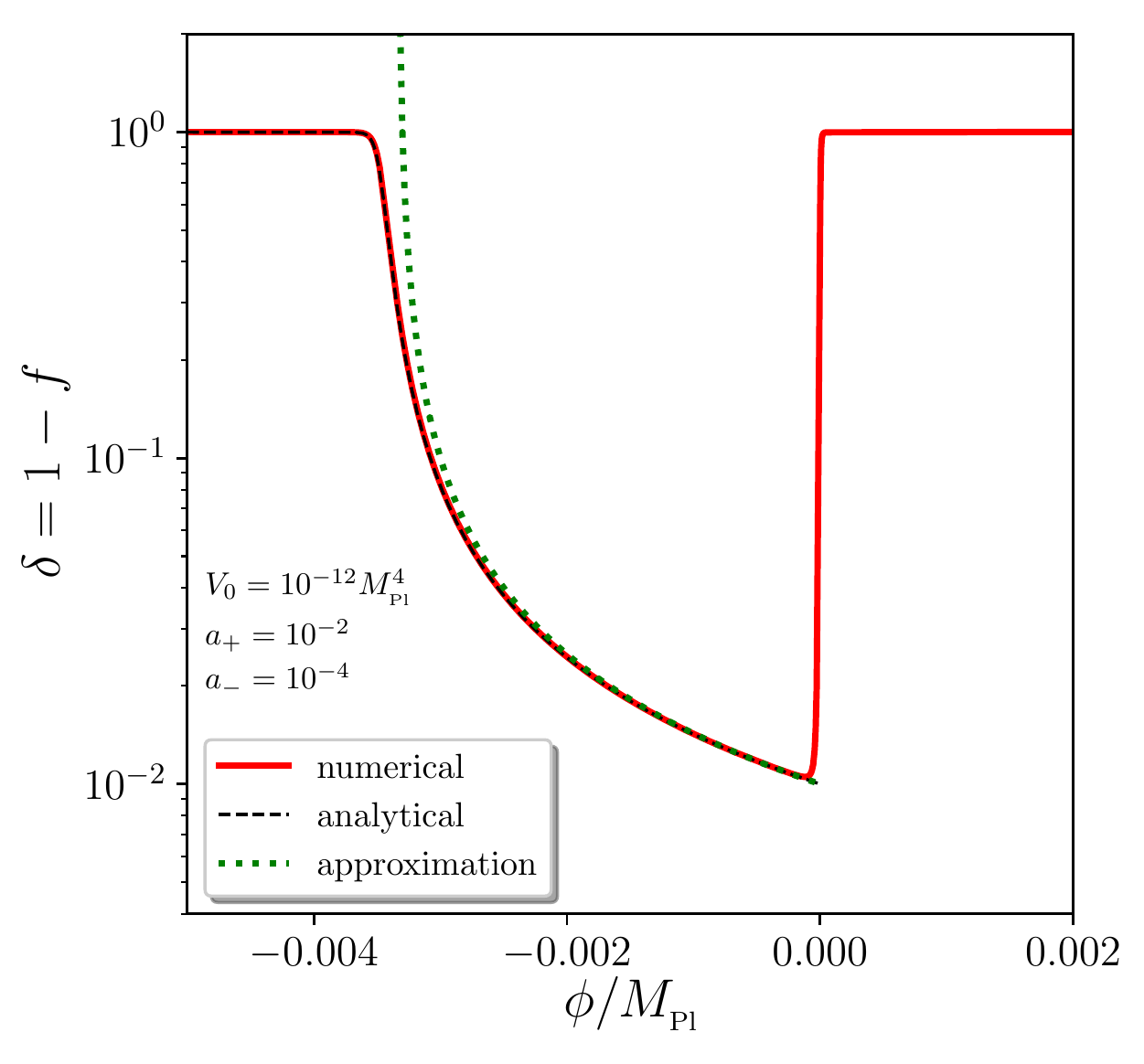}
    \includegraphics[width=0.50\textwidth]{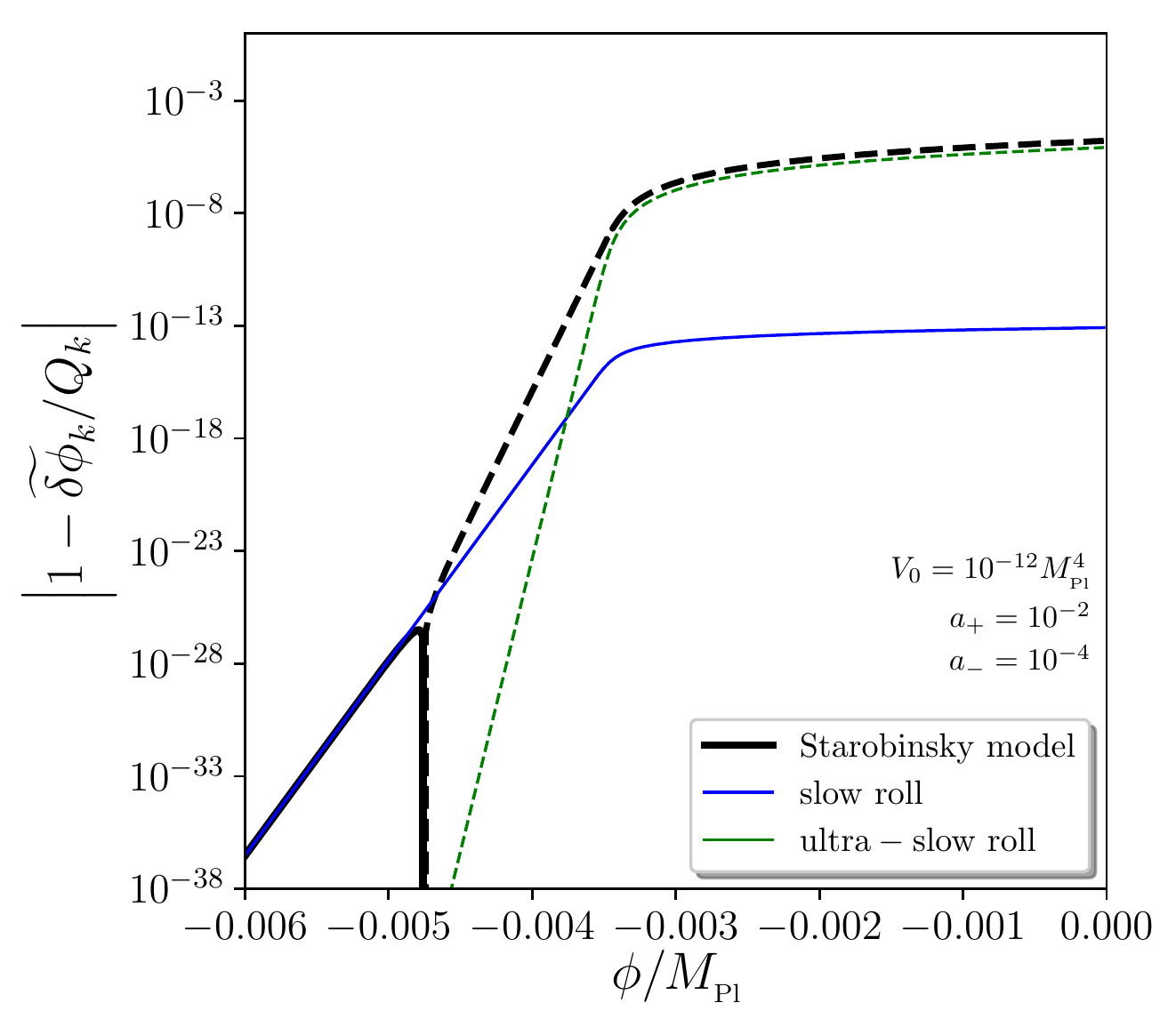}
    \caption[Gauge correction to the field perturbation in the uniform-$N$ gauge in the Starobinsky model of inflation]{Left panel: field acceleration parameter, $f$ defined in \eqref{eq:def:f}, in the Starobinsky model. The red curve corresponds to a numerical integration of the Klein--Gordon equation~(\ref{eq:eom:scalarfield}), the dashed black curve corresponds to the analytical solution~\eqref{eq:f:phi:starobinsky}, while the dotted green line stands for the approximation~(\ref{eq:f:phi:starobinsky:appr}). Right panel: fractional gauge correction to the field perturbation in the uniform-$N$ gauge in the Starobinsky model, for a mode such that $k/aH=10^{-2}$ at the transition time $t=0$. The black line corresponds to the full result~(\ref{eq:Staro:gaugeCorr}), the blue line stands for the slow-roll result~\eqref{eq:GaugeCorr:SR}, and the green line is the ultra-slow-roll result~\eqref{eq:deltaphi:USRtransform}. Solid lines are such that $1-\widetilde{\delta\phi}_k/Q_k>0$ and dashed lines are such that $1-\widetilde{\delta\phi}_k/Q_k<0$. }
    \label{fig:Staro}
\end{figure}
During the USR phase 
the Hubble parameter can be taken as approximately constant, $H\simeq H_0= \sqrt{V_0/(3\Mp^2)}$; the consistency of that assumption will be checked below. The Klein--Gordon equation~(\ref{eq:eom:scalarfield}) then becomes $\ddot{\phi} + 3H_{0}\dot{\phi} + V_0 a_-/\Mp = 0$, and can be solved to give
\bea
\label{eq:phi:t:starobinsky}
\frac{\phi(t)}{\Mp}=\frac{a_+-a_-}{3}\left(\ee^{-3H_0t}-1\right)-a_- H_0 t\, ,
\eea
where we choose $t=0$ to denote the time when $\phi=0$, and the initial velocity is set such that its value at the transition point is given by its slow-roll counterpart in the $\phi>0$ branch of the potential, \ie $\dot{\phi}(\phi=0^-)=\dot{\phi}(\phi=0^+)=-H_0 a_+$. The acceleration parameter defined in \Eq{eq:def:f} is then given by
\bea 
\label{eq:Staro:f(t)}
f(t) &=  1 - \frac{a_-}{a_-+(a_+-a_-)\ee^{-3H_0t}} \, .
\eea 
At the transition time, it reads $f(t=0)= 1-\frac{a_-}{a_+}$, so if $a_-/a_+\ll 1$, $f\simeq 1$ and ultra-slow roll takes place. At late time, however, $f$ is damped so that the system relaxes back to a phase of slow-roll inflation. Note that the solution~\eqref{eq:phi:t:starobinsky} can be inverted, 
\bea 
\label{eq:Staro:traj:inverted}
H_0 t(\phi) = \frac{1}{3}\left(1-\frac{a_+}{a_-}\right) - \frac{\phi}{\Mp a_-} + \frac{1}{3}W_{0}\left[\frac{a_+-a_-}{a_-}\exp\left(
\frac{a_+}{a_-}-1+3 \frac{\phi}{\Mp a_-}\right)\right] \, ,
\eea 
where $W_{0}(x)$ is the $0$-branch of the Lambert function, which leads to the phase-space trajectory
\bea 
\label{eq:Staro:phidot(phi)}
\dot{\phi}(\phi) = -\frac{\Mp }{H} H_0^2 a_- \left\lbrace 1+W_{0}\left[\frac{a_+-a_-}{a_-}\exp\left(\frac{a_+}{a_-}-1+3 \frac{\phi}{\Mp a_-}\right)\right] \right\rbrace \, .
\eea 
In the denominator of the first term in the right-hand side, $H$ is left to vary~\cite{Martin:2011sn}, in such a way that at late time, \ie when $\phi$ goes to $-\infty$, one recovers the slow-roll result $\dot{\phi}=-\Mp H_0^2 a_-/H$. Plugging \Eq{eq:Staro:traj:inverted} into \Eq{eq:Staro:f(t)} also leads to 
\bea \label{eq:f:phi:starobinsky}
f(\phi) = 1 - \frac{1}{1+W_{0}\left[\frac{a_+-a_-}{a_-}\exp\left( 
\frac{a_+}{a_-}-1+3 \frac{\phi}{\Mp a_-}\right)\right]} \, ,
\eea 
which is shown in \Fig{fig:Staro} with the dashed black line and compared to a numerical solution of the Klein--Gordon equation displayed with the solid red line. One can check that $f$ starts from a value close to one at early time and approaches zero at late time. If one expands \Eq{eq:f:phi:starobinsky} around $\phi=0$, one obtains
\bea \label{eq:f:phi:starobinsky:appr}
f\simeq 1-\frac{a_-}{a_++3\frac{\phi}{\Mp}}\, ,
\eea
which matches Eq.~(4.3) of Ref. \cite{Pattison:2018bct}. This approximation is also shown in \Fig{fig:Staro}, with the dotted green line. 

From \Eq{eq:f:phi:starobinsky}, the transition time between USR and $\mathrm{SR}_{-}$, defined as the time when $f=1/2$, is found to be
\bea 
t_{\mathrm{USR}\to\mathrm{SR}} = \frac{1}{3 H_0}\ln\left(\frac{a_+-a_-}{a_-}\right) \, ,
\eea 
which is consistent with Eq.~(4.5) of Ref. \cite{Pattison:2018bct}. Making use of \Eq{eq:phi:t:starobinsky}, the field value at which this happens is given by
\bea 
\phi_{\mathrm{USR}\to\mathrm{SR}} &= -\frac{\Mp }{3}\left[ a_+ - 2 a_-+a_-\ln\left(\frac{a_+-a_-}{a_-}\right)\right]
 \simeq -\frac{a_+}{3}\Mp \, ,
\eea 
where the last expression is derived in the limit $a_-/a_+\ll 1$ and agrees with Eq.~(4.4) of Ref. \cite{Pattison:2018bct}. This allows us to test the assumption made above that the potential, hence the Hubble parameter, does not vary much during the USR phase. The relative shift in the potential value between $\phi=0$ and $\phi_{\mathrm{USR}\to\mathrm{SR}}$ is indeed given by
\bea 
\frac{\Delta V}{V} = \frac{a_-(a_+-a_-)}{3} 
\ll 1 \, ,
\eea
which justifies the above assumption.

Let us now calculate the gauge transformation from the spatially-flat to uniform-$N$ gauge in this model. As explained above, combining \Eq{eq:Staro:phidot(phi)} and~(\ref{eq:phi:t:starobinsky}) leads to
\bea
\dot{\phi}(t) = \frac{H_0^2\Mp}{H}\left[ (a_--a_+)\ee^{-3H_{0}t} -  a_- \right] ,
\eea 
that allows us to both describe the USR and the $\mathrm{SR}_-$ phases, as well as the transition between the two. Making use of the relation $\epsilon_{1}=\dot{\phi}^2/(2\Mp^2H^2)$, one obtains
\bea 
\epsilon_{1}(t) &= \frac{1}{2}\left(\frac{H_0}{H}\right)^4\left[  a_- - (a_--a_+)\ee^{-3H_{0}t}\right]^2 \\
\epsilon_{2}(t) &= -\frac{6(a_--a_+)\ee^{-3H_0t}}{(a_--a_+)\ee^{-3H_0t}-a_-} + 4\epsilon_{1}(t) \, .
\eea 
One can check that, at late times, one recovers $\epsilon_2=4\epsilon_1$, which is indeed satisfied in slow roll for linear potentials, see \Eqs{eq:eps1:V} and~(\ref{eq:eps2:V}).

Since $\mu=0$ in this model, the fact that $\epsilon_1$ remains small implies that \Eq{eq:z''overz:general} is close to its de-Sitter limit. Moreover, one can check that, at early times, the term $Q'_k/Q_k$ in \Eq{eq:sourcefunction:general} provides a subdominant contribution, hence it is sufficient to evaluate $Q'_k/Q_k$ at late time and use the result of \Eq{eq:Q'overQ:SR}, $Q'_k/Q_k\simeq - \epsilon_{2*}/(2\eta)+k^2\eta = a_-^2/\eta+k^2\eta$. One then obtains
\bea 
S &= \frac{i  H_0}{\Mp}\frac{\mathrm{sign}\left(\dot{\phi}\right)}{(2k)^{\frac{3}{2}}\eta}\Bigg\{ 3(a_--a_+)\ee^{-3H_{0}t}\left(1+\frac{a_-^2}{3}\right)
 -  a_-^3  \\
&\hspace{1cm} 
 + \left[{a_-+(a_+-a_-)\ee^{-3H_{0}t}}\right]^3 - k^2\eta^2\left[ \left(a_--a_+\right)\ee^{-3H_{0}t} - a_-\right]\Bigg\} \, .
\eea 
From \Eq{eq:alphaintegral1:general}, we then find the gauge transformation parameter to be
\bea \label{eq:gaugetransformation:starobinsky}
\alpha &\simeq \frac{-i\eta H_0}{3(2k)^{\frac{3}{2}}\Mp}\mathrm{sign}\left(\dot{\phi}\right)
{\Bigg[}  
\frac{(k\eta)^2}{2}a_- 
+ \left(a_--a_+\right)\ee^{-3H_0t}\left(1+\frac{a_-^3}{2}\right)   \\
&\hspace{5mm} 
+ a_-^2\left(a_+-a_-\right)\ee^{-3H_0t} + \frac{a_-\left(a_+-a_-\right)^2}{2}\ee^{-6H_0t} + \frac{\left(a_+-a_-\right)^3}{9}\ee^{-9H_0t} {\Bigg]}  \, ,
\eea 
where only the $(k\eta)^2$-suppressed term that becomes dominant at late times has been kept, \ie there are other $(k\eta)^2$ terms that have been dropped for consistency since they always provide sub-dominant contributions. One can check that at early time, \ie when $t\to 0$, the ultra-slow-roll expression~(\ref{eq:alpha:USR:nu=3/2}) is recovered if $a_-/a_+\ll 1$, while at late time, \ie when $t \to \infty$, the slow-roll expression~(\ref{eq:alphaLO:SR}) is recovered. This gives rise to the gauge correction
\bea
\label{eq:Staro:gaugeCorr}
\widetilde{\delta\phi}_k/ Q_k & \kern-0.1em = \kern-0.1em 1 \kern-0.2em + \kern-0.2em \frac{1}{6}\left(\frac{H_0}{H}\right)^3\left[\left(a_--a_+\right)\ee^{-3H_0 t}-a_-\right]\kern-0.2em
{\Bigg[}  
\frac{(k\eta)^2}{2}a_- 
+ \left(a_--a_+\right)\ee^{-3H_0t}\left(1+\frac{a_-^2}{3}\right) 
\\ &
+ a_-^2\left(a_+-a_-\right)\ee^{-3H_0t} + \frac{a_-\left(a_+-a_-\right)^2}{2}\ee^{-6H_0t} + \frac{\left(a_+-a_-\right)^3}{9}\ee^{-9H_0t} {\Bigg]}\, ,
\eea
which is displayed in the right panel of \Fig{fig:Staro} for a mode such that $k/aH=10^{-2}$ at the transition time $t=0$. Right after the transition point, one can check that the ultra-slow-roll result~\eqref{eq:deltaphi:USRtransform} is recovered (the slight discrepancy is due to the finite value of $a_-/a_+$, \ie the finite initial value of $\delta$, we work with in \Fig{fig:Staro}), and at late time, the slow-roll result~\eqref{eq:GaugeCorr:SR} is obtained. In between, the gauge correction to the noise correlators remains tiny and can therefore be safely neglected.

Thus, from the analysis in this section, we can conclude that the stochastic formalism for inflation in valid well beyond the usual slow-roll approximation. 
We have confirmed the the separate universe approach, which is one of the main pillars that stochastic inflation rests on, is valid and does not require slow roll. 
Furthermore, we have considered the subtle gauge dependence that arises from our choice to use stochastic inflation in the uniform-$N$ gauge, and have found that in practical situations (slow roll, ultra-slow roll, and models that interpolate between them) the gauge transformation is trivial on super-Hubble scales, and stochastic inflation can be applied as usually formulated and without further refinements. 
Note that this does not preclude the existence of situations where these gauge effects might be significant, but in such cases, \Eqs{eq:transform:phi}, \eqref{eq:GaugeTransf:pi}, \eqref{eq:sourcefunction:general} and \eqref{eq:alphaintegral1:general} provide the key formulas to compute them. 

The validity of stochastic inflation beyond slow roll is an important result when one wants to consider the production of primordial black holes, the formation of which is likely to require regimes of inflation that both undergo large stochastic diffusion and violate slow roll.
These implications for primordial black hole formation will be studied in detail in chapters \ref{chapter:quantumdiff:slowroll} and \ref{chapter:USRstochastic}, respectively.

\section{The \texorpdfstring{$\delta N$}{} formalism}
\label{sec:stochasticdeltaN}

In order to calculate the mass fraction of primordial black holes, we need to find the probability distribution of the coarse-grained curvature perturbations, see \Eq{eq:def:beta}.
In this section, we will explain how this can be done using the stochastic-$\delta N$ formalism.

\subsection{The classical-\texorpdfstring{$\delta N$}{} formalism}

The starting point of the stochastic-$\delta N$ formalism is the standard, classical $\delta N$ formalism~\cite{Starobinsky:1982ee, Starobinsky:1986fxa, Sasaki:1995aw, Sasaki:1998ug, Lyth:2004gb, Lyth:2005fi}, which provides a succinct way of relating the fluctuations in the number of \efolds~of expansion during inflation for a family of homogeneous universes with the statistical properties of curvature perturbations. Starting from the unperturbed flat Friedmann-Lema\^{i}tre-Robertson-Walker metric
\bea
\label{eq:metric:FLRW}
\dd s^2 = -\dd t^2 + a^2(t)\delta_{ij}\dd x^{i}\dd x^{j} \, ,
\eea
deviations from isotropy and homogeneity can be added at the perturbative level and contain scalar, vector and tensor degrees of freedom. Gauge redundancies associated with diffeomorphism invariance allow one to choose a specific gauge in which fixed time slices have uniform energy density and fixed spatial worldlines are comoving (in the super-Hubble regime this gauge coincides with the synchronous gauge supplemented by some additional conditions that fix it uniquely). Including spatial perturbations only, one obtains~\cite{Starobinsky:1982ee, Creminelli:2004yq,Salopek:1990jq}
\bea
\dd s^2 = -\dd t^2 + a^2(t)\ee^{2\zeta(t, \bm{x})}\delta_{ij}\dd x^{i}\dd x^{j} \, ,
\eea
where $\zeta$ is the adiabatic curvature perturbation mentioned in \Sec{sec:pbhs}. One can then introduce a local scale factor
\bea
\label{eq:alocal:def}
\tilde{a}(t, \bm{x}) = a(t)\ee^{\zeta(t, \bm{x})} \, ,
\eea
which allows us to express the amount of expansion from an initial flat space-time slice at time $t_\uin$ to a final space-time slice of uniform energy density as
\bea
N(t, \bm{x}) = \ln{\left[ \frac{\tilde{a}(t, \bm{x})}{a(t_{\mathrm{in}})} \right]} \, .
\eea
This is related to the curvature perturbation $\zeta$ via \Eq{eq:alocal:def}, which gives rise to
\bea
\label{eq:zeta:deltaN}
\zeta(t, \bm{x}) = N(t, \bm{x}) - \bar{N}(t) \equiv \delta N \, ,
\eea
where $\bar{N}(t) \equiv \ln{\left[ {a(t)}/{a(t_{\mathrm{in}})} \right]}$ is the unperturbed expansion. 
This expression forms the basis of the $\delta N$ formalism, which follows by making the further simplifying assumption that on super-Hubble scales, each spatial point of the universe evolves independently and this evolution is well approximated by the evolution of an unperturbed universe. 
This is the separate Universe approximation discussed previously, and it allows us to neglect spatial gradients on super-Hubble scales. 
As a consequence, $N(t, \bm{x})$ is the amount of expansion in unperturbed, homogeneous universes, and $\zeta$ can be calculated from the knowledge of the evolution of a family of such universes.

We can also write the $\delta N$ formalism in terms of the inflaton field $\phi$ by perturbing $\phi(\bm{x}) = \phi + \delta\phi(\bm{x})$.
Here, $\phi$ is a homogeneous, unperturbed field value and $\delta\phi$ is a perturbation that we view as originating from vacuum fluctuations. 
We can then rewrite \eqref{eq:zeta:deltaN} as 
\bea 
\zeta(t, \bm{x}) = N(\rho(t), \phi(\bm{x})) - \bar{N}(\rho(t), \phi)\, ,
\eea
in which we now evaluate $N$ in unperturbed universes from an initial field value $\phi$ to a final hypersurface where the energy density has a given value $\rho$.
We can also expand the curvature perturbation as 
\bea \label{eq:zeta:phi}
\zeta(t, \bm{x}) = \delta N \simeq \frac{\partial N}{\partial\phi}\delta\phi \, ,
\eea 
since we know that observed curvature perturbations are approximately Gaussian, and where we can simply evaluate $N$ using the classical formula 
\bea \label{eq:efolds:classicalformula:epsilon}
N(\phi) = \frac{1}{\Mp}\int \frac{\dd\phi}{\sqrt{2\epsilon_1}} \, .
\eea
We can now evaluate the power spectrum of curvature perturbations using the standard $\delta N$ formalism.
Working with the reduced power spectrum $\mathcal{P}_\zeta = k^3/(2\pi^2)P_\zeta$, which can be expressed in term of the power spectrum of $\delta\phi$ through \Eq{eq:zeta:phi}.
Indeed, from \Eq{eq:deSitter:fieldfluctuation} with $\nu\simeq 3/2$, we can calculate the de Sitter power spectrum for a light field to be 
\bea 
\mathcal{P}_{\delta\phi}(k) = \frac{k^3}{2\pi^2}\vert \phi_{\bm{k}}\vert^2 = \left( \frac{H}{2\pi} \right)^2 \, ,
\eea 
where $H$ is to evaluated at horizon crossing ($k=aH$). 
Finally, we can combine this with \Eq{eq:efolds:classicalformula:epsilon} to find 
\bea 
\mathcal{P}_\zeta(k) = \frac{1}{2\Mp^2\epsilon_1}\left( \frac{H}{2\pi} \right)^2 \, ,
\eea 
which matches \Eq{eq:intro:powerspectrum} exactly. 
The approach explained here is the standard $\delta N$ formalism, and in the next section we combine this with the stochastic approach to inflation, to obtain the stochastic-$\delta N$ formalism.

\subsection{The stochastic-\texorpdfstring{$\delta N$}{} formalism}

\begin{figure}[t]
\begin{center}
\includegraphics[width=0.8\textwidth]{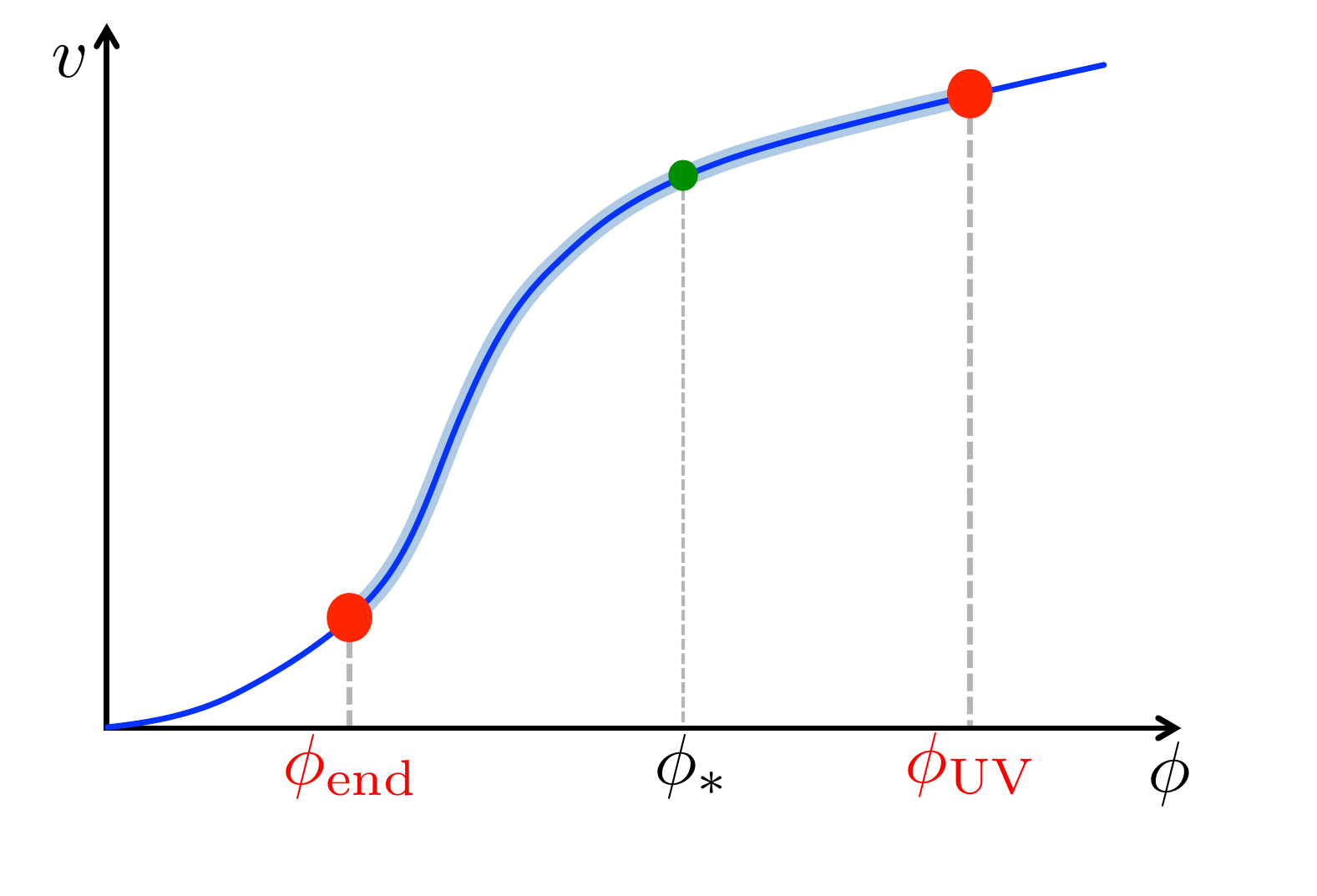}
\caption[Sketch of stochastic dynamics in an arbitrary potential]{Sketch of the single-field stochastic dynamics solved in this work. Starting from $\phi_*$ at initial time, the inflaton field $\phi$ evolved along the potential $v(\phi)$ under the Langevin equation~(\ref{eq:Langevin:slowroll}), until $\phi$ reaches $\phi_\uend$ where inflation ends. A reflective wall is added at $\phiuv$ to prevent the field from exploring arbitrarily large values. The number of \efolds~realised along a family of realisations of the Langevin equation is calculated, and gives rise to the probability distribution of curvature perturbations using the $\delta N$ formalism.}  
\label{fig:sketch}
\end{center}
\end{figure}
The $\delta N$ formalism relies on the calculation of the amount of expansion realised amongst a family of homogeneous universes. When stochastic inflation is employed to describe such a family of universes and to calculate the amount of expansion realised in them, this gives rise to the stochastic-$\delta N$ formalism~\cite{Enqvist:2008kt, Fujita:2013cna, Fujita:2014tja, Vennin:2015hra, Kawasaki:2015ppx, Assadullahi:2016gkk, Vennin:2016wnk}.

This approach is sketched in \Fig{fig:sketch} for the case of the stochastic formalism in slow roll, where the Langevin equation \eqref{eq:KG:efolds:langevin} for $\phi$ (where we have dropped the bar notation but note this equation is only valid on super-Hubble scales) is simply 
\bea
\label{eq:Langevin:slowroll}
\frac{\dd \phi}{\dd N} = -\frac{V'}{3H^2} + \frac{H}{2\pi}\xi\left( N \right) \, ,
\eea
where for ease of notation we now take a prime to mean differentiation with respect to the inflation field $\phi$, \ie $'=\frac{\dd}{\dd\phi}$.
Starting from $\phi=\phi_*$ at an initial time, the inflaton field evolves along the potential $v(\phi)$ under the Langevin equation~(\ref{eq:Langevin:slowroll}), where hereafter we use the rescaled dimensionless potential
\bea
v\equiv \frac{V}{24\pi^2\Mp^4}\, ,
\eea
until it reaches $\phi_\uend$ where inflation ends. 
A reflective wall is added at $\phiuv$ to prevent the field from exploring arbitrarily large values, which can be necessary to renormalise infinities appearing in the theory~\cite{Vennin:2016wnk} (whose results are still independent of $\phiuv$ and of the exact nature of the wall, reflective or absorbing, provided $\phiuv$ lies in some range). 
For example, in Ref. \cite{Assadullahi:2016gkk} it was shown that for a single field monomial potential $v(\phi) = \phi^p$, one can safely take the reflective wall to infinity, $\phi_\mathrm{uv} \to \infty$.
In this case, the probability of the inflaton interacting with the reflective wall becomes zero (for any $p\geq0$), and the physical nature of the wall becomes irrelevant.
This remains true even for $p<1$, despite the fact that in this case the mean number of \efolds~diverges as $\phi_\mathrm{uv} \to \infty$.
However, in the case of a flat potential at the end of inflation, such the case considered in \Sec{sec:PBH:stochastic}, if the value of $\phi_\mathrm{uv}$ is finite then it becomes an important physical parameter that denotes the end of the flat portion of the potential, which we will see also denotes the end of the stochastic dominated region of the potential. 

The amount of expansion realised along a given trajectory is called $\mathcal{N}$, which is a stochastic variable. Thanks to the $\delta N$ formalism, the fluctuation in this number of \efolds, $\mathcal{N}-\langle \mathcal{N} \rangle$, is nothing but the coarse-grained curvature perturbation $\zeta_{\mathrm{cg}}$ defined in \Eq{eq:def:zetacg},
\bea
\delta N_{\mathrm{cg}}\left(\bm{x}\right) = 
\mathcal{N}\left(\bm{x}\right)-\left\langle \mathcal{N} \right\rangle =
\zeta_{\mathrm{cg}}\left(\bm{x}\right) = 
\frac{1}{\left(2\pi\right)^{3/2}}
\int_{k_*}^{k_\uend} \dd \bm{k} \zeta_{\bm{k}} e^{i \bm{k}\cdot \bm{x}}\, ,
\eea
where $k_*$ and $k_\uend$ are the wavenumbers that cross the Hubble radius at initial and final times when $\phi=\phi_*$ and $\phi=\phi_\uend$ respectively. Since, as explained in \Sec{sec:Intro}, the calculation of the PBH mass function relies on the PDF of coarse-grained curvature perturbations, the next step is to calculate the PDF of $\delta N_\mathrm{cg}$.

Before doing so, let us note that quantities related to $\zeta$, and not $\zeta_{\mathrm{cg}}$, can also be calculated in the stochastic-$\delta N$ formalism. For the power spectrum $\calP_\zeta$ for instance, since the coarse-grained $\delta N_{\mathrm{cg}}$ receives an integrated contribution of all modes exiting the Hubble radius during inflation, $\langle \delta N_\mathrm{cg}^2 \rangle = \int_{k_*}^{k_\uend} \calP_\zeta \dd k/k$, one has~\cite{Fujita:2013cna, Vennin:2015hra}
\bea
\label{eq:Pzeta:stochaDeltaN}
\calP_\zeta = \frac{\dd \left\langle \delta N_{\mathrm{cg}}^2 \right\rangle}{\dd \left\langle \mathcal{N}\right\rangle}\, ,
\eea
where we have used the relation $\langle \mathcal{N}\rangle = \ln(a_\uend/a_*) = \ln(k_\uend/k)$, where the last equality is valid at leading order in slow roll only. In the same manner, the local bispectrum can be written as $\mathcal{B}_\zeta \propto \dd^2 \langle \delta N_\mathrm{cg}^3\rangle/\dd \langle \mathcal{N} \rangle^2$, from which the effective $\fnl^\mathrm{local}$ parameter, measuring the ratio between the bispectrum and the power spectrum squared, is given by
\bea
\label{eq:fnl:stochaDeltaN}
\fnl^\mathrm{local} = \frac{5}{72}\frac{\left\langle \delta N_\mathrm{cg}^3\right\rangle}{\dd \left\langle \mathcal{N} \right\rangle^2}\left(\frac{\dd \left\langle \delta N_{\mathrm{cg}}^2 \right\rangle}{\dd \left\langle \mathcal{N}\right\rangle}\right)^{-2}\, .
\eea

\section{First passage time analysis} \label{sec:firstpassage}

In this section, we describe a technique called ``first passage time analysis'', which allows us to calculate the mean number of \efolds~that the inflaton realises between two points, as well as higher order moments of a probability distribution of $\mathcal{N}$, and was first applied to stochastic inflation inflation in Refs. \cite{Starobinsky:1986fx, Vennin:2015hra}.
As well as reviewing this technique, we use this section to extend these methods to include DBI inflation \cite{Silverstein:2003hf,Alishahiha:2004eh} (named for Dirac--Born--Infeld), which is a new result.
This model of inflation includes non-canonical kinetic terms, and is motivated by some models of string theory (see, for example, \cite{Dvali:1998pa,Burgess:2001fx,Kachru:2003sx}), and the inflaton is thought of as the distance between two branes.

\subsection{DBI inflation} \label{sec:DBIinflation}

As before, we consider single scalar field inflation, but in DBI inflation we have a non-canonical kinetic term. 
Since this model is motivated from string theory, it is orignally described by a $10$ dimensional theory.
Once compactified to four dimensions, the classical effective $4D$ action is given by \cite{Lorenz:2010vf}
\bea
S = - \int \dd^4x \sqrt{-g}\left[ \Mp^2 R + V\left(\phi\right) - \left( \frac{\gamma \left(\phi , \partial_{\mu}\phi \right) - 1}{\gamma \left(\phi , \partial_{\mu}\phi \right)} \right) T(\phi) \right] \, ,
\eea
where $R$ is the 4D Ricci scalar, $T(\phi)$ is called the ``warp factor'', and 
\bea
\gamma \left(\phi , \partial_{\mu}\phi \right) = \left( 1 + \frac{g^{\mu \nu}\partial_{\mu} \phi \partial_{\nu}\phi}{T(\phi )}\right) ^{-\frac{1}{2}}
\eea
is defined as the Lorentz factor. 
For an FLRW background, the Friedmann and Klein--Gordon equations then read
\begin{align}
3\Mp^2H^2 &= (\gamma - 1)T(\phi) + V(\phi) \label{eq:DBI:friedmann} \, , \\
\ddot{\phi} + \frac{3H}{\gamma^2}\dot{\phi}  &= -\frac{3\gamma - \gamma^3 - 2}{2\gamma^3}T'(\phi) - \frac{V'(\phi)}{\gamma^3} \, . \label{eq:DBI:kleingordon}
\end{align}
This shows us that $\gamma \to 1$ returns canonical kinetic terms. Also, $\gamma$ describes how the speed of sound $c_{s}$ of this model behaves through $\gamma = \frac{1}{c_s}$.
For DBI inflation in an FLRW background, we have the explicit form \cite{Chen:2006hs,Lorenz:2010vf}
\bea \label{eq:DBI:gamma:general}
\gamma &= \frac{1}{\sqrt{1 - \frac{\dot{\phi}^2}{T(\phi )}}} = \sqrt{1 + \frac{4\Mp^4 }{ T(\phi )}\left( \frac{\dd H}{\dd \phi} \right)^2} \, ,
\eea
where the second equality follows from the Friedmann and Klein--Gordon equations in FLRW DBI inflation.

In the regime $\gamma\to 1$, $\phi$ behaves as usual and obeys the standard Klein--Gordon equation \eqref{eq:eom:scalarfield}, and hence if the potential is sufficiently flat then the inflaton can slow roll. 
However, from \Eq{eq:DBI:gamma:general}, we see that the velocity of the inflaton can never exceed $\sqrt{T(\phi)}$, even if the potential becomes steep\footnote{The limit $\dot{\phi} \to \sqrt{T(\phi)}$ (\ie $\gamma\to \infty$) can therefore be thought of as an extra, ``ultra-relativistic'' regime of inflation.}. 

When we include the non-canonical kinetic term of DBI inflation, the slow-roll Langevin equation \eqref{eq:Langevin:slowroll} is modified to become \cite{Lorenz:2010vf}
\bea \label{eq:intro:DBI:Langevin}
\frac{\text{d}\phi}{\text{d}N} = -\frac{V'}{3\gamma H^2} + \frac{H}{2\pi}\xi (N) \, ,
\eea

This is the stochastic equation of motion for slow-roll DBI inflaton and is the starting point for our following analysis.

\subsection{Mean Number of \texorpdfstring{\efolds}{}}
\label{sec:Mean}

Here we will calculate the mean number of \efolds~the inflaton experiences between an initial and a final field value. 

\subsubsection{Calculating the Mean}

For a given wavenumber $k$, let $\phi_{*}(k)$ be the value of the course-grained field $\phi = \phi_{\text{cl}} + \delta \phi$ when $k$ crosses the Hubble radius.
If inflation ends at $\phiend$, let $\N(k)$ be the number of \efolds~realised between $\phi_{*}(k)$ and $\phiend$.
Here $\N$ is a stochastic quantity, and its variance is given by
\bea
\delta \N^2 (k) = \left< \N ^2(k) \right> - \left< \N (k) \right> ^2 \, .
\eea

Before outlining our computational programme, we shall first derive the so-called It\^{o} lemma, which is a relation obeyed by any smooth function $f$ of $\phi$, for the case of DBI inflation. 
From the Langevin equation \eqref{eq:intro:DBI:Langevin}, we have 
\bea
\dd\phi = -\frac{\Mp^2 V'}{\gamma V} \dd N + \frac{H}{2\pi} \xi \dd N \, .
\eea
and hence if $f(\phi)$ is a smooth function and $\phi$ obeys \eqref{eq:intro:DBI:Langevin}, then by Taylor expansion we have
\bea \label{eq:ito:integrand}
\dd f[\phi (N)] &= f(\phi + \dd\phi ) - f(\phi ) \\
&= f'\dd\phi + \frac{1}{2}f'' \dd\phi^2 + \hdots \\
&\simeq f'[\phi (N)]\sqrt{2v[\phi (N)]}\Mp \xi (N) \dd N - \Mp^2\frac{ f'[\phi (N)]}{ \gamma[\phi(N)]} \frac{v'[\phi(N)]}{v[\phi(N)]} \dd N \\
& \hspace{1cm} + f''[\phi(N)]v[\phi(N)]\Mp^2 \dd N \, ,
\eea
where we have defined the dimensionless potential $v$ to be
\bea
v = \frac{V}{24\pi^2 \Mp^4} \simeq \frac{H^2}{8\pi^2 \Mp^2} \, ,
\eea
with the second approximation
valid in slow-roll. 
Note that, in \Eq{eq:ito:integrand} even though we only keep terms of order $\dd N$, we must work at order $\dd\phi^2$ initially, as some of these terms reduce to linear order when we work in It\^{o} calculus \cite{ItoCalculus}.
We assume the inflation starts at $\phi = \phi_{*}$ and evolves until inflation ends, at $\phi_{1} < \phi_{*}$.
We then have $N = \N$ at $\phi = \phi_{1}$.
Integrating \eqref{eq:ito:integrand} between $N = 0$ (at $\phi = \phi_{*}$) and $N = \N$ then gives a generalised It\^{o} lemma, namely
\bea \label{eq:Ito:lemma} 
f(\phi_{1}) - f(\phi_{*}) &= \int^{\N}_{0} f'[\phi(N)]\sqrt{2v[\phi(N)]}\Mp \xi (N) \dd N \\ 
& \hspace{4mm} + \int^{\N}_{0} \Mp^2\left[ f''[\phi(N)]v[\phi(N)] - \frac{ f'[\phi(N)]}{\gamma [\phi(N)]} \frac{v'[\phi(N)]}{v[\phi(N)]} \right] \dd N \, .
\eea

In order to calculate the mean number of \efolds~$\left< \N \right>$, we look to use the It\^{o} lemma \eqref{eq:Ito:lemma} and define a function $f(\phi )$ such that 
\bea \label{eq:mean:fequation}
\Mp^2\left[ f''[\phi(N)]v[\phi(N)] - \frac{ f'[\phi(N)]}{\gamma[\phi(N)]} \frac{v'[\phi(N)]}{v[\phi(N)]} \right] = -1 \, ,
\eea
with boundary condition $f(\phi_{1} ) = 0$, so that the last integrand in the It\^{o} lemma becomes $-1$.
We can thus evaluate \eqref{eq:Ito:lemma} for our choice of $f$ and find
\bea \label{eq:efolds:N}
\N = f(\phi_{*}) + f'[\phi(N)]\sqrt{2v[\phi(N)]}\Mp \xi (N) \, .
\eea
Taking the stochastic average of this expression over multiple realisations then gives
\bea
\left< {\N} \right> = f(\phi_{*} ) \, ,
\eea
since $\left< \xi \right> = 0$ because $\xi$ is a Gaussian noise. 
Therefore, if we solve \eqref{eq:mean:fequation} and evaluate the result at $\phi_{*}$ we will find the mean value of $\N$.
Doing this yields 
\bea \label{eq:efolds:f}
f(\phi ) = \int^{\phi}_{\phiend} \frac{\dd y}{\Mp^2} \int^{\bar{\phi}}_{y}\frac{\dd x}{v(x)} \exp{\left[ \int ^{y}_{x} \frac{v'(z)}{\gamma (z)v^2(z)} \dd z\right]} \, ,
\eea
where $\bar{\phi} = \bar{\phi}(\phi_{1} )$ is an integration constant allowing us to satisfy our boundary conditions. 
Thus, we can evaluate $f(\phi_{*}) $ and find the mean number of \efolds~to be
\bea \label{eq:efolds:mean}
\left< \N \right> = \int^{\phi_{*}}_{\phiend}\frac{\dd y}{\Mp^2} \int^{\bar{\phi}}_{y} \frac{\dd x}{v(x)} \exp{\left[\int^{y}_{x} \frac{v'(z)}{\gamma (z)v^2(z)} \dd z \right]} \, .
\eea
Note that in the case $\gamma = 1$ this reduces to the standard result for canonical stochastic inflation \cite{Vennin:2015hra}.
This technique is called ``first passage time analysis'', and we now discuss the ``classical limit'' of this expression, before explaining how to extend this technique to calculate higher order moments of the distribution of the number of \efolds.

\subsubsection{Classical Limit}

While \eqref{eq:efolds:mean} is an exact expression, it can be hard to deduce exact general behaviour from this form, and so we seek a ``classical limit'' in which we can check the consistency of our result (\ie if we can recover known canonical results) and to gain some intuition for the leading order behaviour.
As such, we perform a saddle-point approximation around a point $y$, where the exponent is maximal. 
Taylor expanding the exponent around $y$ gives
\bea \label{eq:exponent:taylorexp}
\int^{y}_{x} \frac{v'(z)}{\gamma (z)v^2(z)}\dd z &\simeq  - \frac{v'(y)}{\gamma (y)v^2(y)}(x - y) \\
& \hspace{1mm} + \left( \frac{v'(y)\gamma'(y)}{\gamma^2(y)v^2(y)} + 2\frac{v'^2(y)}{\gamma (y)v^2(y)} - \frac{v''(y)}{\gamma (y)v^2(y)}\right) \frac{(x - y)^2}{2} \, .
\eea
By also Taylor expanding $\frac{1}{v(x)}$ around $y$ 
to first order, and imposing that $(\gamma[y]v[y])^2$ is sufficiently small, gives
\bea \label{cl mean}
\left< \N \right> |_{\text{cl}} = \frac{1}{\Mp^2}\int^{\phi_{*}}_{\phiend} \frac{\gamma(y) v(y)}{v'(y)} \dd y \, .
\eea

This classical limit is valid as long as the Taylor expansion \eqref{eq:exponent:taylorexp} converges, and we quantify this be imposing that the second term must be small, which leads us to define a classicality criterion by 
\bea \label{cl criterion}
\eta_{\text{cl}} \equiv \left| \frac{\gamma v^2}{v'} \left( 2\frac{v'}{v} - \frac{v''}{v'} + \frac{\gamma'}{\gamma} \right) \right| \ll 1 \, .
\eea

\subsection{Number of \texorpdfstring{\efolds}{}~variance}
\label{sec:Variance}

Following a similar calculation, we will now find the variance, or dispersion, in the number of \efolds~realised in this setup.
By squaring \eqref{eq:efolds:N} and averaging the result, we obtain an expression of the number of \efolds~variance as
\bea \label{eq:variance:equation}
\left< \N^2 \right> = f^2(\phi_{*}) + 2\Mp^2\left< \int^{\N}_{0}  f'^2[\phi (N)]v[\phi (N)] \dd N \right> \, .
\eea
In order to use the It\^{o} lemma, we define a function $g[\phi(N)]$ by
\bea \label{eq:variance:gequation}
g''(\phi)v(\phi) - \frac{g'(\phi)v'(\phi)}{\gamma(\phi)v(\phi)} = -2f'^2(\phi)v(\phi) \, ,
\eea
where $f$ is defined by \eqref{eq:efolds:f}.
Now, by using \Eq{eq:Ito:lemma}, imposing the boundary condition $g(\phi_{1}) = 0$, and then taking the stochastic average of the result, we see that
\bea
g(\phi_{*}) &= 2\Mp^2 \left< \int^{\N}_{0}  f'^2[\phi(N)]v[\phi(N)] \dd N \right> \\
&= \left< \N^2 \right> - \left<\N \right>^2 \\
&\equiv \delta \N^2 \, .
\eea
Thus, in order to find the variance of the number of \efolds, we need to solve \eqref{eq:variance:gequation} with $g(\phi_{1}) = 0$ and evaluate the resulting function at $\phi_{*}$.
The formal solution is given by
\bea \label{eq:variance}
\delta \N^2 = g(\phi_{*}) = 2\int_{\phiend}^{\phi_{*}} \dd y \int_{y}^{\bar{\phi}_{2}} \dd x f'^2(x) \exp{\left[ \int^{y}_{x} \frac{v'(z)}{\gamma (z) v^2(z)} \dd z \right]} \, ,
\eea
where once again $\bar{\phi}_{2}$ is an integration constant chosen to realise $g(\phi_{1}) = 0$.

\subsubsection{Classical Limit}

As before, we can derive the classical limit for the variance. 
When $\eta_{\text{cl}} \ll 1$, our saddle point approximation \eqref{eq:exponent:taylorexp} is valid, and we can also expand $f'^2(x)$ around the point $y$ using
\bea \label{eq:taylorexp3}
f'^2(x) \simeq f'^2(y) + 2f'(y)f''(y)(x - y) + \left[ f''(y)^2 + f'(y)f'''(y)\right] (x - y)^2 \, .
\eea
Using these expansions, we find that the classical limit of \eqref{eq:variance} is
\bea \label{eq::variance:classical}
\delta \N^2\big|_{\text{cl}} = \frac{2}{\Mp^4} \int^{\phi_{*}}_{\phiend} \frac{\gamma^3(y)v^4(y)}{v'(y)^3} \dd y \, .
\eea

\subsection{Power Spectrum}
\label{sec:power}

Using the expressions found so far in this section, we can now calculate the power spectrum of curvature fluctuations $\mathcal{P}_{\zeta}(k) = \mathcal{P}_{\delta N}(k) = \frac{\dd}{\dd\left< \N\right> } \delta \N^2\Big|_{\left< \N\right>  = \ln(k_{\text{end}} / k)}$.
It is straightforward to calculate
\bea \label{power}
\mathcal{P}_{\zeta}(\phi_{*}) &= \frac{g'(\phi_{*})}{f'(\phi_{*})} \\
&= 2 \left[ \int^{\bar{\phi}}_{\phi_{*}} \frac{\dd x}{\Mp v(x)} \exp{\left[ \int^{\phi_{*}}_{x} \frac{v'(z)}{\gamma (z) v^2(z)} \dd z\right]} \right]^{-1} \times \nonumber \\
& \hspace{9mm} \int^{\bar{\phi}_{2}}_{\phi_{*}} \frac{\dd x}{\Mp} \left( \int^{\phi_{\infty}}_{x} \frac{\dd y}{\Mp v(y)} \exp{\left[ \int^{\phi_{*}}_{y} \frac{v'(z)}{\gamma (z) v^2(z)} \dd z\right]} \right)^2 \exp{\left[ \int^{\phi_{*}}_{x} \frac{v'(z)}{\gamma (z) v^2(z)} \dd z\right]} \, ,
\eea
where $\mathcal{P}_{\zeta}(\phi_{*})$ stands for the power spectrum calculated at a scale $k$ such that when it crosses the Hubble radius.
We can also calculate the tilt of the power spectrum by noting that, in slow roll, we have
\bea
\frac{\partial}{\partial \ln k} \simeq -\frac{\partial \phi}{\partial \left< \N\right>}  \frac{\partial}{\partial \phi} \, ,
\eea
and hence we have
\bea \label{ns}
n_{s} = 1 - \frac{g''(\phi )}{f'(\phi )g'(\phi )} + \frac{f'' (\phi )}{f'^2(\phi )} \, .
\eea
The full expression is not written out here for brevity, but can be found from \Eqs{eq:efolds:f} and \eqref{eq:variance}.

\subsubsection{Classical Limit} 

Making use of our previous classical expressions for $f$ and $g$, we find the classical limits for the power spectrum and scalar index to be 
\bea \label{cl power}
\mathcal{P}_{\zeta}\big|_{\text{cl}} (\phi_{*}) = \frac{2}{\Mp^2} \frac{v^3(\phi_{*})\gamma ^2 (\phi_{*})}{v'(\phi_{*})^2} \, .
\eea
and
\bea \label{cl ns}
n_{s}\big|_{\text{cl}}(\phi_{*}) = 1 - \Mp^2\left[ \frac{3}{\gamma (\phi_{*})}\left(\frac{v'(\phi_{*})}{v(\phi_{*})}\right)^2 - 2\frac{v''(\phi_{*})}{\gamma (\phi_{*})v(\phi_{*})} + 2\frac{\gamma '(\phi_{*})v'(\phi_{*})}{v(\phi_{*})\gamma^2 (\phi_{*})}\right] \, ,
\eea
respectively.

%
%


We can continue to derive higher order moments of the number of \efolds~realised in this setup, such as the skewness and so on, in similar ways (following the approach of Vennin and Starobinsky \cite{Vennin:2015hra}). 
We can also keep higher order terms in our Taylor expansions in the classical limits to find the next-to-leading order corrections (or as many orders as one requires). 
At this point we note again that including the DBI term $\gamma$ in these expressions is a new result that we present in this thesis. 

In the next chapter, we build on this first passage time analysis and construct a formalism to calculate full probability distributions in the stochastic framework, rather than the individual moments presented here.

\newpage

\chapter{Quantum diffusion during inflation and primordial black holes} \label{chapter:quantumdiff:slowroll}

Having now introduced the stochastic formalism for inflation and studied its validity in a general setting, we now apply this formalism to the specific case of slow-roll inflation. 
Using stochastic inflation, we are able to calculate the full probability density function (PDF) of inflationary curvature perturbations, even in the presence of large quantum backreaction, while making no assumptions of Gaussianity (as is usually done in the literature). 
This is of direct interest for the study of the formation of primordial black holes, which we have seen provide possible solutions to several open problems in modern cosmology. 

In this chapter, making use of the stochastic-$\delta N$ formalism, two complementary methods are developed, one based on solving an ordinary differential equation for the characteristic function of the PDF, and the other based on solving a heat equation for the PDF directly. 
In the classical limit where quantum diffusion is small, we develop an expansion scheme that not only recovers the standard Gaussian PDF at leading order, but also allows us to calculate the first non-Gaussian corrections to the usual result. 
In the opposite limit where quantum diffusion is large, we find that the PDF is given by an elliptic theta function, which is fully characterised by the ratio between the squared width and height (in Planck mass units) of the region where stochastic effects dominate. 
By applying these results to the calculation of the mass fraction of primordial black holes, we show that no more than $\sim 1$ $e$-fold can be spent in regions of the potential dominated by quantum diffusion. 
We then explain how this requirement constrains inflationary potentials with two examples.

This chapter is based on the publication \cite{Pattison:2017mbe} and is organised as follows: 
In \Sec{sec:PDF}, we explain how the full PDF of curvature perturbations can be calculated in stochastic inflation. 
Using the stochastic-$\delta N$ formalism (see \Sec{sec:stochasticdeltaN}), we first derive a set of ordinary differential equations for the moments of this PDF (see \Sec{sec:FPTmoments}), from which two methods of construction of the distribution are proposed, one based on its characteristic function (see \Sec{sec:CharacteristicFunction}) and one based on a heat equation (see \Sec{sec:heatequation}). 
In \Sec{sec:ClassicalLimit}, we derive the classical limit of our formulation, where quantum diffusion is a subdominant correction to the classical field dynamics. 
At leading order, the standard result is recovered, and higher-order corrections allow us to calculate the first non-Gaussian modifications to the PDF of curvature perturbations and to the mass fraction $\beta$. 
In \Sec{sec:StochasticLimit}, we expand our calculation in the opposite limit, where the potential is exactly flat and stochastic effects dominate. In this case, the PDF of curvature perturbations is found to be highly non-Gaussian and is given by an elliptic theta function. 
In \Sec{sec:PBH}, we explain how these two limits enable one to treat more generic inflationary potentials and give a simple calculational programme that updates \Eq{eq:powerconstraint:standard} and allows one to translate PBH observational constraints into constraints on the potential. 
We then illustrate this programme with two examples. 
We finally summarise the main results and present the conclusions of this chapter in \Sec{sec:Conclusion}.

\section{Motivations}
\label{sec:Intro}

As noted in \Sec{sec:pbhs}, the usual calculations for the mass fraction of the Universe contained in primordial black holes (PBHs) relies on two assumptions, namely the use of a Gaussian PDF and the use of the slow-roll approximations. 
In this chapter, we do not make the Gaussian assumption for the PDF of curvature perturbations during inflation, while maintaining the assumption that inflation takes place in a slow-roll regime, and instead calculate these PDFs directly, making use of the stochastic formalism for inflation introduced in chapter \ref{chapter:stochastic:intro}.

As we shall see, producing curvature fluctuations of order $\zeta\sim \zeta_\uc\sim 1$ or higher precisely corresponds to the regime where quantum diffusion dominates the field dynamics over a typical time scale of one \efold. 
The validity of the standard approach that is summarised in \Sec{sec:pbhs} is therefore questionable and this is why here, we present a generic calculation of the PBH abundance from inflation that fully incorporates quantum backreaction effects, and we update \Eq{eq:powerconstraint:standard} to take into account the full quantum dynamics of the inflaton field.

In the slow-roll approximation of the stochastic formalism, recall that the inflaton field $\phi$ follows a Langevin equation of the form
\bea
\label{eq:intro:Langevin}
\frac{\dd \phi}{\dd N} = -\frac{V'}{3H^2} + \frac{H}{2\pi}\xi\left( N \right) \, ,
\eea
where, as in \Sec{sec:firstpassage}, a prime denotes a derivative with respect to the inflaton field. 
The right-hand side of this equation has two terms, the first of which involves $V'$ and is a classical drift term, and the second term involves $\xi$ which is a Gaussian white noise such that $\left\langle \xi \left( N \right) \right\rangle = 0$ and $\left\langle \xi \left( N \right) \xi \left( N' \right)\right\rangle = \delta\left( N - N' \right)$, and which makes the dynamics stochastic.

Over the time scale of one \efold, the ratio between the mean quantum kick $H/(2\pi)$, and the classical drift $V'/(3H^2)$, is of order $\sqrt{\calP_\zeta}$, provided $\calP_\zeta$ follows the classical formula~(\ref{eq:classicalPower}) and where one has made use of the Friedmann slow-roll equation \eqref{eq:friedmann:slowroll}. 
Therefore, if PBHs form when this ratio is of order one or higher, this is precisely when one expects quantum modifications to the standard result to become important.

\section{Probability distribution of curvature perturbations}
\label{sec:PDF}

The calculation of the PBH mass fraction relies on the PDF of the coarse-grained curvature perturbations through \Eq{eq:def:beta}. 
Let us explain how this distribution can be calculated in the stochastic-$\delta N$ formalism introduced in \Sec{sec:stochasticdeltaN}.

\subsection{Statistical moments of first passage times}
\label{sec:FPTmoments}
In order to calculate the PDF of the realised number of \efolds~$\N$, and hence of $\delta N_\mathrm{cg}$ (\ie of $\zeta_\mathrm{cg}$), a first step consists in calculating its statistical moments
\bea
\label{eq:moments}
f_n(\phi) = \left\langle \N^n(\phi) \right\rangle\, ,
\eea
where the dependence on the field value $\phi$ (denoted $\phi_*$ in the discussion around \Fig{fig:sketch}) at which trajectories are initiated is made explicit. 
This can be done using the ``first passage time analysis'' techniques~\cite{Bachelier:1900, Gihman:1972} outlined in \Sec{sec:firstpassage}.
By generalising the methods we explain there, one can derive a hierarchy of ordinary differential equations
\bea
\label{eq:ODE:fn}
f_n'' - \frac{v'}{v^2} f_n' = -\frac{n}{v\Mp^2} f_{n-1}\, ,
\eea
that the moments of the distribution of $\N$ satisfy. 
Note that, as done in the previous section, we have defined $v = {V}/({24\pi^2 \Mp^4)}$ to be the dimensionless potential. 
The hierarchy is initiated at $f_0 = 1$, and for $n \geq 1$ it has to be solved with two boundary conditions, one related to the fact that all trajectories initiated at $\phi_\uend$ realise a vanishing number of \efolds, and the other one implementing the presence of a reflective wall at $\phiuv$, namely
\bea
\label{eq:BC}
f_n(\phiend) = 0\, ,\quad f_n'(\phiuv) = 0\, .
\eea
The formal solution to this problem can be written as
\bea
\label{eq:fn:generalsolution}
f_n(\phi) = n \int_{\phiend}^\phi \frac{\dd x}{\Mp} \int_x^{\phiuv} \frac{\dd y}{\Mp} \ee^{\frac{1}{v(y)}-\frac{1}{v(x)}} \frac{f_{n-1}(y)}{v(y)}\, ,
\eea
which allows one to calculate the moments iteratively. In practice, this relies on performing integrals of increasing dimension, which quickly becomes numerically heavy but provides a convenient way to study the first few moments required to calculate the power spectrum given by \Eq{eq:Pzeta:stochaDeltaN} or the $\fnl^\mathrm{local}$ parameter given by \Eq{eq:fnl:stochaDeltaN}, see Refs. \cite{Vennin:2015hra, Vennin:2016wnk}.

\subsection{The characteristic function approach}
\label{sec:CharacteristicFunction}

In order to relate the PDF of $\N$ to its statistical moments, let us introduce its characteristic function
\bea
\label{eq:characteristicFunction:def}
\chi_\N(t,\phi) \equiv \left\langle \ee^{it \N(\phi)} \right\rangle \, ,
\eea 
which depends on $\phi$ and a dummy parameter $t$, which does not relate to any sort of time coordinate. 
By Taylor expanding $\chi_\N(t,\phi)$ around $t=0$, one has $\chi_\N(t,\phi) = \sum_{n=0}^\infty (it)^n f_n(\phi)/n!$. If one applies the differential operator appearing in the left-hand side of \Eq{eq:ODE:fn} to this expansion, and uses \Eq{eq:ODE:fn}  to replace each term by its right-hand side, one obtains
\bea
\label{eq:ODE:chi}
\left(\frac{\partial^2}{\partial\phi^2} - \frac{v'}{v^2} \frac{\partial}{\partial\phi} + \frac{it}{v\Mp^2}\right)\chi_\N(t,\phi) = 0 \, .
\eea
At fixed $t$, this is an ordinary differential equation in $\phi$, so instead of the hierarchy of coupled differential equations~(\ref{eq:ODE:fn}) one now has a set of uncoupled differential equations to solve, which improves the tractability of the problem. The boundary conditions~(\ref{eq:BC}), together with the fact that $f_0=1$, translate into
\bea
\label{eq:boundary:chi}
\chi_\N\left(t,\phiend\right) = 1\, ,\quad
\frac{\partial\chi_\N}{\partial\phi} \left(t, \phiuv\right) = 0 \, . 
\eea
Let us note that the characteristic function of the fluctuation in the number of \efolds, $\zeta_\mathrm{cg} = \delta N_\mathrm{cg} = \N - \langle \N \rangle = \N -f_1$, can be found by plugging this expression into \Eq{eq:characteristicFunction:def}, which gives rise to 
\bea
\label{eq:chideltaN:chiN}
\chi_{\zeta_\mathrm{cg}}\left(t,\phi\right) = e^{-i f_1(\phi) t} \chi_\N(t,\phi)\, .
\eea
Finally, from  \Eq{eq:characteristicFunction:def}, the characteristic function $\chi_\N$ can be rewritten as
\bea
\label{eq:chi:P}
\chi_{\N}(t,\phi)=\int^{\infty}_{-\infty} \ee^{it\N} P\left(N, \phi\right) \dd N \, ,
\eea
that is to say, the characteristic function is the Fourier transform of the PDF of curvature perturbations. Therefore, the PDF is the inverse Fourier transform of the characteristic function, \ie
\bea
\label{eq:PDF:chi}
P\left(\zeta_{\mathrm{cg}}, \phi\right) = \frac{1}{2\pi} \int^{\infty}_{-\infty} \ee^{-it\left[\zeta_{\mathrm{cg}}+f_1\left(\phi\right)\right]} \chi_{\N}\left(t,\phi\right)\dd t\, ,
\eea
where we have used \Eq{eq:chideltaN:chiN}. 
Thus, the calculational programme is the following: solve \Eq{eq:ODE:chi} with boundary conditions~(\ref{eq:boundary:chi}), calculate $f_1$ by either taking $n=1$ and $f_0=1$ in \Eq{eq:fn:generalsolution} or by noting that $f_1(\phi) = -i \partial\chi_\N /\partial t(t=0,\phi)$, calculate the PDF of curvature perturbations with \Eq{eq:PDF:chi}, and then the mass fraction of PBHs with \Eq{eq:def:beta}.
\subsubsection*{Example: quadratic potential}
\begin{figure}[t]
\begin{center}
\includegraphics[width=0.6\textwidth]{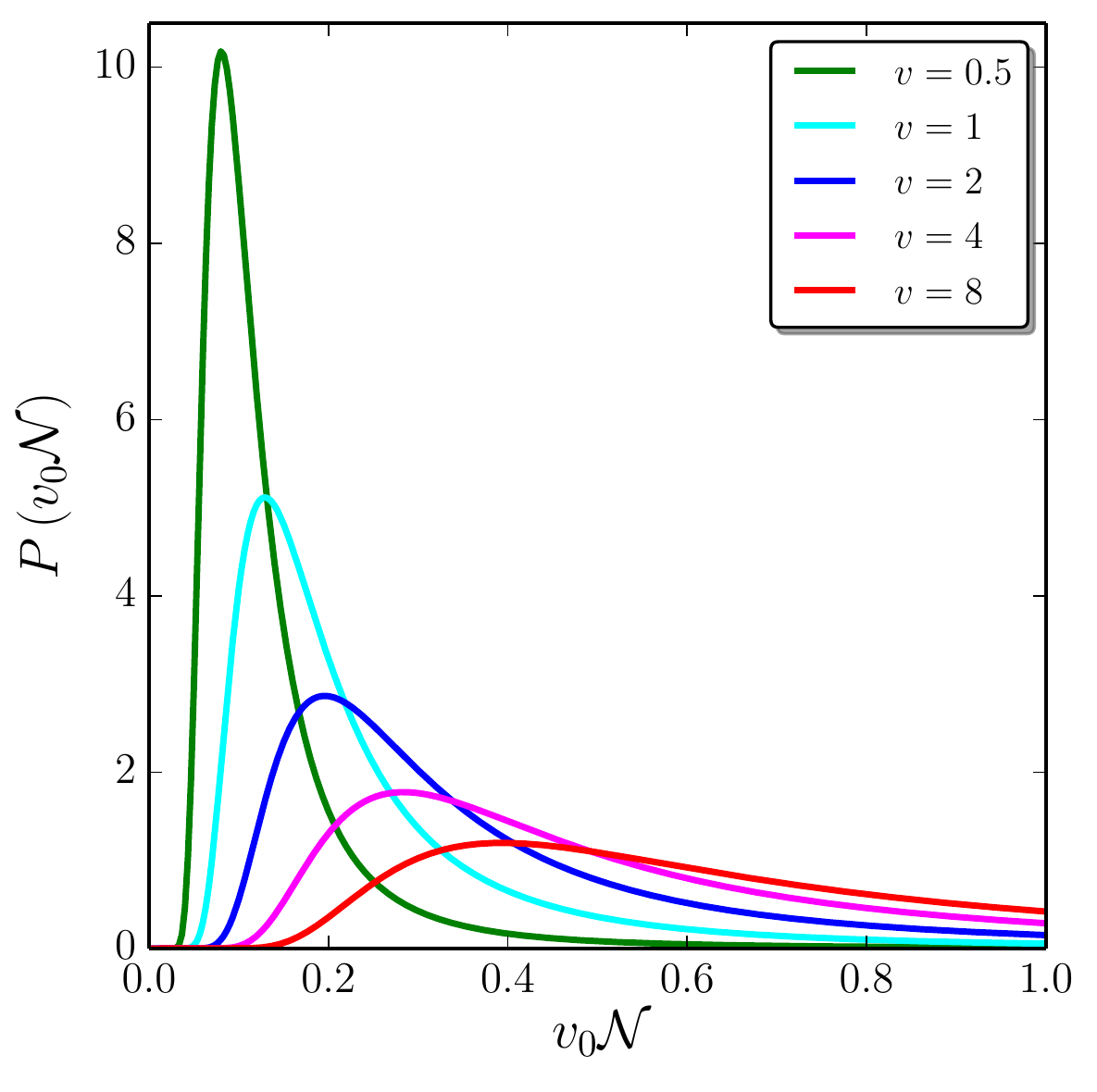}
\caption[Probability distributions of the number of \efolds~$N$ realised in a quadratic potential]{Probability distributions of the number of \efolds~$\N$, rescaled by $v_0$, realised in the quadratic potential~(\ref{eq:quadratic:potential:example}) between an initial field value $\phi$ parametrised by $v(\phi)$ given in the legend, and $\phi_\uend = \sqrt{2}\Mp$ where inflation ends by slow-roll violation. The values displayed for $v$ correspond to very high energies far from the observational window of this model but this is for illustrative purpose only. When $v$ increases, one can see that the PDF has a larger mean value, a larger spread and seems to be less Gaussian, which motivates the need to go beyond Gaussian techniques.} 
\label{fig:pdf_quadratic_solution}
\end{center}
\end{figure}
In order to illustrate this computational programme, let us consider the case of a quadratic potential
\bea
\label{eq:quadratic:potential:example}
v(\phi) = v_{0}\left(\frac{\phi}{\Mp}\right)^2 \, .
\eea
In this case, \Eq{eq:ODE:chi} together with the boundary conditions~(\ref{eq:boundary:chi}) has an exact solution. Taking the $\phiuv \rightarrow \infty$ limit, it is given by 
\bea
\label{eq:chi:quadratic}
\chi_{\N}(t, \phi) =& \left[ \frac{v(\phi)}{v(\phiend)}\right]^{\frac{1 - \alpha(t)}{4}} \frac{{}_{1}F_{1} \left[ \frac{\alpha(t) - 1}{4} ; 1 + \frac{\alpha(t)}{2} ; -\frac{1}{v(\phi)} \right]}{{}_{1}F_{1} \left[ \frac{\alpha(t) - 1}{4} ; 1 + \frac{\alpha(t)}{2} ; -\frac{1}{v(\phiend)} \right]}\, ,
\eea
where $\alpha(t) = \sqrt{1 - \frac{4 i t}{v_{0}}}$ and ${}_1F_1(x;y;z)$ is the Kummer confluent hypergeometric function \cite{Olver:2010:NHM:1830479:hypergeometric, Abramovitz:1970aa:hypergeometric}\footnote{Note that these references use the notation $M(a,b;z)$ for the Kummer confluent hypergeometric function while we use the notation ${}_1F_1$(x,y;z).}. The inverse Fourier transform of \Eq{eq:chi:quadratic} can then be computed numerically, which gives rise to the PDF displayed in \Fig{fig:pdf_quadratic_solution}. At small $v(\phi)$ the PDF is rather peaked and almost Gaussian, while at large $v(\phi)$, it is more spread and deviates more from a Gaussian distribution. In \Secs{sec:ClassicalLimit} and~\ref{sec:StochasticLimit} we will study these two limits one by one, \ie the classical limit where the stochastic corrections are small and the PDF is almost Gaussian, and the stochastic limit where quantum diffusion dominates the inflaton dynamics.
\subsection{The heat equation approach}
\label{sec:heatequation}
Before investigating the classical and stochastic limits, let us note that the problem can be reformulated in terms of a heat equation for the PDF $P(\N,\phi)$. Indeed, if one plugs \Eq{eq:chi:P} into \Eq{eq:ODE:chi}, the two first terms apply on $P(\N,\phi)$ directly, while the third one is given by $i t \chi_\N = \int_{-\infty}^{\infty} \dd N P \partial e^{it\N}/\partial N = - \int_{-\infty}^{\infty} \dd N e^{it\N} \partial P/\partial N $. 
Here, in the first expression, we have simply differentiated \Eq{eq:chi:P} with respect to $\N$, and in the second expression, we have integrated by parts (the boundary terms vanish since $P(\N=\pm\infty,\phi)$ must vanish for the distribution to be normalisable). This gives rise to the heat equation
\bea
\label{eq:heat:phi}
\left(\frac{\partial^2}{\partial\phi^2}  - \frac{v'}{v^2} \frac{\partial}{\partial\phi} - \frac{1}{v\Mp^2} \frac{\partial}{\partial N}\right) P\left( \N, \phi \right) = 0\, .
\eea 
Instead of the infinite set of uncoupled differential equations given in \Eq{eq:ODE:chi}, the problem is now reformulated in terms of a single, but partial, differential equation. Let us note that \Eq{eq:heat:phi} does not have the structure of a Fokker-Planck equation, and should in any case not be confused with the usual Fokker-Planck equation considered in stochastic inflation which governs the PDF of the field value.\footnote{The Langevin equation~(\ref{eq:intro:Langevin}) gives rise to Fokker-Planck equation for the probability density $p(N,\phi )$ of the field to be at $\phi$ at time $N$, which, in the It\^{o} interpretation, is given by
\bea
\label{eq:FokkerPlanck}
\frac{\partial^2}{\partial\phi^2}\left[v p(\N,\phi)\right] +\frac{\partial}{\partial\phi}\left[\frac{v'}{v}p(\N,\phi)\right]-\frac{1}{\Mp^2}\frac{\partial}{\partial N}p(\N,\phi)= 0\, .
\eea
This equation does not coincide with \Eq{eq:heat:phi} for $P(\N,\phi)$ which governs the probability to realise $\N$ \efolds~starting from $\phi$.} When plugging the boundary conditions~(\ref{eq:boundary:chi}) into \Eq{eq:chi:P}, one obtains
\bea
\label{eq:boundary:heat}
P\left( \N, \phiend\right) = \delta(\N)\, ,\quad
\frac{\partial P}{\partial \phi} \left(\N, \phiuv \right) = 0 \, .
\eea
These form the boundary conditions associated to \Eq{eq:heat:phi}.

In order to show that \Eq{eq:heat:phi} has the structure of a heat equation as announced above, one can introduce a change of field variable
\bea
u(\phi) = \int_{\phiend}^\phi \ee^{-\frac{1}{v(\tilde{\phi})}} \frac{\dd \tilde{\phi}}{\Mp}\, ,
\eea
which allows us to rewrite \Eq{eq:heat:phi} as
\bea
\label{eq:heat:u}
\left( v \ee^{-\frac{2}{v}}\frac{\partial^2}{\partial u^2}-\frac{\partial}{\partial N}\right) P(\N,u) = 0 .
\eea
This is a heat equation for a one dimensional medium with diffusivity $v \ee^{-2/v}$, where $\N$ plays the role of time and $u$ the role of space. However, let us stress that heat equations are usually endowed with boundary conditions of a different type as from those in \Eq{eq:boundary:heat}, since in standard heat equations, one usually gives the spatial temperature distribution at an initial time, while \Eq{eq:boundary:heat} involves distributions of times at fixed spatial positions. This is why the numerical methods developed in the literature to solve heat equations would need to be adapted to this kind of boundary conditions but they may provide efficient ways to solve the problem at hand, \eg, in the context of multi-field inflation.
\section{Expansion about the classical limit}
\label{sec:ClassicalLimit}
In the limit of small quantum diffusion, the ``classical'' limit, one needs to check that our formulation allows one to recover the standard results recalled in \Sec{sec:Intro} around \Eq{eq:beta:erfc}. 
This is the goal of this section, where we also calculate the leading order deviation from the standard result in order to best determine its range of validity. 
From the heat equation~(\ref{eq:heat:u}), we saw that the diffusivity increases with $v$, which implies that the classical limit has $v\ll 1$ (this condition is not enough to define the classical regime as we will see below but it constitutes a fair starting point).
We thus perform an expansion in increasing powers of $v$, first in the characteristic function approach introduced in \Sec{sec:CharacteristicFunction}, and then in the heat equation approach introduced in \Sec{sec:heatequation}. 
We will see that the former is much more convenient than the latter which only yields limited results in the classical limit.
\subsection{The characteristic function approach}
\label{sec:classicalAppr:characteristicMethod}
In the ordinary differential equation satisfied by the characteristic function, \Eq{eq:ODE:chi}, an expansion in $v$ is equivalent to an expansion in the diffusion term, involving $\partial^2/\partial\phi^2$.
\subsubsection{Leading order}
\label{sec:classicalAppr:characteristicMethod:LO}
At leading order (LO) in the classical limit, the diffusion term in \Eq{eq:ODE:chi} can be simply neglected, and one has
\bea
\label{eq:ODE:chi:classical}
\left(- \frac{v'}{v} \frac{\partial}{\partial\phi} + \frac{it}{\Mp^2}\right)\chi^{\lo}_\N(t,\phi)=0\, .
\eea 
Making use of the first boundary condition in \Eq{eq:boundary:chi},\footnote{In the expansion about the classical limit, the second boundary condition in \Eq{eq:boundary:chi} cannot be satisfied simultaneously with the first condition. This is why the solutions presented here are, strictly speaking, only valid in the limit $\phiuv \rightarrow \infty$.} this equation can be solved as
\bea
\label{eq:chi:classical:LO}
\chi_\N^\mathrm{\lo}(t,\phi) = \exp\left[it\int_{\phiend}^{\phi} \frac{v(x)}{\Mp^2v'(x)}\dd x\right]\, .
\eea
Note that the integral in the argument of the exponential is the classical number of \efolds, which is also the mean number of \efolds~at leading order in the classical limit, \ie the leading order saddle point expansion of \Eq{eq:fn:generalsolution}~\cite{Vennin:2016wnk},
\bea
\label{eq:f1lo}
 f_1^\lo(\phi)=\frac{1}{\Mp^2}\int_{\phiend}^\phi  \frac{v(x)}{v'(x)} \dd x\, .
\eea 
This is consistent with the formula given below \Eq{eq:PDF:chi}, namely $f_1(\phi) = -i \partial\chi_\N /\partial t(t=0,\phi)$.  As a consequence, \Eq{eq:chideltaN:chiN} implies that $\chi_{\delta N_{\mathrm{cg}}} = 1$, and hence its inverse Fourier transform is $P^\lo\left(\delta N_\mathrm{cg}, \phi\right) = \delta \left(\delta N_\mathrm{cg}\right)$, \ie a Dirac distribution centred around $\delta N_\mathrm{cg} = 0$. 
Thus, at leading order in the classical limit, one simply shuts down quantum diffusion, the dynamics are purely deterministic, $\delta\N \equiv 0$ and there are no curvature perturbations. 
\subsubsection{Next-to-leading order}
\label{sec:nlo}
One thus needs to go to next-to-leading order (NLO) to incorporate curvature perturbations. At NLO, the LO solution~(\ref{eq:chi:classical:LO}) can be used to evaluate the term $\chi^{-1}\partial^2\chi/\partial\phi^2$ in \Eq{eq:ODE:chi}, which then becomes
\bea
\frac{\partial}{\partial\phi}\chi_\N^\nlo -\frac{v^2}{v'} \left(\frac{it}{v\Mp^2}+ \frac{1}{\chi_\N^\lo}\frac{\partial^2 \chi_\N^\lo}{\partial\phi^2} \right)
\chi_\mathcal{N}^\nlo  =0\, .
\eea
Making use of the first boundary condition in \Eq{eq:boundary:chi}, the solution of this first order ordinary differential equation is
\bea
\label{eq:chi:classical:interativeSolution}
\chi_\N^\nlo (t,\phi) = \exp\left\lbrace \int_{\phi_\uend}^\phi \left[ \frac{itv(x)}{\Mp^2v' (x)} + \frac{v^2(x)}{v'(x)} \frac{1}{\chi_\N^\lo(x)}\frac{\partial^2\chi_\N^\lo}{\partial\phi^2}(x) \right] \dd x\right\rbrace\, .
\eea
Notice that if ones replaces $\lo$  by an arbitrary ${n}^\mathrm{th}$ order and $\nlo$ by the ${(n+1)}^\mathrm{th}$ order of the classical expansion, this equation is valid at any order since it is nothing but the iterative solution of \Eq{eq:ODE:chi}. At NLO, plugging \Eq{eq:chi:classical:LO} into \Eq{eq:chi:classical:interativeSolution}, one obtains
\bea
\label{eq:chi:classical:NLO}
\chi_{\N}^\nlo(t,\phi)  = \exp\left[ it f_1^\nlo\left(\phi\right) - \gamma_{1}^\nlo v t^2  \right] \, ,
\eea
where $f_1^\nlo$ is the mean number of \efolds~at NLO~\cite{Vennin:2016wnk},
\bea
f_1^\nlo(\phi)=\frac{1}{\Mp^2}\int_{\phiend}^\phi \dd x \left(\frac{v}{v'} +\frac{v^2}{{v'}}-\frac{v^3 v''}{{v'}^3}\right)\, ,
\eea
and we have defined
\bea
\label{eq:gamma1:nlo:def}
\gamma_{1}^\nlo = \frac{1}{v\Mp^4} \int_{\phiend}^\phi \dd x \frac{v^4}{{v'}^3} \, .
\eea
From this expression, \Eq{eq:chideltaN:chiN} implies that $\chi_{\delta N_\mathrm{cg}}^\nlo\left(t,\phi\right) = \ee^{-\gamma_{1}^\nlo v t^2}$, that is to say $\chi_{\delta\N}^\nlo$ is a Gaussian and hence its inverse Fourier transform $P^\nlo\left(\zeta_\mathrm{cg}, \phi\right)$ is also a Gaussian and is given by
\bea
\label{eq:PDF:classical:NLO}
P^\nlo(\zeta_{\mathrm{cg}}, \phi) = \frac{1}{\sqrt{4\pi\gamma_{1}^\nlo v}} \exp\left(-\frac{\zeta_{\mathrm{cg}}^2}{4\gamma_{1}^\nlo v}\right) \, .
\eea
A crucial remark is that at this order, the power spectrum~(\ref{eq:Pzeta:stochaDeltaN}) is given by~\cite{Vennin:2015hra} \Eq{eq:classicalPower}, so that the variance of the Gaussian distribution~(\ref{eq:PDF:classical:NLO}) reads $2\gamma_{1}^\nlo v = \int_{\phiend}^{\phi} \calP_\zeta^{\nlo} f_1^{'\lo} \dd x$. This precisely matches the standard result recalled above \Eq{eq:beta:erfc}, namely that $P(\zeta_{\mathrm{cg}})$ is a Gaussian PDF with standard deviation given by the integrated power spectrum $\left\langle \zeta_{\mathrm{cg}}^2 \right\rangle = \int_{k}^{k_\uend} \calP_\zeta(\tilde{k})\dd \ln \tilde{k}$, since $\dd \ln k \simeq \dd N = f_1'(\phi) \dd \phi$ at leading order in slow roll.
\subsubsection{Next-to-next-to-leading order}
\label{sec:classical:nnlo}
\begin{figure}[t]
\begin{center}
\includegraphics[width=0.6\textwidth]{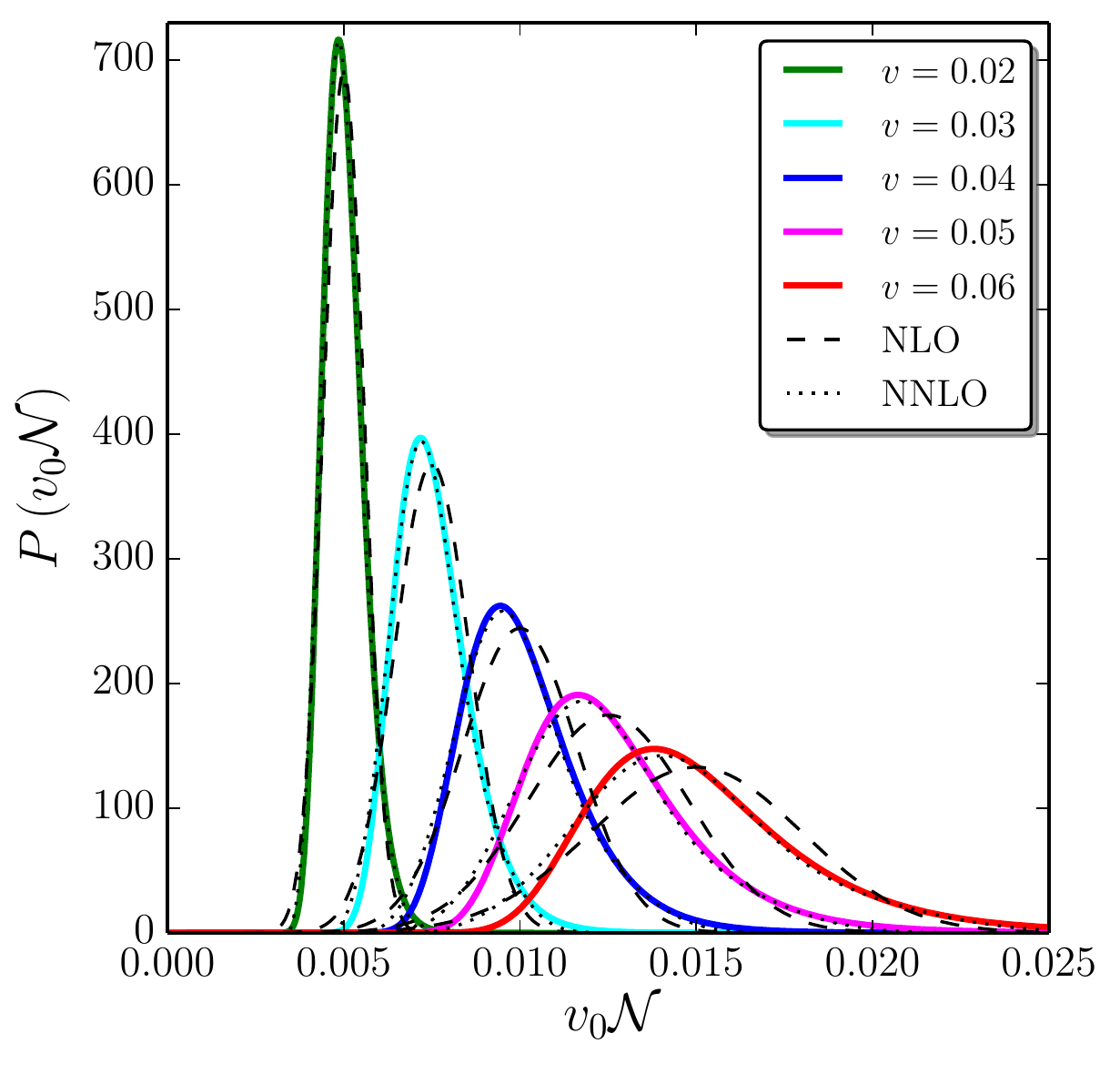}
\caption[Probability distribution of $N$ in a quadratic potential with approximations at NLO and NNLO]{Probability distributions of the number of \efolds~$\N$, rescaled by $v_0$, realised in the quadratic potential~(\ref{eq:quadratic:potential:example}) between an initial field value $\phi$ parametrised by $v(\phi)$ given in the legend, and $\phi_\uend = \sqrt{2}\Mp$ where inflation ends by slow-roll violation, as in \Fig{fig:pdf_quadratic_solution}. The black dashed lines correspond to the NLO (Gaussian) approximation~(\ref{eq:PDF:classical:NLO}), while the dotted lines stand for the NNLO approximation~(\ref{eq:PDF:NNLO}). The smaller $v$ is, the better these approximations are, and the NNLO approximation is substantially better than the NLO one.} 
\label{fig:pdf_quadratic_solution_nnlo}
\end{center}
\end{figure}

In order to study the first non-Gaussian corrections to the standard result, one needs to go to next-to-next-to-leading order (NNLO). As explained in \Sec{sec:nlo}, one can simply increment the order of the iterative relation~(\ref{eq:chi:classical:interativeSolution}), \ie replace $\lo$ by $\nlo$ and $\nlo$ by $\nnlo$. Plugging in \Eq{eq:chi:classical:NLO}, and making use of \Eq{eq:chideltaN:chiN}, this gives rise to
\bea
\label{eq:chiNNLO}
\chi_{\delta N_{\mathrm{cg}}}^\nnlo\left(t,\phi\right)  = \exp\left( - \gamma_{1}^\nnlo v t^2 - i\gamma_{2}^\nnlo v^2t^3 
 \right)  \, ,
\eea
where we have only kept the terms that are consistent at that order and where we have defined
\bea
\label{eq:gamma:nnlo:def}
\gamma_{1}^\nnlo &= \frac{1}{v\Mp^4 }  \int_{\phiend}^\phi\dd x\left(\frac{v^4 }{v'^3}+6\frac{v^5 }{v'^3}-5\frac{v^6 v''}{v'^5}\right) \, ,\\
\gamma_{2}^\nnlo &= \frac{2}{v^2 \Mp^6}  \int_{\phiend}^\phi\dd x \frac{v^7}{v'^5} \, .
\eea
One can already see that since the characteristic function is not a Gaussian; the PDF is not a Gaussian distribution. Using \Eq{eq:PDF:chi}, it is given by
\bea
\label{eq:PDF:NNLO:chi}
P^{\nnlo}\left( \delta N_{\mathrm{cg}}, \phi \right) = \frac{1}{2\pi}\int_{-\infty}^{\infty} \dd t \exp\left( -it\delta N_{\mathrm{cg}} - \gamma_{1}^\nnlo vt^2 + i\gamma_{2}^\nnlo v^2t^3  \right) \, .
\eea
In this integral, the second term in the argument of the exponential makes the integrand become negligible when $\gamma_1^\nnlo v t^2 \gg 1$, \ie for $\vert t \vert \gg t_\uc$ where $t_\uc = (\gamma_{1}^\nnlo v )^{-1/2}$. When $t=\pm t_\uc$, the ratio between the third and the second terms in the argument of the exponential of \Eq{eq:PDF:NNLO:chi} is of order $(\gamma_2^\nnlo / \gamma_1^\nnlo) \sqrt{v/\gamma_1^\nnlo}$, \ie of order $\sqrt{v}$ in an expansion in $v$ since the $\gamma_i$ parameters have been defined to carry no dimension of $v$ (at least at their leading orders). This is why, over the domain of integration where most of the contribution to the integral comes from, the third term is negligible and can be Taylor expanded. One obtains 
\bea
\label{eq:PDF:NNLO}
P^{\nnlo}\left( \zeta_\mathrm{cg}, \phi \right)  = \frac{1}{\sqrt{4 \pi \gamma_{1}^\nnlo v}} \exp\left(-\frac{\zeta_\mathrm{cg}^2}{4\gamma_{1}^\nnlo v}\right)\left[ 1 - \frac{\gamma_{2}^\nnlo}{8\left({\gamma_{1}^\nnlo}\right)^3 v} \zeta_\mathrm{cg} \left( 6 \gamma_{1}^\nnlo v - \zeta_\mathrm{cg}^2 \right) \right] \, .
\eea 
For the quadratic potential example discussed in \Sec{sec:CharacteristicFunction}, in \Fig{fig:pdf_quadratic_solution_nnlo} we have reproduced \Fig{fig:pdf_quadratic_solution} (for different values of $v$ to better illustrate the behaviours of the classical approximations) where we have superimposed the NLO approximation~(\ref{eq:PDF:classical:NLO}) and the NNLO approximation~(\ref{eq:PDF:NNLO}). One can check that these approximations become better at smaller values of $v$ as expected, and that the NNLO approximation always provides a better fit than the NLO one. 

As a consistency check, one can verify that the distribution~(\ref{eq:PDF:NNLO}) yields the same moments at NNLO order as the ones derived in Ref. \cite{Vennin:2015hra} by calculating the integrals~(\ref{eq:fn:generalsolution}) with a saddle-point approximation technique at NNLO. For the second moment, one has $\langle \delta N_{\mathrm{cg}}^2 \rangle = \int^{\infty}_{-\infty} \zeta_{\mathrm{cg}}^2 P^{\nnlo} ( \zeta_\mathrm{cg}, \phi ) \dd \zeta_{\mathrm{cg}} = 2 \gamma_1^\nnlo v$, which coincides with Eq.~(3.35) of Ref. \cite{Vennin:2015hra}. Similarly for the third moment, $\langle \delta N_{\mathrm{cg}}^3 \rangle = \int^{\infty}_{-\infty} \zeta_{\mathrm{cg}}^3 P^{\nnlo} ( \zeta_\mathrm{cg}, \phi ) \dd \zeta_{\mathrm{cg}} = 6 \gamma_2^\nnlo v^2$, which coincides with Eq.~(3.37) of Ref. \cite{Vennin:2015hra}. The two methods, \ie the iterative solution~(\ref{eq:chi:classical:interativeSolution}) of the characteristic function equation and the saddle-point expansion of the integrals~(\ref{eq:fn:generalsolution}), are therefore equivalent.

Let us also note that the characteristic function, $\chi_{\cal N}(t,\phi)$ defined in \Eq{eq:characteristicFunction:def}, is closely related to the cumulant generating function for the probability distribution
\bea
\label{defKtau}
K_{\cal N}(\tau,\phi) = \ln \langle e^{\tau {\cal N}(\phi)} \rangle = \sum_{n=1}^\infty \frac{\kappa_n(\phi)}{n!} \tau^n \,.
\eea
By comparing \Eqs{eq:characteristicFunction:def} and~(\ref{defKtau}) indeed, one simply has $\chi_{\cal N}(t,\phi)=\exp\left[K_{\cal N}(it,\phi)\right]$. If we now compare \Eqs{eq:chiNNLO} and~(\ref{defKtau}), we can read off the first cumulants
\bea
\kappa_2(\phi) = 2 v \gamma_{1} \,,  \quad \kappa_3(\phi) = 6v^2 \gamma_{2} \,.
\eea
One measure of the deviation from a Gaussian distribution is the skewness of the distribution which is determined by the ratio of these cumulants
\bea
\label{eq:gamma_skew}
\gamma_{\mathrm{skew}} \equiv \frac{\kappa_3}{\kappa_2^{3/2}} = \frac{3v^{1/2}\gamma_{2}}{\sqrt{2}(\gamma_{1})^{3/2}}
\,.
\eea
Since $\gamma_2$ is non-vanishing at next-to-next-to-leading (and higher) order only, the NNLO term thus represents the first non-Gaussian correction to the standard Gaussian result obtained at NLO in the classical limit. 

At this order, the distribution is positively skewed, which is indeed the case for all the distributions displayed in \Figs{fig:pdf_quadratic_solution} and~\ref{fig:pdf_quadratic_solution_nnlo}. One can also note that the parameter introduced below \Eq{eq:PDF:NNLO:chi}, that must be small in order for  the classical expansion to be valid at NNLO, exactly coincides with $\gamma_{\mathrm{skew}}$. The above formulae are therefore correct in the limit $\gamma_{\mathrm{skew}}\ll 1$ only. Finally, the correcting term in the brackets of \Eq{eq:PDF:NNLO} can be expressed as $\gamma_{\mathrm{skew}} \left(\zeta_\mathrm{cg}^2/\left\langle \zeta_\mathrm{cg}^2 \right\rangle\right)^{3/2}$, where we have used the relation $\langle \delta N_{\mathrm{cg}}^2 \rangle =  2 \gamma_1^\nnlo v$ given above together with \Eq{eq:gamma_skew}. This shows that $\gamma_{\mathrm{skew}}\ll 1$ only ensures the correcting term to be small when $\zeta_\mathrm{cg}^2$ is of order $\left\langle \zeta_\mathrm{cg}^2 \right\rangle$, \ie around the maximum of the distribution. The classical approximation is therefore an expansion about the maximum of the distribution that should be expected to fail in the tail. Since the PBH threshold $\zeta_\uc$ is usually in the far tail of the distribution (in the standard calculation recalled in \Sec{sec:Intro}, at the level of the observational bounds, one has $\left\langle \zeta_\mathrm{cg}^2 \right\rangle \sim 10^{-2} \ll \zeta_\uc^2\sim 1$), one may need to go beyond the classical approximation in such cases.
\subsection{The heat equation approach}
Before moving on to the stochastic limit, let us briefly explain how the heat equation approach proceeds in the classical limit. At LO, neglecting the diffusion term in \Eq{eq:heat:phi}, one has to solve $\Mp^2 v'/v \, \partial P/\partial \phi + \partial P/\partial N =0$, with the first boundary condition of \Eq{eq:boundary:heat}. Using the method of characteristics to solve first-order partial differential equations, one obtains
\bea
\label{eq:heat:classical:LO:solution}
P^{\lo} \left( \N, \phi \right) = \delta \left[ \N - f_1^{\lo}\left(\phi\right) \right] \, ,
\eea
where $ f_1^{\lo}$ has been defined in \Eq{eq:f1lo} and corresponds to the classical number of \efolds, which is also the mean number of e-folds at leading order in the classical limit. One therefore recovers the result of \Sec{sec:classicalAppr:characteristicMethod:LO}. At NLO, one can use \Eq{eq:heat:classical:LO:solution} to calculate the diffusive term in \Eq{eq:heat:phi} and iterate the procedure. However, by doing so, one has to solve a first-order partial differential equation with a source term that involves derivatives of the Dirac distribution. This makes the solving procedure technically complicated, and we therefore do not pursue this direction further since a simpler way to obtain the solution was already presented in \Sec{sec:classicalAppr:characteristicMethod}. One can already see the benefit of having two solving procedures at hand, which will become even more obvious in what follows.
\section{The stochastic limit}
\label{sec:StochasticLimit}
\begin{figure}[t]
\begin{center}
\includegraphics[width=0.6\textwidth]{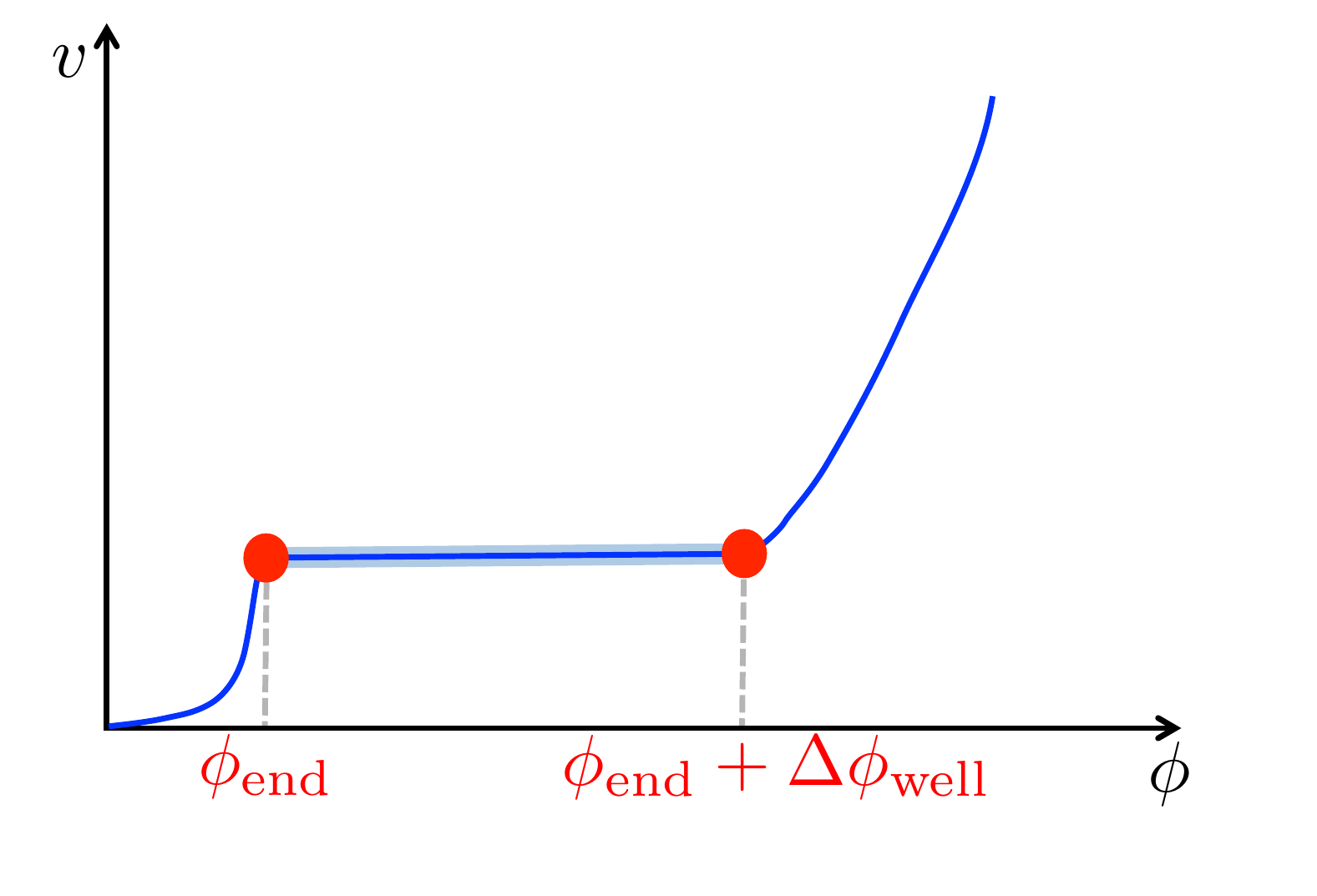}
\caption[Sketch of the ``quantum well'' for an exactly flat potential]{
Schematic representation of the single-field stochastic dynamics solved in \Sec{sec:StochasticLimit}, where the potential may be taken to be 
exactly constant over the ``quantum well'' regime delimited by $\phi_\uend$ and $\phi_\uend+\Delta\phiwell$. Inflation terminates at $\phi_\uend$, where either the potential becomes very steep or a mechanism other than slow-roll violation ends inflation, and a reflective wall is placed at $\phi_\uend+\Delta\phiwell$, which can be seen as the point where the dynamics become classically dominated and the classical drift prevents the field from escaping the quantum well.}  
\label{fig:sketch2}
\end{center}
\end{figure}
We now consider the opposite limit where the inflaton field dynamics are dominated by quantum diffusion. This is the case if the potential is exactly flat, since then the slow-roll classical drift vanishes. We thus consider a potential that is constant between the two values $\phi_\uend$ and $\phi_\uend+\Delta\phiwell$, where $\Delta\phiwell$ denotes the width of this ``quantum well''. Inflation terminates when the field reaches $\phi_\uend$ (where either the potential is assumed to become very steep, or a mechanism other than slow-roll violation must be invoked to end inflation), and a reflective wall is located at  $\phi_\uend+\Delta\phiwell$, which can be seen as the point where the dynamics become classically dominated so that the probability for field trajectories to climb up this part of the potential and escape the quantum well can be neglected. The situation is depicted in \Fig{fig:sketch2}, and in \Sec{sec:PBH} we will see why these assumptions allow one to study most cases of interest. 
\subsection{The characteristic function approach}
If the potential $v=v_0$ is constant, the potential gradient term vanishes in \Eq{eq:ODE:chi} and making use of the boundary conditions~(\ref{eq:boundary:chi}), where $\phiuv$ is replaced by $\phi_\uend+\Delta\phiwell$, one obtains
\bea
\label{eq:chiN:cosh}
\chi_\N\left(t, \phi \right) = \frac{\cosh\left[\alpha \sqrt{t} \mu \left(x-1\right)\right]}{\cosh\left(\alpha \sqrt{t} \mu\right)}\, .
\eea
In this expression, $x\equiv (\phi-\phi_\uend)/\Delta\phiwell$, $\alpha\equiv (i-1)/\sqrt{2}$, and we have introduced the parameter
\bea
\label{eq:def:mu}
\mu^{2} = \frac{(\Delta\phiwell)^2}{v_{0}\Mp^2} 
\eea
which is the ratio between the squared width of the quantum well and its height, in Planck mass units, and which is the only combination through which these two quantities appear.

\begin{figure}[t]
\begin{center}
\includegraphics[width=0.6\textwidth]{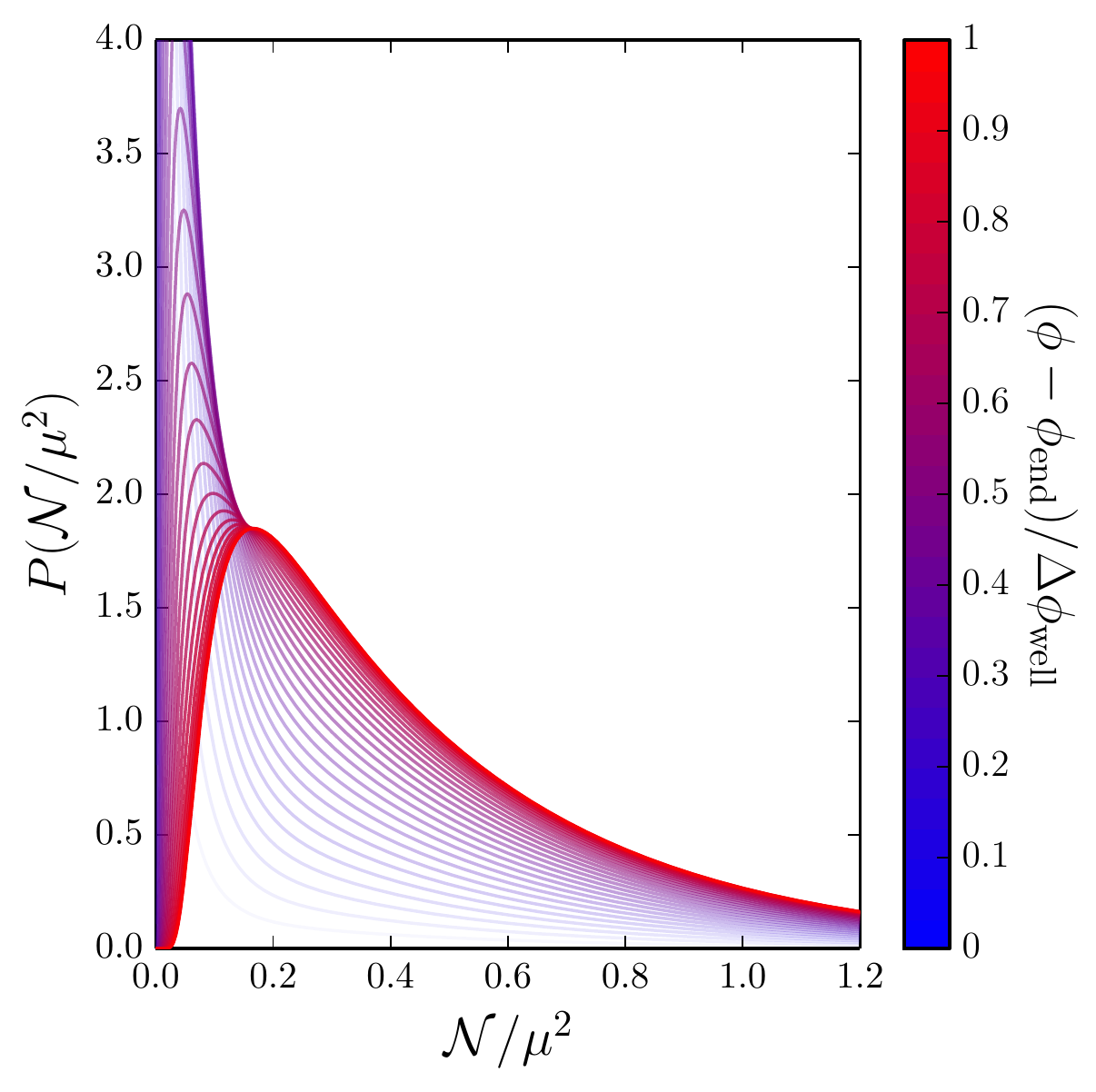}
\caption[Probability distribution of the number of \efolds~$N$ in a flat potential]{Probability distributions of the number of \efolds~$\N$, rescaled by $\mu^2$, realised in the constant potential depicted in \Fig{fig:sketch2} between $\phi$ and $\phi_\uend$, where different colours correspond to different values of $\phi$. When $\phi$ approaches $\phi_\uend$, the distribution becomes more peaked and the transparency of the curves is increased for displayed purposes.} 
\label{fig:pdf_stochastic}
\end{center}
\end{figure}

The PDF can be obtained by inverse Fourier transforming \Eq{eq:chiN:cosh}, see \Eq{eq:PDF:chi}, which can be done after Taylor expanding the characteristic function~(\ref{eq:chiN:cosh}) and inverse Fourier transforming each term in the sum. This leads to
\bea
\label{eq:stocha:CharacteristicFunctionMethod:PDF}
& P\left(\N, \phi \right) = \frac{1}{2\sqrt{\pi}}\frac{\mu}{\N^{3/2}} \times \\
& \quad \left\lbrace
 \sum_{n=0}^\infty (-1)^n \left[2(n+1) - x \right]
 \ee^{- \frac{\mu^2}{4\N} \left[2(n+1) - x \right]^2} 
+ \sum_{n=0}^\infty (-1)^n \left[2n + x \right]
 \ee^{- \frac{\mu^2}{4\N} \left[2n + x \right]^2} 
\right\rbrace\, .
\eea
This PDF is displayed in \Fig{fig:pdf_stochastic} for different values of $x$. Interestingly, \Eq{eq:stocha:CharacteristicFunctionMethod:PDF} can be resummed to give a closed form when combined with the result from the heat equation approach presented below in \Sec{sec:StochasticLimit:HeatEquationApproach}. For now, we can derive closed form expressions at both boundaries of the quantum well, \ie in the two limits $\phi\simeq \phi_\uend+\Delta\phiwell$ and $\phi\simeq\phiend$.
\subsubsection*{Reflective boundary of the quantum well}
In the case where $\phi =  \phi_\uend+\Delta\phiwell$, or $x=1$, \ie at the reflective boundary of the quantum well, \Eq{eq:stocha:CharacteristicFunctionMethod:PDF} reduces to $P(\N, \phiwell) =\mu/\sqrt{\pi} \N^{-3/2}  \sum_{n=0}^\infty (-1)^n (2n+1) \ee^{- \frac{\mu^2}{4\N}  (2n+1)^2}$. Making use of the elliptic theta functions~\cite{Olver:2010:NHM:1830479:theta, Abramovitz:1970aa:theta} introduced in \App{appendix:identities}, this can be rewritten as 
\bea
\label{eq:PDF:phiwall:thetaElliptic}
P\left(\N, \phi =  \phi_\uend+\Delta\phiwell\right) = \frac{\mu}{2 \sqrt{\pi} \N^{3/2}}  
\vartheta^\prime_1\left(0,\ee^{-\frac{\mu^2}{\N}}\right) \, ,
\eea
where $\vartheta_1^\prime$ is the derivative (with respect to the first argument) of the first elliptic theta function, see \Eq{eq:theta1prime:def}.
\subsubsection*{Absorbing boundary of the quantum well}
\begin{figure}[t]
\begin{center}
\includegraphics[width=0.496\textwidth]{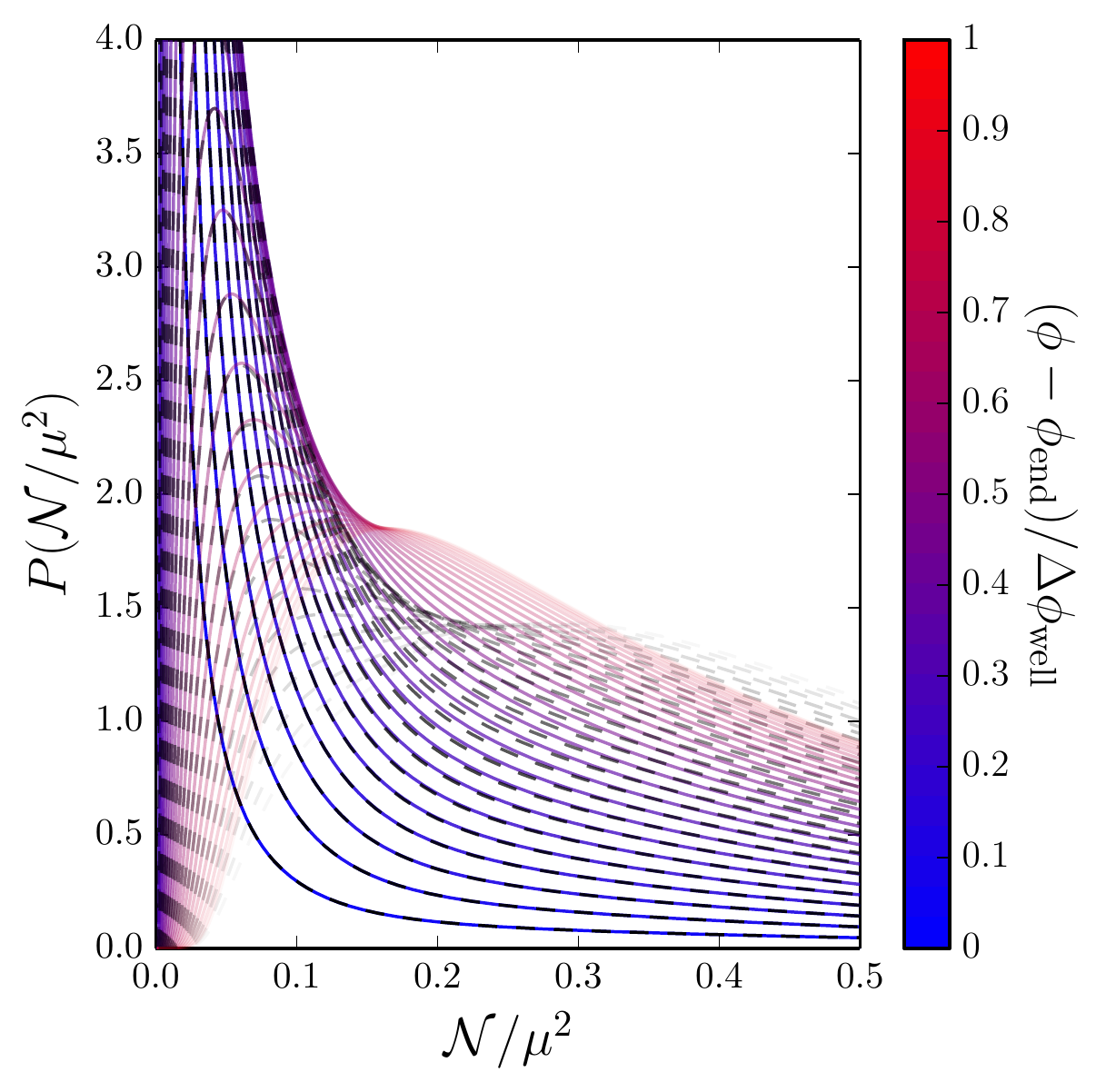}
\includegraphics[width=0.496\textwidth]{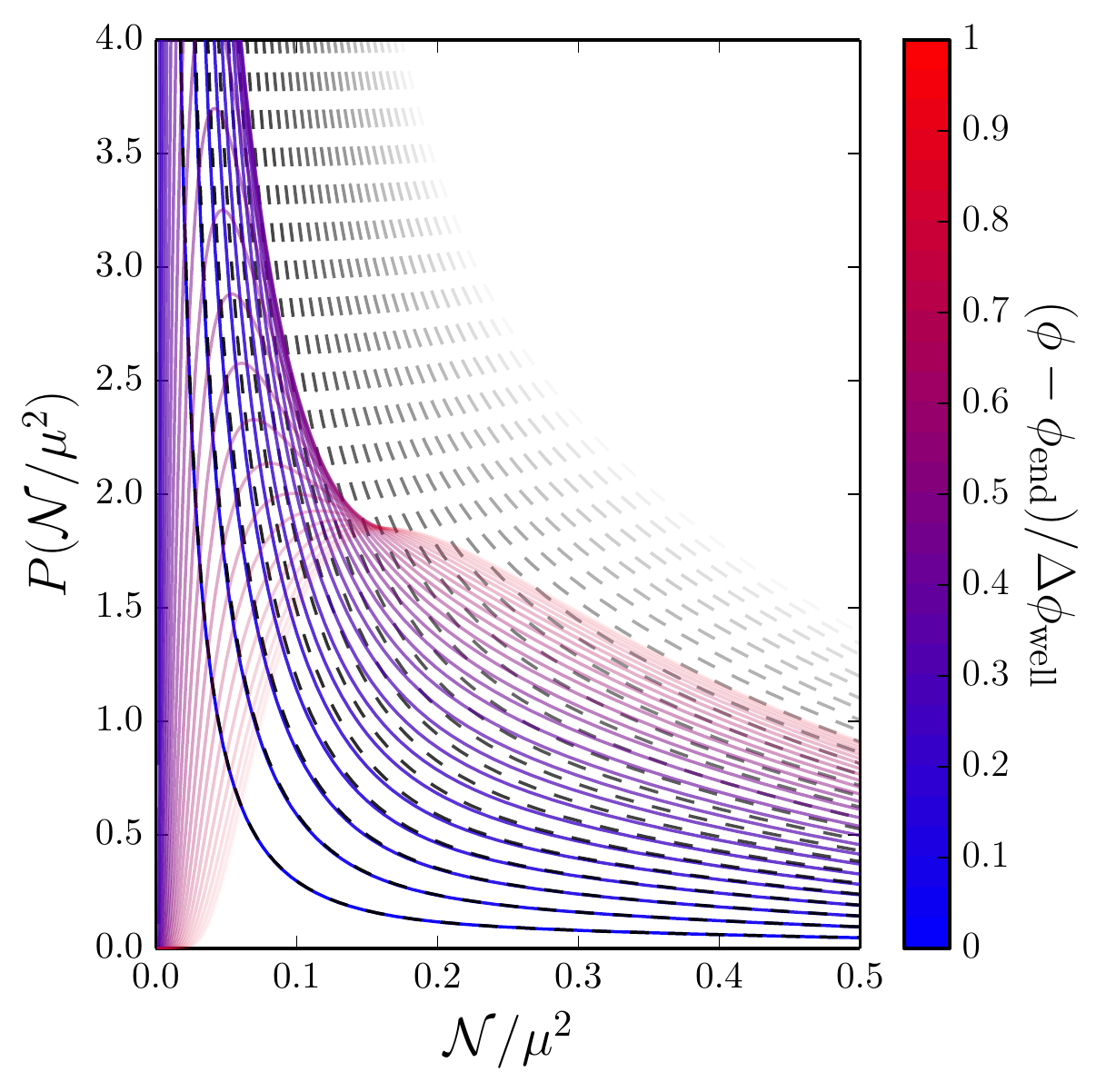}
\caption[Probability distribution for $N$ in a flat potential, compared to different approximation schemes]{Probability distributions of the number of \efolds~$\N$, rescaled by $\mu^2$, realised in the constant potential depicted in \Fig{fig:sketch2} between $\phi$ and $\phi_\uend$. In both panels, different colours correspond to different values of $\phi$, and the black dashed lines correspond to approximations. Left panel: the approximation~(\ref{eq:PDF:stocha:phiend:appr}) is displayed with the black dashed lines. Right panel: the approximation~(\ref{eq:PDF:stocha:phiend:appr:2}) is displayed with the black dashed lines. These approximations are valid close to the absorbing boundary of the quantum well where inflation ends. When $\phi$ increases, the approximation becomes worse, and the transparency of the curves is increased for displayed purposes, but one can see that the approximation~(\ref{eq:PDF:stocha:phiend:appr}) is excellent up to $(\phi-\phiend)/\Delta\phiwell \sim 0.3$, and slightly better than the approximation~(\ref{eq:PDF:stocha:phiend:appr:2}).} 
\label{fig:pdf_stochastic_appr}
\end{center}
\end{figure}

In the case where $\phi \simeq  \phi_\uend$, or $x\ll 1$, \ie at the absorbing boundary of the quantum well, an approximated formula can be obtained by noting that \Eq{eq:stocha:CharacteristicFunctionMethod:PDF} can be rewritten as $P(\N, \phi) = \mu /(2\sqrt{\pi}\N^{3/2})[x \ee^{-\mu^2 x^2/(4\N)}+F(-x)-F(x)]$, with $F(x)\equiv \sum_{n=0}^\infty (-1)^n [2(n +1)+ x]
 \ee^{- \frac{\mu^2}{4 \N} [2(n+1)+x]^2}$. In the limit where $x\ll 1$, $F(-x)-F(x)\simeq -2 x F'(0)$, where $F'(0)=1/2 - 1/2 \vartheta_4(0,\ee^{-\mu^2/\N})-\mu^2/(4\N)\vartheta_4''(0,\ee^{-\mu^2/4})$, see \Eq{eq:theta4primeprime:def}. This gives rise to
\bea
\label{eq:PDF:stocha:phiend:appr}
P\left(\N, \phi \simeq \phi_\uend \right) \simeq \frac{\mu x}{2\sqrt{\pi}\N^{3/2}}
\left[ \ee^{- \frac{\mu^2 x^2}{4 \N }} 
- 1 + \vartheta_4\left(0,\ee^{- \frac{\mu^2}{\N} } \right) + \frac{\mu^2}{2 \N} \vartheta_4^{\prime\prime}\left(0,\ee^{- \frac{\mu^2}{\N} } \right) 
\right] \, .
\eea
This approximation is superimposed to the full result~(\ref{eq:PDF:phiwall:thetaElliptic}) in the left panel of \Fig{fig:pdf_stochastic_appr}, where one can check that the agreement is excellent even up to $x\sim 0.3$.

As a brief aside, let us note how this setup is altered when we include the non-canonical kinetic terms of DBI inflation (see \Sec{sec:DBIinflation} for a brief introduction).
The equation for the characteristic function is modified to give 
\bea
\label{eq:DBI:ODE:chi}
\left(\frac{\partial^2}{\partial\phi^2} - \frac{v'}{\gamma v^2} \frac{\partial}{\partial\phi} + \frac{it}{v \Mp^2}\right)\chi_\N(t,\phi) = 0 \, .
\eea
In the case of a flat potential at the end of inflation we neglect the term $v'/(\gamma v^2)$ and can then solve \eqref{eq:ODE:chi} exactly, giving us the exact  same equation we find in the canonical case.
This is because the DBI correction is a correction to the classical drift, but does not affect the amplitude of the quantum fluctuations.
This means quantum diffusion proceeds as it does in the absence of a DBI term, and the solution for the characteristic function is simply given by \Eq{eq:chiN:cosh} and our conclusions for a flat potential do not change for DBI inflation, \ie in ``quantum wells" of a potential we have stochastic effects dominating.

%
%
\subsection{The heat equation approach}
\label{sec:StochasticLimit:HeatEquationApproach}
Let us now move on to the heat equation approach, since combined with the results of the characteristic function approach, this will allow us to derive a closed form for the PDF at arbitrary values of $x$. In the case of a constant potential, the heat equation~(\ref{eq:heat:phi}) becomes $(v_0\Mp^2\partial^2/\partial\phi^2 - \partial/\partial N)P(\N, \phi ) = 0$. The second boundary condition of \Eq{eq:boundary:heat},  $\partial P/\partial\phi (\N, \phi_\uend+\Delta\phiwell) = 0$, leads to a Fourrier decomposition of the form
\bea
\label{eq:stochastic:heat:FourierDecomposition}
P\left(\N, \phi \right) = \sum_{n=0}^\infty \left\lbrace A_n\left(\N\right) \sin\left[\left(\frac{\pi}{2}+ n\pi\right)x\right]
+B_n\left(\N\right) \cos\left(n\pi x \right) \right\rbrace \, ,
\eea
where, by plugging \Eq{eq:stochastic:heat:FourierDecomposition} into the heat equation~(\ref{eq:heat:phi}), the coefficients $A_n$ and $B_n$ must satisfy
\bea
\frac{\partial A_n}{\partial N} = -\frac{\pi^2}{\mu^2} \left(n+\frac{1}{2}\right)^2 A_n\, , \quad
\frac{\partial B_n}{\partial N} = -\frac{\pi^2}{\mu^2} n^2 B_n\, .
\eea
This leads to
\bea
A_n(\mathcal{N})  = a_n \exp\left[ -\frac{\pi^2}{\mu^2} \left(n+\frac{1}{2}\right)^2 \mathcal{N}\right]\, , \quad
B_n(\mathcal{N})  = b_n  \exp\left(-\frac{\pi^2}{\mu^2} n^2 \mathcal{N} \right)\, ,
\eea
where $a_n$ and $b_n$ are coefficients that depend only on $n$. They can be calculated by identifying \Eqs{eq:stocha:CharacteristicFunctionMethod:PDF} and~(\ref{eq:stochastic:heat:FourierDecomposition}) in the $\N \to 0$ limit. In this limit, in \Eq{eq:stocha:CharacteristicFunctionMethod:PDF}, the term with $n=0$ of the second sum is the dominant contribution, and using the fact that $\ee^{-x^2/(4\sigma)}/ (2\sqrt{\pi \sigma}) \to \delta(x)$ when $\sigma\to 0$, hence $- x\ee^{-x^2/(4\sigma)}/ (4\sigma \sqrt{\pi \sigma}) \to \delta'(x)$ when $\sigma\to 0$, one has 
\bea
\label{eq:stochastic:PDF:Neq0}
P(\N,\phi)\underset{\N \to 0}{\longrightarrow} -2 v_0 \Mp^2 \delta'(\phi-\phi_\uend)\, .
\eea
In passing, one notes that this expression implies that $P(\N=0,\phi)=0$ when $\phi \neq \phi_\uend$, which is consistent with the continuity of the distribution when $\N=0$ and with the fact that the probability to realise a negative number of \efolds~obviously vanishes. The case $\phi=\phiend$ is singular because of the first boundary condition of \Eq{eq:boundary:heat}, which explains the singularity in \Eq{eq:stochastic:PDF:Neq0}. The coefficients $a_n$ and $b_n$ can then be expressed as $a_n = \int_{-1}^1 \dd x P(\N=0,\phi) \sin[(n+1/2)\pi x]$ for $n\geq 0$ and $b_n = \int_{-1}^1 \dd x P(\N=0,\phi) \cos[n\pi x]$ for $n\geq 0$, where we recall that the link between $\phi$ and $x$ is given above \Eq{eq:def:mu}. This gives rise to $a_n=2\pi(n+1/2)/\mu^2$ and $b_n=0$, hence 
\bea
\label{eq:stocha:HeatMethod:PDF:expansion}
P\left(\N, \phi \right) = & \frac{2 \pi}{\mu^2} \sum_{n=0}^\infty \left( n + \frac{1}{2} \right) \exp\left[ -\frac{\pi^2}{\mu^2} \left(n+\frac{1}{2}\right)^2 \N\right] \sin\left[x \pi\left(n+\frac{1}{2}\right)\right] \, ,
\eea
which can be written as
\bea
\label{eq:PDF:thetatwo}
P\left( \N, \phi \right) = & -\frac{\pi}{2 \mu^2} \vartheta_{2}' \left( \frac{\pi}{2}x, \ee^{-\frac{\pi^2}{\mu^2} \N} \right) \, ,
\eea
see \Eq{eq:theta2prime:def}. A few comments are in order.

First, let us stress that the results from both methods, the characteristic function one and the heat equation one, have been necessary to derive this closed form, since the expression coming from the characteristic function has allowed us to calculate the coefficients $a_n$ and $b_n$ in the heat equation solution. This further illustrates how useful it is to have two approaches at hand.

Second, the expansion~(\ref{eq:stocha:HeatMethod:PDF:expansion}) is an alternative to the one given in \Eq{eq:stocha:CharacteristicFunctionMethod:PDF} for the PDF. One can numerically check that they are identical, and in \Fig{fig:pdf_stochastic}, $P(\N,\phi)$ is displayed as a function of $\N$ for various values of $\phi$. The difference between \Eqs{eq:stocha:CharacteristicFunctionMethod:PDF} and~(\ref{eq:stocha:HeatMethod:PDF:expansion}) is that they correspond to expansions around different regions of the PDF. In \Eq{eq:stocha:CharacteristicFunctionMethod:PDF}, since one is summing over increasing powers of $\ee^{-1/\N}$, one is expanding around $\N=0$, \ie on the ``left'' tail of the distribution. In \Eq{eq:stocha:HeatMethod:PDF:expansion} however, since one is summing over increasing powers of $\ee^{-\N}$, one is expanding around $\N=\infty$, \ie on the ``right'' tail of the distribution. Therefore, if one wants to study the PDF by truncating the expansion at some fixed order $n$, one should choose to work with the expression that better describes the part of the distribution one is interested in, so that both expressions can a priori be useful (let us stress again that, in the limit where all terms in the sums are included, both expressions match exactly for all values of $\N$). 

Third,  by plugging $x=1$ in \Eq{eq:PDF:thetatwo}, one obtains an expression for $P(\N,\phi=\phiend+\Delta\phiwell)$ that is an alternative to \Eq{eq:PDF:phiwall:thetaElliptic} even if both formulae involve elliptic theta functions. In \App{appendix:identities}, we show that both expressions are equivalent, due to identities satisfied by the elliptic theta functions. In fact, a third expression for $P(\N,\phi=\phiend+\Delta\phiwell)$ can even be obtained by plugging $x=1$ into \Eq{eq:stocha:HeatMethod:PDF:expansion} and the consistency with the two other ones is also shown in \App{appendix:identities}.

Fourth, an approximated formula for the PDF in the limit $\phi\sim\phiend$ can be derived by Taylor expanding \Eq{eq:PDF:thetatwo}, 
\bea
\label{eq:PDF:stocha:phiend:appr:2}
P\left(\N, \phi \simeq \phiend \right) \simeq -\frac{\pi^2}{4 \mu^2} x \vartheta_{2}'' \left(0, \ee^{-\frac{\pi^2}{\mu^2}\N} \right )\, ,
\eea
see \Eq{eq:theta2primeprime:def}. This provides an alternative to the approximation~(\ref{eq:PDF:stocha:phiend:appr}), that is displayed in the right panel of \Fig{fig:pdf_stochastic_appr}. Numerically, one can check that \Eq{eq:PDF:stocha:phiend:appr} is slightly better.

Fifth, the PDF of coarse-grained curvature perturbations decays exponentially as $\ee^{-\zeta_{\mathrm{cg}}}$, \ie much slower than the Gaussian decay $\ee^{-\zeta_{\mathrm{cg}}^2}$. Since PBHs form along the tail of these distributions, we expect their mass fraction to be greatly affected by this highly non-Gaussian behaviour. More precisely, on the tail, one has 
\bea
\label{eq:PDF:stoch:tail}
P(\zeta_{\mathrm{cg}},\phi)\propto \ee^{-\frac{\pi^2}{4\mu^2}\zeta_{\mathrm{cg}}}\, ,
\eea
which is given by the dominant mode $n=0$ in the expansion~(\ref{eq:stocha:HeatMethod:PDF:expansion}). Interestingly, the decay rate of the distribution is independent of $\phi$. Let us also note that another case where the PDF decays exponentially is in presence of large local non-Gaussianities, when the PDF is a $\chi^2$ distribution~\cite{Young:2013oia, Young:2015cyn}.
\section{Primordial black holes}
\label{sec:PBH}
The formalism developed so far allows one to derive the PDF of coarse-grained curvature perturbations produced during a phase of single-field slow-roll inflation. Let us now apply this result to the calculation of the mass fraction of PBHs discussed in \Sec{sec:Intro}.
\subsection{Classical limit}
\label{sec:PBH:classical}
In the classical limit detailed in \Sec{sec:ClassicalLimit}, the PDF is approximately Gaussian, see \Eq{eq:PDF:classical:NLO}, so that the considerations presented in the introduction apply. 
By plugging \Eq{eq:PDF:classical:NLO} into \Eq{eq:def:beta}, one has $\beta = \erfc[\zeta_\uc/(2\sqrt{v\gamma_1})]$, which is consistent with \Eq{eq:beta:erfc} as noted below \Eq{eq:PDF:classical:NLO}. In the $\beta\ll 1$ limit, this leads to 
\bea 
\label{eq:constraint:classical}
v \gamma_1 \simeq -\frac{\zeta_\uc^2}{4 \ln \beta}\, ,
\eea
where from now on, the order at which the $\gamma_i$ parameters are calculated is omitted for simplicity. Approximating $\gamma_1$ given in \Eq{eq:gamma1:nlo:def} by $\gamma_1\simeq (v/v')^3 \Delta\phi/\Mp^4$, where $\Delta\phi = \vert \phi-\phiend\vert$ is the field excursion, one obtains
\bea
\label{eq:Vconstraint:standard}
\left\vert  \frac{\Delta \phi v^4}{ {v^{\prime}}^3 \Mp^4} \right\vert \simeq - \frac{\zeta_\uc^2}{4\ln\beta(M)}\, .
\eea
In this expression, let us recall that the left-hand side must be evaluated at a value $\phi$ which is related to the PBH mass $M$ by identifying the wavenumber that exits the Hubble radius during inflation at the time when the inflaton field equals $\phi$, with the one that re-enters the Hubble radius during the radiation-dominated era when the mass contained in a Hubble patch equals $M$. For instance, with $\zeta_\uc=1$, the bound $\beta<10^{-22}$ leads to the requirement that the left-hand side of \Eq{eq:Vconstraint:standard} be smaller than $0.005$, which constrains the inflationary potential.

In passing, let us see how the first non-Gaussian correction derived in \Sec{sec:classical:nnlo} affects this result. 
Plugging \Eq{eq:PDF:NNLO} into \Eq{eq:def:beta}, one obtains 
\bea
\beta(M) = \erfc\left(\frac{\zeta_\uc}{2\sqrt{v\gamma_1}}\right) + \frac{\gamma_2}{4\sqrt{v\pi\gamma_1^5}}\ee^{-\frac{\zeta_\uc^2}{4 v \gamma_1}}\left(\zeta_\uc^2-2v\gamma_1\right)\, .
\eea
In the $\beta\ll 1$ limit, \ie in the $\zeta_\uc^2 \gg v \gamma_1$ limit, this reads $\beta\simeq 2 \ee^{-\zeta_\uc^2/(4 v \gamma_1)}\sqrt{v\gamma_1/\pi}/\zeta_\uc[1+\gamma_2 \zeta_\uc^3/(8 v \gamma_1^{3})]$. In this regime, one can see that the non-Gaussian correction is in fact larger than the Gaussian leading order, which signals that the non-Gaussian expansion breaks down on the far tail of the distribution. This also suggests that non-Gaussianities cannot be simply treated at the perturbative level when it comes to PBH mass fractions~\cite{Young:2015cyn}.
\subsection{Stochastic limit}
\label{sec:PBH:stochastic}
\begin{figure}[t]
\begin{center}
\includegraphics[width=0.496\textwidth]{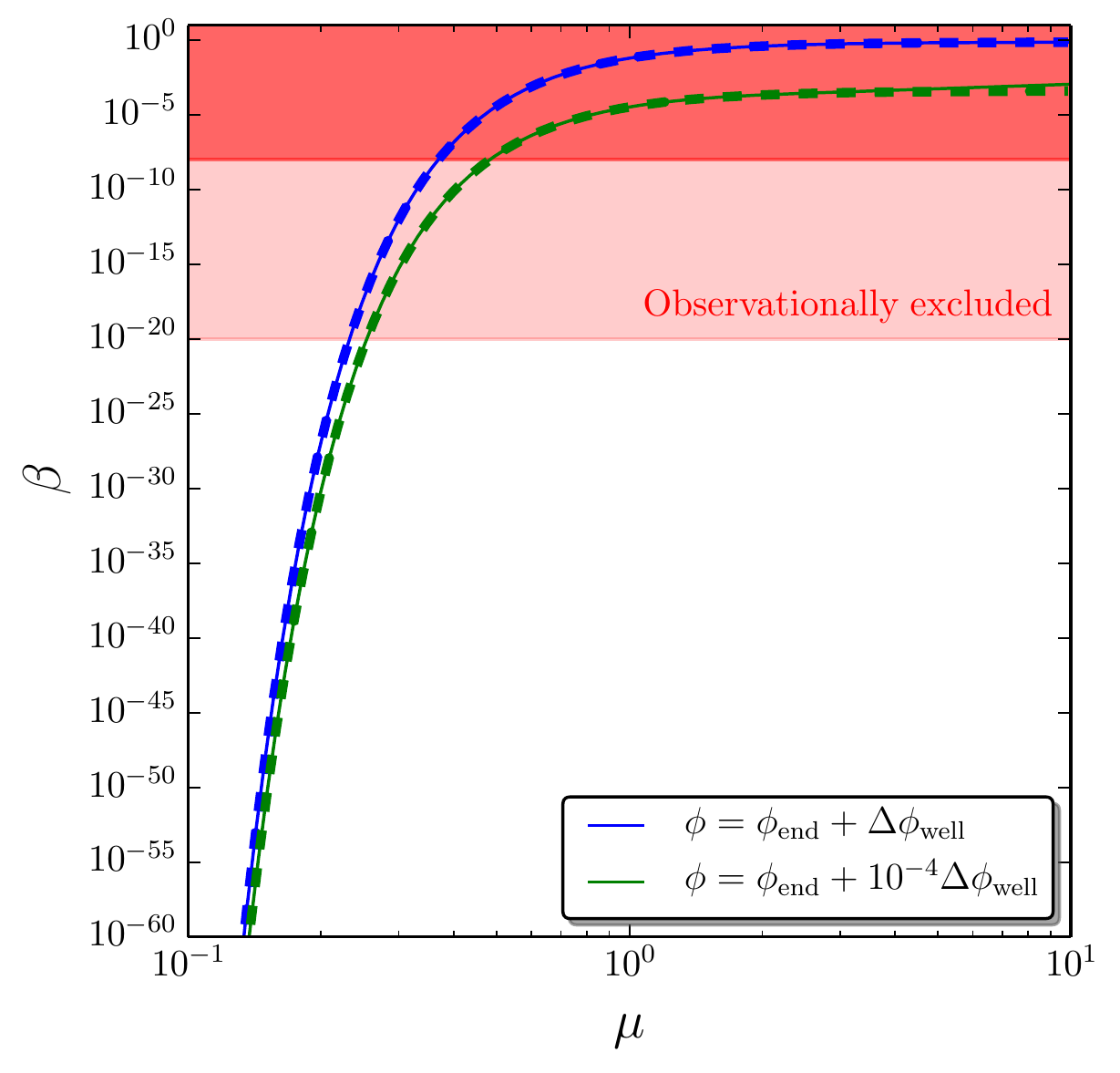}
\includegraphics[width=0.496\textwidth]{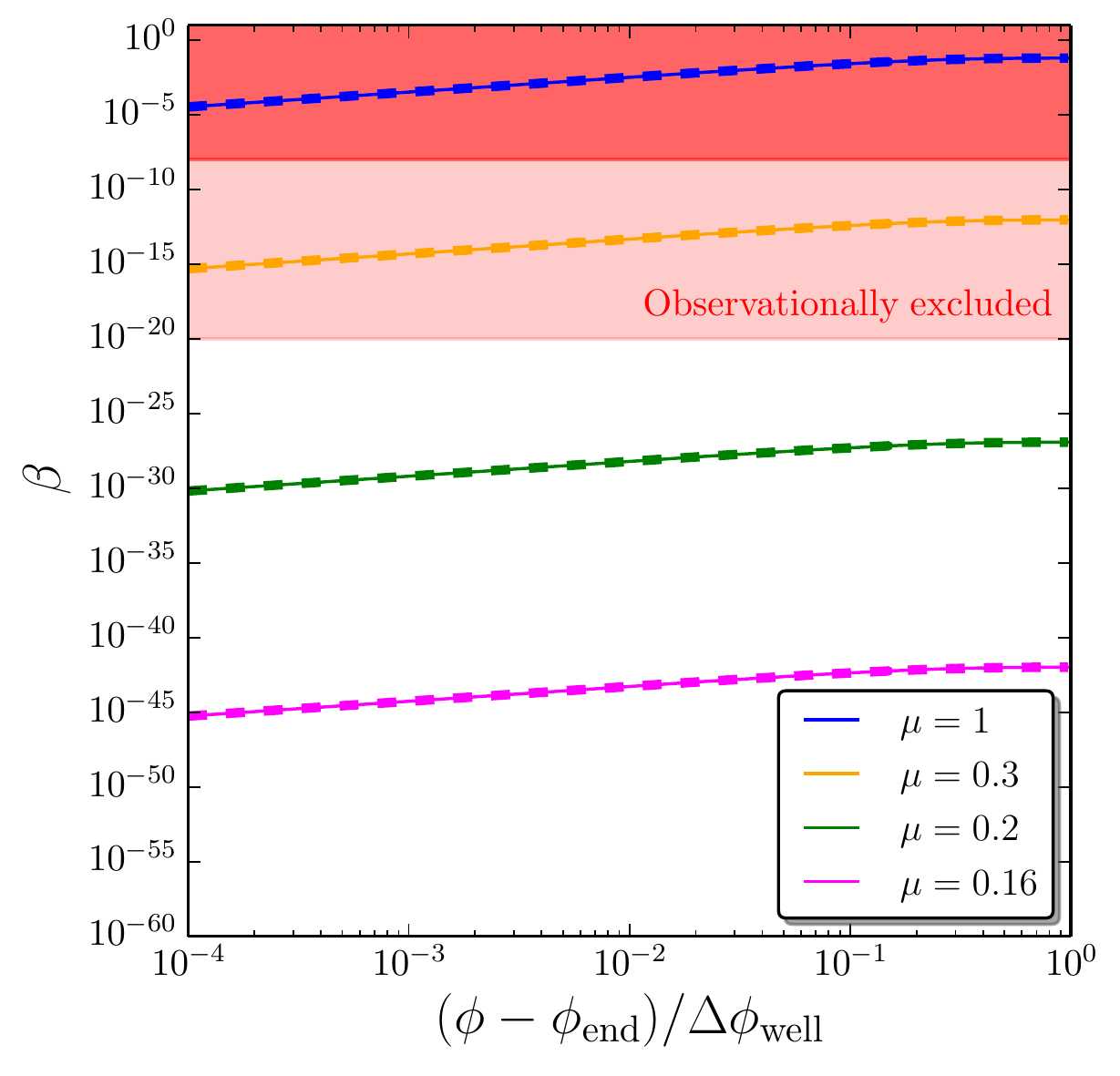}	
\caption[Mass fraction of PBHs in a flat potential]{Mass fraction $\beta$ of primordial black holes in the quantum diffusion dominated regime. The left panel displays $\beta$ evaluated at $\phi=\phi_\uend+\Delta\phiwell$ (blue), \ie at the reflective boundary of the quantum well, and at $\phi=\phi_\uend+10^{-4}\Delta\phiwell$, \ie close to the absorbing boundary of the quantum well, as a function of $\mu=\Delta\phiwell/(\sqrt{v_0}\Mp)$. In the right panel, $\beta$ is plotted as a function of $\phi$ for a few values of $\mu$. One can see that the mass fraction depends very weakly on $\phi$ but very strongly on $\mu$. In both panels, we have taken $\zeta_{\uc} = 1$, the solid lines correspond to the full expression~(\ref{eq:beta:full}) and the dashed line to the approximation~(\ref{eq:beta:stocha:appr}). The shaded region is excluded by observations, the light shaded area roughly corresponds to constraints for PBH masses between $10^{9}\mathrm{g}$ and $10^{16}\mathrm{g}$, the dark shaded area for PBH masses between $10^{16}\mathrm{g}$ and $10^{50}\mathrm{g}$ (see discussion in \Sec{sec:Intro}).} 
\label{fig:beta}
\end{center}
\end{figure}
Let us now see how the constraint~(\ref{eq:Vconstraint:standard}) changes in the presence of large quantum diffusion, as considered in \Sec{sec:StochasticLimit}. In this case, the PDF of coarse-grained curvature perturbations $\zeta_{\mathrm{cg}} = \delta N_{\mathrm{cg}} = \N-\langle \N \rangle$ can be obtained from \Eq{eq:stocha:HeatMethod:PDF:expansion},
\bea
\label{eq:beta:full}
\beta(M) = \frac{4}{\pi} \sum^{\infty}_{n=0} \frac{1}{\left( n+\frac{1}{2} \right)}\sin{\left[ \pi \left( n + \frac{1}{2} \right)x \right]} \exp{\left\lbrace -\pi^2 \left( n + \frac{1}{2}\right)^2 \left[ x \left( 1 - \frac{x}{2} \right) + \frac{\zeta_{c}}{\mu^2}\right] \right\rbrace} \, .
\eea 
In this expression, we have replaced $\langle \N \rangle = f_1 = \mu^2 x(1-x/2)$ which can be obtained by setting the potential to a constant in \Eq{eq:fn:generalsolution}. Let us recall that $x=(\phi-\phiend)/\Delta\phiwell$ and that $M$ and $\phi$ are related as explained below \Eq{eq:Vconstraint:standard}. When $x=0$, \ie when $\phi=\phi_\uend$, \Eq{eq:beta:full} yields $\beta=0$, which is consistent with the fact that the PDF of $\zeta_{\mathrm{cg}}$ is a Dirac distribution in this case.

The mass fraction~(\ref{eq:beta:full}) depends only on $\phi$, $\mu$ and $\zeta_\uc$. It is displayed in \Fig{fig:beta} for $\zeta_\uc=1$, as a function of $\mu$ for $x=1$, \ie $\phi=\phi_\uend+\Delta\phiwell$, and $x=10^{-4}$, \ie $\phi=\phi_\uend+10^{-4}\Delta\phiwell$, in the left panel, and as a function of $\phi$ for a few values of $\mu$ in the right panel. One can see that $\beta$ depends only weakly on $\phi$ but very strongly on $\mu$, which is constrained to be at most of order one. More precisely, if one assumes that $\zeta_\uc\gg \mu^2$ so that $\zeta_\uc $ is well within the tail of the distribution and one can keep only the mode $n=0$ in \Eq{eq:beta:full}, as was done when deriving \Eq{eq:PDF:stoch:tail}, one has
\bea
\label{eq:beta:stocha:appr}
\beta(M) \simeq \frac{8}{\pi}\sin\left(\frac{\pi x}{2}\right)\ee^{-\frac{\pi^2}{8}\left[x(2-x)\right]+\frac{2\zeta_\uc}{\mu^2}}\, .
\eea
This expression is superimposed to the full result~(\ref{eq:beta:full}) in \Fig{fig:beta} where one can see that it provides a very good approximation even when the condition $\zeta_\uc\gg \mu^2$ is not satisfied. This is because, in \Eq{eq:beta:full}, higher terms in the sum are not only suppressed by higher powers of $\ee^{-\zeta_\uc^2/\mu^2}$ but also by higher powers of $\ee^{-\pi^2x(1-x/2)}$, so that \Eq{eq:beta:stocha:appr} is an excellent proxy for all values of $\mu$ except if $x$ is tiny. With $x=1$, it gives rise to
\bea
\label{eq:stocha:constraint:mu}
\mu^2 = \frac{\Delta\phi_{\mathrm{well}}^2}{v_0\Mp^2} = -\frac{2\zeta_\uc}{1+\frac{8}{\pi^2}\ln\left(\frac{\pi}{8}\beta\right)}\, .
\eea

Several comments are in order regarding this result. First, with $\zeta_\uc=1$, $\beta<10^{-24}$ gives rise to $\mu<0.21$ and $\beta<10^{-5}$ gives rise to $\mu<0.47$. The requirement that $\mu$ be smaller than one is therefore very generic and rather independent of the level of the constraint on $\beta$ or the precise value chosen for $\zeta_\uc$. Since $v_0$ needs to be smaller than $10^{-10}$ to satisfy the upper bound~\cite{Ade:2015lrj} on the tensor-to-scalar ratio in the CMB observational window, this also means that $\Delta\phiwell$ cannot exceed $\sim 10^{-5}\Mp$.

Second, \Eq{eq:stocha:constraint:mu} should be compared with its classical equivalent, \Eq{eq:Vconstraint:standard}. In the left-hand sides of these formulae, the scalings with $\Delta\phi$ and $v$ are not the same. In particular, while the PBH mass fraction increases with the energy scale $v$ in the classical picture, in the stochastic limit, it goes in the opposite direction. One should also note that when the potential is exactly flat, $v'=0$, the classical result diverges, but the stochastic one remains finite. In the right-hand sides, the scaling with $\zeta_\uc$ is also different, since the shape of the PDF $P(\zeta_{\mathrm{cg}})$ is not the same (it has a Gaussian decay in the classical case and an exponential decay in the stochastic one). The expressions~(\ref{eq:Vconstraint:standard}) and~(\ref{eq:stocha:constraint:mu}) are therefore very different, and thus translate into very different constraints on the inflationary potential. 

Third, as mentioned below \Eq{eq:beta:full}, the mean number of \efolds~realised across the quantum well is of order $\mu^2$, 
\bea
\langle \N \rangle = \mu^2 x \left(1-\frac{x}{2}\right)\, .
\eea
The conclusion one reaches is therefore remarkably simple: either the region dominated by stochastic effects is much less than one \efold~long and PBHs are not overproduced ($\mu\ll 1$), or it is much more than one \efold~long and PBHs are overproduced ($\mu\gg 1$). Interestingly, heuristic arguments lead to a similar conclusion in Ref. \cite{GarciaBellido:1996qt}, in the context of hybrid inflation.

Fourth, in terms of the power spectrum, since \Eq{eq:Pzeta:stochaDeltaN} gives $\calP_\zeta = f_2'/f_1'- 2f_1$, with $f_1$ given above and $f_2 = \mu^4 x(1 - x^2/2 + x^3/8)/3$ as can be obtained by setting the potential to a constant in \Eq{eq:fn:generalsolution}, one has 
\bea
\label{eq:Pzeta:stochasticLimit}
\calP_\zeta = \frac{\mu^2}{3}\left(2x^2-4x+2\right)\, ,
\eea
so $\mu^2$ is also the amplitude of the power spectrum. With $\beta<10^{-22}$, the constraint~(\ref{eq:stocha:constraint:mu}) on $\mu$ translates into $\calP_\zeta<1.6\times 10^{-2}$ for the value of the power spectrum close to the end of inflation. However, contrary to the classical condition $\calP_\zeta \Delta N<10^{-2}$ recalled below \Eq{eq:powerconstraint:standard}, this constraint does not involve the number of \efolds~since here, a single parameter, $\mu$, determines everything: the mean number of \efolds, the power spectrum amplitude, and the mass fraction. 
\subsection{Recipe for analysing a generic potential}
So far, we have calculated the PBH mass fraction produced in the classical limit and when the inflaton field dynamics are dominated by quantum diffusion. 
In order to analyse a generic potential, it remains to determine where both limits apply. This can be done by comparing the NLO and NNLO results in the classical limit to estimate the conditions under which the classical expansion is under control. For instance, comparing \Eqs{eq:gamma1:nlo:def} and~(\ref{eq:gamma:nnlo:def}) for $\gamma_1$, which gives the mass fraction $\beta$ at NLO as explained in \Sec{sec:PBH:classical}, one can see that $\vert \gamma_1^\nlo - \gamma_1^\nnlo \vert \ll \gamma_1^\nlo $ if $v\ll 1$ and $\vert v^2 v''/{v'}^2 \vert \ll 1$. The first condition is always satisfied, since as already pointed out, $v$ needs to be smaller than $10^{-10}$ to satisfy the upper bound~\cite{Ade:2015lrj} on the tensor-to-scalar ratio in the CMB observational window. The second condition defines our ``classicality criterion''~\cite{Vennin:2015hra}
\bea
\label{eq:eta_class}
\eta_{\mathrm{class}} \equiv \left \vert \frac{v^2 v^{''}}{{v'}^2} \right\vert\, .
\eea
When $\eta_{\mathrm{class}}\ll 1$, the classical expansion is under control, at least at NNLO, and one can use the results of  \Sec{sec:PBH:classical}. 
When $\eta_{\mathrm{class}}\gg 1$, one is far from the classical regime, quantum diffusion dominates the inflaton field dynamics and the results of \Sec{sec:PBH:stochastic} apply. When $\eta_{\mathrm{class}}$ is of order one, a full numerical treatment is required. The ``recipe'' for analysing a generic potential is therefore the following:
\begin{itemize}
\item calculate $\eta_{\mathrm{class}}$ given by \Eq{eq:eta_class} and identify the regions of the potential where $\eta_{\mathrm{class}}\ll 1$ and  $\eta_{\mathrm{class}}\gg 1$;
\item in the regions where $\eta_{\mathrm{class}}\ll 1$, make use of the constraint from \Eq{eq:constraint:classical};
\item in the ``quantum wells'' defined by $\eta_{\mathrm{class}}\gg 1$, make use of the constraint from \Eq{eq:stocha:constraint:mu}.
\end{itemize}

In the following, we illustrate this calculational programme with two examples and check its validity.
\subsection{Example 1: \texorpdfstring{$V\propto 1+\phi^p$}{}}
\label{sec:example:1_plus_phi_to_the_p}
We first consider the case where PBHs can form at scales that exit the Hubble radius towards the end of inflation, where the potential can be approximated by a Taylor expansion around $\phi=0$ where inflation is assumed to end ($\phi_\uend=0$), so
\bea
\label{eq:pot:expansEnd}
v=v_0\left[1+\left(\frac{\phi}{\phi_0}\right)^p\right]\, .
\eea
In this model, inflation does not end by slow-roll violation but another mechanism must be invoked~\cite{Linde:1991km, Linde:1993cn, Copeland:1994vg, Renaux-Petel:2015mga, Renaux-Petel:2017dia}. We also assume that the potential is in the vacuum-dominated regime for the range of field values relevant for PBH formation, so that $\phi\ll \phi_0$. A comprehensive study of this potential is performed in \App{appendix:cases} where all cases of interest are systematically identified and investigated. Here, we simply check that the calculational programme sketched above allows us to recover the main results.

In order to describe the model~(\ref{eq:pot:expansEnd}) in terms of the situation depicted in \Fig{fig:sketch2}, one has to assess $\Delta\phiwell$, which marks the boundary between the classical and the stochastic regimes. In the vacuum-dominated approximation, \Eq{eq:eta_class} gives rise to $\eta_{\mathrm{class}} \simeq (p-1) v_0 (\phi/\phi_0)^{-p}/p$, which is of order one when $\phi = \Delta\phiwell$ with
\bea
\label{eq:phiwell:expansEnd}
\Delta\phiwell \simeq  \phi_0 v_0^{\frac{1}{p}}\, .
\eea
Since $(\Delta\phiwell/\phi_0)^p=v_0\ll 1$, the vacuum-dominated condition is always satisfied at this transition point. However, the slow-roll conditions are not always met, and in \App{appendix:cases} it is shown that slow roll is indeed violated at $\phi = \Delta\phiwell$ if $\phi_0/\Mp < v_0^{(p-2)/(2p)}$, unless $p=1$ for which slow-roll is violated if $\phi_0< \Mp$. In such cases, the expansion~(\ref{eq:pot:expansEnd}) fails to cover the whole quantum well and higher-order terms in the potential must be included for a consistent analysis. Otherwise, we can keep following the recipe given above.

In the classical regime, $\phi\gg \Delta\phiwell$, \Eq{eq:constraint:classical} applies, where $v \gamma_1$ is given by \Eq{eq:gamma1:nlo:def}. In the vacuum-dominated approximation, it reads $v\gamma_1 \simeq v_0 (\phi_0/\Mp)^4/(4p^3-3p^4)[(\phi/\phi_0)^{4-3p}-(\phi_\uend/\phi_0)^{4-3p}]$. Neglecting the contribution from $\phi_\uend$, which lies outside the validity range of the classical formula anyway, one can evaluate this expression at $\phi = \Delta\phiwell$ where the power spectrum is maximal, and combining this with \Eq{eq:constraint:classical} leads to
\bea
\label{eq:expansEnd:classConstraint}
\frac{v_0^{\frac{2}{p}-1}}{\sqrt{\vert 4p^3-3p^4 \vert }}\left(\frac{\phi_0}{\Mp}\right)^2 \simeq  \frac{\zeta_\uc}{2\sqrt{\vert \ln \beta \vert}}\, .
\eea
In the stochastic regime, combining \Eqs{eq:stocha:constraint:mu} and~(\ref{eq:phiwell:expansEnd}), one has
\bea
\label{eq:expansEnd:stochConstraint}
v_0^{\frac{2}{p}-1}\left(\frac{\phi_0}{\Mp}\right)^2 \simeq  \frac{2\zeta_\uc}{\left\vert 1+\frac{8}{\pi^2}\ln\left(\frac{\pi}{8}\beta\right)\right\vert}\, .
\eea
It is interesting to notice that up to an overall factor of order one, the two constraints~(\ref{eq:expansEnd:classConstraint}) and~(\ref{eq:expansEnd:stochConstraint}) are very similar, even though they are obtained in very different regimes that yield very different PDFs for the curvature perturbations. 

It is also important to note that the slow-roll conditions given above imply that $v_0^{2/p-1}(\phi_0/\Mp)^2\gg 1$ except if $p=1$. Therefore, if $p$ is different from $1$, either PBHs are too abundant and the model is ruled out, or slow roll is strongly violated before one exits the classical regime and one needs to go beyond the present formalism to calculate PBH mass fractions. The case $p=1$ is subtle, since \Eq{eq:eta_class} gives $\eta_{\mathrm{class}}=0$. One would thus have to extend the classical expansion of \Sec{sec:ClassicalLimit} to next-to-next-to-next to leading order (NNNLO) to determine what the first stochastic correction is and under which condition the classical approximation holds, and investigate numerically the regime where it does not. 
This study is left for future work.

\begin{figure}[t]
\begin{center}
\includegraphics[width=0.6\textwidth]{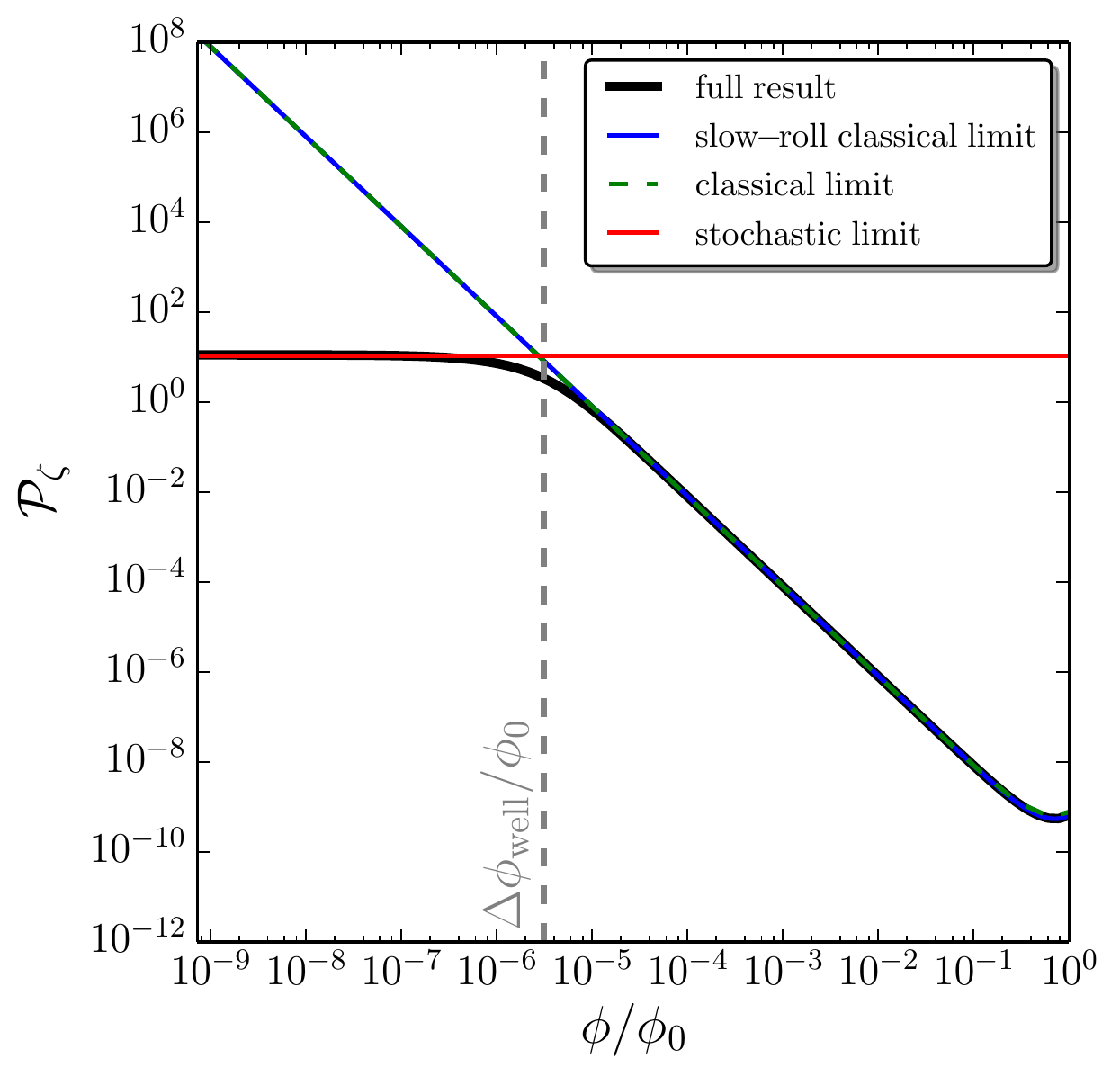}
\caption[Power spectrum of curvature perturbations in a quadratic potential]{Power spectrum of curvature perturbations $\calP_\zeta$ produced in the potential~(\ref{eq:pot:expansEnd}) with  $p=2$, $v_0 = 10^{-11}$, $\phi_0=4 \Mp$ and $\phiuv=10^{4}\phi_0$ (solid black line). The blue line corresponds to the slow-roll classical limit~(\ref{eq:classicalPower}), while the green dashed line is obtained from solving the full Klein-Gordon equation. The red line corresponds to the stochastic limit assuming the potential is exactly flat for $\phi<\Delta\phiwell$ and that a reflective wall is located at $\phi = \Delta\phiwell$.  The value of $\Delta\phiwell$ obtained from requiring $\eta_{\mathrm{class}}=1$ is displayed with the grey vertical dotted line and delimitates the classical and stochastic regimes.} 
\label{fig:quad:power:spectrum}
\end{center}
\end{figure}
In passing, let us check that approximating the full potential~(\ref{eq:pot:expansEnd}) as a piecewise function consisting of a constant piece and a classical one, separated at $\phi=\Delta\phiwell$, is numerically justified. In \Fig{fig:quad:power:spectrum}, we show the power spectrum computed numerically from \Eqs{eq:fn:generalsolution} and~(\ref{eq:Pzeta:stochaDeltaN}), which gives $\calP_\zeta = f_2'/f_1'- 2f_1$, in the potential~(\ref{eq:pot:expansEnd}) with $p=2$, $v_0 = 10^{-11}$, $\phi_0=4 \Mp$ and $\phiuv=10^{4}\phi_0$ (solid black line). The blue line corresponds to the slow-roll classical limit~(\ref{eq:classicalPower}), and the green dashed line is obtained from solving the full Klein-Gordon equation. The agreement of this solution with the slow-roll formula confirms that the slow-roll conditions are satisfied for the parameters used in this example. The red line corresponds to the stochastic limit~(\ref{eq:Pzeta:stochasticLimit}) $\calP_\zeta = 2\mu^2/3$ at $\phi=0$, where $\mu$ is given by \Eqs{eq:def:mu} and~(\ref{eq:phiwell:expansEnd}), which yields $\calP_\zeta \sim 2 (\phi_0/\Mp)^2 v_0^{2/p-1}/3$. One can see that both limits are correctly reproduced, and that the value of $\Delta\phiwell$ obtained in \Eq{eq:phiwell:expansEnd} from our classicality criterion $\eta_{\mathrm{class}}<1$, and displayed with the grey vertical dotted line, indeed separates the two regimes. In \App{appendix:cases}, an analytical expression for $\calP_\zeta$ in the regime $\phi\ll\Delta\phiwell$ is derived, and one finds $\calP_\zeta = 2\Gamma^2(1/p)v_0^{2/p-1}(\phi_0/\Mp)^2/p^2$, see \Eq{stochasticpower}, where $\Gamma$ is the gamma function. Up to an overall numerical constant of order one, one recovers the result obtained from simply assuming the potential to be exactly flat until $\phi=\Delta\phiwell$, where $\eta_{\mathrm{class}}=1$, and setting a reflective wall there. This confirms the validity of this approach.
\subsection{Example 2: running-mass inflation}
\begin{figure}[t]
\begin{center}
\includegraphics[width=0.6\textwidth]{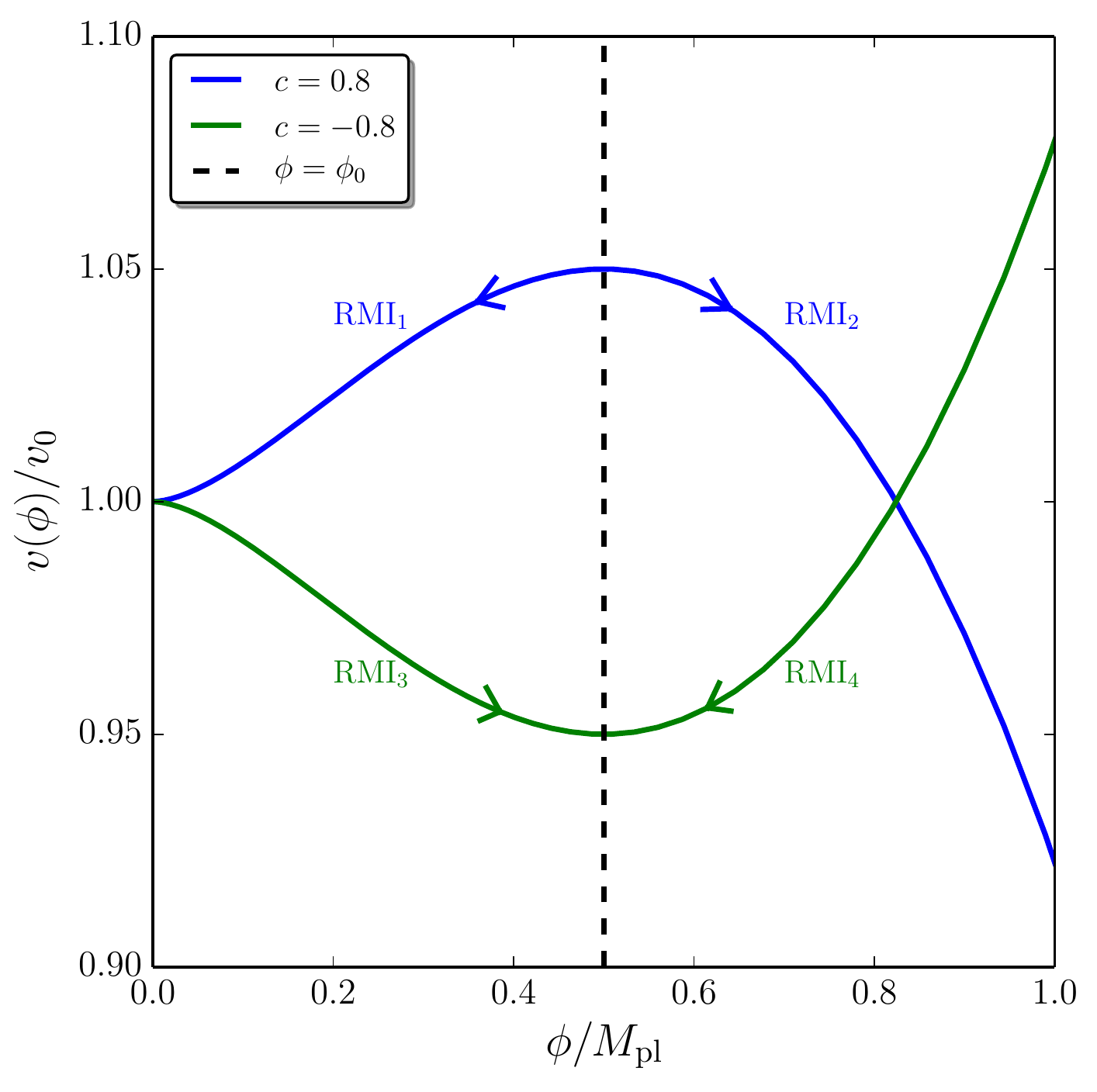}		
\caption[The potential for running-mass inflation]{The potential~(\ref{eq:RMI:potential}) for running-mass inflation (RMI) with $\phi_{0} = 0.5\Mp$. The blue curve takes $c = 0.8$ and the green curve takes $c = -0.8$ (these values may not be physical but they have been chosen to produce a clear plot). RMI is shown to have four possible realisations (RMI${}_1$, RMI${}_2$, RMI${}_3$ and RMI${}_4$), depending on the sign of $c$ and on whether $\phi$ is initially smaller or larger than $\phi_0$. Except for RMI${}_2$, the potential flattens as inflation proceeds, which can lead to the formation of PBHs for scales exiting the Hubble radius towards the end of inflation.} 
\label{fig:RMI:Potential}
\end{center}
\end{figure}
Let us now consider another example, running-mass inflation (RMI) \cite{Stewart:1996ey}, where the inflationary potential is given by
\bea
\label{eq:RMI:potential}
v\left( \phi \right) = v_{0} \left\lbrace 1 - \frac{c}{2}\left[ -\frac{1}{2} + \ln{\left( \frac{\phi}{\phi_{0}}\right)} \right] \frac{\phi^2}{\Mp^2}\right\rbrace\, .
\eea
In this expression, $c$ is a dimensionless coupling constant assumed to be much smaller than one, $c\ll 1$ (more precisely, as discussed in Ref. \cite{Martin:2013tda}, within supersymmetry, natural values of $c$ are $c \simeq 10^{-2}$ to $10^{-1}$ for soft masses values matching the energy scale of inflation), and $\phi_0$ must be sub-Planckian, $\phi_0\ll \Mp$. 

The potential~(\ref{eq:RMI:potential}) is displayed in \Fig{fig:RMI:Potential}, where one can see that depending on the sign of $c$ and on whether $\phi$ is initially smaller or larger than $\phi_0$, inflation can proceed in four regimes~\cite{Covi:1998mb}, that we denote RMI${}_1$, RMI${}_2$, RMI${}_3$ and RMI${}_4$. In RMI${}_1$, $c > 0$, $\phi < \phi_{0}$ and $\phi$ decreases towards $\phi = 0$ as inflation proceeds. RMI${}_2$ also has $c > 0$ but in this case $\phi > \phi_{0}$ and $\phi$ increases away from $\phi_{0}$ throughout inflation. RMI${}_3$ and RMI${}_4$ both have $c < 0$, but RMI${}_3$ has $\phi < \phi_{0}$ with $\phi$ increasing towards $\phi_0$ during inflation, while RMI${}_4$ has $\phi > \phi_{0}$ and $\phi$ decreases towards $\phi_0$ during inflation. In RMI${}_1$, RMI${}_3$ and RMI${}_4$, the potential flattens as inflation proceeds, which may lead to the production of PBHs at scales that exit the Hubble radius towards the end of inflation, as studied in Refs. \cite{Leach:2000ea, Drees:2011hb, Akrami:2016vrq}. 
The width of the ``quantum well'' in these cases is determined by the condition $\eta_{\mathrm{class}}>1$, where $\eta_{\mathrm{class}}$ is given by \Eq{eq:eta_class}. In the vacuum-dominated regime, it reads
\bea
\eta_{\mathrm{class}} \simeq \frac{v_0}{\vert c \vert}\frac{\Mp^2}{\phi^2}\frac{\left\vert 1+\ln\left(\frac{\phi}{\phi_0}\right)\right\vert}{\ln^2\left(\frac{\phi}{\phi_0}\right)}
\, .
\eea

For RMI${}_1$, the equation $\eta_{\mathrm{class}}(\phiwell)=1$ yields $\phiwell/\phi_0 \sqrt{\vert \ln(\phiwell/\phi_0 ) \vert} = \sqrt{v_0/c} \Mp/\phi_0$, where we have assumed that $\phiwell \ll \phi_0$ so that $\vert \ln(\phiwell/\phi_0)\vert \gg 1$. This can be solved as
\bea
\label{eq:phiwell:RMI1}
\phiwell = \phi_0 \exp\left[\frac{1}{2 }W_{-1}\left( -2 \frac{v_0}{ c  }\frac{\Mp^2}{\phi_0^2}\right)\right]\, ,
\eea
where $W_{-1}$ is the $-1$ branch of the Lambert function~\cite{Olver:2010:NHM:1830479:lambert}. The approximation $\vert \ln(\phiwell/\phi_0)\vert \gg 1$ is satisfied when the argument of the Lambert function in \Eq{eq:phiwell:RMI1} is much smaller than one, which is typically the case for the values of $v_0$, $c$ and $\phi_0$ considered in the literature~\cite{Martin:2013tda, Martin:2013nzq}. In this limit, one can Taylor expand the Lambert function according to $W_{-1}(-x)\simeq \ln x$ when $x\ll 1$, which gives rise to $\phiwell \simeq \sqrt{2v_0/c} \Mp$, and hence
\bea
\label{eq:Deltaphiwell:RMI1}
\Delta\phiwell = \left\vert \phiwell \right\vert \simeq   \sqrt{\frac{2 v_0}{c}}\Mp\, .
\eea
In this expression, we have assumed that inflation terminates at $\phi=0$ (otherwise, $\phiend$ must be subtracted from the right-hand side). Making use of \Eq{eq:def:mu}, this leads to
\bea
\label{eq:MU:RMI1}
\mu^2\simeq \frac{2}{c} \gg 1\, .
\eea
The result is remarkably simple since it depends only on the coupling constant $c$, and on neither $v_0$ nor $\phi_0$. As explained at the beginning of this section, $c$ is always much smaller than one, which implies that $\mu\gg 1$ and according to the discussion of \Sec{sec:PBH:stochastic}, PBHs are too abundant in this case. One concludes that the stochastic regime of the potential cannot be explored in RMI${}_1$, \ie $\phiend$ should be at least of the order of $\Delta\phiwell$ given in \Eq{eq:Deltaphiwell:RMI1}.

For RMI${}_2$, the potential does not flatten as inflation proceeds so the inflaton field dynamics is never dominated by quantum diffusion for scales smaller than those probed in the CMB.

For RMI${}_3$ and RMI${}_4$, assuming that $\phiwell$ is very close to $\phi_0$ so that $\vert \ln(\phiwell/\phi_0)\vert \ll 1$, the equation $\eta_{\mathrm{class}}(\phiwell)=1$ reduces to $\phiwell/\phi_0 \vert \ln(\phiwell/\phi_0 ) \vert = \sqrt{v_0/\vert c \vert} \Mp/\phi_0$. This can be solved as
\bea
\label{eq:phiwell:RMI34}
\phiwell = \phi_0 \exp\left[W_0\left(\mp \sqrt{\frac{v_0}{\vert c \vert}}\frac{\Mp}{\phi_0}\right)\right]\, ,
\eea
where $W_0$ is the principal branch of the Lambert function, and its argument comes with a minus sign in RMI${}_3$ and with a plus sign in RMI${}_4$. The approximation $\vert \ln(\phiwell/\phi_0)\vert \ll 1$ is satisfied when the argument of the Lambert function in \Eq{eq:phiwell:RMI34} is much smaller than one, which is the same condition as the one coming from $\phiwell\ll\phi_0$ in RMI${}_1$. In this limit, one can Taylor expand the Lambert function as $W_0(x)\simeq x$ when $x\ll 1$. One obtains $\phiwell \simeq \phi_0\mp \Mp\sqrt{v_0/\vert c \vert}$, and hence
\bea
\label{eq:Deltaphiwell:RMI34}
\Delta\phiwell = \left\vert \phiwell-\phi_0 \right\vert \simeq  \Mp\sqrt{\frac{v_0}{\left\vert c \right\vert}}\, .
\eea
Up to a factor $\sqrt{2}$, this expression is the same as \Eq{eq:Deltaphiwell:RMI1}. This leads to
\bea
\label{eq:MU:RMI34}
\mu^2\simeq \frac{1}{\left\vert c \right\vert} \gg 1\, ,
\eea
and the same conclusions as the ones drawn for RMI${}_1$ apply, namely that one cannot explore the quantum well of the potential without producing too many PBHs, so $\vert \phiend-\phi_0\vert $ should be at least of order $\Delta\phiwell$ given in \Eq{eq:Deltaphiwell:RMI34}.

If this is indeed the case, the classical approximation is valid throughout the entire period of inflation, and \Eq{eq:classicalPower} gives rise to
\bea
\calP_\zeta \simeq 2\frac{v_0}{c^2}\frac{\Mp^2}{\phi^2}\ln^{-2}\left(\frac{\phi}{\phi_0}\right)\, .
\eea
When this expression is evaluated at $\phiwell$, given by \Eq{eq:phiwell:RMI1} for RMI${}_1$ and by \Eq{eq:phiwell:RMI34} for RMI${}_3$ and RMI${}_4$, one finds
\bea
\calP_\zeta\left(\phiwell\right) \simeq
\begin{cases}
\frac{4}{c} &\mathrm{in}\ \mathrm{RMI}_1\\
\frac{2}{\left\vert c\right\vert } &\mathrm{in}\ \mathrm{RMI}_3\ \mathrm{and}\ \mathrm{RMI}_4
\end{cases}\, .
\eea
It is interesting to notice that, up to an overall numerical constant of order one, this also corresponds to the stochastic limit~(\ref{eq:Pzeta:stochasticLimit}), $\calP_\zeta \sim \mu^2 \sim 1/c$, where one makes use of \Eqs{eq:MU:RMI1} and~(\ref{eq:MU:RMI34}). This is similar to what was found in \Sec{sec:example:1_plus_phi_to_the_p}. Since $\vert c \vert \ll 1$, this means that the classical power spectrum is already larger than one when one enters the quantum well. This implies that, for this model, analyses relying on the classical formalism only should exclude the quantum well (even if not for the correct reason) and should therefore be valid. 
However, let us stress that the approach developed in this work was necessary in order to check the consistency of the standard results.

\section{Conclusions}
\label{sec:Conclusion}

Let us now summarise the main finding of this chapter. 
Making use of the stochastic-$\delta N$ formalism, we have developed a calculational framework in which the PDF of the coarse-grained curvature perturbations produced during inflation can be derived exactly, even in the presence of large quantum backreaction on the inflaton field dynamics.  
More precisely, we have proposed two complementary methods, one based on solving an ordinary differential equation for the characteristic function of the PDF, and the other based on solving a heat equation for the PDF directly. We have shown that depending on the problem one considers, the method to be preferred can vary. 
We have then derived a classicality criterion that determines whether the effects of quantum diffusion are important or not. When this is not the case, \ie in the classical limit, we have developed an expansion scheme that not only recovers the standard Gaussian PDF at leading order, but also allows one to calculate the first non-Gaussian corrections to the usual result. In the opposite limit, \ie when quantum diffusion plays the dominant role in the field dynamics, we have found that the PDF follows an elliptic theta function, whose tail decays only exponentially, and which is fully characterised by a single parameter, given by $\mu^2 = (\Delta\phiwell)^2/(v_0\Mp^2)$. This parameter measures the ratio between the squared width of the quantum well and its height, in Planck mass units. The mean number of \efolds~realised across the quantum well, the amplitude of the power spectrum, and, if $\zeta_\uc \sim 1$, the inverse log of the PBH mass fraction, are all of order $\mu^2$. Therefore, observational constraints on the abundance of PBHs put an upper bound on $\mu^2$ that is of order one, and imposes that one cannot spend more than $\sim 1$ \efold~in regions of the potential dominated by quantum diffusion. For a given potential, one must therefore determine whether a diffusion dominated quantum well exists, and check that its width squared is smaller than its height. Finally, we have illustrated our calculational programme with two examples.

However, it is important to remember that in this chapter we have worked exclusively in the slow-roll regime. 
There are cases where slow roll is violated when scales smaller than those probed in the CMB exit the Hubble radius and the standard PBH calculation does not apply, as recently pointed out in Refs. \cite{Kannike:2017bxn, Germani:2017bcs, Motohashi:2017kbs}. 
This, for instance, happens when the inflationary potential has a flat inflection point \cite{Garcia-Bellido:2017mdw}, around which slow roll is transiently violated and one can enter an ultra slow-roll phase. 
Close to such an inflection point, quantum diffusion plays an important role and this also needs to be included. 
This requires the formalism presented in this chapter to be extended beyond the slow-roll approximation~\cite{Grain:2017dqa}, and in the next chapter we will extend the methods presented here into the ultra-slow-roll regime.

%

%
%
%


\newpage

\chapter{Ultra-slow-roll inflation with quantum diffusion} \label{chapter:USRstochastic}

In this chapter, we will develop the stochastic formalism of inflation (see \Sec{chapter:stochastic:intro}) into the regime of ultra-slow-roll (USR) inflation.
Since the submission of this thesis, the work in this chapter work has been continued and (eventually) submitted for publication in \cite{Pattison:2021oen}.
This was done in collaboration with Vennin, Wands, Assadullahi.

In slow roll, the stochastic-$\delta N$ formalism has been used to reconstruct the full probability distribution function for the primordial density field~\cite{Pattison:2017mbe, Ezquiaga:2019ftu} finding large deviations from a Gaussian distribution in the nonlinear tail of the distribution, as described in chapter \ref{chapter:quantumdiff:slowroll}.
However, it has recently become clear that there are physical situations of interest in which stochastic effects are important but also where slow roll is violated. 
This is commonly seen in the discussion of primordial black holes (PBHs) from inflationary models with a flat portion of the potential near the end of inflation~\cite{Garcia-Bellido:2017mdw, Germani:2017bcs, Biagetti:2018pjj, Ezquiaga:2018gbw, Firouzjahi:2018vet}.
When the potential is very flat, the inflaton can be subject to large stochastic diffusion~\cite{Pattison:2017mbe}, as well being kicked off the slow-roll attractor and violating the associated approximations. 
This can lead to a phase of USR inflation~\cite{Inoue:2001zt, Kinney:2005vj}, which we have seen can be stable~\cite{Pattison:2018bct} if certain conditions are met (see Chapter \ref{chapter:USRstability}).

Here we will extend the characteristic function approach introduced in chapter \ref{chapter:quantumdiff:slowroll} to the (USR) regime. 
We will reconstruct the probability distribution of curvature perturbations in specific cases, and discuss possible consequences for PBH production during such a phase of USR inflation. 

This chapter is organised as follows: 
In \Sec{sec:stochUSR}, we discuss the USR system we are considering and explain how we build a stochastic formalism for this regime. 
In \Sec{sec:characteristicUSR}, we develop the characteristic function formalism that we originally introduced in Chapter \ref{chapter:quantumdiff:slowroll} to be used in USR, and in \Sec{sec:classicallimit} we show how this can be solved iteratively in the ``classical limit'', where quantum diffusion provides a subdominant contribution to the dynamics of the inflaton. 
In \Sec{sec:latetimelimit}, we solve the USR system in the late-time limit, where the classical velocity of the inflaton has completely decayed and the dynamics are purely quantum diffusion.
Then, in \Sec{sec:stochasticlimit}, we solve the stochastic limit, where there is a small, but non-vanishing, classical velocity, and also discuss the implications this can have on primordial black hole production. 
Finally, in \Sec{sec:USRdiscussion}, we present some concluding remarks. 

\section{Stochastic ultra-slow-roll inflation} \label{sec:stochUSR}

Let us briefly review the USR regime for inflation, before we discuss the stochastic formalism in this regime.
USR is realised when the gradient of the inflaton potential becomes negligible with respect to the Hubble damping in the equation of motion (\ref{eq:eom:scalarfield}), so $V'(\phi)/3H\dot\phi\to 0$. 

In this limit, the classical equation of motion \eqref{eq:eom:scalarfield} becomes friction-dominated and can be simplified to 
\bea \label{eq:usr:eom}
\ddot{\phi} + 3H\dot{\phi} \simeq 0 \, ,
\eea 
which gives the classical USR solution $\dot{\phi} = \dot{\phi}_{\mathrm{in}}\ee^{-3N}$, and hence
\bea
\label{eq:usr:dotphisol}
\dot{\phi}=\dot{\phi}_\uin+3H(\phi_\uin-\phi)
\,, 
\eea
This USR evolution is dependent on the initial conditions of the inflaton field and its velocity, unlike in slow roll where the slow-roll trajectory is an attractor independent of the initial velocity 

These solutions are classical background solutions, and do not yet include the effect of quantum diffusion on the inflaton's evolution.
We assume that the inflaton classically decreases ($\dot\phi<0$) towards a minimum field value $\phi_\mathrm{end}$ where inflation ends. 
However quantum diffusion can in principle probe arbitrarily large field values. We will assume that at sufficiently large field values the potential becomes sufficiently steep that the potential gradient cannot be neglected and the resulting classical drift towards $\phi_\mathrm{end}$ dominates over the quantum diffusion.
We therefore impose a cut-off at large field values by imposing a reflective boundary at a maximum value $\phi_\mathrm{well}$. We refer to the flat portion of a potential, as a ``quantum well" \cite{Pattison:2017mbe}, which has a finite width 
\bea \label{eq:def:deltaphiwell}
\Delta\phi_\mathrm{well} = \phi_\mathrm{well} - \phi_\mathrm{end} \, .
\eea 

We can use the classical equation (\ref{eq:usr:dotphisol}) to calculate the critical velocity, which is the minimum velocity the system needs at $\phi_\mathrm{in}$ in order to exactly reach the endpoint $\phi_\mathrm{end}$ purely through classical USR motion. 
Hence from Eq.~(\ref{eq:usr:dotphisol}), with $\dot{\phi}\to0$ as $\phi\to\phi_\mathrm{end}$, we have 
\bea
\dot{\phi}_\mathrm{in}\bigg|_{\mathrm{crit}} = -3H\left(\phi_\mathrm{in}-\phi_\mathrm{end}\right) \, . 
\eea 
This is the classical velocity needed to reach $\phi_\mathrm{end}$ from a given initial point in the quantum well. 
Let us define $p_\mathrm{crit}$ to be the classical velocity needed to traverse the \textit{entire} quantum well, from $\phi_\mathrm{well}$ to $\phi_\mathrm{end}$, under classical motion alone, so that
\bea  \label{eq:criticalvelocity}
p_\mathrm{crit} = -3H \Delta \phi_\mathrm{well} \, . 
\eea 

\begin{figure}
    \centering
    \includegraphics[width=0.9\textwidth]{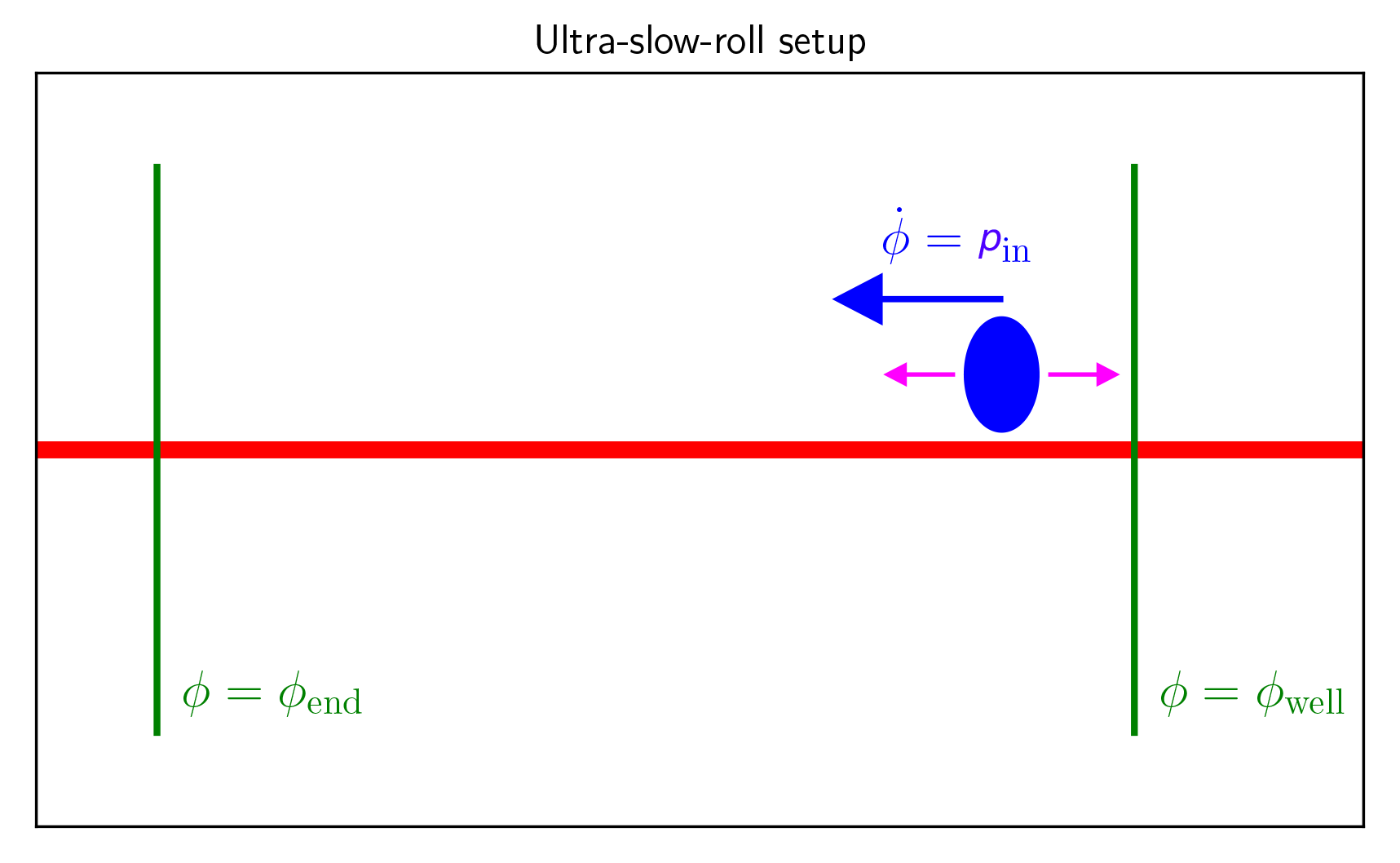}
    \caption[Sketch of the ultra-slow-roll scenario]{A sketch of the ultra-slow-roll scenario. The inflaton moves on the flat potential through a classical velocity, denoted by the blue arrow, and by quantum diffusion, denoted by the magenta arrows. Inflation ends at $\phi=\phi_\mathrm{end}$ and there is a reflective wall at $\phi=\phi_\mathrm{well}$, which in practice should be thought of as a point where the potential becomes steep enough that the classical drift pushes us towards $\phi_\mathrm{end}$. \label{fig:usr:sketch}}
\end{figure}

To take into account quantum kicks, we use stochastic inflation, where in USR we can evaluate the general Langevin equations \eqref{eq:conjmomentum:langevin} and \eqref{eq:KG:efolds:langevin} for a flat potential \cite{Firouzjahi:2018vet} to find
\begin{align}
\label{eq:eom:phi:stochastic} \frac{\dd \phi}{\dd N} &= \frac{p}{H} + \frac{H}{2\pi}\xi(N) \, , \\
\frac{\dd p}{\dd N} &= -3p \label{eq:eom:v:stochastic} \, ,
\end{align}
where $p=\dd \phi/\dd t$ (in the background) and $H$ is the Hubble parameter. 
This is valid for a flat potential $V'=0$ and where we  neglect the noise associated with $p$ as it is second order in the coarse-graining parameter, see \cite{Firouzjahi:2018vet} for details of this.
We will treat this as a $2-$dimensional problem, and hence regard $\phi$ and $p$ as independent variables. 
A sketch of the setup we are considering is displayed in \Fig{fig:usr:sketch}.

From here, in order to reduce the number of parameters needed to describe this USR system, we make the change of variables
\bea \label{eq:xy:variabletransforms}
&\phi \to x = \frac{\phi - \phi_\mathrm{end}}{\Delta \phi_\mathrm{well}} \, , &  p \to y = \frac{p}{p_\mathrm{crit}} \, ,
\eea 
where $p_\mathrm{crit}$ is given by \eqref{eq:criticalvelocity}.
Using these variables, the Langevin equations \eqref{eq:eom:phi:stochastic} and \eqref{eq:eom:v:stochastic} become
\bea \label{eq:langevin:xy}
\frac{\dd x}{\dd N} &= -3 y + \frac{\sqrt{2}}{\mu} \xi(N) \\
\frac{\dd y}{\dd N} &= -3y \, ,
\eea 
where we have assumed $H^2 \simeq V_0/(3\Mp^2)$ is constant\footnote{In our $(x,y)$ variables, the Hubble parameter is given by 
\bea 
H^2 = \frac{V_0}{3\Mp^2 - \frac{9}{2}\Delta\phi_\mathrm{well}{}^2y^2} \, ,
\eea 
and hence the assumption $H\simeq constant$ is equivalent to $y < \Mp/\Delta\phi_\mathrm{well}$, \ie we assume the quantum well is not ``too" wide.
} and defined the dimensionless parameter
\bea \label{eq:def:mu:usr}
\mu^2 = \frac{8\pi^2}{H^2}\Delta \phi_\mathrm{well}{}^2 \, ,
\eea 
as in Chapter \ref{chapter:quantumdiff:slowroll}.
In these variables, our entire problem is specified with only one parameter, namely $\mu$, along with the initial state $(x_\mathrm{in}, y_\mathrm{in})$. 
Note that $x \in [0,1]$, and for $y_\mathrm{in}<1$ the field would not cross the quantum well under purely classical motion and will become dominated by the quantum diffusion at some time, while for $y_\mathrm{in} \geq 1$ the field would cross the well classically and for large enough\footnote{In fact, if one is only interested in a USR regime that is inflating (as opposed to USR non-inflation), then there exists a maximum value for the variable $y$.
This can be seen by first noting that $\dot{H} = \dot{\phi}^2/(2\Mp^2)$, and hence 
\bea 
H^2 = \frac{\dot{\phi}^2}{2\epsilon_1\Mp^2} \, .
\eea 
This in turn allows us to write 
\bea 
y = \frac{\sqrt{2\epsilon_1}\Mp}{3\Delta\phi_\mathrm{well}} \, ,
\eea 
where we have taken a minus sign when taking $\sqrt{H^2}$ (since $\dot{\phi} < 0$ and $H>0$), and hence imposing inflation ($\epsilon_1 < 1$) gives a maximum value of $y$ to be 
\bea \label{eq:ymax:generic}
y_\mathrm{max} = \frac{\sqrt{2}\Mp}{3\Delta\phi_\mathrm{well}} \, .
\eea 
In particular, we see that if one considers the extreme case an infinitely wide, flat potential then $y_\mathrm{max}\to 0$, and hence this case is described by free diffusion in \eqref{eq:langevin:xy}.
This makes sense, since the classical velocity needed to traverse an infinitely wide well is infinite, $p_\mathrm{crit}\to\infty$ in \eqref{eq:criticalvelocity}.
} values of $y_\mathrm{in}$ it may never become diffusion dominated.

\section{Characteristic function formalism}
\label{sec:characteristicUSR}

In this section we discuss the characteristic function formalism developed in chapter \ref{chapter:quantumdiff:slowroll} and extend the analysis for USR inflation. 

\subsection{First passage time analysis}

While we use the same first passage time analysis techniques as in chapter \ref{chapter:quantumdiff:slowroll}, here we present an alternative (but equivalent) way of deriving the relevant equations \cite{Vennin:2015hra}. 
We begin by deriving a $2D$ Fokker--Planck equation for the system \eqref{eq:eom:phi:stochastic}, which is the equation governing the moments of the probability distribution of the number of \efolds~$P(N)$ (these calculations are demonstrated in \App{appendix:FokkerPlanck:USR}).
and we find this to be
\bea \label{eq:FP:USR}
\frac{\partial P}{\partial N} &= \left[ 3 + 3y\left(\frac{\partial}{\partial x} + \frac{\partial}{\partial y}\right) + \frac{1}{\mu^2}\frac{\partial^2}{\partial x^2}\right] P \\
&\equiv \mathcal{L}_{\mathrm{FP}} \cdot P \, .
\eea 
As detailed in \cite{Assadullahi:2016gkk}, the equation that gives the moments of the PDF, defined as $f_n(x_i, y) = \langle \mathcal{N}^n \rangle (x_i^{\mathrm{in}} = x_i, y)$, is then given by 
\bea 
\mathcal{L}^\dagger_{\mathrm{FP}} \cdot f_n(x_i, y) = -n f_{n-1}(x_i, y) \, ,
\eea 
where $\mathcal{L}^\dagger_{\mathrm{FP}}$ is the adjoint Fokker--Planck operator, and hence the moments of $P$ are given by the recursive relation
\bea \label{eq:usr:moments:pde}
\left[ \frac{1}{\mu^2}\frac{\partial^2}{\partial x^2} - 3y\left(\frac{\partial}{\partial x} + \frac{\partial}{\partial y}\right) \right]f_n = -nf_{n-1} \, ,
\eea 
with $n \geq 1$ and $f_0 = 1$, and boundary conditions 
\bea \label{eq:moments:initialconditions}
&f_n(0, y) = 0 \, , & \frac{\partial f_n}{\partial x}(1, y) = 0 \, ,
\eea 
\ie all trajectories initiated at $x_\mathrm{in}=0$ realise a vanishing
number of e-folds, and we implement the presence of a reflective wall at $x=1$.

\subsection{Characteristic function}

We again make use of the characteristic function $\chi_{\mathcal{N}}(t;x,y)$, given by \Eq{eq:characteristicFunction:def}, which is the (inverse) Fourier transform of the probability distrution of the number of \efolds, see \Eq{eq:PDF:chi}, where we see that the dummy variable $t$ has become the number of \efolds~$\mathcal{N}$ in this transformation.
The characteristic function of curvature perturbations, defined by $\zeta_\mathrm{cg} = \delta N_\mathrm{cg} = \mathcal{N} - \left< \mathcal{N} \right>$, can then be found via \Eq{eq:chideltaN:chiN},
where the mean number of \efolds~$\left<\mathcal{N}\right>(x,y)$ is given by
\bea \label{eq:meanefolds:charfunction}
\langle \mathcal{N}\rangle(x,y) &= f_1(x,y) = -i\left.\frac{\partial \chi_\mathcal{N}}{\partial t}\right|_{t=0} \, ,
\eea 
and the PDF of curvature fluctuation through the Fourier transform
\bea
\label{eq:PDF:chideltaN}
P\left(\zeta_\mathrm{cg}; x, y\right) = \frac{1}{2\pi} \int^{\infty}_{-\infty} \ee^{-it\left[\zeta_\mathrm{cg} - \left<\mathcal{N}\right>(x)\right]} \chi_{\mathcal{N}}\left(t; x, y \right)\dd t\, .
\eea
In fact, the characteristic function generates all the moments of the associated probability distribution and the $n^{\mathrm{th}}$ moment is given by
\bea \label{eq:char:generatemoments}
f_n(x,y) &= i^{-n}\frac{\partial^n}{\partial t^n}\chi_\mathcal{N}(t; x,y)\bigg|_{t=0} \, .
\eea 
If we combine the definition of the characteristic function \eqref{eq:characteristicFunction:def} with the equation of motion for the moments \eqref{eq:usr:moments:pde}, we find an equation for the characteristic function in USR, namely
\bea \label{eq:char:pde}
\left[ \frac{1}{\mu^2}\frac{\partial^2}{\partial x^2} - 3y\left(\frac{\partial}{\partial x} + \frac{\partial}{\partial y}\right) + it \right]\chi_\mathcal{N}(t;x,y) = 0 \, .
\eea 
This is still a second order PDE, but is now uncoupled and must be solved at each fixed $t$, rather than being a tower of coupled PDEs for each moment of the distribution that becomes increasingly more difficult to solve, as \eqref{eq:usr:moments:pde} does. 
The system \eqref{eq:char:pde} also comes endowed with initial conditions given by
\bea \label{eq:char:initialconditions}
& \chi_\mathcal{N}(t, 0, y) = 1 \, , & \frac{\partial \chi_\mathcal{N}}{\partial x}(t, 1, y) = 0 \, .
\eea 
We do not have the general analytic solution for the USR characteristic function given by \eqref{eq:char:pde}, but we can study this equation in certain simplifying limits. We will first give the ``classical limit" in which the classical drift dominates over the quantum diffusion in \Sec{sec:classicallimit}, and we will then go on to consider the ``stochastic limit" in which the quantum diffusion dominates. In \Sec{sec:latetimelimit} we solve \eqref{eq:char:pde} analytically in the ``late-time limit" in which the velocity $y$ has completely decayed and we simply have the inflaton freely diffusing on its potential, and in \Sec{sec:stochasticlimit} we include first-order corrections due to a small classical velocity $y$.

\section{Classical limit}
\label{sec:classicallimit}

In this section, we expand the characteristic function about its classical limit, in which the stochastic diffusion in negligible in comparison to the classical drift. 

\subsection{Leading order}

Let us first solve the leading order classical limit of equation \eqref{eq:char:pde}. 
This corresponds to neglecting the second derivative term, which originates from the quantum diffusion, and so the classical limit of the equation is 
\bea \label{eq:char:classicallimit}
\left[ - 3y\left(\frac{\partial}{\partial x} + \frac{\partial}{\partial y}\right) + it \right]\chi_\mathcal{N}(t;x,y) = 0 \, ,
\eea 
along with the initial condition $\chi_{\mathcal{N}} (t, 0, y) = 1$.
This PDE can then be solved (see \App{app:classicalLO:charfunction} for details) and gives the classical limit of the characteristic function to be
\bea \label{eq:char:classicalLO:solution}
\chi_\mathcal{N}\big|_\mathrm{cl}(t;x,y) = \left( 1 - \frac{x}{y}\right)^{-\frac{it}{3}} \, .
\eea 
Note that this solution does not satisfy the second boundary condition $x=1$, but this is not necessary in the classical limit as there is no diffusion. 
We can use now this to calculate the leading order of  specific moments $f_n(x,y)$ of the PDF $P(\mathcal{N};x,y)$. 
For instance, the mean number of \efolds, given by \eqref{eq:meanefolds:charfunction}, is found to be 
\bea \label{eq:meanefolds:classical:LO}
\langle \mathcal{N}\rangle(x,y)\big|_\mathrm{cl} 
&= -\frac{1}{3}\ln\left[ 1 - \frac{x}{y}\right] \, ,
\eea 
which we note matches the classical limit found in \cite{Firouzjahi:2018vet} (see equation $(4.5)$ in their paper).
Note that from \eqref{eq:chideltaN:chiN}, we find that $\chi_{\zeta_\mathrm{cg}} = 1$ and hence from \eqref{eq:PDF:chideltaN} we have $P(\zeta_\mathrm{cg}) = \delta(\zeta_\mathrm{cg})$, or in terms of \efolds 
\bea 
P(\mathcal{N};x,y) = \delta\left( \mathcal{N} - \langle \mathcal{N}\rangle(x,y)\big|_\mathrm{cl}\right) \, .
\eea 
This is exactly as one would expect at leading order in the classical limit, since we have completely neglected quantum diffusion and so any trajectory in the system will realise the same, deterministic number of \efolds. 

\subsection{Next-to-leading order}

We can now use our classical solution to perform an expansion about the classical solution, \ie for small quantum diffusion. 
We do this by using the leading-order classical solution as the source term for the next-to-leading order (NLO) solution, \ie
\bea 
\left[ - 3y\left(\frac{\partial}{\partial x} + \frac{\partial}{\partial y}\right) + it \right]\chi_\mathcal{N}(t;x,y)\Big|_{\mathrm{NLO}} = -  \frac{1}{\mu^2}\frac{\partial^2}{\partial x^2} \chi_{\mathcal{N}}(t;x,y)\Big|_{\mathrm{cl}} \, .
\eea  
Inputting the classical solution \eqref{eq:char:classicalLO:solution}, the differential equation we need to solve becomes
\bea 
\left[ - 3y\left(\frac{\partial}{\partial x} + \frac{\partial}{\partial y}\right) + it \right]\chi_\mathcal{N}(t;x,y)\Big|_{\mathrm{NLO}} = -  \frac{1}{y^2}\frac{it}{3\mu^2}\left(1+\frac{it}{3}\right)\left( 1- \frac{x}{y}\right)^{-2-\frac{it}{3}} \, ,
\eea 
subject to the initial conditions \eqref{eq:char:initialconditions}.
Solving this PDE using the same methods as before (see \App{app:classicalNLO:charfunction}), we find the next-to-leading order solution to be
\bea \label{eq:characteristic:NLO}
\chi_\mathcal{N}\bigg|_\mathrm{NLO} &= \left( 1 - \frac{x}{y}\right)^{-\frac{it}{3}}\left[ 1 - \frac{it}{3}\left(1+\frac{it}{3}\right)\frac{\frac{1}{3}\ln\left[ 1-\frac{x}{y}\right]}{\mu^2(y-x)^2} \right] \\ 
&= \left( 1 - \frac{x}{y}\right)^{-\frac{it}{3}}\left[ 1 + \frac{it}{3}\left(1+\frac{it}{3}\right)\frac{\left< \mathcal{N} \right> \big|_\mathrm{cl}}{\mu^2(y-x)^2} \right] \, .
\eea 
This procedure can be iterated to higher orders, where one can use the expression of the characteristic function at order $(\mathrm{N})^n$LO to evaluate the second derivative term in \eqref{eq:char:pde} and then solve this equation for the characteristic function at order $(\mathrm{N})^{n+1}$LO.
Once again we can calculate the mean number of \efolds~using \eqref{eq:meanefolds:charfunction} and we find 
\bea \label{eq:meaefolds:classical:NLO}
\langle \mathcal{N}\rangle(x,y)\Big|_{\mathrm{NLO}} 
&= -\frac{1}{3}\ln\left[ 1 - \frac{x}{y}\right]\left[ 1 - \frac{1}{3\mu^2\left(y-x\right)^2} \right] \\
&= \left< \mathcal{N} \right> \big|_\mathrm{cl}\left[ 1 - \frac{1}{3\mu^2\left(y-x\right)^2} \right] \, ,
\eea 
where the correction to the classical solution is of order $\mu^{-2}$, and once again $\mu$ is the only parameter that controls the expression.
 
We can also calculate the power spectrum at leading order using 
\bea 
P_{\zeta} = \frac{\dd \delta \mathcal{N}^2}{\dd \langle \mathcal{N}\rangle} = \frac{(\delta\mathcal{N}^{2})'}{\langle\mathcal{N}\rangle'} \, ,
\eea 
where $\delta \mathcal{N}^2 = \langle\mathcal{N}^2\rangle - \langle\mathcal{N}\rangle^2$, and in the second equality where a prime is a derivative with respect to the classical number of \efolds. 
Making use of \eqref{eq:char:generatemoments} and noting that $\langle\mathcal{N}^2\rangle(x,y) = f_2(x,y)$, we have 
\bea 
\langle\mathcal{N}^2\rangle &= \frac{2\left< \mathcal{N} \right> \big|_\mathrm{cl}}{9\mu^2\left(y-x\right)^2} + \left< \mathcal{N} \right> \big|_\mathrm{cl}{}^2\left[ 1 - \frac{2}{3\mu^2(y-x)^2} \right] \, , \\
\eea 
and hence
\bea 
 \delta \mathcal{N}^2 &= \frac{2}{9\mu^2(y-x)^2}\left< \mathcal{N} \right> \big|_\mathrm{cl}\left[ 1 - \frac{\left< \mathcal{N} \right> \big|_\mathrm{cl}}{\mu^2(y-x)^2} \right] \, . \\
\eea 
Thus, we find the power spectrum to be
\bea 
P_{\zeta}(x,y) \simeq \frac{2}{9\mu^2\left(y-x\right)^2} \, , 
\eea 
at leading order in $\mu^{-2}$.
Here we see that the correction to the power spectrum is of the same order of the correction to the mean number of \efolds~\eqref{eq:meaefolds:classical:NLO}, and hence in the classical limit of our system, the power spectrum remains small.

\section{Late-time limit} \label{sec:latetimelimit}

If we consider the late-time limit of our USR system, which relates to trajectories that experience a long time in the USR regime, then we expect the (classical) velocity $y$ to become negligible since $y = y_\mathrm{in}\ee^{-3N}$ decays exponentially.
The late-time limit thus corresponds to $y\approx 0$, and \eqref{eq:char:pde} becomes
\bea \label{eq:char:latetime}
\frac{1}{\mu^2}\frac{\partial^2}{\partial x^2}\chi_{\mathcal{N}}(t; x, y) = -it\chi_{\mathcal{N}}(t;x,y) \, .
\eea 
This is equivalent to free diffusion on a flat potential (\ie the inflaton enters the quantum well with no classical velocity), which has previously been studied in chapter \ref{chapter:quantumdiff:slowroll}, and we note that in this zero velocity limit, we recover the slow-roll Langevin equations for a flat potential. 
This means that, despite what one may be concerned with, it is not inconsistent to use the slow-roll Langevin equation on an exactly flat potential.
In fact, one should consider this as working in the small velocity, or late time, limit of the full USR system. 
Classically this limit is not well-defined, since the inflaton will remain at rest if there is no classical velocity present, but when one considers quantum diffusion this effectively ``regularises" the problem and gives a well-defined system. 

The solution for the characteristic function defined by \eqref{eq:char:latetime} is then given by \Eq{eq:chiN:cosh}, where it is interesting to note that this late-time ($y\to0$) limit of the USR problem is equivalent to the slow-roll problem studied in \Sec{sec:StochasticLimit}.
The PDF of the number of \efolds~is then the inverse Fourier transform of the characteristic function, and is given by \Eq{eq:stocha:HeatMethod:PDF:expansion}, which we rewrite here for convenience
\bea \label{eq:pdf:usr:analyitic}
P\left(\N, \phi \right) = & \frac{2 \pi}{\mu^2} \sum_{n=0}^\infty \left( n + \frac{1}{2} \right) \exp\left[ -\frac{\pi^2}{\mu^2} \left(n+\frac{1}{2}\right)^2 \N\right] \sin\left[x \pi\left(n+\frac{1}{2}\right)\right] \, .
\eea
\begin{figure} 
    \centering
    \includegraphics[width=0.49\textwidth]{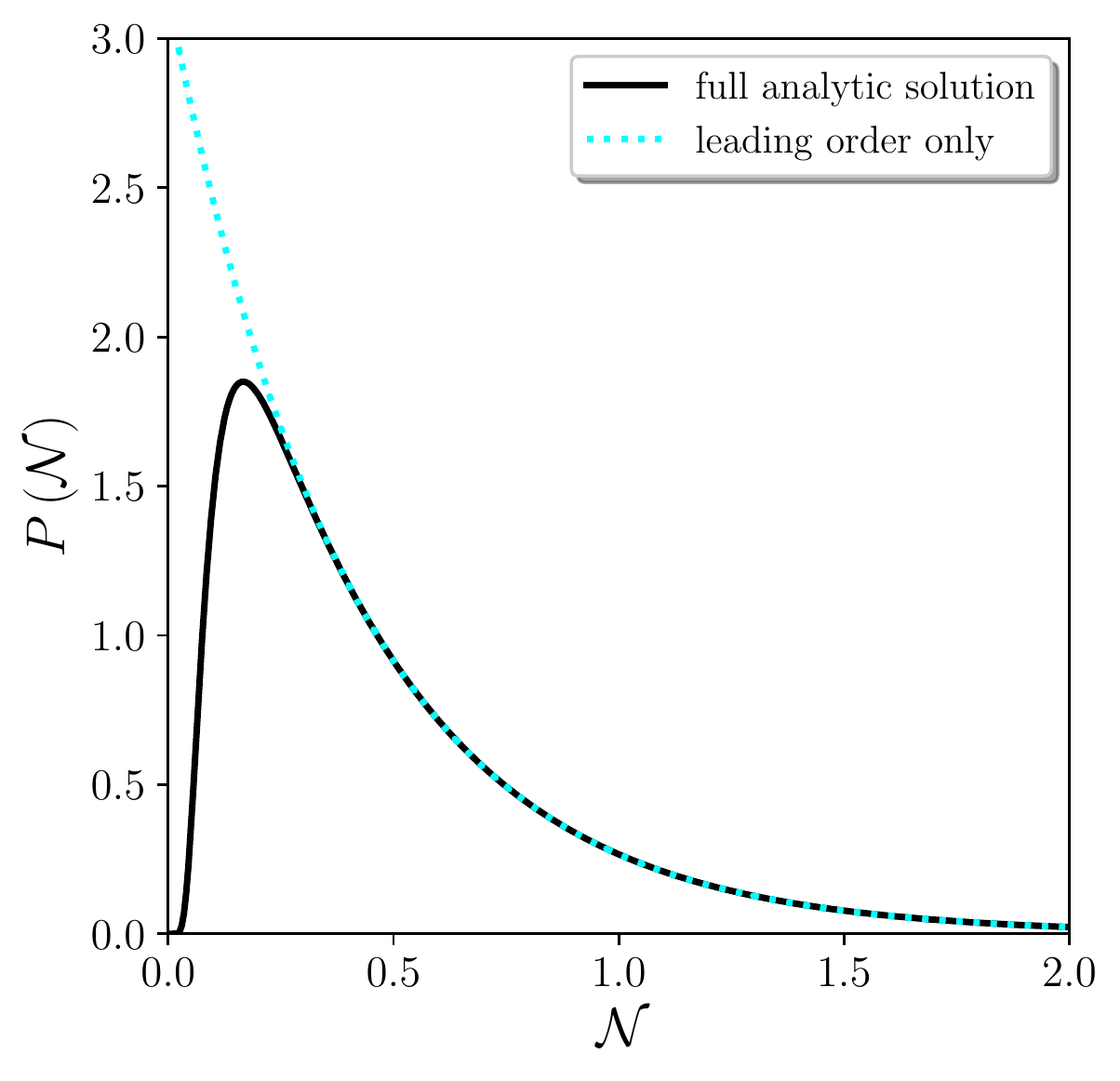}
    \includegraphics[width=0.49\textwidth]{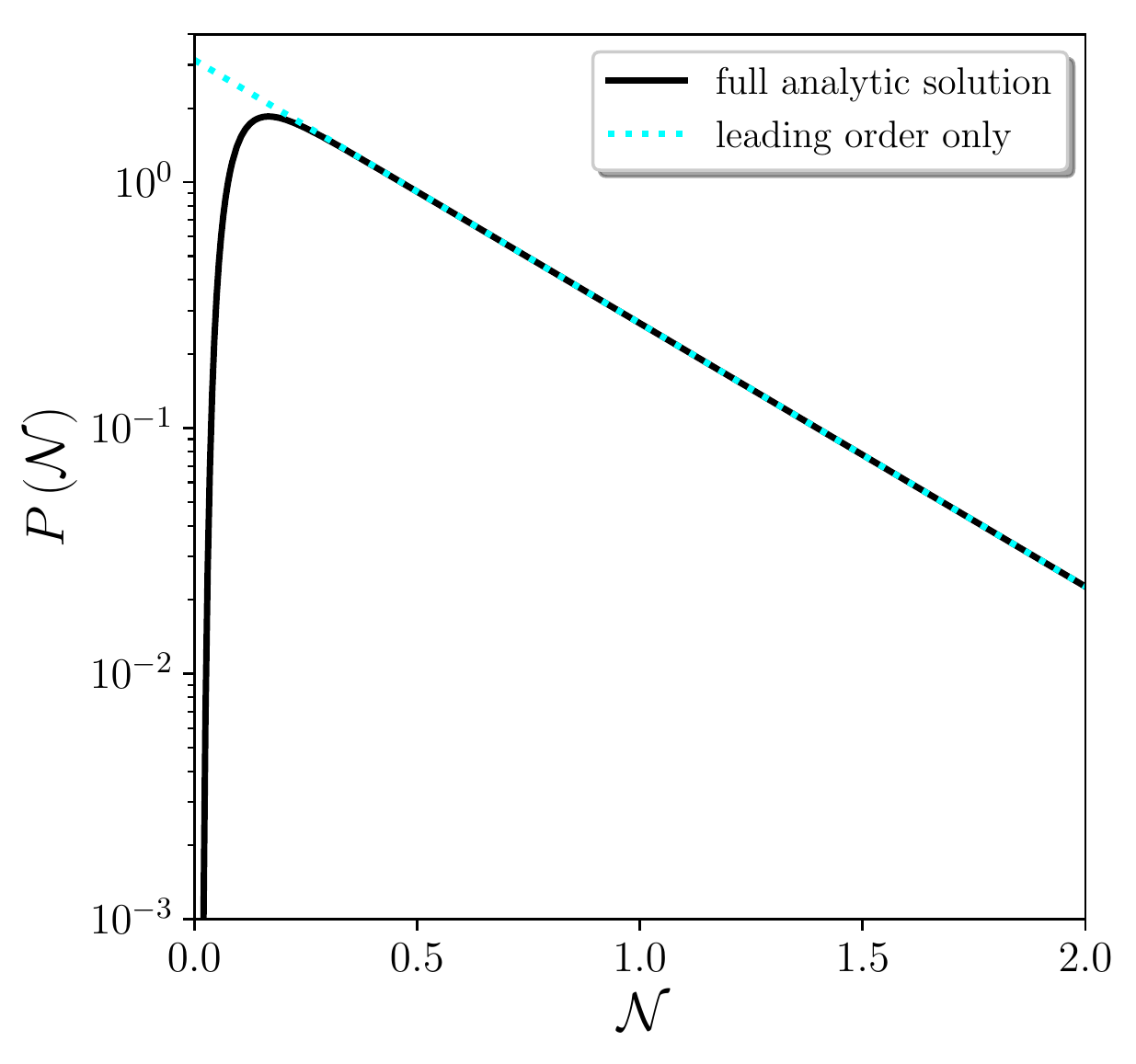}
    \caption[Probability distribution of $N$ in the late time, ultra-slow roll limit]{The PDF \eqref{eq:pdf:usr:analyitic} is plotted in black. The cyan dashed line is only the leading order term in \eqref{eq:pdf:usr:analyitic}, corresponding to the $n=0$ term, and we see the exponential decay of the tail of the distribution. The left hand plot uses a linear scale for the $y$-axis while the right hand plot uses a log scale.  \label{fig:pdf:analytic}}
\end{figure}
This PDF is plotted in the left hand side of \Fig{fig:pdf:analytic} for $\mu=1$. 

Let us note that \eqref{eq:pdf:usr:analyitic} is an expansion in powers of $\propto \ee^{-\mathcal{N}}$, that is, we are expanding about $\mathcal{N}=\infty$ which is the tail on the right-hand side of the distribution (describing large values of $\mathcal{N}$) in \Fig{fig:pdf:analytic}. 
In particular, we see that the tail of the distribution decays exponentially, and not as a Gaussian, which is the approximation that is often used in the study of the formation of primordial black holes. 
To illustrate this point, we also plot just the leading order exponential of the PDF in the right hand plot of \Fig{fig:pdf:analytic}, which corresponds to the $n=0$ term in \eqref{eq:pdf:usr:analyitic}.
That is, we plot
\bea 
P\left(\mathcal{N}, \phi\right)\bigg|_{n=0} = \frac{\pi}{\mu^2}\exp\left[ -\frac{\pi^2}{4\mu^2}\mathcal{N} \right] \sin\left[ \frac{\pi x}{2} \right] \, ,
\eea 
and we see that on the tail this is perfectly sufficient to explain the decay of the distribution.

An interesting observation about the late time limit is that it is independent of the initial velocity $y_\mathrm{in}$ and hence all trajectories tend to the same behaviour in the large-$\mathcal{N}$ limit.
This means that once the classical velocity has become negligible, we can model the late-time behaviour as free diffusion using \eqref{eq:pdf:usr:analyitic} from the field value at which the velocity vanishes. 
That is, the decay rate of the PDF at late times is given by the width of the USR region once $y\to0$.

We can also calculate the mean number of \efolds~from \eqref{eq:chiN:cosh} to be 
\bea \label{eq:usr:efolds:analytic}
\left< \mathcal{N}(x)\right> = \mu^2x\left(1-\frac{x}{2}\right) \, ,
\eea 
which, in USR,  gives $\left< \mathcal{N}\right>$ for an initial field value $\phi$ and zero initial velocity.

\section{Stochastic USR limit} \label{sec:stochasticlimit}

In order to understand the dynamics of the inflaton in the stochastic limit, we will expand \Eq{eq:char:pde} in the limit of small classical velocity $y$.
This corresponds to the quantum diffusion dominating over the classical drift during USR inflation. 

\subsection{Stochastic limit expansion}

\begin{figure} 
    \centering
    \includegraphics[width=0.49\textwidth]{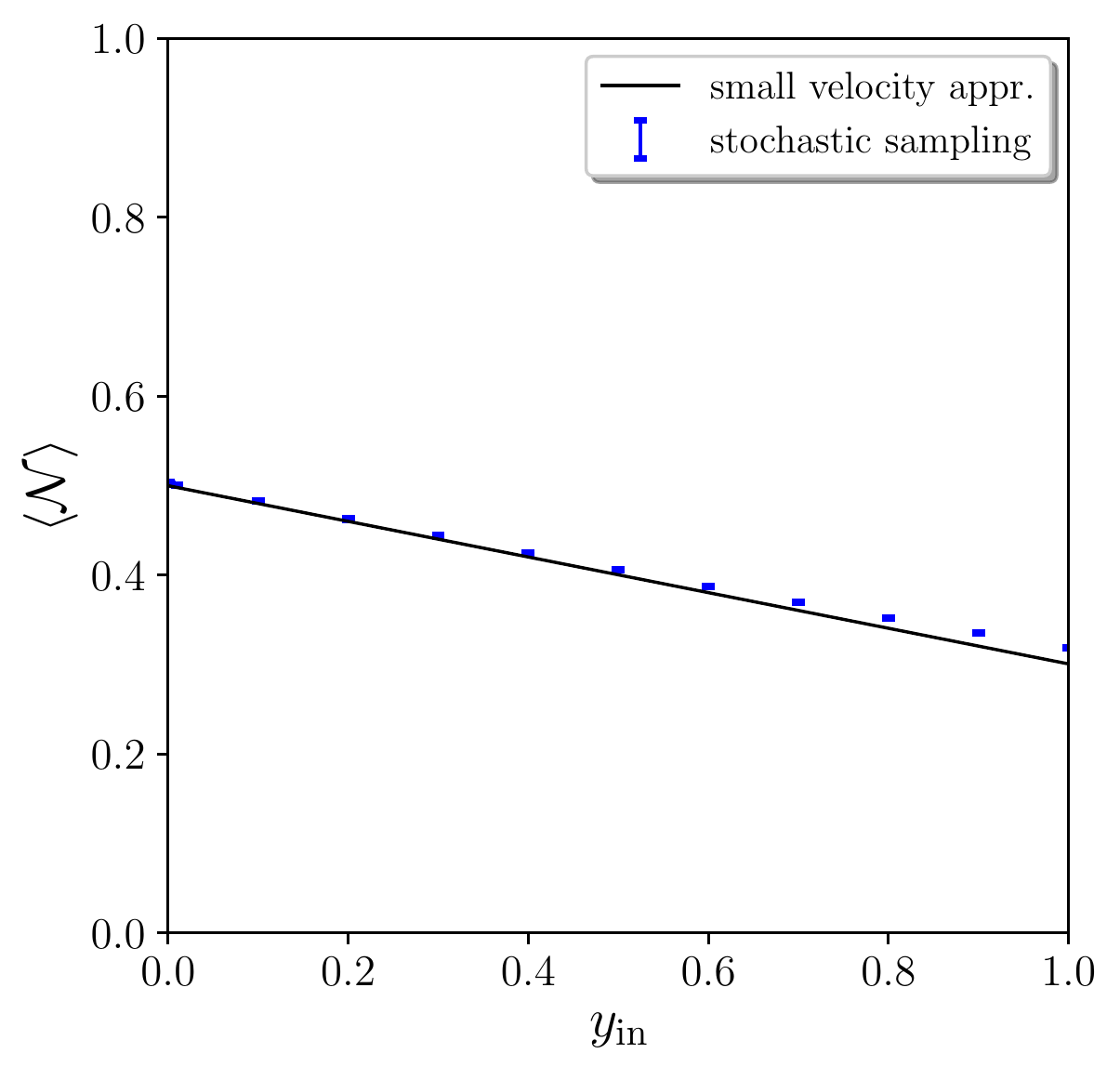}
    \includegraphics[width=0.49\textwidth]{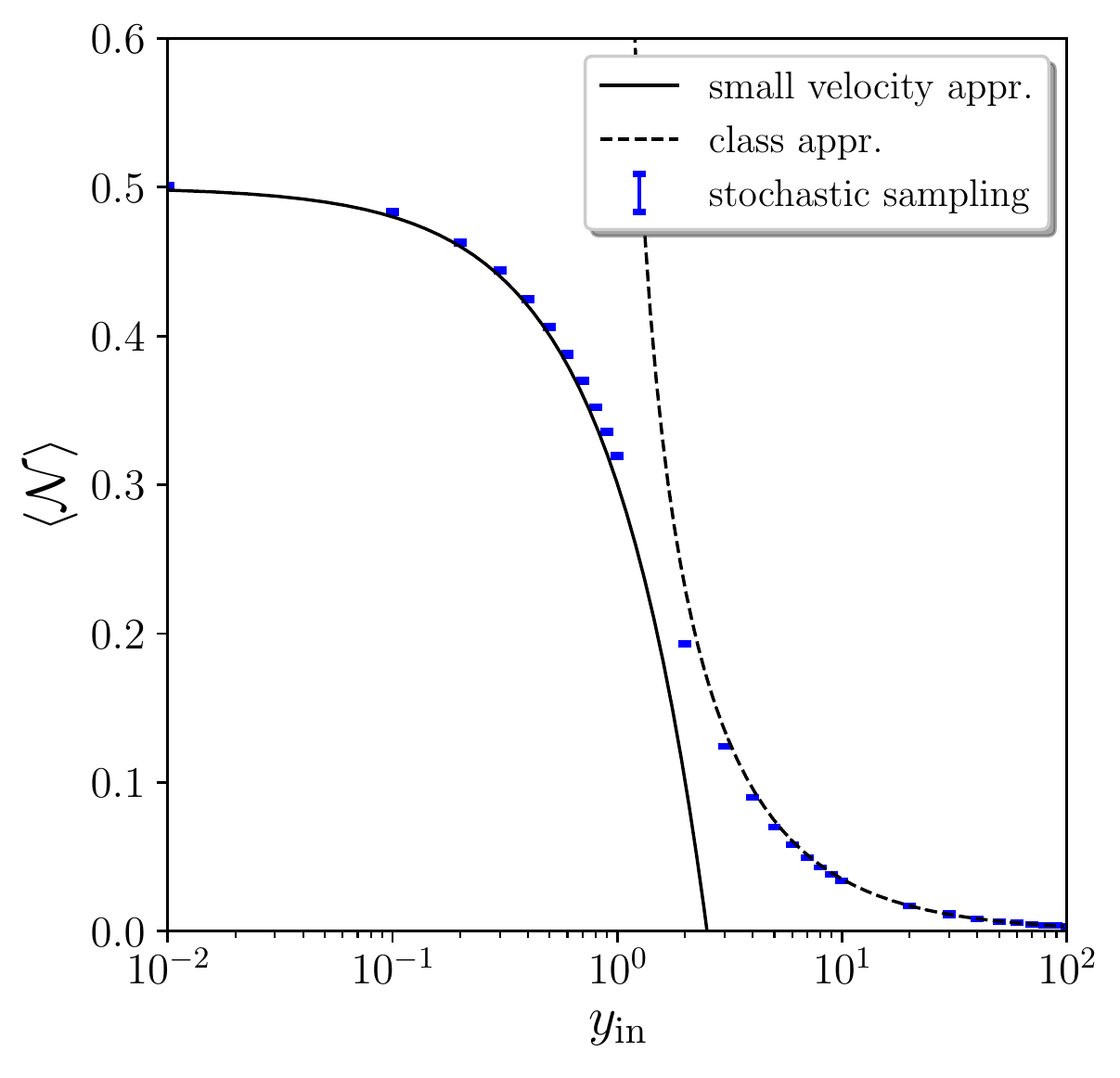}
    \caption[The mean number of \efolds~in the ultra-slow-roll limit]{The mean number of \efolds~is plotted as a function of the initial veloicty $y_\mathrm{in}$, for $\mu=1$. The solid black curves are the small-$y$ approximation \eqref{eq:meanefolds:smally}, the dashed black curve corresponds to the classical result \eqref{eq:meanefolds:classical:LO}, and the blue bars are reconstructed from a large number of realisations of the Langevin equations \eqref{eq:langevin:xy}. The size of the bars is the $2\sigma$ estimate for the statistical error, which are found using a jackknife procedure explained in the main text. In the right hand plot we see that the transition between the two different limiting cases that we have considered happens at around $y_\mathrm{in}=1$, as we expect. \label{fig:meanefolds:smally}}
\end{figure}

In the stochastic, small-$y$ limit, let us Taylor expand the characteristic function as 
\bea \label{eq:chi:smally:taylorexp}
\chi_\mathcal{N}(x,y,t) = \chi_\mathcal{N}(x,0,t) + yf(x,t) \, , 
\eea 
where $\chi_\mathcal{N}(x,0,t)$ is given by \eqref{eq:chiN:cosh}. 
At leading order in the stochastic USR (small-$y$) limit we have
\bea 
\left[ \frac{1}{\mu^2}\frac{\partial^2}{\partial x^2} - \left( 3 - it \right) \right]f(x,t) &= 3\frac{\partial}{\partial x}\chi_\mathcal{N}(x,0,t) \\
&= 3\alpha\sqrt{t}\mu \frac{\sinh\left[ \alpha\sqrt{t}\mu(x-1)\right]}{\cosh\left[ \alpha \sqrt{t}\mu\right]} \, , 
\eea 
with boundary conditions 
\bea \label{eq:boundaryconditions:smallylimit} 
& f(0,t) = 0 \, , & \frac{\partial f}{\partial x}(1,t) = 0 \, . 
\eea 
One can check that this is solved via
\bea 
f(x,t) &= C_1(t)\ee^{\sqrt{3-it}\mu x} + C_2(t)\ee^{-\sqrt{3-it}\mu x} - \alpha \sqrt{t}\mu\frac{\sinh\left[ \alpha\sqrt{t}\mu(x-1)\right]}{\cosh\left[ \alpha \sqrt{t}\mu\right]} \, , 
\eea 
for some $C_1(t)$ and $C_2(t)$ to be found using the boundary conditions \eqref{eq:boundaryconditions:smallylimit}.
Doing this allows us to calculate $f$ completely, which we find to be 
\bea 
f(x,t) &= \frac{-i\sqrt{t}\mu}{\sqrt{3-it}\cosh\left(\alpha\sqrt{t}\mu\right)\cosh\left(\sqrt{3-it}\mu\right)}\left\{ \sqrt{t}\sinh\left(\sqrt{3-it}\mu x\right) \right. \\
& \hspace{1cm} \left. - \alpha^*\sqrt{3-it}\sinh\left(\alpha\sqrt{t}\mu\right)\cosh\left[\sqrt{3-it}\mu(x-1)\right] \right. \\
& \hspace{1cm} \left. - \alpha^*\sqrt{3-it}\cosh\left(\sqrt{3-it}\mu\right)\sinh\left[\alpha\sqrt{t}\mu(x-1)\right] \right\} \, ,
\eea 
where $\alpha^*$ is the complex conjugate of $\alpha$. 
Recall at this point that 
\bea 
\chi_\mathcal{N}(x,y,t) = \chi_\mathcal{N}(x,0,t) + yf(x,t) \, , 
\eea 
where each of these terms is given above. 
As we have done before, we can find the mean number than \efolds~using \eqref{eq:meanefolds:charfunction}, and then find 
\bea \label{eq:meanefolds:smally}
\left< \mathcal{N} \right>(x,y) &= \frac{\mu^2}{2}\left( 2x - x^2 - 2y 
+2xy \right) \\
& \hspace{1cm} + \frac{\mu y}{\cosh\left( \sqrt{3}\mu \right)}\left\{ \mu\cosh\left[ \sqrt{3}\mu\left(x-1\right) \right] - \frac{1}{\sqrt{3}}\sinh\left[ \sqrt{3}\mu x \right] \right\} \, ,
\eea 
which, as expected, reduces to \eqref{eq:usr:efolds:analytic} when $y=0$. 
Here, $x$ and $y$ denote the initial values of these variables when the USR regime begins. 
The expression \eqref{eq:meanefolds:smally} is displayed as a function of the initial velocity $y_\mathrm{in}$ in \Fig{fig:meanefolds:smally} and compared with a numerical simulation of many realisation of the Langevin equations \eqref{eq:eom:phi:stochastic}-\eqref{eq:eom:v:stochastic} for $\mu=1$. 
We use this value for $\mu$ as we have seen in Chapter \ref{chapter:quantumdiff:slowroll} that this provides an upper limit for $\mu$ if we want to satisfy PBH constraints. 
The number of numerical simulation used varies between $10^6$ and $10^8$ depending on the value of the initial velocity, and the ensemble average of the number of \efolds~elapsed in each of the realisations is displayed with the blue bars. 
These bars correspond to an estimate of the $2\sigma$ statistical error, which is presence due to using a finite number of realisations, and we estimate this error using the jackknife resampling method, which we now describe.

For a sample of $n$ trajectories, the central limit theorem states that the statistical error on this sample scales as $1/\sqrt{n}$, and so we can write the standard deviation as $\sigma=\lambda/\sqrt{n}$.
We now need to find $\lambda$, and to do this we divide our set of realisations into $n_\mathrm{sub}$ subsamples, each of size $n/n_\mathrm{sub}$.
In each subsample, we then compute the mean number of \efolds~$\left<\mathcal{N}\right>_{(n_\mathrm{sub})}$ for that subsample and then compute the standard deviation $\sigma_{n/n_\mathrm{sub}}$ across the set of values of $\left<\mathcal{N}\right>_{(n_\mathrm{sub})}$ that are obtained. 
This give us $\lambda=\sqrt{n/n_\mathrm{sub}}\sigma_{n/n_\mathrm{sub}}$, and hence $\sigma_n=\sigma_{n_\mathrm{sub}}/\sqrt{n_\mathrm{sub}}$.
In practice, we take $n_\mathrm{sub}=100$ and evaluate the statistical error using this formula to give the error bars displayed in subsequent figures. 

We see in \Fig{fig:meanefolds:smally} that our small-$y$ approximation matches the simulations very well for $y_\mathrm{in} < 1$ and our classical formula \eqref{eq:meanefolds:classical:LO}, which is displayed with the black dashed line in the right hand panel, matches the simulation very well for $y_\mathrm{in} \gg 1$, so both limits here are well understood.

\subsection{Implications for primordial black holes}

In the stochastic (small-$y$) limit, we have expanded to find the characteristic function, and our goal now is to find the PDF associated to this characteristic function.
Following \cite{Ezquiaga:2019ftu}, we now want to find the poles of $\chi_\mathcal{N}$, given as $\Lambda_n = it$, so that we can expand $\chi_\mathcal{N}$ as 
\bea 
\chi_\mathcal{N} = \sum_n \frac{a_n(x)}{\Lambda_n-it} + g(t,x) \, , 
\eea 
where $g(t,x)$ is some regular function, and $a_n(x)$ is the residual associated to $\Lambda_n$ which can be found via 
\bea 
a_n(x) = -i\left[\frac{\partial}{\partial t}\chi_\mathcal{N}^{-1}\left(t=-i\Lambda_n, x, y\right)\right]^{-1} \, . 
\eea 
After doing this expansion, it will then be simple to find the PDF, which is given by 
\bea 
P\left(\mathcal{N}\right) = \sum_n a_n(x)\ee^{-\Lambda_n\mathcal{N}} \, .
\eea 
Both terms in $\chi_\mathcal{N}$ (\ie in \Eq{eq:chi:smally:taylorexp}) have the term $\cosh\left(\alpha\sqrt{t}\mu\right)$ in their denominator, which has associated poles at
\bea 
\cosh\left(\alpha\sqrt{t}\mu\right) = \cos\left(i\alpha\sqrt{t}\mu\right) = \cos\left(\sqrt{it}\mu\right) = 0 \, ,
\eea 
\ie when 
\bea 
it = \frac{\pi^2}{\mu^2}\left(n+\frac{1}{2}\right)^2 \equiv \Lambda_n^{(1)} \, . 
\eea 
The corresponding residual is then given by
\bea 
a_n^{(1)} &= \frac{\pi}{\mu^2}\left(n+\frac{1}{2}\right)\left\{ 2\sin\left[\pi\left(n+\frac{1}{2}\right)x\right] - \pi(2n+1)y\cos\left[ \pi\left(n+\frac{1}{2}\right)x\right] \right\} \\
& \hspace{1cm} + \frac{\left(-1\right)^n 2\pi^2\left(n+\frac{1}{2}\right)^2 y}{\mu^2\sqrt{3\mu^2-\pi^2\left(n+\frac{1}{2}\right)^2}\cosh\left[ \mu\sqrt{3-\frac{\pi^2\left(n+\frac{1}{2}\right)^2}{\mu^2}} \right]} \\ 
& \hspace{1cm} \times \left\{(-1)^n \sqrt{3\mu^2-\pi^2(\left(n+\frac{1}{2}\right)^2}\cosh\left[ \mu(x-1)\sqrt{3-\frac{\pi^2\left(n+\frac{1}{2}\right)^2}{\mu^2}} \right] \right. \\ 
& \hspace{2cm} \left. - \pi\left(n+\frac{1}{2}\right)\sinh\left[ \mu x\sqrt{3-\frac{\pi^2\left(n+\frac{1}{2}\right)^2}{\mu^2}} \right] \right\} \, . 
\eea 
The NLO term $f(x,y)$ also features the term $\cosh(\sqrt{3-it}\mu)$ in its denominator, which has a pole at
\bea 
it = 3+ \frac{\pi^2}{\mu^2}\left(n+\frac{1}{2}\right)^2 \equiv \Lambda_n^{(2)} = \Lambda_n^{(1)} + 3 \, , 
\eea 
with associated residue
\bea 
a_n^{(2)} &= \frac{2(-1)^n y}{\mu^2}\frac{\sin\left[ \left(n+\frac{1}{2}\right)\pi x\right]}{\cos\left[ \sqrt{3\mu^2 + \pi^2\left(n+\frac{1}{2}\right)^2} \right]} \left\{ -3\mu^2 - \pi^2\left(n+\frac{1}{2}\right)^2 \right. \\ 
& \hspace{1cm} \left. +\pi(-1)^n \left(n+\frac{1}{2}\right)\sqrt{3\mu^2 + \pi^2\left( n+\frac{1}{2} \right)^2}\sin\left[ \sqrt{3\mu^2 + \pi^2\left( n+\frac{1}{2} \right)^2} \right] \right\} \, . 
\eea 
Then the PDF is given by 
\bea \label{eq:PDF:smallylimit}
P(\mathcal{N}; x, y) &= \sum_{i=1,2}\sum_{n=0}^{\infty}a_n^{(i)}(x,y)\ee^{-\Lambda_n^{(i)}\mathcal{N}} &= \sum_{n=0}^{\infty}\left[ a_n^{(1)} + a_n^{(2)}\ee^{-3\mathcal{N}} \right] \ee^{-\Lambda_n^{(1)}\mathcal{N}} \, ,
\eea
where each of these terms is given above. 

\begin{figure} 
    \centering
    \includegraphics[width=0.49\textwidth]{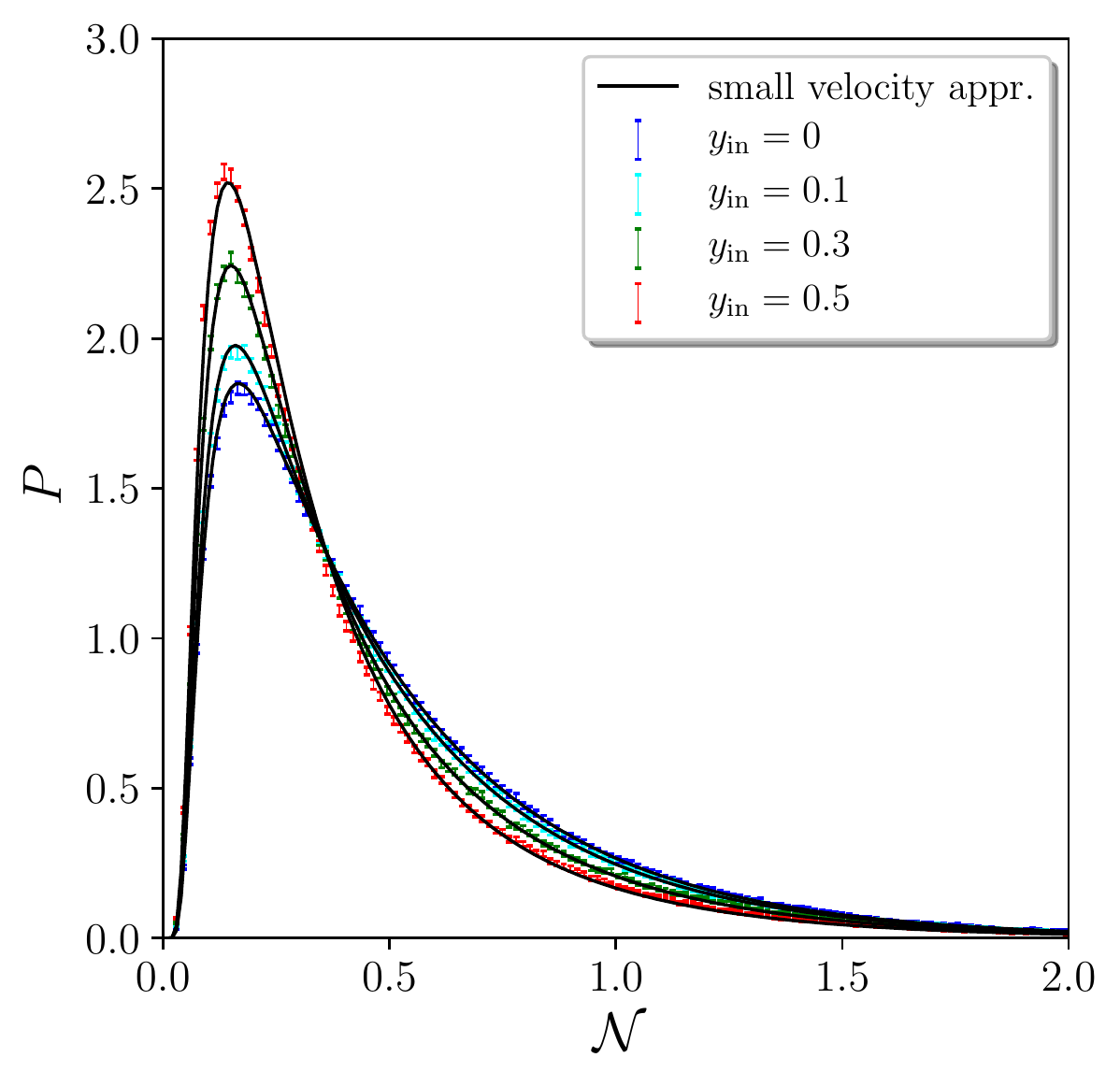}
    \includegraphics[width=0.49\textwidth]{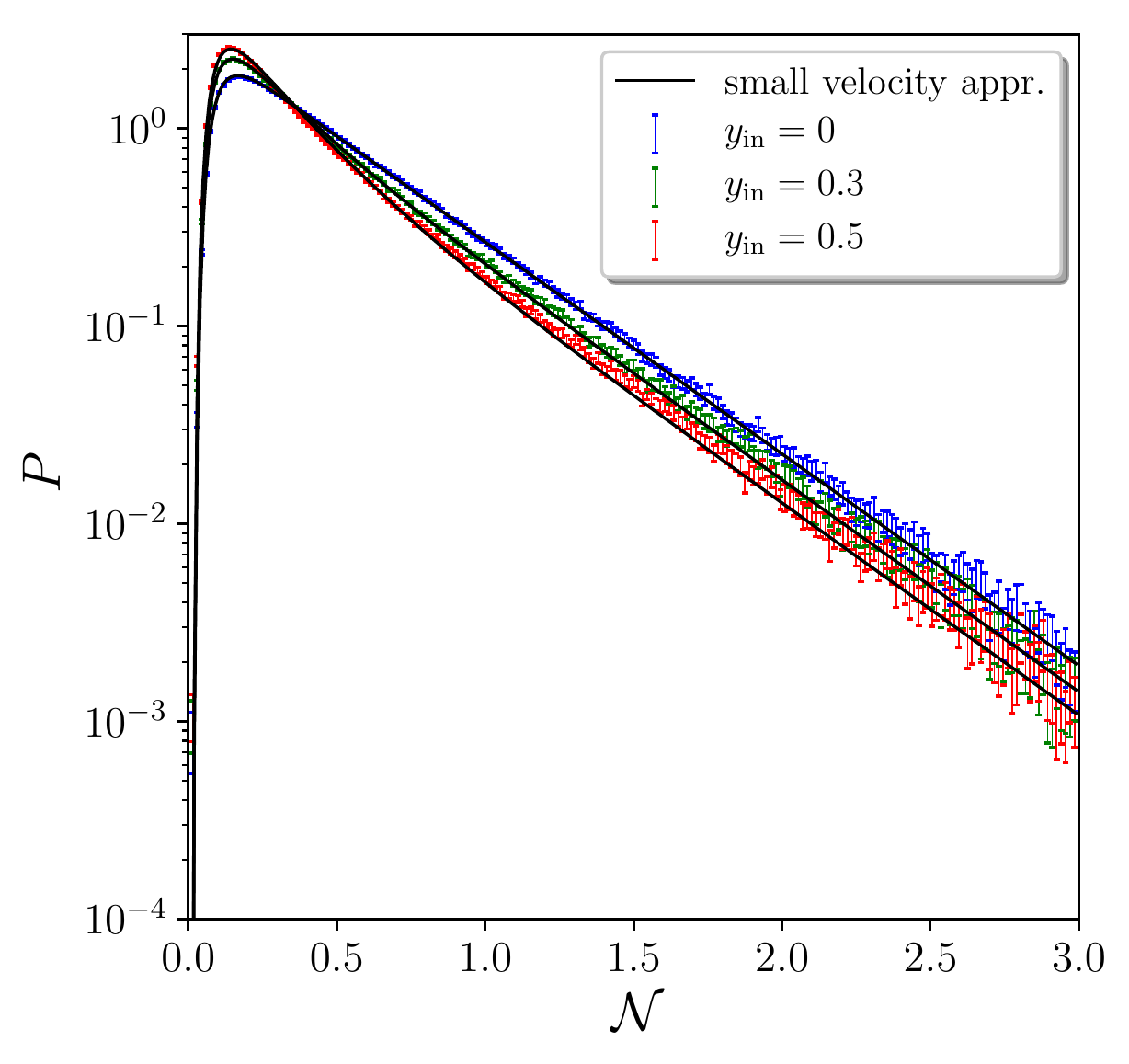}
    \caption[Probability distribution of $N$ in ultra-slow roll]{The probability distribution $P$ of the number of \efolds~$\mathcal{N}$ for several different values of the initial velocity $y_\mathrm{in}$. In the left panel we use a linear scale of the vertical axis, and in the right panel we use a logarithmic axis (the value $y_\mathrm{in}=0.1$ is not shown in the right panel for display convenience), and in both plots we take the value $\mu=1$. The black curves correspond to the small-$y$ expression \eqref{eq:PDF:smallylimit}, which for $y_\mathrm{in}=0$ is equivalent to the slow-roll solution \eqref{eq:pdf:usr:analyitic}, while the error bars are reconstructed from a large number of realisations of the Langevin equations \eqref{eq:langevin:xy}, with Gaussian kernel density of width $N=0.005$. The size of the bars is the $2\sigma$ estimate for the statistical error, which are found using a jackknife procedure explained in the main text. We see that our analytic approximation provide a very good fit to the simulations for all values of $y_\mathrm{in}$ taken. \label{fig:pdf:smally}}
\end{figure}

Note that the presence of the classical velocity does not change the location of the poles that are present at leading order in this calculation, but rather adds a second set of poles. 
This new set of poles do not contribute at leading order since their associated residuals vanish when $y=0$, and so their contribution to \eqref{eq:PDF:smallylimit} vanishes. 
As such, a classical velocity does not change the decay rate of the PDF, and only provides a small correction to the amplitude of the PDF (as the residuals are suppressed by the factor of $\ee^{-3\mathcal{N}}$).
Our formula \eqref{eq:PDF:smallylimit} is compared to numerical simulations of the USR Langevin equation in \Fig{fig:pdf:smally}, where we can see that it provides a very good fit of the PDF, even for reasonably large values of $y_\mathrm{in}$.
On the tail of the distribution we see that the error bars become larger, and this is because realisations that experience the larger numbers of \efolds~of inflation become rarer, and hence the statistics are based on a fewer numbers of trajectories. 
However, we still see that the decay rate of the tail of the distribution is still independent of the initial velocity $y_\mathrm{in}$.

\begin{figure} 
    \centering
    \includegraphics[width=0.49\textwidth]{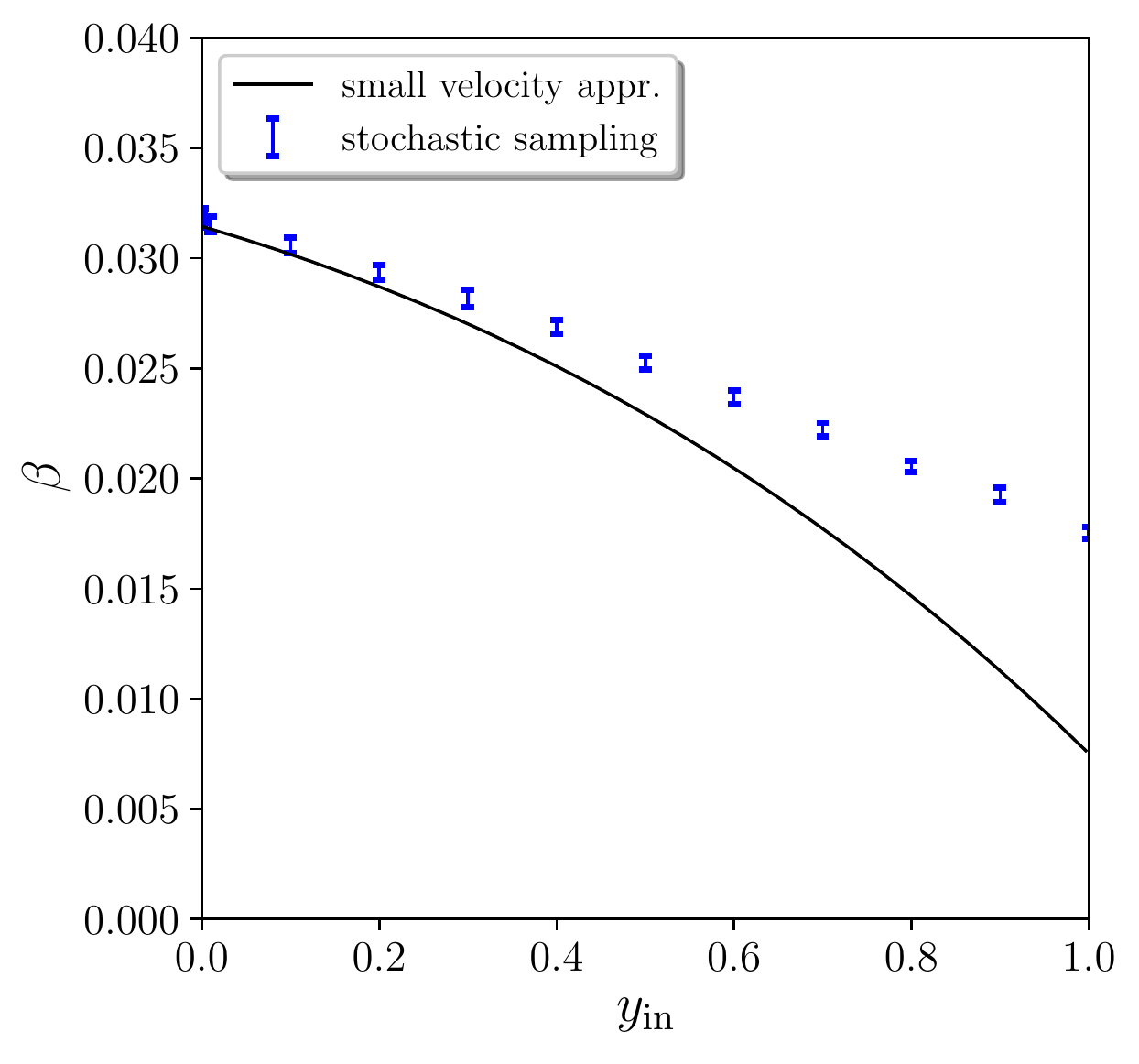}
    \includegraphics[width=0.49\textwidth]{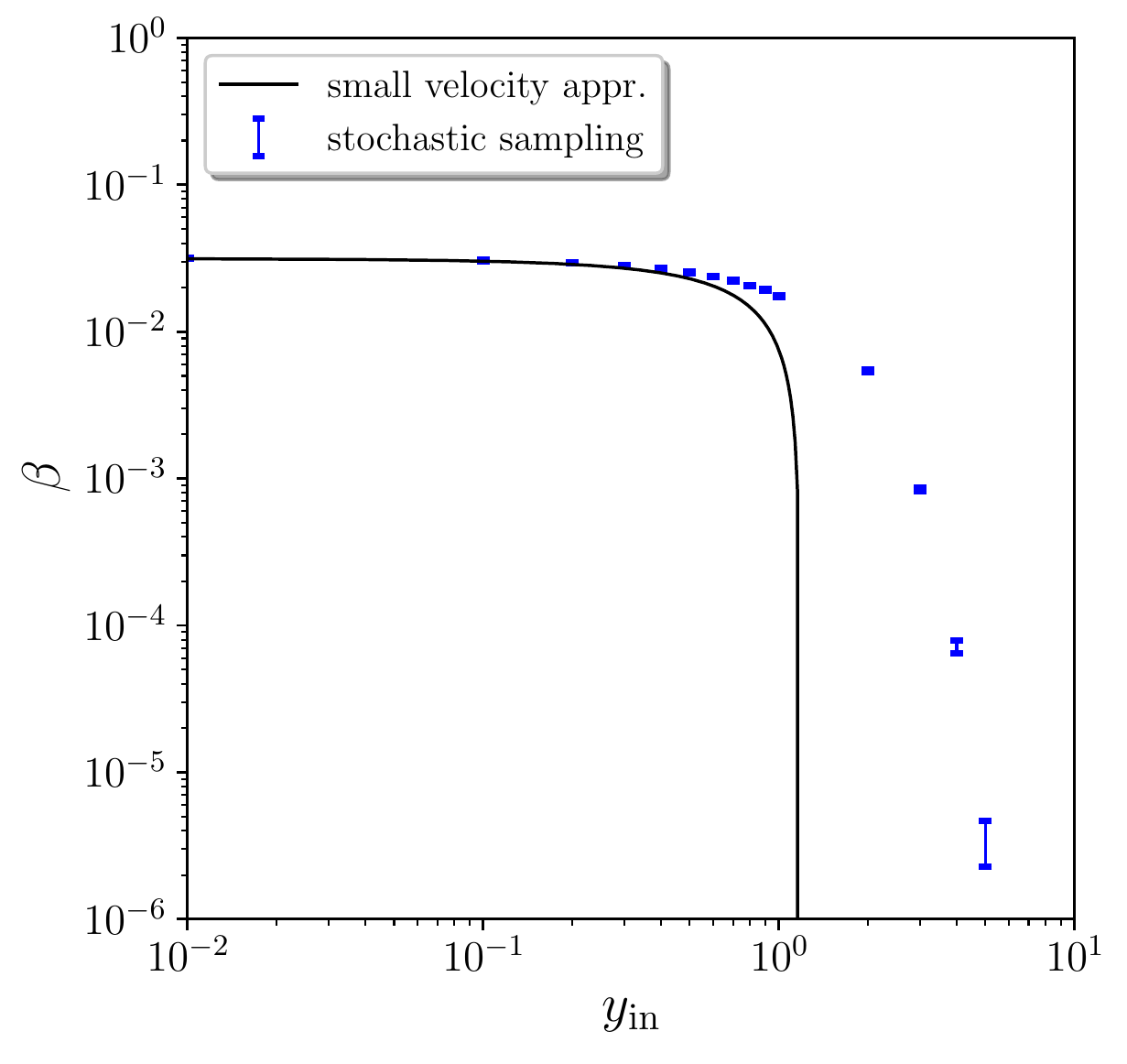}
    \caption[Mass fraction of PBHs formed in an ultra-slow-roll regime]{The PBH mass fraction $\beta$ in a flat potential is plotted as a function of the inital velocity $y_\mathrm{in}$, for $\mu=1$. The black curves are the small-$y$ approximation \eqref{eq:beta:smally}, while the blue bars are reconstructed from a large number of numerical simulations of the Langevin equations \eqref{eq:langevin:xy}. The size of the bars is the $2\sigma$ estimate for the statistical error, estimated using a jackknife procedure. \label{fig:beta:smally}}
\end{figure}

From here, we can calculate the mass fraction $\beta$ of primordial black holes using \Eq{eq:def:beta}, and here it is found to be 
\bea \label{eq:beta:smally}
\beta &= \sum_{i=1,2}\sum_{n=0}^{\infty}\frac{1}{\Lambda_n^{(i)}}a_n^{(i)}(x,y)\ee^{-\Lambda_n^{(i)}\left[\left<\mathcal{N}\right>(x)+\zeta_c\right]} \\
&= \sum_{n=0}^{\infty} \left[ \frac{a_n^{(1)}}{\Lambda_n^{(1)}} + \frac{a_n^{(2)}}{\Lambda_n^{(1)}+3}\ee^{-3\left[\left<\mathcal{N}\right>(x)+\zeta_c\right]} \right] \ee^{-\Lambda_n^{(1)}\left[\left<\mathcal{N}\right>(x)+\zeta_c\right]} \, . 
\eea 

This expression is compared to the results of numerical realisations of our USR Langvin equations \eqref{eq:langevin:xy} in \Fig{fig:beta:smally}. 
In practice, for the numerical realisations, we take a conservative value for the critical value of the curvature perturbation $\zeta_\mathrm{c}=1$, and then to obtain $\beta$ we take the number of realisations that produce more \efolds~than the mean value plus $\zeta_\mathrm{c}$ and divide this by the total number of realisations.
When $\beta$ becomes very small we must compute a very large number of realisations in order to avoid being dominated by statistical noise (in practice we must produce many more realisations than $1/\beta$), and hence this becomes very numerically expensive and we therefore do not calculate values of $\beta$ smaller than $\sim 10^{-6}$.
In our numerical simulations, for $y_\mathrm{in} > 6$ we produced $10^7$ realisations and exactly none of these experienced more than $\left< \mathcal{N} \right> + 1$ \efolds~of inflation, and hence we can only place an upper bound of $\sim 10^{-7}$ for these values of the initial velocity.

\section{Discussion} \label{sec:USRdiscussion}

We have seen that in USR inflation we can find the PDF of curvature fluctuations by solving for the characteristic function from \Eq{eq:char:pde}, along with initial conditions \eqref{eq:char:initialconditions}, and then perform an inverse Fourier transform to find the PDF. 
In turn, through \eqref{eq:def:beta}, this tells us how much of the Universe is contained in PBHs at the time of their formation. 
However, solving this system with the given initial conditions is not possible in general, and so we inform our study by considering limiting cases of this system. 

In the classical limit, \ie when the classical velocity dominates over the stochastic diffusion, the system becomes solvable in an iterative way, to as high an order as one requires.  
In the opposite limit, when the stochastic term dominates, we have performed an expansion in the small classical velocity and found the NLO solution in this limit.
In this case, both the PDF and the PBH mass fraction can be compared to numerical simulations, and we see consistent results. 

This work demonstrates that we have a good understanding of these two limiting cases for USR inflation, but the question of which of these limits is physically relevant, or if both can be realised in practice, remains.
If we consider the Starobinsky model of inflation, see for example \Sec{sec:Staro}, then the initial velocity in the USR phase is given by the slow-roll attractor that proceeds it. 
On the other hand, as we have seen in \Sec{sec:example:inflectionPoint}, for a cubic inflection point model a transition from slow roll into ultra-slow roll actually ends inflation, and hence an inflating USR regime must start in USR. 
This means that we can, \textit{a priori}, give the inflaton any initial velocity that we want (provided it is less than the limit given by \eqref{eq:ymax:generic}), and hence both of the limits we have considered may be applicable. 
We leave further study of these specific examples for future work. 

In the next, and final, chapter, we present a brief summary of the work in this thesis, as well as outlining some directions of future work which build on these results.


\newpage

\chapter{Conclusions}

\section{Summary}

In this thesis, we have covered several aspects of cosmological inflation. 
We began by introducing the necessary background cosmology, including the FLRW Universe and the Hot Big Bang model of the Universe. 
We explained how an early period of accelerated expansion of the Universe can solve the classical problems of this Big Bang model, and introduced the slow-roll approximations of inflation, which are both simplifying of the dynamics and physically motivated. 
We then showed how large curvature fluctuations during inflation can later collapse into dense objects, such as primordial black holes, once inflation has ended. 
This is interesting because we saw that primordial black holes might solve several open problems that persist in modern cosmology, including the nature of dark matter and the origin of supermassive black holes at high ($z\sim 7$) redshifts. 
We also showed that in situations where a significant number of primordial black holes can be produced, the usual slow-roll approximations break down, and we may enter a phase so ultra-slow-roll inflation. 

In Chapter \ref{chapter:USRstability} we performed a stability analysis of ultra-slow-roll inflation, since it is often stated in the literature that this regime is always unstable and short-lived.
We performed this analysis in terms of the dimensionless field acceleration parameter, $f=1-\delta$ in \Eq{eq:def:f}, that quantifies the acceleration of the inflaton field relative to the Hubble friction term in the Klein-Gordon equation. 
We saw that ultra-slow-roll inflation is stable ($\dd |\delta|/ \dd t<0$) for a scalar field rolling down a convex potential ($\dot{V}<0$ and $V_{,\phi\phi}>0$) if the condition 
\bea
\Mp \frac{\dd^2 V}{\dd\phi^2}>\left\vert \frac{\dd V}{\dd \phi} \right\vert\, 
\eea
is satisfied. 

We then compared our analytical results against numerical examples. 
For the Starobinsky model \eqref{eq:def:pot:strobinsky} where the potential is made of two linear pieces, the condition~\eqref{eq:USR:stable:criterion:conclusion} is never fulfilled, since $V_{,\phi\phi}=0$, and ultra-slow-roll inflation is never stable. 
It lasts for a number of \efolds~of order one. 
We also analysed the case of a potential (\ref{eq:pot:inflectionPoint:cubic}) with a flat inflection point at $\phi=0$, $V\propto 1+(\phi/\phi_0)^3$, where we found that ultra-slow roll is stable in the range $0<\phi<\Mp$, \ie when approaching the inflection point. 
When $\phi_0\ll\Mp$ the ultra-slow-roll regime is always short lived, but when $\phi_0\gg \Mp$ it can last for a large number of \efolds. 
In fact, if one considers for instance a potential of the form $V\propto 1+\alpha\ee^{\beta\phi/\Mp}$, with $\alpha\beta\ll 1$ and $\beta\gg 1$, there is even an infinite phase of classical ultra-slow roll for $\phi<0$.

However, this was a purely classical analysis and we expect a very flat potential to require a complete stochastic treatment to be fully understood. 
This stochastic formalism was introduced in Chapter \ref{chapter:stochastic:intro}, and we discussed the requirements for this formalism to be valid. 
These requirements are well-known to be satisfied in slow-roll inflation, but had not previously been explored beyond the slow-roll regime. 
This motivated us to explicitly check these conditions in Chapter \ref{chapter:stochastic:intro}. 

One of the key assumptions made by the stochastic formalism for non-test fields is the separate universe approach, which pictures the universe on super-Hubble scales as causally disconnected regions that evolve under local FLRW dynamics.
While the separate universe approach is known to be valid within slow roll, its status once slow roll is violated was not so clear.
By showing that the dynamics of super-Hubble fluctuations can be recovered by perturbing the background FLRW equations of motion, we demonstrated that this approach is in fact valid and does not require slow roll. 

Subtleties also arise regarding the gauge in which the stochastic equations are written. 
The time variable that is usually used in the Langevin equation is the logarithmic expansion, \ie the number of \efolds~$N$, and this variable is left unperturbed, which means we are implicitly working in a gauge in which the expansion is uniform.
However, the field fluctuations, which determine the correlations of the noises, are usually quantised in the spatially-flat gauge, where they coincide with the gauge-invariant Sasaki--Mukhanov scalar field perturbations, $Q$. 
We therefore have to perform a gauge transformation from the spatially-flat to the uniform-$N$ gauge before evaluating the stochastic noise due to quantum field fluctuations. 
We calculated this transformation and showed that it is proportional to the non-adiabatic pressure perturbation. 
Since this vanishes on large scales in the presence of a dynamical attractor, such as in slow roll, the gauge transformation becomes trivial in that case (\ie the two gauges coincide on super-Hubble scales). 
We also examined the case of ultra-slow roll, where we found that the gauge transformation is also trivial, despite the fact that ultra-slow roll in this case is not a dynamical attractor in phase space. 
Finally, we studied the (linear piece-wise) Starobinsky model, where the dynamics interpolates between a phase of ultra-slow-roll and slow-roll inflation, and found that the same conclusions apply.
Thus, in all three cases, the gauge transformation that is required prior to evaluating the noise correlators in stochastic inflation turns out to be trivial on super-Hubble scales, and stochastic inflation as usually formulated can be applied without further refinements.

In order to consider the formation of primordial black holes, we require the full probability distribution of curvature perturbations, and this can be found using the stochastic-$\delta N$ formalism. 
We described how the first few moments of this distribution can be calculated in an iterative manner, but further work was required to calculate the full distribution, including the tail, accurately. 
In Chapter \ref{chapter:quantumdiff:slowroll}, we introduced a characteristic function formalism, which is built upon the stochastic framework, and used this to calculate the full PDF of curvature perturbations in slow-roll inflation. 
We considered both the classical limit and the stochastic limit and derived expressions for the PDF in both cases, as well as defining a classicality criterion to determine which limit is appropriate for each region of a potential. 

We found that the tail of the distribution decays exponentially \cite{Ezquiaga:2019ftu} and not as a Gaussian, and this decay is fully characterised by a single parameter, given by $\mu^2 = 
(\Delta\phiwell)^2/(v_0\Mp^2)$, where we define the quantum well to be the part of the potential that is dominated by stochastic effects. 
The parameter $\mu^2$ also determines the mean number of \efolds~realised across the quantum well, the amplitude of the power spectrum, and the PBH mass fraction. 
We then found that observational constraints on the abundance of PBHs put an upper bound on $\mu^2$ that is of order one, which in turn constrains the inflaton to spend no more than $\sim 1$ \efold~in the quantum well. 

Finally, in Chapter \ref{chapter:USRstochastic}, we extended our methods to consistently use stochastic inflation in a USR regime to derive full PDFs for curvature fluctuations.
We considered the two separate limits, one dominated by a classical drift, and one dominated by quantum diffusion. 
In the latter case, we saw that exponential decay of the PDFs (which we also saw in slow roll) is present in USR, and this seems to be a universal property that exists when stochastic effects are correctly accounted for. 
The study of whether the classical or stochastic limit is the physically relevant one is left for future work, and in the next section we outline other directions of research that emerge as a result of the work we have presented in this thesis. 

\section{Future work}

Here we discuss some new and interesting research directions that open up as a consequence of the results obtained in this thesis:

First, in slow roll, we have shown that the effects of quantum diffusion on the PDF of curvature perturbations and on the mass fraction of PBHs can be dramatic in regions of the potential where the classical approximation breaks down. 
This implies that some of the constraints on inflationary models derived in the literature, from non-observations of PBHs and using only the classical approximation, may have to be revised. 
This could have important consequences for these models.


Next, it has been shown~\cite{Assadullahi:2016gkk, Vennin:2016wnk} that in the presence of multiple fields, the effects of quantum diffusion can be even more drastic. 
Formation of PBHs in multi-field models of inflation, such as hybrid inflation~\cite{GarciaBellido:1996qt, Bugaev:2011qt, Bugaev:2011wy, Clesse:2015wea, Kawasaki:2015ppx}, would therefore be interesting to study with the stochastic techniques developed in this thesis.

There are also other astrophysical objects for which the knowledge of the full probability distribution of cosmological perturbations produced during inflation is important, such as ultra-compact mini-halos~\cite{Shandera:2012ke}. 
Using our results to calculate the abundance of such objects is another interesting prospect.

Warm inflation \cite{Berera:1995wh, Berera:1995ie, Berera:2008ar} is a phenomenological model of the early universe that describes the decay of the inflaton into the standard model particles during inflation, rather than during a separate reheating phase after inflation. 
The decay of the inflaton is described as a thermal dissipation, which can be studied using standard stochastic techniques (see, for example, Ref. \cite{Bellini:1999ug}).
It would therefore be interesting to apply our new stochastic methods to study quantum diffusion in warm inflation, both in slow roll and USR, and also to consider the effects of the thermal dissipation on the production of curvature perturbations, and hence on the formation of PBHs. 

One can also incorporate stochastic effects in collapsing models of the universe (see, eg, \cite{Miranda:2019ara}), and the conditions we explained in Chapter \ref{chapter:stochastic:intro} also need to be tested in such a scenario. 
This has not been explored before and it would therefore also be interesting to explicitly test the validity of the stochastic formalism in collapsing models. 

Finally, as discussed in Chapter \ref{chapter:USRstochastic} and mentioned above, it is important to apply the USR stochastic formalism that we have developed to realistic models that lead to PBH production, such as \cite{Garcia-Bellido:2017mdw, Ezquiaga:2018gbw}.
This will be interesting to see the effects of stochastic USR inflation on the formation of PBHs, although we expect it to greatly increase the formation rate of these rare objects, similarly to the slow-roll case, and limit the length of time the inflaton can spend on a flat potential without violating observational constraints. 

We can see there are still many exciting areas to explore regarding perturbations from inflation, the stochastic formalism, and primordial black holes. 
We have made substantial progress understanding curvature fluctuations using the non-perturbative techniques of the stochastic-$\delta N$ formalism, as well as extending our understanding of stochastic inflation and the formation of primordial black holes when quantum effects are fully accounted for.
We have opened up many exciting areas to continue to research in the future.




%
\phantomsection
\addcontentsline{toc}{chapter}{Appendices}
\appendix 

\chapter{FLRW Christoffel symbols and Einstein tensor}
%
\label{appendix:flrwchristoffel}

Recall the the FLRW metric, given in \eqref{eq:FLRWmetric}, is 
\bea \label{eq:app:flrwmetric}
\dd s^2 = g_{\mu \nu}\dd x^{\mu}\dd x^{\nu} = -\dd t^2 + a^2(t)\left[ \frac{\dd r^2}{1-Kr^2} + r^2\dd\theta^2 + r^2\sin^2\theta\dd\phi^2 \right]  \, .
\eea
In order to define the Ricci tensor of this metric we must first define the Christoffel symbols to be 
\bea \label{eq:def:christoffel}
\Gamma^\lambda_{\mu\nu} = \frac{1}{2}g^{\lambda \sigma}\left( \partial_\nu g_{\sigma\mu} + \partial_\mu g_{\sigma\nu} - \partial_\sigma g_{\mu\nu}  \right) \, ,
\eea 
which are symmetric in their lower indices, so $\Gamma^\lambda_{\mu\nu} = \Gamma^\lambda_{\nu\mu}$.
Plugging in the metric components from \eqref{eq:app:flrwmetric} we find the non-vanishing Christoffel symbols for the FLRW metric to be
\bea
	&\Gamma _{11}^{0}=\frac{a\dot{a}}{1-K{{r}^{2}}};\quad
	 \Gamma _{22}^{0}= a\dot{a}{r^2};\quad
	\Gamma _{33}^{0}=a\dot{a}{r^2}\sin^2\theta\\
	&\Gamma_{10}^1=\Gamma _{01}^1 =\frac{\dot a}{a};
	\quad
	\Gamma _{11}^1 =\frac{Kr}{1-Kr^2};\quad
	\Gamma _{22}^1=-r(1-Kr^2);\quad
	\Gamma _{33}^1=-r(1-Kr^2)\sin^2\theta;\\
	&\Gamma^{2}_{20}=\Gamma^{2}_{02}
	=  \frac{\dot{a}}{a};\quad
	\Gamma^{2}_{21}=\Gamma^{2}_{12}=\frac 1 r ;\quad
	\Gamma^{2}_{33}=-\sin\theta\cos\theta;\\
	&\Gamma^{3}_{30}=\Gamma^{3}_{03}
	=\frac{\dot{a}}{a};\quad
	\Gamma^{3}_{31}=\Gamma^{3}_{13}=\frac 1 r ;\quad
	\Gamma^{3}_{32}=\Gamma^{3}_{23}
	=\cot\theta \, ,
\eea
where the indices denote the coordinates through $(0,1,2,3) = (t,r,\theta,\phi)$.
The Ricci tensor is then defined in terms of Christoffel symbols as
\bea \label{eq:def:riccitensor}
R_{\mu \nu } = \partial_{\lambda}\Gamma _{\mu \nu}^\lambda
	- \partial_{\nu}\Gamma _{\mu \lambda }^\lambda
	+\Gamma_{\mu \nu }^{\rho}
		\Gamma _{\rho \lambda }^\lambda
	- \Gamma _{\mu \lambda }^\rho
		\Gamma _{\rho \nu }^{\lambda} \, ,
\eea 
and in the FLRW metric the non-zero components can be computed to be 
\bea
  &R_{00}  =  - 3\frac{\ddot a}{a} \, , \\
  &R_{11}  = \frac{\left( {\ddot aa + 2\dot a^2  + 2k} \right)}{1 - Kr^2 }\, ,\\
 &R_{22}  = \left( \ddot aa + 2\dot a^2  + 2K \right)r^2 \, , \\
&R_{33}  = \left( \ddot aa + 2\dot a^2  + 2K \right)r^2 \sin ^2 \theta \, .
\eea
We can also compute the Ricci scalar $R = g^{\mu\nu}R_{\mu\nu} = R^\mu{}_{\nu}$ to be 
\bea 
R =  - 6\left( \frac{\ddot a}{a} + \left(\frac{\dot a}{a}\right)^2 + \frac{K}{a^2} \right) \, .
\eea 
Finally, let us explicitly give the non-zero components for the Einstein tensor $G_{\mu\nu} = R_{\mu\nu} - (R/2)g_{\mu\nu}$, which are found to be 
\bea 
&G_{00}  =  3 \left[ \left(\frac{\dot a}{a}\right)^2 + \frac{K}{a^2} \right] \, , \\
&G_{11}  = -\left[ \left(\frac{\dot a}{a}\right)^2 + \frac{2\ddot a}{a} + \frac{K}{a^2} \right] \frac{a^2}{1 - Kr^2 }\, ,\\
&G_{22}  = -\left[ \left(\frac{\dot a}{a}\right)^2 + \frac{2\ddot a}{a} + \frac{K}{a^2} \right] a^2r^2 \, , \\
&G_{33}  = -\left[ \left(\frac{\dot a}{a}\right)^2 + \frac{2\ddot a}{a} + \frac{K}{a^2} \right] a^2r^2 \sin ^2 \theta \, .
\eea 

\newpage

\chapter{Sasaki--Mukhanov equation} \label{appendix:MSequation}
In this appendix, we  derive the general expression \eqref{eq:z''overz:general} for $z''/z$ in the Sasaki--Mukhanov equation and discuss both the slow-roll and the ultra-slow-roll limits.

\subsection*{Deriving the general expression}

We start with the Sasaki--Mukhanov variable
\bea \label{eq:def:v}
v_{\bm{k}} = z\zeta_{\bm{k}} \, ,
\eea 
where 
\bea 
z = a \sqrt{2 \epsilon_{1}}\Mp \, ,
\eea 
and $v_k$ obeys the Sasaki--Mukhanov equation
\bea 
v_{\bm{k}}'' + \left( k^2 - \frac{z''}{z} \right) v_{\bm{k}} = 0 \, .
\eea 
Note that in the spatially flat gauge, we have $v_k = a \delta \phi_k =aQ_k $.
Let us also reiterate the notation we use here: $\dot{ } = \frac{\dd}{\dd t}$ ($t$ is proper time) and $' = \frac{\dd}{\dd \eta}$ ($\eta$ is conformal time), so $\frac{\dd}{\dd \eta} = a \frac{\dd}{\dd t}$.
Combining
\bea 
\epsilon_{1} = -\frac{\dot{H}}{H^2} = 1 - \frac{\mathcal{H}'}{\mathcal{H}^2}
\eea 
with the Friedmann equation~(\ref{eq:friedmann}), one obtains $\epsilon_{1} =  \dot{\phi}^2/(2\Mp^2H^2)$, and $z$ can be rewritten as 
\bea 
z  = \frac{a \dot{\phi}}{H} \, .
\eea 
This allows us to calculate 
\bea \label{eq:z'/z}
\frac{z'}{z} = \frac{a}{z}\dot{z} = a\left( \frac{\dot{a}}{a} + \frac{\ddot{\phi}}{\dot{\phi}} - \frac{\dot{H}}{H} \right) \, .
\eea 
In terms of the slow-roll parameter $\epsilon_{1}$ and the relative acceleration parameter $f$ defined in \Eq{eq:def:f}, this reads 
\bea \label{eq:z'/z:exact}
\frac{z'}{z}= aH\left(1 - 3f + \epsilon_{1} \right) \, .
\eea 
In order to calculate $z''/z$, notice that
\bea \label{eq:intermediate}
\left(\frac{z'}{z}\right)' = \frac{z''}{z} - \frac{z'^2}{z^2} \, ,
\eea 
where the left-hand side can be derived from \Eq{eq:z'/z:exact},
\bea \label{eq:intermediate2}
\left(\frac{z'}{z}\right)' = a\frac{\dd}{\dd t}\left( \frac{z'}{z} \right) = a^2H\left(1 - 3f + \epsilon_{1} \right) \left( \frac{\dot{a}}{a} + \frac{\dot{H}}{H} + \frac{ \dot{\epsilon_{1}}-3\dot{f}}{1 - 3f + \epsilon_{1}} \right) \, .
\eea 
Derivating $\epsilon_{1}  = \dot{\phi}^2/(2\Mp^2H^2)$ with respect to time, and making use of the Klein--Gordon equation~(\ref{eq:eom:scalarfield}), we have
\bea \label{eq:epsilondot}
\dot{\epsilon_{1}} = 2H\epsilon_{1} \left( \epsilon_{1} - 3f \right) \, ,
\eea 
and from \Eq{eq:def:f} we can calculate
\bea 
\frac{\dot{f}}{H} &= \mu + \left(f - 1\right)\left( \epsilon_{1} + 3f \right) \, .
\eea 
where the dimensionless mass parameter $\mu$ is defined in \Eq{mu:def}. Thus we can evaluate \Eq{eq:intermediate2} as
\bea 
\left(\frac{z'}{z}\right)' &= a^2H^2 \left( 1+6f+3\epsilon_{1} - 3\mu + \epsilon_{1}^2 - 9f^2 - 6f\epsilon_{1} \right) \, .
\eea 
Combining this with \Eqs{eq:z'/z:exact} and \eqref{eq:intermediate}, we find
\bea \label{eq:z''}
\frac{z''}{z} &= a^2H^2 \left( 2 + 5\epsilon_{1} - 3\mu - 12f\epsilon_{1} + 2\epsilon_{1}^2 \right)\, ,
\eea 
which is an exact expression that makes no approximations. 
\subsection*{Slow-roll limit}
\label{sec:z''/z:SR}
In the slow-roll regime,
\begin{align}
\label{eq:eps1:V}
\epsilon_{1}&\simeq \epsilon_{1}^V \equiv \frac{\Mp^2}{2} \left( \frac{V_{,\phi}}{V} \right)^2 \, , \\
 \label{eq:eps2:V}
\epsilon_{2}&\simeq \epsilon_{2}^V \equiv 2\Mp^2\left[ \left( \frac{V_{,\phi}}{V} \right)^2 - \frac{V_{,\phi\phi}}{V} \right] \, , \\
H^2 &\simeq \frac{V}{3\Mp^2} \, ,
\end{align}
and hence
\begin{align}
\mu &\simeq \Mp^2 \frac{V_{,\phi\phi}}{V} \\
\epsilon_{2} &\simeq 4\epsilon_{1} - 2 \mu \, .
\end{align}
At leading order in the slow-roll parameters, we therefore see that \eqref{eq:z''} reduces to 
\bea 
\frac{z''}{z} &\simeq a^2H^2 \left[ 2 - \epsilon_{1} + \frac{3}{2}\epsilon_{2}  +\order{\epsilon^2}\right] \, , 
\eea 
where we note that, since $f = \frac{2\epsilon_1-\epsilon_2}{6} $ \cite{Pattison:2018bct}, terms of order $\mathcal{O}(f\epsilon_{1})$ are neglected at first order.
In order to write \eqref{eq:z''} as an explicit function of conformal time, note that 
\bea \label{eq:eta:integral}
\eta &= \int \frac{\dd t}{a} = \int \frac{\dd a}{a^2H(a)} = -\frac{1}{aH} + \int \frac{\epsilon_{1} \dd a}{a^2H} \, ,
\eea 
where we have integrated by parts to get the last equality.
From \Eq{eq:epsilondot}, we have
\bea 
\frac{\dot{\epsilon_{1}}}{\epsilon_{1}} &= 2H \left( \epsilon_{1} - 3f \right) \, ,
\eea 
and thus we can again integrate by parts to find 
\bea 
\int \frac{\epsilon_{1} \dd a}{a^2H} &= -\frac{\epsilon_{1}}{aH} + \int \frac{\dd a}{a^2H}\frac{\dot{\epsilon_{1}}}{\epsilon_{1}}\frac{\epsilon_{1}}{H} + \mathcal{O}\left( \epsilon_{1}^2 \right)
 = -\frac{\epsilon_{1}}{aH} + \mathcal{O}\left(\epsilon_{1}^2, f\epsilon_{1} \right) \, .
\eea 
Therefore, from \Eq{eq:eta:integral},
\bea \label{eq:tau:slowroll}
\eta \simeq -\frac{1}{aH}\left( 1 + \epsilon_{1} \right)
\eea 
at first order in slow roll, and \Eq{eq:z''} becomes
\bea \label{eq:z'':slowrolllimit}
\frac{z''}{z} &\simeq \frac{2}{\eta^2} \left( 1 + \frac{3}{2}\epsilon_{1} + \frac{3}{4}\epsilon_{2} \right)\, ,
\eea 
as announced in the main text.

\subsection*{Near ultra-slow-roll limit}
In the ultra-slow-roll regime, the field acceleration parameter is close to one and it is convenient to parameterise
\bea 
f = 1 - \delta \, ,
\eea 
where $\vert \delta \vert \ll 1$. In terms of $\delta$, \Eq{eq:z''} becomes
\bea 
\frac{z''}{z} &= a^2H^2 \left( 2 - 7\epsilon_{1} + 2 \epsilon_{1}^2 + 12\delta\epsilon_{1} - 3\mu \right) \, .
\eea 
In order to derive an explicit expression in terms of the conformal $\eta$ using \Eq{eq:eta:integral}, we note from \Eq{eq:epsilondot} that we have
\bea 
\frac{\dot{\epsilon_{1}}}{\epsilon_{1}} &= 2H \left( \epsilon_{1} - 3f \right)
 = -6H \left( 1 - \delta - \frac{\epsilon_{1}}{3} \right) \, .
\eea 
This allows us to again integrate by parts in \Eq{eq:eta:integral} and find
\bea \label{eq:integratebyparts}
\int \frac{\epsilon_{1} \dd a}{a^2H} &= -\frac{1}{7}\frac{\epsilon_{1}}{aH} + \mathcal{O}\left( \epsilon_{1}\delta, \epsilon_{1}^2 \right) \, ,
\eea 
and hence
\bea \label{eq:tau:USR}
\eta = -\frac{1}{aH}\left( 1 + \frac{1}{7}\epsilon_{1} \right)  + \mathcal{O}\left( \epsilon_{1}\delta, \epsilon_{1}^2 \right) \, .
\eea 
Thus, in ultra-slow roll, \Eq{eq:z''} becomes
\bea 
\label{eq:z''/z:usr}
\frac{z''}{z} & = \frac{1}{\eta^2} \left[ 2 - 3 \mu - \frac{3}{7}\epsilon_{1}\left( 15 + 2\mu \right) \right] + \mathcal{O}\left( \epsilon_{1}\delta, \epsilon_{1}^2 \right) \, .
\eea 
If the effective mass parameter $\mu$ is small, the leading-order behaviour is the same as in conventional slow roll, but it differs if $\mu$ is of order one or larger.

\chapter{First slow-roll correction in ultra-slow roll}
\label{app:usr:epscorrections}
In this section, we solve \Eq{eq:MSequn} perturbatively in $\epsilon_{1*}$, in order to check that the leading-order solutions given in \Sec{sec:USR} are indeed consistent. We still consider the case of an exactly flat potential, so that $\delta=\mu=0$, and \Eq{eq:z''/z:usr} reads
\bea 
\frac{z''}{z} = \frac{1}{\eta^2}\left( 2 - \frac{45}{7}\epsilon_{1}\right) = \frac{1}{\eta^2}\left[  2 - \frac{45}{7}\epsilon_{1*}\left(\frac{\eta}{\eta_{*}}\right)^6\right] \, .
\eea
At first order in $\epsilon_{1*}$, the comoving Hubble parameter is given by \Eq{eq:tau:USR}, namely
\bea \label{eq:mathcalH:USR}
\mathcal{H} = -\frac{1}{\eta}\left[ 1 + \frac{1}{7}\epsilon_{1} + \mathcal{O}(\epsilon_{1}^2) \right] \, .
\eea 
Unlike the slow-roll case, this cannot be integrated to find $a(\eta)$, but we can instead perform an expansion in powers of $\epsilon_{1*}$ to find
\bea \label{eq:a:USR}
a(\eta) = -\frac{1}{H_* \eta}\left[ 1 - \frac{1}{42}\left(\epsilon_{1}-\epsilon_{1*}\right) + \mathcal{O}(\epsilon_{1}^2) \right] \, .
\eea

In order to solve \Eq{eq:MSequn} perturbatively, we first note that the leading order solution is simply the $\nu=\frac{3}{2}$ solution already given in \Eq{eq:v:nu=3/2}.
To find the first correction to this, we introduce 
\bea \label{eq:v:firstordereps}
v_{\bm{k}} = \frac{\ee^{-ik\eta}}{\sqrt{2k}}\left(1 - \frac{i}{k\eta}\right)\left[ 1+\epsilon_{1*}f_{\bm{k}}(\eta)\right] \, ,
\eea
for some function $f_{\bm{k}}(\eta)$.
If we substitute \Eq{eq:v:firstordereps} back into \Eq{eq:MSequn} and solve the resultant differential equation for $f(\eta)$, we find that 
\bea \label{eq:f':usr}
f'_{\bm{k}}(\eta) &=  -\frac{45}{28}\frac{1}{k^3}\frac{\eta^2}{\eta_{*}^6}\frac{\ee^{2ik\eta}}{(k\eta-i)^2}\left[ \ee^{-2ik\eta}\left(7i - 14k\eta - 14ik^2\eta^2+8k^3\eta^3 + 2ik^4\eta^4\right) \right. \\
& \hspace{2cm} \left. - \ee^{-2ik\eta_{\mathrm{start}}}\left(7i - 14k\eta_{\mathrm{start}} - 14ik^2\eta_{\mathrm{start}}^2+8k^3\eta_{\mathrm{start}}^3 + 2ik^4\eta_{\mathrm{start}}^4\right) \right] \, ,
\eea 
where the integration constant $\eta_{\mathrm{start}}$ must be chosen such that
\bea 
\epsilon_{1\mathrm{start}} = \epsilon_{1*}\left(\frac{\eta_{\mathrm{start}}}{\eta_{*}}\right)^6 < 1 \, .
\eea 
Combining \Eqs{eq:a:USR} and~(\ref{eq:v:firstordereps}) at leading order in $\epsilon_{1*}$, recalling that $v_{\bm{k}} = a Q_{\bm{k}}$, we can then calculate
\bea 
\frac{Q'_{\bm{k}}}{Q_{\bm{k}}} &=  - \frac{ik\eta}{\eta - \frac{i}{k}} + \epsilon_{1*}\left[ \frac{1}{7\eta}\left(\frac{\eta}{\eta_*}\right)^6 - f'_{\bm{k}}\right] \, ,
\eea
where $f'_{\bm{k}}$ is given by \Eq{eq:f':usr}.
Note that at leading order in $\epsilon_{1*}$, this reduces to \Eq{eq:Q'overQ:USR:nu=3over2}, as expected. We also see that the source function \eqref{eq:sourcefunction:general} becomes
\bea \label{eq:S:USReps1correction}
S_{\bm{k}} &= - \sqrt{\frac{\epsilon_1}{2}}\frac{Q_{\bm{k}}}{\Mp}\left\lbrace \frac{3}{\eta} - \frac{ik\eta}{\eta-\frac{i}{k}} - \epsilon_{1*}\left[ \frac{3}{7\eta}\left(\frac{\eta}{\eta_{*}}\right)^6  - f'_k\right] + \mathcal{O}(\epsilon_{1*}^2) \right\rbrace \, .
\eea
This implies that
\bea 
Q_{\bm{k}} & = -\frac{H_*}{\sqrt{2k}}\ee^{-ik\eta}\left[ \eta - \frac{i}{k} + \mathcal{O}(\epsilon_{1*})\right] \\
S_{\bm{k}} &= -\sqrt{\frac{\epsilon_{1*}}{2}}\frac{Q}{\Mp}\left(\frac{\eta}{\eta_*}\right)^3\left[ \frac{3}{\eta} - \frac{ik\eta}{\eta-\frac{i}{k}} + \mathcal{O}(\epsilon_{1*}) \right] \, .
\eea 
Making use of \Eq{eq:alphaintegral1:general}, we thus find 
\bea 
\alpha_{\bm{k}} &
\simeq 
 -\frac{iH_{*}\sqrt{\epsilon_{1*}}}{6\Mp}{k^{-\frac{5}{2}}}(k\eta)^4
 \left[ 1 + \mathcal{O}(\epsilon_{1*})\right]   \, ,
\eea
of which \Eq{eq:alpha:USR:nu=3/2} indeed captures the leading order. The situation is therefore different than in slow roll where the leading order result vanishes and the dominant contribution comes from the decaying mode. This is because, as stressed above, the presence of a dynamical attractor in slow roll makes the non-adiabatic pressure perturbation vanish, which is not the case in ultra-slow roll.

\chapter{Elliptic theta functions}
\label{appendix:identities}
In \Sec{sec:StochasticLimit}, the PDF of coarse-grained curvature perturbations is expressed in terms of elliptic theta functions. In this appendix, we define these special functions and give some of their properties that are relevant for the considerations of \Sec{sec:StochasticLimit}. There are four elliptic theta functions, defined as~\cite{Olver:2010:NHM:1830479:theta, Abramovitz:1970aa:theta}
\bea
\vartheta_1\left(z,q\right) &= 2 \sum_{n=0}^\infty (-1)^n q^{\left(n+\frac{1}{2}\right)^2}\sin\left[\left(2n+1\right)z\right]\, ,\\
\vartheta_2\left(z,q\right) &= 2 \sum_{n=0}^\infty  q^{\left(n+\frac{1}{2}\right)^2}\cos\left[\left(2n+1\right)z\right]\, ,\\
\vartheta_3\left(z,q\right) &= 1+2 \sum_{n=1}^\infty q^{n^2}\cos\left(2nz\right)\, ,\\
\vartheta_4\left(z,q\right) &= 1+2 \sum_{n=1}^\infty (-1)^nq^{n^2}\cos\left(2nz\right)\, .
\eea
By convention, $\vartheta_i^\prime$ denotes the derivative of $\vartheta_i$ with respect to its first argument $z$. For instance, one has
\bea
\label{eq:theta1prime:def}
\vartheta_1^\prime\left(z,q\right) = 2 \sum_{n=0}^\infty (-1)^n q^{\left(n+\frac{1}{2}\right)^2}\left(2n+1\right)\cos\left[\left(2n+1\right)z\right]\, ,
\eea
which appears in \Eq{eq:PDF:phiwall:thetaElliptic}. As another example, one has
\bea
\label{eq:theta4primeprime:def}
\vartheta_4^{\prime\prime}\left(z,q\right) = -8\sum_{n=1}^\infty (-1)^nq^{n^2}n^2\cos\left(2nz\right)\, ,
\eea
which is used in \Eq{eq:PDF:stocha:phiend:appr}. As a third example, one has
\bea
\label{eq:theta2prime:def}
\vartheta_2^{\prime}\left(z,q\right) = -2 \sum_{n=0}^\infty  q^{\left(n+\frac{1}{2}\right)^2}\left(2n+1\right)\sin\left[\left(2n+1\right)z\right]\, ,
\eea
which appears in \Eq{eq:PDF:thetatwo}.  As a last example, one has
\bea
\label{eq:theta2primeprime:def}
\vartheta_2^{\prime\prime}\left(z,q\right) = -2 \sum_{n=0}^\infty  q^{\left(n+\frac{1}{2}\right)^2}\left(2n+1\right)^2\cos\left[\left(2n+1\right)z\right]\, ,
\eea
which is used in \Eq{eq:PDF:stocha:phiend:appr:2}. The function $\vartheta_i(z,q)$ is noted \texttt{EllipticTheta[i,z,q]} in Mathematica and $\vartheta_i^\prime(z,q)$ is noted \texttt{EllipticThetaPrime[i,z,q]}.

Let us now show that the different expressions for $P(\N,\phi=\phiend+\Delta\phiwell)$ obtained in \Sec{sec:StochasticLimit} in the stochastic dominated regime are equivalent. A first expression is given by \Eq{eq:PDF:phiwall:thetaElliptic}, a second expression can be derived by plugging $x=1$ into \Eq{eq:stocha:HeatMethod:PDF:expansion} and making use of \Eq{eq:theta1prime:def}, and a third expression is given by plugging $x=1$ in \Eq{eq:PDF:thetatwo}. The three formulae are equivalent if
\bea
\label{eq:theta:identity}
\left(\frac{\mu}{\sqrt{\pi\N}}\right)^3 \vartheta^\prime_1\left(0,\ee^{-\frac{\mu^2}{\N}}\right) = 
\vartheta_{1}'\left( 0, \ee^{ -\frac{\pi^2}{\mu^2} \N} \right)  = 
-\vartheta_{2}'\left( \frac{\pi}{2}, \ee^{ -\frac{\pi^2}{\mu^2} \N } \right) \, .
\eea
The first equality in \Eq{eq:theta:identity} can be shown from the Jacobi identity for a modular transformation of the first elliptic theta function, see Eq.~(20.7.30) of Ref. \cite{NIST:DLMF},
\bea
\label{app:identity:modulartrans}
\left( -i \tau \right)^{\frac{1}{2}} \vartheta_{1} \left( z, \ee^{i\pi \tau} \right) = -i \ee^{ -\frac{z^2}{\pi \tau}} \vartheta_{1} \left( -\frac{z}{\tau}, -\ee^{-\frac{i\pi}{\tau}} \right) \, .
\eea
By taking $\tau = i/(a \pi)$ and differentiating \Eq{app:identity:modulartrans} with respect to $z$, one obtains
\bea
\label{app:id:diff}
\left( \pi a \right)^{\frac{1}{2}} \vartheta_{1}' \left( z, \ee^{-\frac{1}{a}} \right) = - \frac{2iz}{a}\ee^{az^2} \vartheta_{1} \left( - i \pi a z, \ee^{-\pi^2 a} \right) + a \pi \ee^{az^2} \vartheta_{1}' \left( - i \pi a z, \ee^{-\pi^2 a} \right) \, .
\eea
Taking $z = 0$, one recovers the first equality in \Eq{eq:theta:identity}. The second equality in \Eq{eq:theta:identity} simply follows from \Eqs{eq:theta1prime:def} and~(\ref{eq:theta2prime:def}).

\chapter{Detailed analysis of the model \texorpdfstring{$V\propto 1+\phi^p$}{}}
\label{appendix:cases}
In this appendix, we present a detailed analysis of the model discussed in \Sec{sec:example:1_plus_phi_to_the_p}, where the inflationary potential is of the form
\bea
\label{app:eq:def:potential}
v(\phi) = v_{0}\left[ 1 + \left( \frac{\phi}{\phi_{0}} \right)^{p} \right] \, .
\eea

In order to use the slow-roll approximation, one needs to check that the slow-roll conditions~\cite{liddle_lyth_2000}, $\Mp^{2}(v'/v)^2 \ll 1$, $\Mp^{2}|v''/v| \ll 1$, and $\Mp^{4}|v'''v'/v^{2}| \ll 1$, are satisfied. Here, we use the three first slow-roll conditions only, since these are the only ones currently constrained by observations. In order to satisfy the third condition, one requires a condition involving the position of $\phi$ with respect to
\bea
\label{constraint3}
\phi_{\text{sr1}} \equiv {\phi_{0}} \left( \frac{\phi_{0}}{\Mp}\right) ^{\frac{2}{p - 2}} \, ,
\eea
\ie $\phi \ll \phi_{\text{sr1}}$ if $p > 2$ and $\phi \gg \phi_{\text{sr1}}$ if $p<2$. The second slow-roll condition reduces to an equivalent condition. If $p=1$, there is no such condition and if $p=2$, it reduces to $\phi_0\gg \Mp$. The first condition constrains the position of $\phi$ with respect to
\bea
\label{constraint1}
\phi_{\text{sr2}} \equiv  \phi_0 \left(\frac{\phi_{0}}{\Mp}\right)^{\frac{1}{p - 1}} \, ,
\eea
namely $\phi\ll \phi_{\text{sr2}} $ if $p>1$ and $\phi\gg \phi_{\text{sr2}} $ if $p<1$ (if $p=1$, it reduces to $\phi_0\gg \Mp$).

As explained in \Sec{sec:example:1_plus_phi_to_the_p}, quantum diffusion plays an important role when $\eta_{\mathrm{class}}\gg 1$, where $\eta_{\mathrm{class}}$ is given in \Eq{eq:eta_class}, which leads to $\phi \ll \Delta\phiwell$,  where
\bea
\label{app:example:Deltaphiwell}
\Delta\phiwell = \phi_0 v_0^{\frac{1}{p}}\, ,
\eea
see \Eq{eq:phiwell:expansEnd}. Hereafter, we work in the vacuum-dominated regime, in which $\phi \ll \phi_0$ and $v \simeq v_{0}$. Making use of \Eq{eq:stocha:constraint:mu}, this gives rise to
\bea
\label{eq:app:example:mu}
\mu^2 = \left(\frac{\phi_0}{\Mp}\right)^2 v_0^{\frac{2}{p}-1}\, .
\eea

In the classical regime, \ie when $\phi\gg \Delta\phiwell $, the power spectrum is given by \Eq{eq:classicalPower}, which gives rise to
\bea
\label{clpower}
\mathcal{P}_{\zeta} |_{\text{cl}} = \frac{2v_{0}}{p^2}\left(\frac{\phi_{0}}{\Mp}\right)^{2}\left(\frac{\phi_{0}}{\phi}\right)^{2p - 2} \, .
\eea
Thus we see that the classical power spectrum is larger than unity when $\phi < \phicl$ if $p>1$, and $\phi >\phicl$ if $p<1$, where
\bea 
\label{clpower>1}
\phicl = {\phi_{0}} \left[ \frac{2v_{0}}{p^2}\left(\frac{\phi_{0}}{\Mp}\right)^2 \right]^{\frac{1}{2p - 2}}\, .
\eea
The number of \efolds~realised between $\phi$ and $\phiend$ can also be calculated in the classical regime using \Eq{eq:classicalPower}, and one obtains
\bea 
\label{clefolds}
N_{\text{end}} - N \simeq \frac{\phi_{0}^2}{p(p - 2)\Mp^{2}}\left[ \left( \frac{\phi_{0}}{\phiend}\right)^{p - 2} - \left( \frac{\phi_{0}}{\phi}\right)^{p - 2}\right]\, .
\eea
Note that this expression is singular for $p=2$, and this case is treated separately in \Sec{app:V_eq_1_plus_phi_to_the_p:p_eq_2}. Combining \Eqs{clpower>1} and~(\ref{clefolds}), one can rewrite the classical power spectrum as
\bea
\label{clPzeta}
\mathcal{P}_{\zeta} |_{\text{cl}} = \frac{2v_{0}}{p^2}\left(\frac{\phi_{0}}{\Mp}\right)^{2}\left[ \frac{p(2 - p)\Mp^{2}}{\phi_{0}^2} (N_{\text{end}} - N) + \left( \frac{\phi_{0}}{\phiend}\right)^{p - 2} \right]^{\frac{2p - 2}{p - 2}} \, .
\eea

In the stochastic regime,  \ie when $\phi\ll \Delta\phiwell $, the mean number of \efolds~can be computed from \Eq{eq:fn:generalsolution} for $f_1$. In the limit $\phiuv \rightarrow \infty$, and taking $\phi \ll \Delta\phiwell$, one obtains
\bea 
\label{stochasticefolds}
\left< \N \right> = \Gamma \left( \frac{1}{p} \right) \frac{\phi_{0}}{p \Mp^2v_{0}^{1 - \frac{1}{p}}} \left( \phi - \phiend \right) \, .
\eea
Similarly, using \Eq{eq:fn:generalsolution} for $f_2$, and the formula $\calP_\zeta = f_2'/f_1'- 2f_1$ from \Eq{eq:Pzeta:stochaDeltaN}, the power spectrum is given by
\bea 
\label{stochasticpower}
\mathcal{P}_{\zeta} (\phi) = \frac{2}{p^2}\Gamma^{2}\left( \frac{1}{p} \right) v_{0}^{\frac{2}{p} - 1} \left(\frac{\phi_{0}}{\Mp}\right)^2 \, .
\eea
\section{Case \texorpdfstring{$p=2$}{}}
\label{app:V_eq_1_plus_phi_to_the_p:p_eq_2}
We first consider the case of $p = 2$, where the slow-roll conditions reduce to $\phi_0 \gg \Mp$. Let us also note that \Eq{clefolds} is  singular for $p=2$, and that it should be replaced by
\bea
N_\uend - N \simeq \frac{\phi_0^2}{2\Mp^2} \ln\left(\frac{\phi}{\phi_\uend}\right)\, .
\eea
As noted in \Sec{sec:example:1_plus_phi_to_the_p}, the constant value found for the power spectrum in the stochastic limit, \Eq{stochasticpower}, corresponds (up to an order one prefactor) to the classical power spectrum~(\ref{clPzeta}) evaluated at $\phisto$. Therefore, when $\phi$ decreases, the stochastic and the classical result coincide until $\phi$ becomes smaller than $\phisto$, where the stochastic power spectrum saturates to a constant value and the classical power spectrum continues to increase. Since the slow-roll condition implies that  $\phi_{0} \gg \Mp$, the power spectrum is always larger than one in this regime. This can be also seen from $\phicl < \phisto$, as can be checked explicitly from \Eqs{app:example:Deltaphiwell} and~(\ref{clpower>1}). 

Furthermore, the number of \efolds~(\ref{stochasticefolds}) spent in the stochastic regime is of order $\mean{\N} = \sqrt{\pi}/2(\phi_0/\Mp)^2$, which is larger than unity. It is interesting to note that both the amplitude of the power spectrum and the number of \efolds~during which the stochastic regime depend only on $\phi_0$. PBHs are therefore overproduced in this case.
\section{Case \texorpdfstring{$p > 2$}{}}
We now consider values of $p$ such that $p > 2$. In this case, the slow-roll condition that is the strictest depends on whether $\phi_0$ is sub-Planckian or super-Planckian. More precisely, since $2/({p-2}) > 1/({p-1})$ for $p>2$, if $\phi_{0} < \Mp$, then $\phi_{\text{sr1}} < \phi_{\text{sr2}}$ and the slow-roll condition is given by $\phi \ll \phi_{\text{sr1}}$, and if $\phi_{0} > \Mp$ then $\phi_{\text{sr2}} < \phi_{\text{sr1}}$ and it is given by $\phi \ll \phi_{\text{sr2}}$.
\subsection{Case \texorpdfstring{$p > 2$}{} and \texorpdfstring{$\phi_{0} > \Mp$}{}}
\label{app:example:p_gt_2_and_phi0_gt_Mp}
In this case, the slow-roll condition reads $\phi \ll \phi_{\text{sr2}}$. Making use of \Eqs{constraint1}, \eqref{app:example:Deltaphiwell} and \eqref{clpower>1}, one can show that
\bea
\phisto < \phicl < \phi_{\text{sr2}}\, .
\eea
As a consequence, when stochastic effects become important, the classical power spectrum is already larger than one and so quantum diffusion cannot ``rescue'' the model in this case. Stochastic effects do reduce the amount of power, but not soon enough to keep the amount of PBH below the observationally constrained level.
\subsection{Case \texorpdfstring{$p > 2$}{} and \texorpdfstring{$\phi_{0} < \Mp$}{}}
In this case, the slow-roll condition reads $\phi \ll \phi_{\text{sr1}}$. Two sub-cases need to be distinguished.
\paragraph{Case $p>2$ and $v_{0}^{\frac{p - 2}{2p}}<\phi_{0}/\Mp<1 $}
$ $\\
In this case, \Eqs{constraint3}, \eqref{app:example:Deltaphiwell} and \eqref{clpower>1} lead to
\bea
\phisto < \phicl < \phi_{\text{sr1}}\, .
\eea
The situation is therefore very similar to the case $p>2$ and $\phi_{0} > \Mp$, and quantum diffusion does not sufficiently suppress PBH production. 
\paragraph{Case $p>2$ and $\phi_{0}/\Mp < v_{0}^{\frac{p - 2}{2p}}$}
$ $\\
In this case, \Eqs{constraint3}, \eqref{app:example:Deltaphiwell} and \eqref{clpower>1} give a reversed hierarchy, namely
\bea
\phi_{\text{sr1}} < \phicl < \phisto \,.
\eea
In the region where the slow-roll approximation applies, $\phi\ll \phi_{\text{sr1}}$, the classical power spectrum is therefore always larger than one. However this region is dominated by stochastic effects since $\phi_{\text{sr1}} < \phisto$. In the stochastic regime, the expressions we have previously derived receive a contribution from $\phi > \phi_{\text{sr1}}$ since we are integrating beyond $\phisto$, and are therefore inconsistent in this case. Intuitively, one may think that the violation of slow roll induces the suppression of the noise amplitude in the Langevin equation~(\ref{eq:intro:Langevin}) so that $\Delta\phiwell$ should in fact be replaced with $\phi_{\text{sr1}}$. The situation is then similar to the one sketched in \Fig{fig:sketch2}, and from \Eqs{eq:def:mu} and~(\ref{constraint3}), one has
\bea
\mu^2 \sim \frac{\phi_{\text{sr1}}^2}{v_0\Mp^2} = \frac{1}{v_0}\left(\frac{\phi_0}{\Mp}\right)^{\frac{2p}{p-2}}\, .
\eea
Since the condition under which this case is defined is $\phi_{0}/\Mp < v_{0}^{\frac{p - 2}{2p}}$, one has $\mu^2<1$, and PBHs are not overproduced in this case.
\section{Case \texorpdfstring{$1 < p < 2$}{}}
In this case, slow roll is valid in the range
\bea
\phi_{\text{sr1}} \ll \phi \ll \phi_{\text{sr2}}\, .
\eea
Note that this condition implies that $\phi_{\text{sr1}} \ll \phi_{\text{sr2}}$, which is the case only if $\phi_0\gg \Mp$. One then has $\phi_{\text{sr2}}\gg \phi_0$, so this range extends beyond the vacuum-dominated regime, and the interval of interest is in fact
\bea
\label{eq:app:example:case_1_lt_p_lt_2:sr_cond}
\phi_{\text{sr1}} \ll \phi \ll \phi_0\, .
\eea
Two sub-cases need to be distinguished.

\subsection{Case \texorpdfstring{$1 < p < 2$}{} and \texorpdfstring{$\phi_0/\Mp < v_{0}^{\frac{p - 2}{2p}}$}{}}

In this case, \Eqs{constraint3}, \eqref{constraint1}, \eqref{app:example:Deltaphiwell} and \eqref{clpower>1} give rise to
\bea
\phicl < \phisto < \phi_{\text{sr1}} < \phi_0< \phi_{\text{sr2}}\, .
\eea
This means that in the region of interest given by \Eq{eq:app:example:case_1_lt_p_lt_2:sr_cond}, the classical approximation is valid and predicts that PBHs are not overproduced.

\subsection{Case \texorpdfstring{$1 < p < 2$}{} and \texorpdfstring{$\phi_0/\Mp > v_{0}^{\frac{p - 2}{2p}}$}{}}

In this case, \Eqs{constraint3}, \eqref{constraint1}, \eqref{app:example:Deltaphiwell} and \eqref{clpower>1} give rise to
\bea
 \phi_{\text{sr1}} < \phisto < \phicl < \phi_0< \phi_{\text{sr2}}\, .
\eea
The situation is therefore similar to the case $p > 2$ and $\phi_{0} > \Mp$ described in \Sec{app:example:p_gt_2_and_phi0_gt_Mp}, and quantum diffusion does not sufficiently suppress PBH production.
\section{Case \texorpdfstring{$0 < p < 1$}{}}
If one takes $p<1$, contrary to the previous cases, the condition for the classical power spectrum to be larger than one is $\phi > \phicl$, where $\phicl$ is given by \Eq{clpower>1}. Furthermore, in this case, the slow-roll conditions read $\phi \gg \phi_{\mathrm{sr1}}$ and $\phi \gg \phi_{\mathrm{sr2}}$, and which of these conditions is the strictest depends on whether $\phi_0$ is sub- or super-Planckian.

\subsection{Case \texorpdfstring{$0 < p < 1$}{} and \texorpdfstring{$\phi_0<\Mp$}{}}

In this case the slow-roll condition reads
\bea
\phi \gg \phi_{\mathrm{sr}1}\, .
\eea
Since $p<1$, $\phi_0<\Mp$ implies that $\phi_0/\Mp < v_0^{\frac{p-2}{2p}}$ and from \Eqs{constraint3}, \eqref{app:example:Deltaphiwell} and \eqref{clpower>1}, one has
\bea
\phisto < \phi_{\text{sr1}}  < \phicl   \, .
\eea
In this case, stochastic effects cannot play an important role in the slow-roll region of the potential, where the power spectrum is smaller than one provided $\phi< \phicl$.
\subsection{Case \texorpdfstring{$0 < p < 1$}{} and \texorpdfstring{$\phi_0>\Mp$}{}}
In this case the slow-roll condition reads
\bea
\phi \gg \phi_{\mathrm{sr}2}\, ,
\eea
and two sub-cases need to be distinguished.
\paragraph{Case $0 < p < 1$ and $\frac{\phi_{0}}{\Mp} > v_{0}^{\frac{p - 2}{2p}}$}
$ $\\
From \Eqs{constraint1}, \eqref{app:example:Deltaphiwell} and \eqref{clpower>1}, one has
\bea
\phicl < \phi_{\text{sr2}} < \phisto \, .
\eea
In the classical region of the potential, $\phi>\phisto$, the power spectrum is much larger than one and PBHs are overproduced. In the stochastic region of the potential, $\phi_{\text{sr2}} <\phi< \phisto$, assuming $\phi_{\text{sr2}} \ll \phisto$, $\mu^2$ is given by \Eq{eq:app:example:mu}, which is much larger than one for $\phi_0>\Mp$ and $p<2$. Therefore PBHs are also too abundant in this part of the potential, and this case is observationally excluded.
\paragraph{Case $0 < p < 1$ and $1<\frac{\phi_{0}}{\Mp} < v_{0}^{\frac{p - 2}{2p}}$}
$ $\\
From \Eqs{constraint1}, \eqref{app:example:Deltaphiwell} and \eqref{clpower>1}, one has
\bea
\phisto < \phi_{\text{sr2}} < \phicl \, .
\eea
This case is similar to the one where $0<p<1$ and $\phi_0<\Mp$, and stochastic effects do not play an important role in the slow-roll region of the potential. 
\section{Case \texorpdfstring{$p=1$}{}}
Finally, let us consider the case $p=1$. The slow-roll approximation is valid throughout the entire vacuum-dominated region of the potential provided
\bea
\phi_0\gg\Mp.
\eea
This case is however more subtle than the previous ones, since from \Eq{eq:eta_class}, one has 
\bea
\eta_{\mathrm{class}}\equiv 0\, .
\eea
This means that the first stochastic correction in the classical expansion presented in \Sec{sec:ClassicalLimit} vanishes. This does not imply that quantum diffusion never plays a role since higher-order terms can still spoil the classical result, but this suggests that our classicality criterion fails in this case to identify where stochastic effects become important.

This is why no clear conclusion can be drawn in this case. In practice, one should extend the classical expansion of \Sec{sec:ClassicalLimit} to next-to-next-to-next to leading order (NNNLO) at least to determine what the first stochastic correction is and under which condition the classical approximation holds, and investigate numerically the cases where it does not. We leave these considerations for future work.

\chapter{Deriving the Fokker--Planck equation} \label{appendix:FokkerPlanck:USR}

\section{Ultra-slow roll}

In order to find the Fokker--Planck equation related to the USR system outlined here, we use the general derivation that is explained in \cite{risken1989fpe, 2007cond.mat..1242G}.
For a general multi-variate system of Langevin equations given by 
\bea 
\frac{\dd y_i}{\dd t} = A_i(y_1,...,y_n, t) + \sum_k B_{ik}(y_1,...,y_n, t)\xi_k(t) \, ,
\eea 
where $\langle\xi_i(t)\rangle=0$ and $\langle\xi_i(t_1)\xi_j(t_2)\rangle=2D\delta_{ij}\delta(t_1-t_2)$, the general Fokker--Planck equation  for $P(y_1,...,y_n, t)$ is given by
\bea \label{eq:FP:general}
\frac{\partial }{\partial t}P &= -\sum_i\frac{\partial}{\partial y_i}\left\{ \left[ A_i + D\sum_{jk}B_{jk}\frac{\partial B_{ik}}{\partial y_j} \right] P \right\} \\
& \hspace{2cm} + D\sum_{ij}\frac{\partial^2}{\partial y_i\partial y_j}\left\{ \left[ \sum_k B_{ik}B_{jk} \right] P \right\} \, .
\eea 
For the system given by \eqref{eq:eom:phi:stochastic} and \eqref{eq:eom:v:stochastic}, we have 
\bea 
&A_\phi = \frac{v}{H} \, , &A_v = -3v \, , & B_{\phi\phi} = \frac{H}{2\pi} \, , & D=\frac{1}{2} \, ,
\eea 
all other coefficients are zero, and we take $t=N$. 
Hence, in the case of USR inflation, the general equation \eqref{eq:FP:general} becomes
\bea \label{app:eq:FP:USR}
\frac{\partial P}{\partial N} &= \left[ 3 - \frac{v}{H}\frac{\partial}{\partial\phi} + 3v\frac{\partial}{\partial v} + \frac{H^2}{8\pi^2}\frac{\partial^2}{\partial \phi^2}\right] P \\
&\equiv \mathcal{L}_{\mathrm{FP}} \cdot P \, ,
\eea 
where $\mathcal{L}_{\mathrm{FP}}$ is the Fokker--Planck operator and we have used the fact that $H$ is independent of $\phi$ from
\bea 
3\Mp^2 H^2 = V_0 + \frac{1}{2}v^2 \, .
\eea 

As detailed in App A of \cite{Assadullahi:2016gkk}, the equation that gives the moments of the PDF, defined as $f_n(\phi_i) = \langle \mathcal{N}^n \rangle (\phi_i^{\mathrm{in}} = \phi_i)$, is given by 
\bea 
\mathcal{L}^\dagger_{\mathrm{FP}} \cdot f_n(\phi_i) = -n f_{n-1}(\phi_i) \, ,
\eea 
where $\mathcal{L}^\dagger_{\mathrm{FP}}$ is the adjoint FP operator. 
Hence, we need to find the adjoint of \eqref{app:eq:FP:USR}.

In general, for an operator given by 
\bea 
L\cdot u(x) = \sum_n a_n(x) D^n u(x) \, ,
\eea 
the adjoint operator is given by
\bea 
L^\dagger\cdot u(x) = \sum_n (-1)^n D^n\left[a_n(x)u(x)\right] \, .
\eea 
Applying this to \eqref{app:eq:FP:USR}, we find that
\bea 
\mathcal{L}^\dagger_{\mathrm{FP}} \cdot P = \left[ \frac{v}{H}\frac{\partial}{\partial\phi} - 3v\frac{\partial}{\partial v} + \frac{H^2}{8\pi^2}\frac{\partial^2}{\partial\phi^2} \right] P \, ,
\eea 
Hence, for USR inflation, the moments of $P$ are given by
\bea \label{app:eq:usr:moments:pde}
 \left[\frac{H^2}{8\pi^2}\frac{\partial^2}{\partial\phi^2} + \frac{v}{H}\frac{\partial}{\partial\phi} - 3v\frac{\partial}{\partial v} \right] f_n(\phi, v) = -n f_{n-1}(\phi, v) \, ,
\eea 
with $n \geq 1$ and $f_0 = 1$, and boundary conditions 
\bea \label{app:eq:moments:initialconditions}
&f_n(\phi_\mathrm{end}, v) = 0 \, , & \frac{\partial f_n}{\partial \phi}(\phi_{\mathrm{uv}}, v) = 0 \, ,
\eea 
\ie all trajectories initiated at $\phi_\mathrm{end}$ realise a vanishing
number of e-folds, and the other one implementing the presence of a reflective wall at $\phi_{\mathrm{uv}}$.

\section{Slow roll} \label{appendix:slowrollFP}

Let us also, for comparison to the ultra-slow-roll case, briefly outline the method of deriving the Fokker--Planck equation for the multi-field, slow-roll system
\bea 
\frac{\dd\phi_i}{\dd N} = -\frac{V_{,\phi_i}}{3H^2} + \frac{H}{2\pi}\xi(N) 
\eea 
where we have $D$ fields (so $0\leq i \leq D$) moving in a potential $V(\phi_i)$.
Following the method outlined in the previous section, we have 
\bea 
&A_i = -\frac{V_{,\phi_i}}{3H^2} \, , & B_{ii} &= \frac{H}{2\pi} \, ,
\eea 
and all $B_{ij} = 0$ for $i\neq j$. 
Hence, the Fokker--Planck equation corresponding to this system is then
\bea 
\frac{\partial P}{\partial N} &= \sum_i\frac{\partial}{\partial\phi_i}\left[\frac{V_{,\phi_i}}{3H^2}P\right] + \frac{1}{2}\sum_i\frac{\partial^2}{\partial\phi_i^2}\left[\left(\frac{H}{2\pi}\right)^2P\right] \\
&= \sum_i\frac{\partial}{\partial\phi_i}\left[\frac{V_{,\phi_i}}{V}\Mp^2 P\right] + \sum_i\frac{\partial^2}{\partial\phi_i^2}\left[\frac{V}{3\Mp^2}\frac{1}{8\pi^2}P\right] \, ,
\eea 
where in the last equality we have used $3\Mp^2 H^2 = V$ in slow roll. 
Finally, for simplicity, we define the dimensionless potential 
\bea 
\upsilon = \frac{V}{24\pi^2\Mp^4} \, ,
\eea 
and hence the Fokker-Planck equation becomes 
\bea 
\frac{\partial P}{\partial N} &= \sum_i\frac{\partial}{\partial\phi_i}\left[\frac{\upsilon_{,\phi_i}}{\upsilon}\Mp^2 P\right] + \sum_i\frac{\partial^2}{\partial\phi_i^2}\left[\Mp^2\upsilon P\right] \, .
\eea 

\section{Leading order classical characteristic function} \label{app:classicalLO:charfunction}

In order to use the method of characteristics, we parameterise with a parameter $u$ such that $\phi=\phi(u)$ and $v=v(u)$, and hence $\chi_\mathcal{N}(t,u) = \chi_\mathcal{N}(t,\phi(u),v(u))$.
By then comparing the total derivative 
\bea 
\frac{\dd \chi_\mathcal{N}}{\dd u} = \frac{\partial \chi_\mathcal{N}}{\partial \phi}\frac{\partial \phi}{\partial u} + \frac{\partial \chi_\mathcal{N}}{\partial v}\frac{\partial v}{\partial u} \, , 
\eea 
with our known classical PDE \eqref{eq:char:classicallimit} 
\bea 
\frac{1}{H}\frac{\partial \chi_\mathcal{N}}{\partial\phi} - 3\frac{\partial \chi_\mathcal{N}}{\partial v} + \frac{it}{v} \chi_{\mathcal{N}} = 0 \, ,
\eea 
we are motivated to choose 
\bea 
&\frac{\partial \phi}{\partial u} = \frac{1}{H} \, , 
&\frac{\partial v}{\partial u} = -3 \, .
\eea 
Hence we have 
\bea  \label{eq:characteristiccurves}
&\phi = \phi_0 +\frac{u}{H} \, ,
&v = v_0 - 3u \, ,
\eea 
for some parameters $\phi_0$ and $v_0$ that set the characteristic lines we are following. 
Note that we have also assumed the $H$ can be approximated as constant for now, as done in \cite{Firouzjahi:2018vet}.
Under this change of variables, our classical equation becomes 
\bea 
\frac{\dd \chi_\mathcal{N}}{\dd u} = -\frac{it}{v_0 - 3u}\chi_\mathcal{N} \, .
\eea 
This is solved by 
\bea 
\chi_\mathcal{N} = \chi_0(\phi_0,v_0)\left[ v_0 - 3u  \right]^{\frac{it}{3}} \, , 
\eea 
where $\chi_0$ is an arbitrary function of our parameters $\phi_0$ and $v_0$.
We are free to choose one of these parameters, so let us simply choose $v_0=0$, and hence for each given $(\phi,v)$, we have 
\bea 
&\phi_0 = \phi + \frac{v}{3H} \, , & u = -\frac{v}{3} \, .
\eea 
Using this, we can transform back to our original variables and find 
\bea 
\chi_\mathcal{N}(t,\phi,v) = \chi_0\left( \phi + \frac{v}{3H} \right)v^{\frac{it}{3}} \, .
\eea 
To find the unknown function $\chi_0$ we use our initial condition $\chi_{\mathcal{N}} (t, \phi_{\mathrm{end}}, v) = 1$. With this, we obtain our final solution to be
\bea \label{eq:char:classicalLO:App}
\left. \chi_\mathcal{N}(t,\phi,v)\right|_{\mathrm{cl}} = \left[ v + 3H(\phi - \phi_\mathrm{end}) \right]^{-\frac{it}{3}}v^{\frac{it}{3}}
\eea 

\subsection{Classical NLO solution}
\label{app:classicalNLO:charfunction}

By again choosing the same characteristic curves \eqref{eq:characteristiccurves}, and choosing $v_0=0$ as before, the solution is given by 
\bea 
\chi_{\mathcal{N}}(t,\phi,v)\Big|_{\mathrm{NLO}} = \left[ \chi_1\left(\phi+\frac{v}{3H}\right) + \ln v \frac{iH^4t}{8\pi^2}\left(1+\frac{it}{3}\right)\left[ v + 3H\left(\phi - \phi_\mathrm{end}\right)\right]^{-2-\frac{it}{3}} \right] v^{\frac{it}{3}} \, ,
\eea 
where $\chi_1$ is some arbitrary function to be determined by the initial conditions. 
This initial condition gives us 
\bea 
\chi_1\left( \phi_\mathrm{end}+\frac{v}{3H} \right) = v^{-\frac{it}{3}}\left[ 1 - \frac{iH^4t}{8\pi^2}\left(1+\frac{it}{3}\right) \frac{\ln v}{v^{2}}\right]
\eea 
Setting $X=\phi_\mathrm{end} + \frac{v}{3H}$ lets us find $\chi_1(X)$, and hence 
\bea 
\chi_{\mathcal{N}}(t,\phi,v)\Big|_{\mathrm{NLO}} = v^{\frac{it}{3}}\left[ v + 3H\left(\phi - \phi_\mathrm{end}\right)\right]^{-\frac{it}{3}}\left[ 1 - \frac{iH^4t}{8\pi^2}\left(1+\frac{it}{3}\right)\frac{\ln \left[ 1 + \frac{3H}{v}\left(\phi-\phi_\mathrm{end}\right)\right]}{\left[ v + 3H\left(\phi - \phi_\mathrm{end}\right)\right]^{2}} \right]  \, .
\eea

\newpage
\bibliographystyle{JHEP}
\phantomsection
\addcontentsline{toc}{chapter}{Bibliography}
\bibliography{Thesis}

\end{document}